\documentclass[11pt]{article}
\usepackage{amsmath,amssymb}
\usepackage{amsfonts}
\usepackage{scalerel}
\usepackage[dvipsnames]{xcolor}
\usepackage{graphicx}
\usepackage{float}
\usepackage{subfig}
\usepackage{cite}
\usepackage{tikz}\usetikzlibrary{shapes,arrows,chains}
\usetikzlibrary[calc]
\usepackage[colorlinks=true,
linkcolor=Blue, 
citecolor=Blue,
filecolor=Blue,
urlcolor=Blue,
linktoc=page, %%%
pdfstartview=FitV,
bookmarksopen=true]{hyperref}

\topmargin - 0.5 in

\def\a{\alpha}
\def\b{\beta}
\def\s{\sigma}
\def\l{\lambda}
\def\p{\partial}
\def\d{\delta}

\def\g{\gamma}
\def\th{\theta}
\def\om{\omega}
\def\r{\rightarrow}
\def\O{\mathcal{O}}
\def\N{\mathcal{N}}

\def\B{\mathcal{B}}
\def\E{\mathcal{E}}
\def\L{\Lambda}
\newcommand{\bi}{\begin{itemize}}
\newcommand{\ei}{\end{itemize}}
\newcommand{\ap}{\alpha'}
\newcommand{\apt}{\tilde{\alpha}'}

\newcommand{\vep}{\varepsilon}

\newcommand{\be}{\begin{equation}}
\newcommand{\ee}{\end{equation}}
\newcommand{\bea}{\begin{eqnarray}}
\newcommand{\eea}{\end{eqnarray}}

\usepackage{color}

\usepackage[normalem]{ulem}  % \sout{old text} for strikeout

\numberwithin{equation}{section}
\makeatletter
\renewcommand{\@seccntformat}[1]{%
  \csname the#1\endcsname.\ }
\makeatother

\title{
\vskip-5mm
Ascending  the attractor flow in the D1-D5 system
\vspace{0.3cm}}

\author{ Silvia Georgescu$^{\S, \dag}$,
Monica Guica$^{\S,\ddag,\sharp }$ and Nicolas Kovensky$^{\S}$\vspace{2mm}\\
\\\vspace{1mm}
${}^\S$\emph{\small Universit\'e Paris-Saclay, CNRS, CEA,} 
\emph{\small Institut de Physique Th\'eorique, 91191 Gif-sur-Yvette, France} \\ %\vspace{1mm}
 \vspace{1mm}
${}^\dag$\emph{\small 
CPHT, CNRS, \'Ecole polytechnique, Institut Polytechnique de Paris, 91120 Palaiseau, France} \\  \vspace{1mm}
${}^\ddag$\emph{\small Institute of Physics, Ecole Polytechnique Fed\'erale de Lausanne, CH-1015 Lausanne, Switzerland} \\ 
${}^\sharp$\emph{\small Theoretical Physics Department, CERN, CH-1211 Geneva 23, Switzerland }}

\date{}

\begin{document}

\maketitle

\abstract{
\vskip4mm

\noindent We study maximally supersymmetric irrelevant deformations of the D1-D5 CFT that correspond to following the  attractor flow in reverse in the dual half-BPS black string solutions of type IIB supergravity  on K3. When a single, quadratic condition is imposed  on the parameters of the $22$ such irrelevant deformations, the asymptotics of the solution degenerate to a linear dilaton-like spacetime. We identify each such degeneration limit with a known decoupling limit of string theory, which yields little string theory or  deformations thereof (the so-called open brane LST, or OD$p$ theories), compactified to two dimensions. This  suggests that a $21$-parameter family of the above deformations leads to  UV-complete theories, which are string theories decoupled from gravity that are continuously connected to each other. %} \com{I mean some (perhaps weak) form of our LST moduli space idea}.
%
% that at least in these cases, the irrelevant deformations should lead to a well-defined, UV-complete non-gravitational (string) theory. \emph{More?} 
All  these theories have been argued to display Hagedorn behaviour; we show that including the F1 strings leads to an additional Cardy term. The resulting entropy formula  closely resembles that of single-trace $T\bar T$-deformed CFTs, whose generalisations  could provide   possibly tractable effective  two-dimensional descriptions of the above web of theories. % that is akin to $T\bar T$.

\vskip2mm

\noindent We also consider the asymptotically flat black strings. At fixed temperature, the partition function is dominated by thermodynamically stable, `small' black string solutions, similar to the ones in the decoupled backgrounds. %\new{stable solutions with small horizons,} whose size may nevertheless be made much larger than the \old{AdS} \new{AdS$_3$ }scale by tuning the coupling. 
We   show that certain asymptotic symmetries of these black strings bear a striking resemblance with the state-dependent symmetries of single-trace $T\bar T$, and  break down precisely when the background solution reaches the `large' black string threshold. This suggests that small, asymptotically flat black strings may also admit a $T\bar T$ - like effective description. 

% The asymptotic symmetris turn out to be field-dependent generalisations of conformal transformations, which are highly reminiscent  of the field-dependent coordinate transformations in $T\bar T$ - deformed CFTs. We also comment on the interpretation of non-extremal black holes in this setup.
}

\tableofcontents

\section{Introduction}

Finding a microscopic explanation for the entropy of \emph{general} black holes has been a main driving force of research in quantum gravity, ever since Bekenstein and Hawking's discovery that black holes are thermodynamic systems \cite{Bekenstein:1973ur,Bardeen:1973gs,Hawking:1975vcx}. The AdS/CFT correspondence \cite{Maldacena:1997re} marked a significant advance in this direction, as it provided a microscopic explanation of the entropy of large black holes in anti de-Sitter spacetimes   in terms of the thermal entropy in the dual CFT \cite{Gubser:1996de}. In AdS$_3$/CFT$_2$, the coefficient of the entropy can be matched exactly \cite{Strominger:1997eq}.

Central to the derivation of the AdS/CFT correspondence is the existence of a decoupling limit, which ensures a separation between the open and closed string descriptions of the brane system. This limit is a low-energy one, in which the open string description flows to a CFT and the closed string modes are confined to an AdS throat, where their  redshift is nearly infinite.  There are, however, many other decoupling limits known within string theory, which relate non-local field theories \cite{Seiberg:1999vs,Maldacena:1999mh} or string theories decoupled from gravity \cite{Seiberg:1997zk,Losev:1997hx,Aharony:1998ub,Seiberg:2000ms,Harmark:2000wv,Gopakumar:2000na,Gopakumar:2000ep,Harmark:2000ff}  to space-times with asymptotics different from AdS. It is all but natural, if one seeks to understand the entropy black holes with non-AdS asymptotics - such as, for example, asymptotically flat black holes - to consider these more general instances of holography. 

In this article, we concentrate on the D1-D5 system, which consists of a stack of D5 branes wrapped on K3 (for specificity; we could have equally considered $T^4$) and D1 branes parallel to their remaining worldvolume direction.   This system has been very extensively studied
%\cite{Strominger:1996sh,Callan:1996dv,Horowitz:1996ay} 
(see \cite{David:2002wn} and references therein). At zero temperature and angular potential, its supergravity description is as a half-BPS black string in type IIB supergravity compactified on K3.  We will be studying both  this supersymmetric solution and its non-extremal version.  

The supersymmetric black string is well-known to exhibit an attractor mechanism \cite{Ferrara:1995ih,Ferrara:1996dd,Sen:2005wa},   whereby a subset of the moduli at infinity must reach prespecified values at the horizon, which are  determined by the charges of the black string. More specifically, for type IIB supergravity on K3, the massless scalars at infinity  locally parametrise  the coset 

\be
\mathcal{M}_\infty = \frac{SO(5,21)}{SO(5) \times SO(21)}  \label{minfty}
\ee
  Upon flowing to the near-horizon region,  $21$ of the initial $105$ massless scalar fields acquire a mass from the AdS$_3$ point of view \cite{Maldacena:1998bw} and their values become fixed. From the perspective of the D1-D5 CFT that describes the near-horizon region, the $21$ attracted scalars correspond to  single-trace  irrelevant operators of left/right conformal dimension  $(2,2)$; 
  one additional such irrelevant operator  is related to changes in the size of the AdS$_3$ and $S^3$. The moduli space of the near-horizon region, parametrised by the remaining massless scalars, is locally  the coset
\be
\mathcal{M}_{hor} =  \frac{SO(4,21)}{SO(4) \times SO(21)}  \label{mhor}
\ee  
and coincides with the moduli space of the D1-D5 CFT \cite{Dijkgraaf:1998gf}.   
  
Given the standard identification between the radial direction in the bulk and the energy scale in the boundary theory, as well as  between the supergravity moduli and the coupling constants on the brane worldvolume, the attractor flow may be seen, from the boundary perspective, as  a flow towards low energies that ends at an IR fixed point \cite{deBoer:2008ss}. $22$ of the couplings of the original theory become irrelevant in the IR.  The corresponding picture is sketched in the figure below\footnote{As discussed at length in  \cite{deBoer:2008ss}, the asymptotic supergravity moduli space $\mathcal{M}_\infty$ is expected to be identified with the space of couplings of the dual worldvolume theory. However, since the latter is in general not under control, the only rigorous identification can be between the near-horizon moduli space and the conformal manifold of the D1-D5 CFT, and small departures away from this manifold in the effective low-energy worldvolume theory and the corresponding supergravity perturbations.  One goal of this article is to argue that the boundary moduli space can be made sense of beyond this infinitesimal approximation. }. 
 
 %
%\begin{figure}[!htb]
%   \begin{minipage}{0.45\textwidth}
%     \centering
%     \includegraphics[width=.9\linewidth]{afbs}
%     \caption{\small  }\label{}
%   \end{minipage}\hfill
%   \begin{minipage}{0.4\textwidth}
%     \centering
%     \includegraphics[width=.9\linewidth]{attrflowinv2}
%     \caption{\small .}\label{}
%   \end{minipage}
%\end{figure}

\begin{figure}[h] 
    \centering
     \captionsetup{width=.8\linewidth}
    \subfloat{{\includegraphics[width=6.7cm]{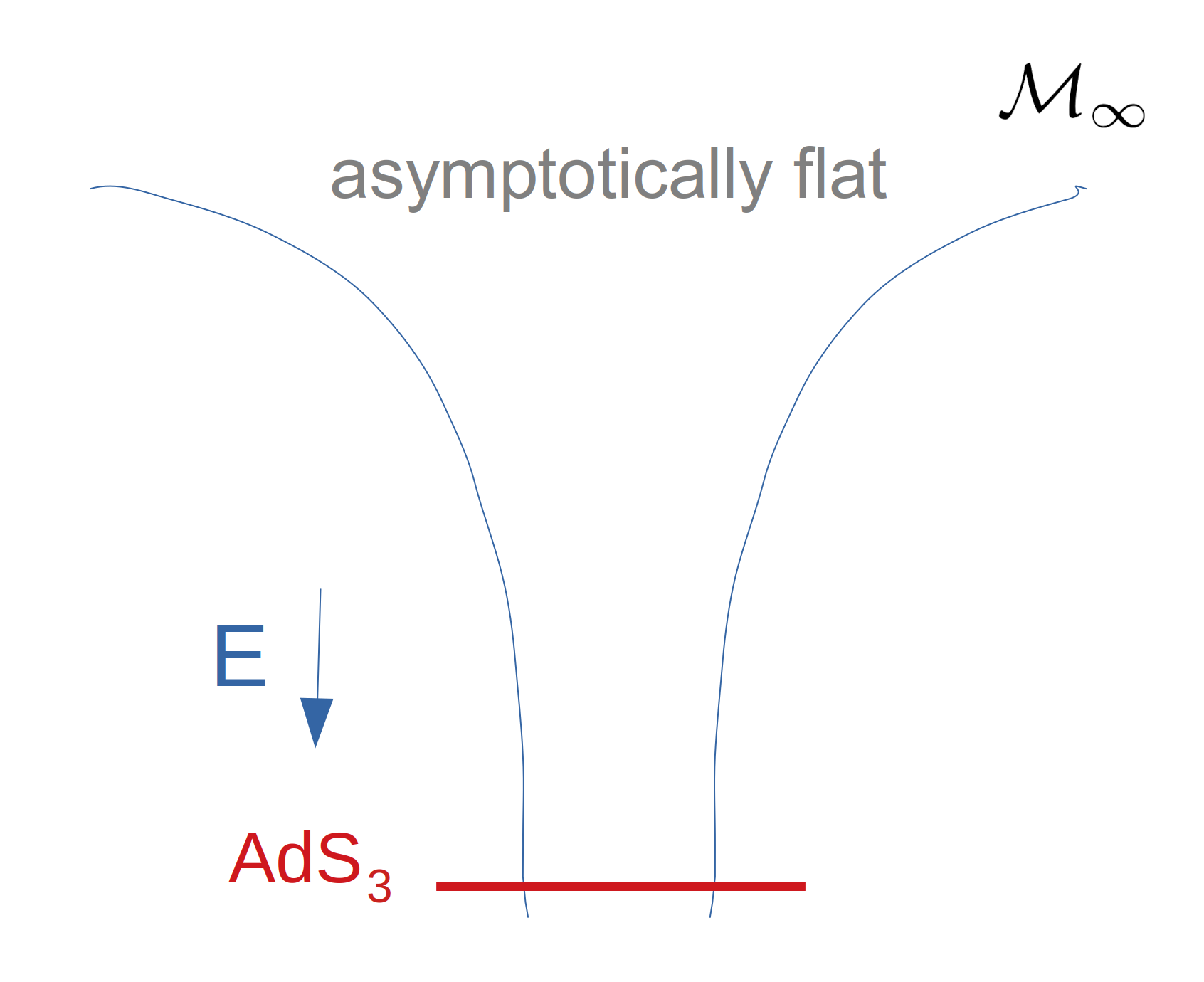} }}%
    \qquad\qquad\qquad
    \subfloat{{\includegraphics[width=5.4cm]{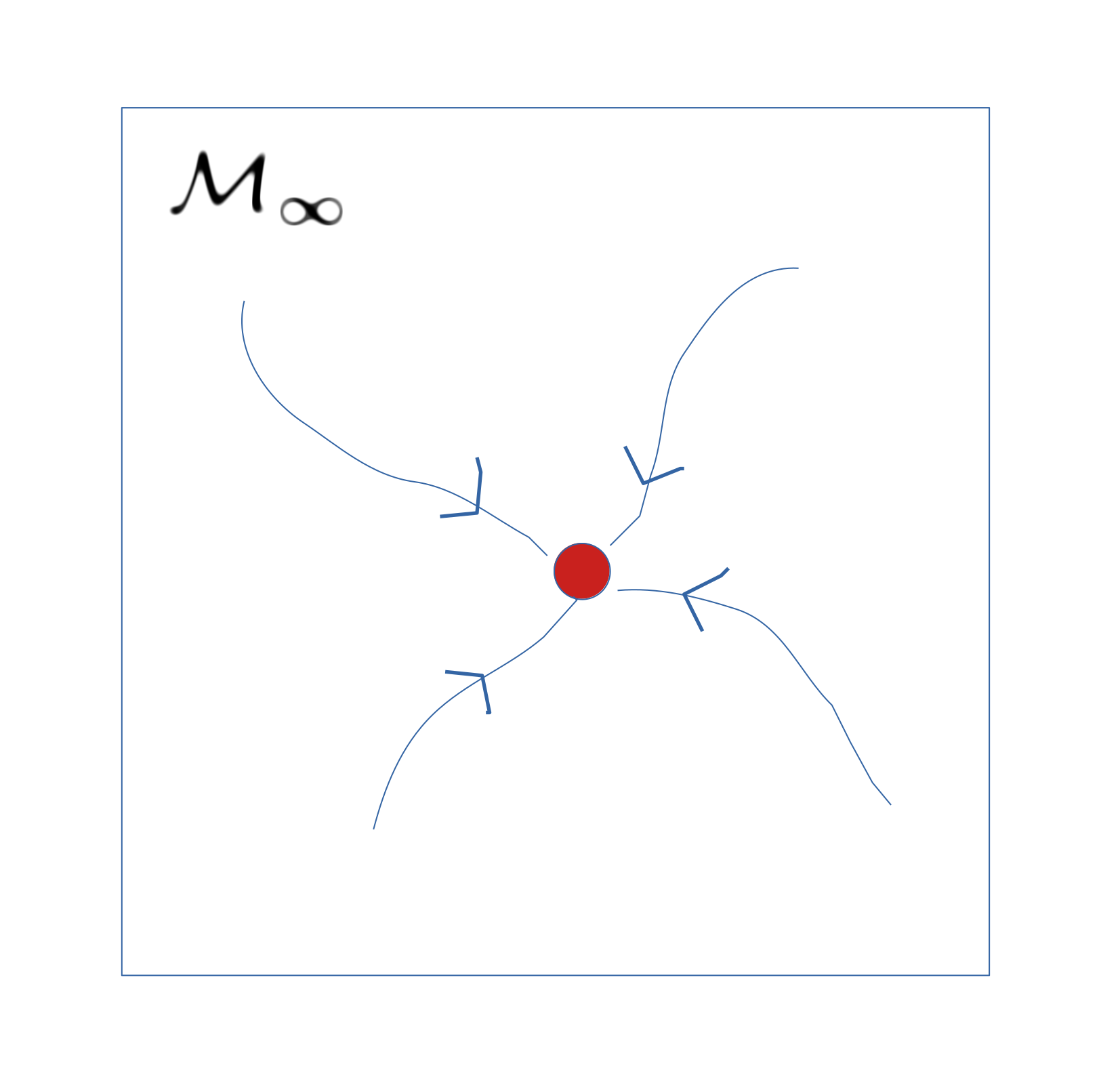} }}%
    \caption{
\footnotesize{The attractor flow corresponds to a flow towards low energies, both in the gravity picture (left), where one nears the AdS$_3$ throat of the black string, and in the boundary one (right), where one flows towards an IR CFT fixed point (red dot). The moduli spaces of the near-horizon region (red line) and of the CFT  are identified and are given by \eqref{mhor}.  }}%
    \label{fig:1}%
\end{figure}

 % correspond to sources for exactly marginal operators that move one along the conformal manifold of the D1-D5 CFT. 

 %Consequently, the holographic picture for the half-BPS black string solution is \emph{Place figure attractor flow}. 

\noindent One may verify this picture infinitesimally around the attractor point, by matching the maximally supersymmetric single-trace irrelevant deformations to the leading departure of the black string solution away from the near-horizon one%\textcolor{blue}{, as well as the \textcolor{red}{connection for these operators} over the CFT moduli space to the connection on the normal bundle of \eqref{mhor} inside \eqref{minfty} }
\cite{deBoer:2008ss}.

 %infinitesimal coefficients of the irrelevant operators  AdS$_3 \times S^3$. % The identification of the operators depends on the point in moduli space. 

\begin{figure}[h]
    \centering
     \captionsetup{width=.8\linewidth}
    \subfloat{{\includegraphics[width=6.7cm]{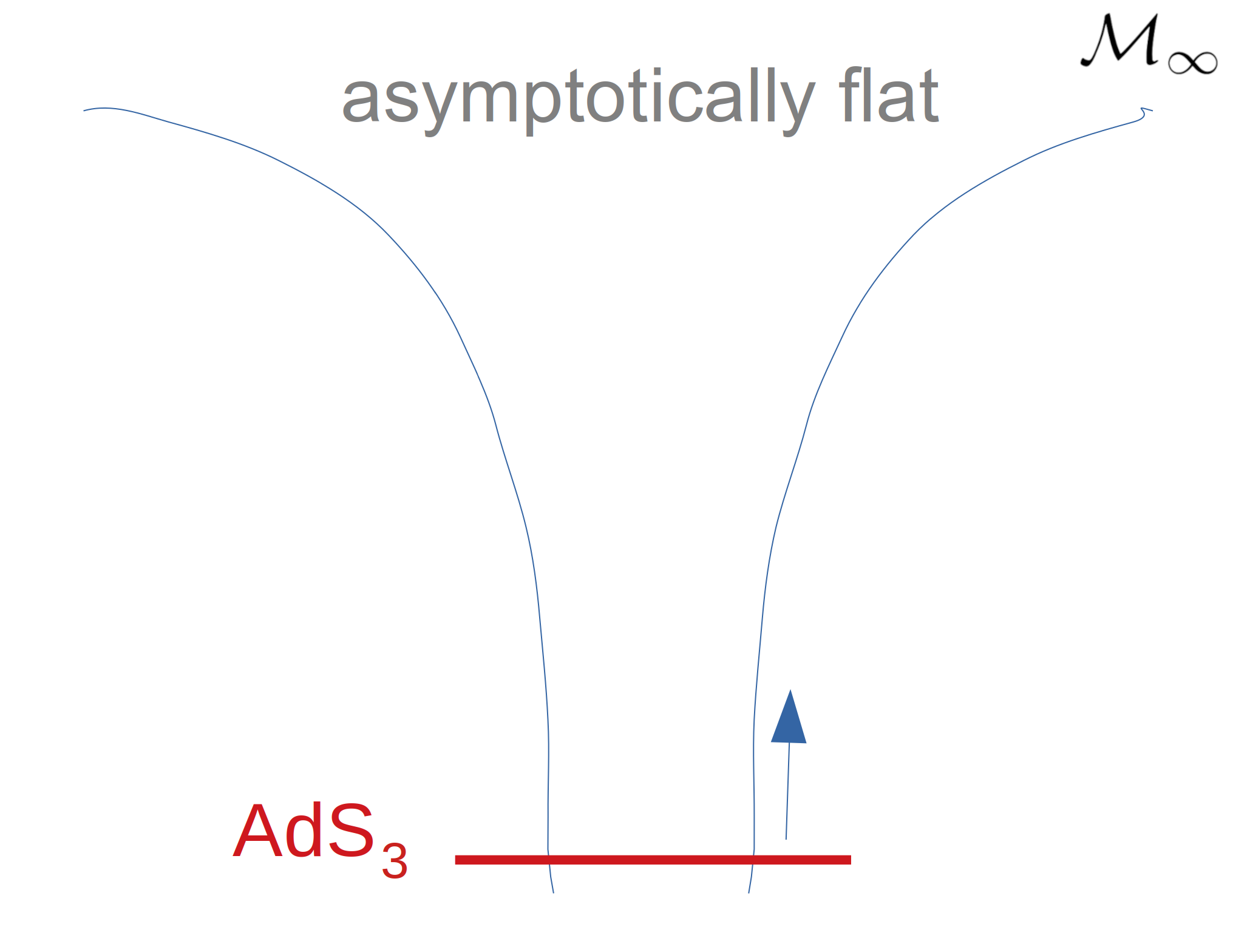} }}%
    \qquad\qquad\qquad
    \subfloat%[\centering label 2]
    {{\includegraphics[width=5.4cm]{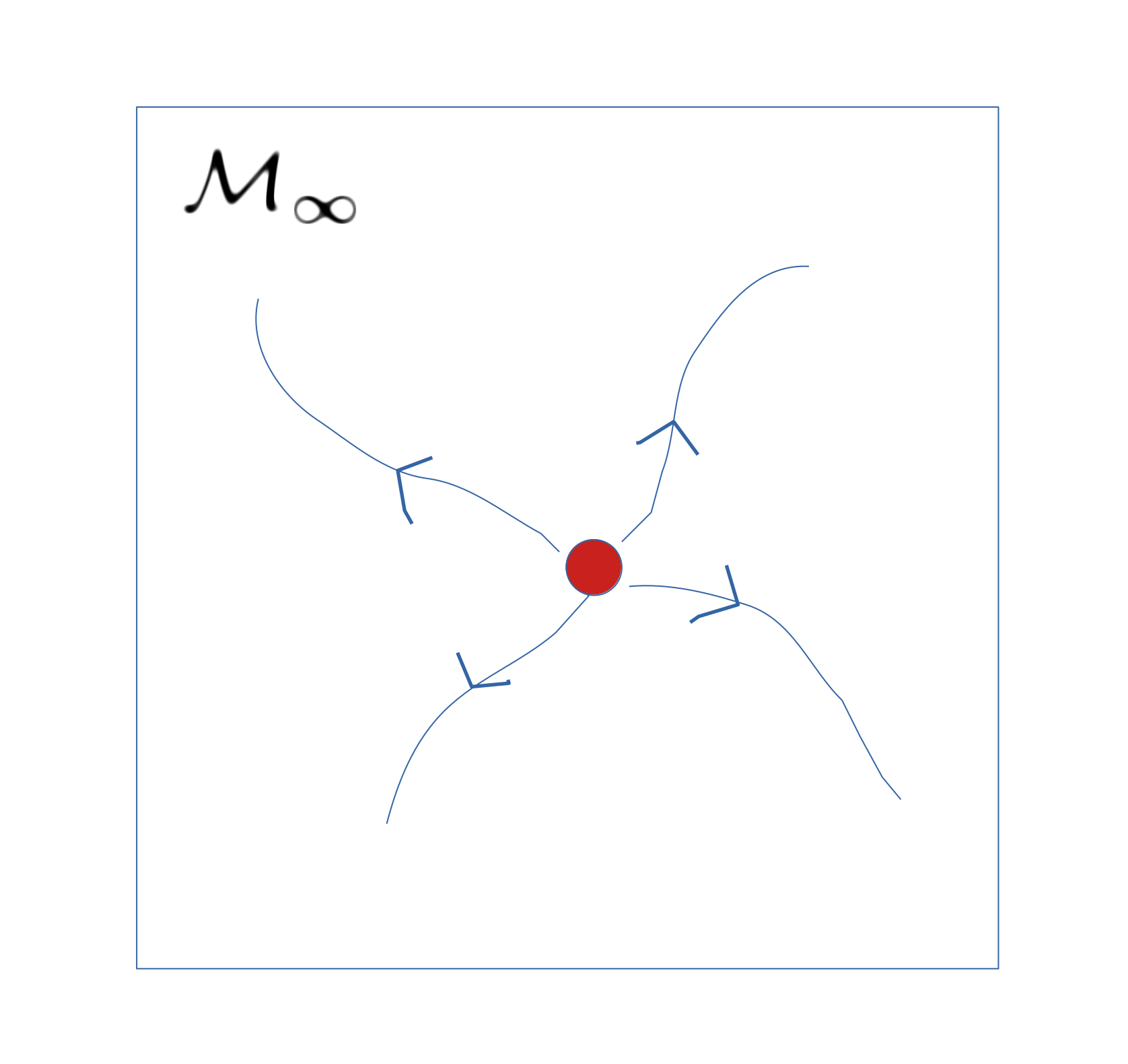} }}%
    \caption{\small{The reverse attractor flow is mapped to turning on irrelevant deformations in the D1-D5 CFT. Reaching out a finite distance outside the throat corresponds to understanding the boundary theory at \emph{finite} values of the irrelevant couplings.  }}%
    \label{fig:2}%
\end{figure}

The framework of this article consists of looking at this picture in reverse (figure \ref{fig:2}), whereby we map the inverse attractor flow out of the AdS$_3 \times S^3$ near-horizon region - which should eventually bring us  to flat space -  to a $22$ - parameter family of maximally supersymmetric irrelevant deformations of the D1-D5 CFT. As already explained, these deformations start as $(2,2)$ irrelevant operators near the IR fixed point and may, in principle, receive many corrections as one flows upwards in energy. The question we would like to address herein is:

\bigskip

\parbox{0.87\linewidth}{Q:~ \emph{What type of theory does one obtain for \underline{finite} values of the irrelevant couplings?}}

\bigskip

\noindent In general, following an irrelevant deformation  to higher orders  in (conformal) perturbation theory is a very difficult and possibly ill-defined task, due to the proliferation of unfixed counterterms. However, there is some hope that the particular deformations above may be tractable, as maximal supersymmetry severely restricts the other operators that may appear, at least in the supergravity limit. Such an approach was put forth in \cite{Intriligator:1999ai} for D3 branes, and has been recently revisited in \cite{Caetano:2020ofu}, but has yet to yield non-trivial predictions.

%
%\begin{figure}[!htb]
%   \begin{minipage}{0.45\textwidth}
%     \centering
%     \includegraphics[width=.9\linewidth]{afbsout}
%     \caption{\small  }\label{}
%   \end{minipage}\hfill
%   \begin{minipage}{0.4\textwidth}
%     \centering
%     \includegraphics[width=.9\linewidth]{attrfloutv2}
%     \caption{\small .}\label{}
%   \end{minipage}
%\end{figure}

\noindent The recent examples of exactly solvable \emph{finite} irrelevant deformations of two-dimensional QFTs  \cite{Smirnov:2016lqw} have given renewed hope of solving this problem. The resulting theories  %constructed from bilinears of conserved currents and 
%can be tracked for \emph{finite} values of the deformation parameter and 
appear to be UV-complete \cite{Dubovsky:2013ira} and intrinsically non-local at the scale set by the irrelevant coupling.  The best-studied examples of such theories are $T\bar T$ - deformed CFTs \cite{Smirnov:2016lqw,Cavaglia:2016oda}, which are Lorentz invariant,  and $J\bar T$ - deformed CFTs \cite{Guica:2017lia}, which are Lorentz-breaking, but preserve the standard  conformal symmetries and locality along one lightlike direction. Both  classes of theories share many features with standard, \emph{local} two-dimensional CFTs, such as the presence of Virasoro generators \cite{Guica:2020uhm,Guica:2021pzy,Guica:2022gts,Guica:2020eab} and correlation functions that are related in a simple, universal, albeit non-local manner to  
standard CFT$_2$ correlators \cite{Guica:2021fkv,Cui:2023jrb,Aharony:2023dod}. As argued in \cite{Guica:2022gts}, the natural basis for the extended symmetry generators is a Fourier one, in which the symmetry algebra becomes a $T\bar T$/$J\bar T$ - specific non-linear   deformation of the Virasoro $\times $ Virasoro algebra. There also exists a `single-trace' generalisation of $T\bar T$-
  and $J\bar T$-deformed CFTs \cite{Giveon:2017nie,Chakraborty:2023wel,Apolo:2018qpq,Chakraborty:2018vja}, which simply denotes the symmetric product orbifold of the above theories. The universal structure of the deformed spectrum and correlation functions, as well as the infinite symmetries, carries over from the double-trace case, in a completely predictable fashion \cite{Chakraborty:2023wel} (for the spectrum, see also \cite{Apolo:2023aho}). 
  
An interesting connection between $T\bar T$ deformations and non-AdS holography has been put forth in \cite{Giveon:2017nie}, who showed that the asymptotically linear dilaton  (henceforth ALD) spacetime  obtained in the NS5 decoupling limit (namely $g_s \r 0$, keeping $\a'$ fixed) of the NS5-F1 system shares many features with a single-trace  $T\bar T$ - deformed CFT. To be more precise, this decoupled spacetime  interpolates between an AdS$_3$ region in the IR and a linear dilaton one in the UV, and thus its holographic description is precisely that of an irrelevant flow out of a CFT; on the boundary side, the decoupling limit yields LST, a non-local, non-gravitational theory with string-like excitations \cite{Seiberg:1997zk}, compactified on K3 or $T^4$. %{\color{red}whose size is assumed to be much smaller than that of the IR AdS$_3$.  } {\color{ForestGreen}(from the D1D5 review "Strictly speaking, we should wrap the D1-D5 string on a large circle to avoid the Gregory-Laflamme instability", from Aharony "The NS-NS description is invariant under a T-duality transformation inverting the four cycles of the torus" so in what sense $v$ is small?)}
 Research in this subfield has followed two main strands: the first one concerns  the worldsheet string theory description of the decoupled spacetime, which exhibits a perfect match between the spectrum and correlation functions of long string states/operators in this background and their $T\bar T$ - deformed counterparts. Thus, there is very good evidence that the long string subsector of string theory in this background is described by a symmetric product orbifold if $T\bar T$ - deformed CFTs. This subsector does not, however, dominate the entropy \cite{Giveon:2005mi}.  On the other hand, the short string spectrum and correlators do not match the $T\bar T$ ones, thus precluding an exact duality between string theory in the ALD blackground and single-trace $T\bar T$ (except perhaps for the case of a single NS5-brane \cite{
Gaberdiel:2018rqv,Eberhardt:2018ouy}, when the size of the AdS factor is string scale). 

The  second strand of results connect \emph{universal} quantities in the decoupled linear dilaton spacetime - which are, in principle, defined throughout the moduli space and carry information about the entire theory   -   and single-trace $T\bar T$. One such observable is the entropy.  In \cite{Giveon:2017nie},   
it was shown that the entropy of black holes in the ALD spacetime precisely matches the single-trace $T\bar T$ - deformed entropy, which has a characteristic Cardy $\r$ Hagedorn behaviour 
%\textbf{\emph{Check factors!}}

\be
S = 2\pi \sqrt{\frac{ c}{3}(p E R + \mu  E^2)}
\ee
where $c$ is the central charge of the seed $CFT$, $p$ is the number of copies in the symmetric orbifold and $\mu$ is the dimensionful coupling. We have set the momentum to zero, for simplicity.  

Another important piece of evidence for a universal connection between the two is that the asymptotic symmetry group  of the ALD spacetime \emph{precisely matches} \cite{Georgescu:2022iyx} the particular non-linear modification of the Virasoro algebra  that represents the symmetry algebra of single-trace $T\bar T$ - deformed CFTs \cite{Chakraborty:2023wel}.  Given that the holographic dual of the ALD spacetime cannot be the \emph{exact} symmetric product orbifold of $T\bar T$ - deformed CFTs, these results point towards the existence of \emph{generalisations} of single-trace $T\bar T$ - deformed CFTs that share the same behaviour of the entropy and the extended symmetries. How to define the appropriate generalisations is is an interesting question for future research.%\footnote{It is tempting  to speculate that the standard single-trace $T\bar T$ deformation may bear the same relation to the more general $T\bar T$ - like ones  as an exactly marginal $J\bar J$ deformation to general exactly marginal deformations of a CFT. The over-regularity of the $T\bar T$ - deformed spectrum and the ability to generate most deformed observables via a field-dependent coordinate transformations whould  be expected to disappear in the more general case, which is a desirable feature.  \cite{}  }. %, (though the compactified LST dual of the ALD spacetimes suggests that they should contain strings. \emph{Keep?} )

%These observables are expected to be robust, \emph{Why?} unlike the long strings, which only exist at singular points \cite{} in the D1-D5 moduli space. 

The NS5-F1 system discussed above is just a particular case of the general half-BPS black string solutions mentioned at the beginning of this section, where the only moduli that are turned on are the dilaton and the overall volume of the internal manifold. The six-dimensional dilaton is an attracted scalar, and thus the black string geometry corresponds to turning on the corresponding irrelevant deformation, denoted $\l_-$, in the D1-D5 CFT. The size of the $AdS_3$ also grows when moving out of the near-horizon region, and its associated deformation parameter is $\l_+ \geq |\l_-|$. %; for the genral black string solution, we have $\l_+ \geq \l_-$. 
 In a convenient normalisation, the NS5 decoupling limit then corresponds to the particular case in which

\be \label{irrvcouppurens}
\l_+ = \l_- = \a' k
\ee 
where $\a' k$ is the effective irrelevant coupling, with $k$ the number of NS5 branes. Even though this result is extracted from an infinitesimal analysis (at small $\a'$), the existence of the
UV-complete  LST  description for finite  $\a'$ 
indicates that there should be a sense in which  the irrelevant deformations  \eqref{irrvcouppurens} can be turned on a finite amount. This viewpoint is supported by the fact that the deformation is nearly exact in the supergravity regime, and that its $T\bar T$ effective description also exists at the corresponding finite value of the irrelevant parameter\footnote{We will not assume that only the  maximally supersymmetric single-trace irrelevant operators in short multiplets  are turned on, but we will assume that 
%
% to finite values of $\l_\pm$ above that corresponds to compactified  LST with finite $\a'$. Even though we do not know which operators are turned on, in principle, we will assume, using the supergravity intuitions, that
  the coefficients of all operators that are present only depend on the corresponding $\l$ parameter.}.
  %
%  will all be determined by the $\l$ parameters. As we discuss in section \ref{section24} the supersymmetry implies that this deformation is almost exact at supergravity level.  The finite coupling thus refers to the sugra description, having convinced ourselves a field theory counterpart should exist. It is also the finite effective  $T\bar T$ coupling. \emph{Explain better!} 
%  We will therefore not make a distinction, as far as terminology is concerned, between  the irrelevant  couplings with the corresponding sugper parameter interchangeably. 
In the following, we will be using the terminology ``irrelevant coupling'' and ``supergravity parameter'' interchangeably.

\begin{figure}[h]
    \centering
     \captionsetup{width=.8\linewidth}
        \subfloat{{\includegraphics[width=6.7cm]{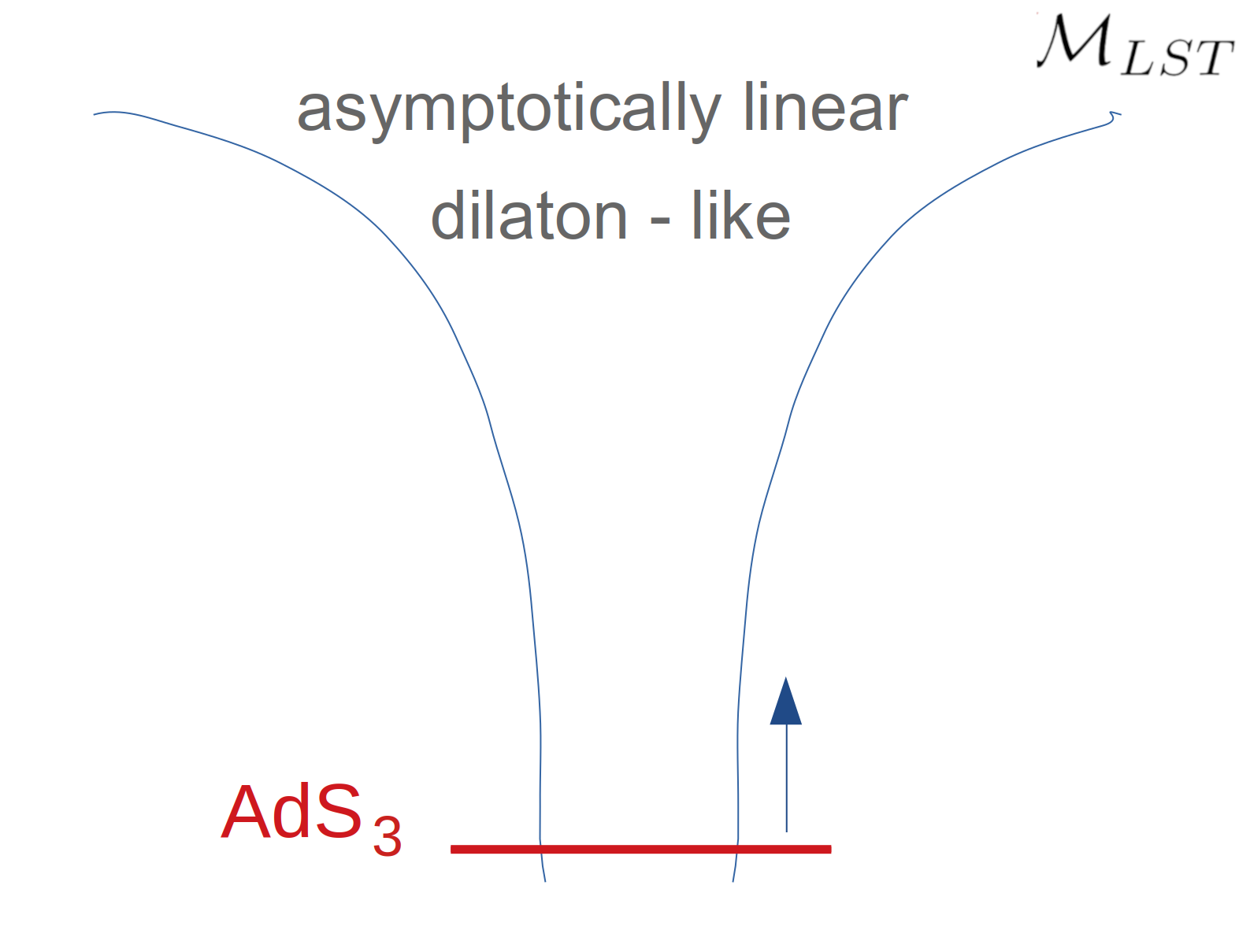} }}%
    \qquad\qquad\qquad
    \subfloat{{\includegraphics[width=4.9cm]{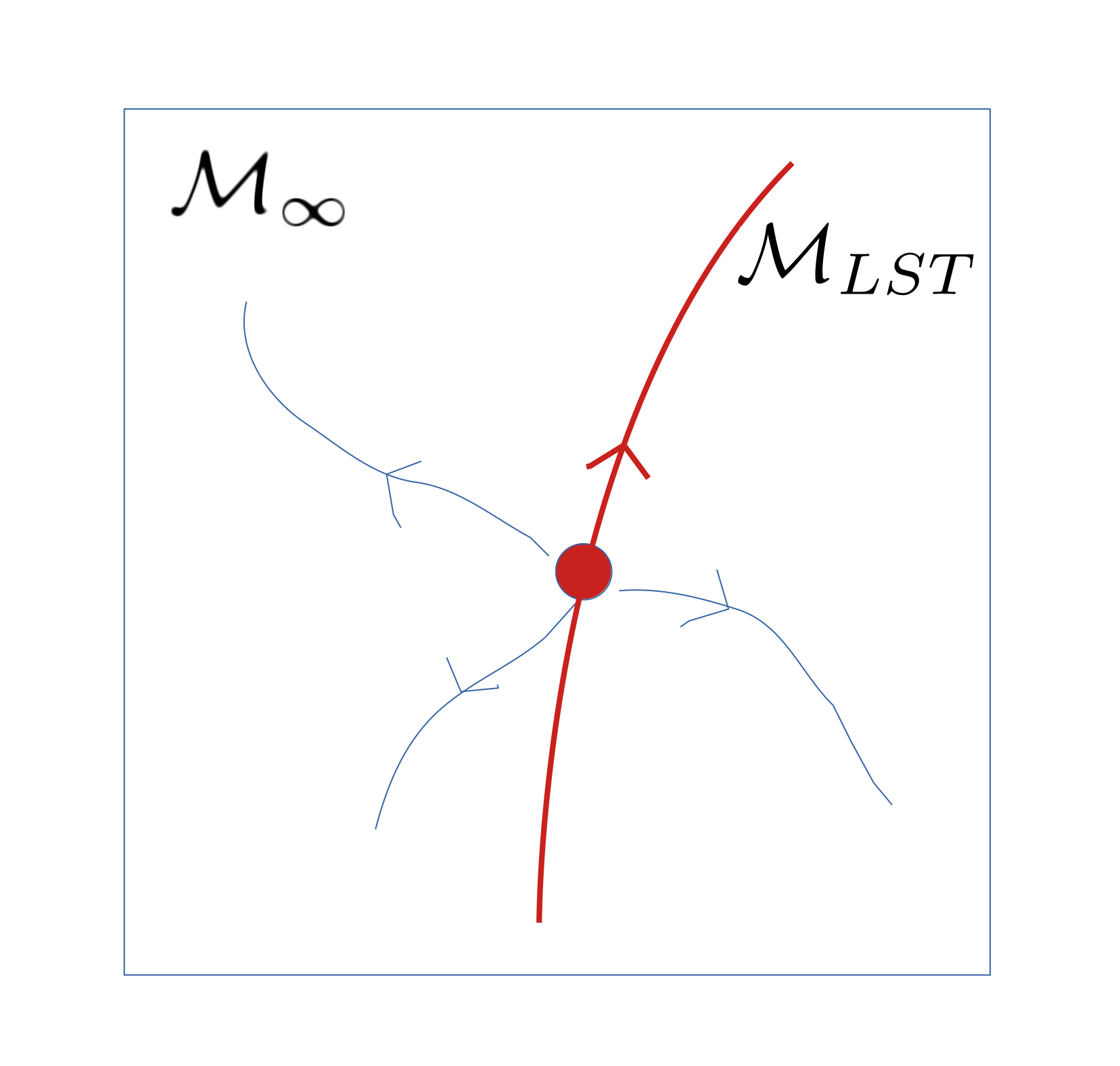} }}%
    \caption{\footnotesize{The asymptotics of the black string solutions degenerate on a codimension one subspace of the moduli space at infinity - $\mathcal{M}_{LST}$ - which is spanned by known decoupling limits of string theory that yield little string theory and  deformations thereof. It corresponds to turning on $21$ of the possible $22$ maximally supersymmetric irrelevant couplings. }}%
    \label{fig:3}%
\end{figure}

 One of the main goals of this article is to argue that this picture extends to the full moduli space of the D1-D5 system. Namely, we show that all instances in which the asymptotics of the general half-BPS black string solution degenerate to an ALD-like spacetime - which happens on a codimension one subspace of the moduli space at infinity - correspond to a known decoupling limit of string theory. We thus expect that the irrelevant deformations whose coefficients satisfy the corresponding field-theoretical constraint  yield a  \emph{UV-complete} theory (see figure \ref{fig:3}). 
These deformations span a $21$-dimensional subspace of the $22$ possible maximally supersymmetric irrelevant deformations of the system. The remaining irrelevant direction turns on the deformation to six-dimensional asymptotically flat space, for which a rigorous decoupling limit is not known. %\emph{Figure!} 

%
%\begin{figure}[!htb]
%   \begin{minipage}{0.45\textwidth}
%     \centering
%     \includegraphics[width=.9\linewidth]{aldv2}
%     \caption{\small  }\label{}
%   \end{minipage}\hfill
%   \begin{minipage}{0.4\textwidth}
%     \centering
%     \includegraphics[width=.8\linewidth]{mlstv2}
%     \caption{\small .}\label{}
%   \end{minipage}
%\end{figure}

The corresponding UV-complete theory is LST for the pure NS5-F1 system;  in the remaining cases, it will be a deformed version thereof known as open brane LST (OBLST, or OD$p$, for $p=1,3$) \cite{Gopakumar:2000ep,Harmark:2000ff}, all compactified on K3.  These theories were discovered in the context of understanding non-trivial decoupling limits of D-branes in a background Kalb-Ramond field. If the $B$-field is magnetic, then the decoupling limit yields  a non-commutative deformation of the super Yang-Mills theory that describes the low-energy dynamics of the brane \cite{Seiberg:1999vs}. If, on the other hand, it is electric, then the resulting theory is a space-time non-commutative open string theory (NCOS) that is decoupled from closed strings \cite{Seiberg:2000ms,Gopakumar:2000na}.  The NCOS description is only valid at weak coupling; in particular, its associated Hagedorn behaviour cannot be seen in supergravity \cite{Seiberg:2000ms,Harmark:2000wv}. The strong coupling limit of the D5 NCOS of interest to us is obtained via type IIB S-duality and is known as the  
 OD$1$ theory; the OD$3$ ones are obtained by two further T-dualities \cite{Gopakumar:2000ep}. 
 %
%  is S-dual to the six-dimensional non-commutative open string theory (NCOS) -  a theory of open strings living on the worldvolume of a D5 brane in a critical electric field, which is decoupled from closed strings and exhibits space-time non-commutativity. 
  It follows from this construction that the OD$p$/OBLST theories correspond to decoupling limits of NS5 branes in a background critical RR field, which in our case is either a two-form gauge field along the black string worldvolume (for OD$1$) or a four-form gauge field with two legs along one of the 19 anti-self-dual two-cycles of the K3 (for OD$3$).
   
   All these decoupled theories are non-local and have closed (little) string excitations and generically  also open D-branes;  it is sometimes said they are space-time non-commutative, in the sense that they are S-dual to the space-time non-commutative NCOS.  They  are connected to each other via the web of dualities we briefly sketched; specific representatives may be obtained in particular limits of the coupling space, denoted as $\mathcal{M}_{LST}$ in the figure above.  These couplings are related to ratios of the various irrelevant deformation coefficients that can be read off the  supergravity description. Thus, our (partial) answer to the question raised at the beginning is 
    
\vskip5mm
    
 A: ~ ~  \parbox{0.85\linewidth}{ \emph{When restricted to a codimension one subspace of the moduli space at infinity, the irrelevant deformations lead to a UV-complete theory which is, generically, a deformation of little string theory compactified on K3.} }

\vskip5mm

\noindent  Infinitesimally away from this subspace, we can identify the remaining supergravity parameter with the coefficient of the irrelevant deformation that turns on the deformation to flat space; however, we still do not have a clear picture of what the resulting worldvolume theory is at finite values of this parameter.

Most of the above-mentioned theories are strongly-coupled and do not have a tractable worldvolume description%\footnote{While NCOS may be weakly coupled, its associated Hagedorn behaviour is not visible in the supergravity regime (which corresponds, as usual, to string coupling), where it is washed out by LST. }
. One may use, however, the dual gravitational description to infer some of their properties. In LST, it is well-known that the little strings lead to Hagedorn behaviour at high energies \cite{Aharony:1998ub,Maldacena:1996ya,Kutasov:2001uf} which, as reviewed above, becomes Cardy $\r$ Hagedorn behaviour in presence of the F1s. In the literature, it has been argued that all the $ODp$/OBLST theories also exhibit Hagedorn behaviour at high energies; working under the same assumptions, we show that this behaviour becomes Cardy $\r$ Hagedorn upon the inclusion of the F1 strings, with a dependence on the parameters that is, again, very similar to that of single-trace $T\bar T$. Thus, all the  decoupled theories that span the codimension one subspace of the moduli space  appear to exhibit ``$T\bar T$ - like'' behaviour in their entropy, which is presumably associated to the universal presence of little strings. %It would be very interesting to also compute the asymptotic symmetries of these decoupled backgrounds and check whether it, too, show $T\bar T$ - like featues.
Whether their connection to $T\bar T$ is deeper than this may be investigated by - for example - computing the asymptotic symmetry groups of the associated spacetimes. 
 
 \medskip

With respect to the original asymptotically flat region, all the black strings responsible for the behaviour above are near-extremal. It is interesting to also discuss non-extremal asymptotically flat black strings from this perspective. One problematic aspect of such objects is, of course, that they are thermodynamically unstable. If one works in the canonical ensemble it is nonetheless possible to show, following the entirely analogous analysis  \cite{kutd3} of non-extremal D3-branes, that at any fixed temperature below a certain 
 maximum one, the pure NS5-F1 system contains
  two possible black string solutions: one whose horizon is  larger than a certain radius $r_0^*$,  and one that is smaller (see figure \ref{Tvsr0h}). The large black string is thermodynamically unstable, but also has higher free energy than the small one, which is stable. The canonical ensemble is thus dominated by the small black string. %\textcolor{blue}{ One interesting feature we find in the NS5-F1 system with respect to the D3 one is that now the maximum size of the small black string can be made much larger than the size of the near-horizon AdS$_3$, thanks to a tunable coupling. }
It appears reasonable to conjecture that the entropy of the small black string 
may be reproduced by a boundary theory with Hagedorn-like behaviour, as suggested by the existence of a maximal temperature in this system.  This %\textcolor{blue}{As they do not dominate the thermal ensemble, large black strings are in a way similar to the small black holes in global AdS$_{D>3}$,  %- whose entropy has been notoriusly difficult to understand microscopically -
% but differ from them in that 
%their  entropy  is larger than the maximum}
 entropy can be  associated with  just the  open string excitations on the D-branes. 

%The status of the large black hole appears similar to that of small black holes in AdS - which are unstable, but never dominate the canonical ensemble - and may be equally/ their entropy is notoriously  difficult to understand microscopically.  Of course, this sheds no light on the microcanonical understanding of these black holes/remains as puzzling as ever.

Motivated by the possibility to describe at least the small asymptotically flat black strings holographically,  we  study their asymptotic symmetries  from this ``$T\bar T$ perspective'', which suggests that the dual theory lives along the  directions parallel to the string. We therefore 
concentrate on the dependence of the asymptotic Killing vectors on the black string coordinates, ignoring  their dependence on the asymptotic $S^3$ ones. The resulting problem can  be studied within a very simple three-dimensional  consistent truncation of the non-extremal black string. Remarkably, we find that the asymptotic symmetry generators are extremely similar to those of the ALD spacetime \cite{Georgescu:2022iyx} and thus single-trace $T\bar T$, as they correspond to field-dependent generalisations of conformal transformations.
Interestingly, their background-dependent expressions break down %(become imaginary) 
\emph{precisely}  when the maximum size of the small black strings is reached, suggesting that a $T\bar T$ - like  effective description is valid only below the associated maximum energy.

 %This suggests that the theory obtained by turning on the irrelevant deformation that takes us from ALD-like moduli space to flat space may itself display $T\bar T$ - like features. 

To summarize, the global picture that we arrrive at is that  turning on $21$ of the $22$ possible maximally supersymmetric irrelevant deformations of the D1-D5 system for a \emph{finite} amount leads to a UV-complete theory, which is in general  a deformation of little string theory. The entropy of black holes in these theories appears to correspond to  the thermal entropy of (little) strings, displaying a Cardy $\to$ Hagedorn behaviour that is also characteristic of single-trace $T\bar T$ - deformed CFTs. 
 Thus, all the theories on this submanifold of the moduli space at infinity  display $T\bar T$ - like features, without exactly being a symmetric product orbifold of $T\bar T$ - deformed CFTs. This suggests the existence of a ``$T\bar T$ universality class'' of two-dimensional QFTs, which is yet to be understood. We also provided some evidence
  that the stable branch of asymptotically flat black strings may also fit within such a description for energies below a certain threshold, despite the fact that open and closed string modes in asymptotically flat space  may not be fully decoupled.  

This article is organised as follows. In section \ref{sugrareview}, we review the relevant black string solutions and the relationship between the attractor flow  and  irrelevant deformations of the D1-D5 CFT. In section \ref{section3}, we link the asymptotic degeneration limits of various black string backgrounds to known decoupling limits of string theory, for which we provide a brief review. Finally, in section \ref{section4} we analyse the thermodynamics and asymptotic symmetries of the non-extremal asymptotically flat backgrounds, as well as the thermodynamics of the decoupled ones. We end with a discussion and future directions in section \ref{discussion}. Many of the technical details of our various calculations are relegated to the  appendices. 
%
%The motivation goes as:
%
%\bi
%\item non-AdS holography $\r$ can try via irrelevant deformmations $\r$ UV complete
%\item generally very hard, but recently solvable $T\bar T$
%\item the decoupled ALD background presents $T\bar T$ - like features (entropy, ASG) $\r$ universality class of theories of which this is a tractable example?
%\ei
%or: AdS/CFT = most famous decoupling limit. But there are others, less well-known/harder to study. They are however relevant for non-AdS holography. 

%The extremal NS-F1 solution is a supersymmetric flow from AF to AdS$_3$, where the leading irrelevant operator in the IR is $(2,2)$. In the special case $c_5=0$, Kutasov showed that there exists an intermediate decoupling limit. In this article, we study general inverse attractor flow in 1/2 BPS black string and show that

%\begin{enumerate}
%\item there exists a codimension one subspace of the moduli space at infinity where the theory is decoupled 
%\item bring (thermo) evidence that all decoupled theories possess effective $T\bar T$ - like dynamics
%\end{enumerate}
%Interpretation:
%\bi
%\item attractor flows in reverse (finite amount)
%\item same theory? (20 of 21 extra moduli at infinity? )
%\item these decoupled ALD-like backgrounds are physically different from flat (e.g. self-dual)
%\ei

\section{Supergravity, attractors and irrelevant deformations \label{sugrareview}}

In this section, we review the construction of the  half-BPS black string solutions of type IIB supergravity compactified on K3, as well as their relation to irrelevant deformations of the D1-D5 CFT. 

%\new{Here we work mostly in the six-dimensional framework. }
More precisely, in  subsection \ref{6dsupergravity} 
we  briefly review  six-dimensional supergravity coupled to tensor multiplets.  Then, in \ref{section22n1} we review the construction of half-BPS black string solutions in $(1,0)$ six-dimensional supergravity, mostly as a warm-up for the construction, in subsection \ref{section23n2}, of the $(2,0)$ attractor solutions, whose parametrization we explain in detail. Finally, we review the relation between the attracted scalars and irrelevant deformations of the D1-D5 CFT. We also explain how to generate, in type IIB supergravity, a basis of solutions of interest from the simple  D1-D5 or F1-NS5  solution.

\subsection{Six-dimensional supergravity
\label{6dsupergravity}}

Let us consider six-dimensional $(1,0)$ or $(2,0)$ supergravity coupled to $n$ tensor multiplets.  In  the $(1,0)$  theory, the gravity multiplet consists of the graviton, two left gravitini and one self-dual three-form field, while the tensor multiplets consist of one anti-self-dual tensor, two right  fermions (`tensorini') % \emph{Is this the standard name?} 
and one scalar. 
In $(2,0)$ supergravity, the gravity multiplet contains the graviton, four gravitini and five self-dual three-form fields, whereas the tensor multiplet contains one anti-self-dual tensor, four fermions and five scalars%\footnote{One may reduce the $(2,0)$ tensor multiplet to a (1,0) tensor multiplet and one hyper (with 2 fermions and four scalars); however, there is no such decomposition of the $(2,0)$ gravity multiplet into $(1,0)$ ones. \emph{Correct? More comments?}}
. The resulting theories thus contain $P$ self-dual three-form fields and $Q$ anti-self-dual ones,  where  for $(1,0)$ theories, $P=1$ and\footnote{One should also add other multiplets in order to cancel the gravitational and other anomalies \cite{Alvarez-Gaume:1983ihn,Schwarz:1995zw}.}  $Q=n$%(arbitrary \emph{True?}) {\color{ForestGreen}(the gravitational anomaly is canceled if $H − V + 29T = 273$, so we can have any $n$ provided that we compensate with the other multiplets. But there are also other anomaly constraints from gauge and mixed, see the factorization requirements in 2.2-2.7 in \href{https://arxiv.org/abs/2106.10839}{$https://arxiv.org/abs/2106.10839$} )}
, whereas for $(2,0)$ ones $P=5$ and $Q=21$, as required by anomaly cancellation \cite{Alvarez-Gaume:1983ihn}. % {\color{ForestGreen} (the anomaly polynomials are written in eq 118 of the Alvarez-Gaume Witten paper. We can see that we have anomaly cancellation for 21 tensor multiplets (take into account that SD and ASD forms contribute with opposite signs).)}
In both cases - and equally so for the non-chiral supersymmetric six-dimensional theories - the scalars are described by a nonlinear sigma model whose target space is locally of the form $SO(P,Q)/SO(P) \times SO(Q)$.  %\emph{Should we introduce a special notation for the coset?}%\emph{Is it O or SO?} {\color{ForestGreen}(in the simple case of only plus signature we have that $O(n+1)/O(n)$ is the the n-sphere, while $SO(n+1)/SO(n)$ is the oriented n-sphere, see for ex wiki page for homogeneous spaces. I couldn't find a proof for this, but it seems natural to generalize to the arbitrary signature case. Topologically speaking I think they are the same and we probably don't care about orientation. I will ask some maths people. ) }
For more details, see \cite{ferraraUduality}.

The description of the non-linear model of scalars taking values in a quotient space $G/H$, with $G$ non-compact and $H$ its maximal compact subgroup, is well known  \cite{Gaillard:1981rj}. %\textbf{\emph{Please cite the original papers. I used the review in  Gaillard-Zumino}}. 
Assuming that $H$ acts by left multiplication, we require that the theory be invariant under the following gauge transformations  %\emph{May need to put on the left to agree with $X$.}{\color{ForestGreen}(do we want to keep this notation with $g$? it might be confusing, we use $g$ for ex in 2.13)}
\be
g(x) \r  h (x) g(x) \, ,
\ee
for any $h \in H$, and with $g \in G$. We then use the  gauge connection $Q_\mu$, valued in $\mathfrak{h} = {\rm Lie}(H)$, to introduce the covariant derivative
\be
\label{eq: def covariant derivative}
D_\mu g = \p_\mu g -  Q_\mu g \,, 
\ee
which transforms in the same way as $g$, provided $Q_\mu \r  h Q_\mu h^{-1} + \p_\mu h h^{-1}$. The gauge-invariant Lagrangian is then simply given by 
\be
\label{eq: Lcoset}
\mathcal{L}_{\rm coset} = - \frac{1}{2} Tr\, [( D_\mu g\, g^{-1})^2] %= - \frac{1}{2}  Tr [P_\mu P^\mu] \;, \;\;\;\;\; P_\mu \equiv g^{-1} D_\mu g 
\, .
\ee
%\old{The field $Q_\mu$ does not appear with derivatives in the Lagrangian. Varying with respect to it yields the constraint $Tr [ D^\mu g \,  g^{-1}  \d Q_\mu] =0$, implying  that $P_\mu \equiv \frac{1}{\sqrt{2}} D_\mu g\, g^{-1}  $}
The variation of the Lagrangian with respect to the field $Q_\mu$, which appears with no derivatives, yields the constraint ${\rm Tr} [ D^\mu g \,  g^{-1}  \d Q_\mu] =0$, implying  that 
\begin{equation} \label{Pcosetrep}
    P_\mu \equiv \frac{1}{\sqrt{2}} D_\mu g\, g^{-1} 
\end{equation}
is in the orthogonal complement of $\mathfrak{h}$ inside $\mathfrak{g} = {\rm Lie}(G)$. In other words, decomposing $\p_\mu g\, g^{-1} \in \mathfrak{g}$ as
\be
\p_\mu g\,  g^{-1}  = Q_\mu + P_\mu \sqrt{2}
\ee 
provides a concrete way to express the gauge connection $Q_\mu$ and the physical fields $P_\mu$ in terms of the group elements and their derivatives\footnote{In appendix \ref{sec: Appendix B - consistent truncation}, we work out in full detail this decomposition for the case $P=Q=2$.}. 
%
%and can be thought of as giving a definition of $Q_\mu$ and the physical 'fields' $P_\mu$ in terms of $g$ and its derivative. The theory also has a \emph{global} symmetry, given by right multiplication, with conserved currents xxx. \emph{Need?} {\color{ForestGreen}(since $H$ is a direct product, we can split the connection into one from $SO(P)$ and one for $SO(Q)$, which are the $Q$ and $S$ connections in Romans; these connections seem to be flat by the Maurer-Cartan eqs, but not trivial.)}  \emph{Is the connection already flat at this stage (e.g., following from the e.o.m.), or this is an extra assumption?}
Note the above decomposition is perfectly compatible with the gauge transformations. %for $Q_\mu$, which always keeps it in the Lie algebra of $H$. The scalar kinetic term is $Tr P_\mu^2$, and only involves the physical degrees of freedom.] 

%\textbf{\emph{Still need: send also to \cite{ferraraUduality} for details?}}
%
The Lagrangian \eqref{eq: Lcoset} also has a global symmetry, corresponding to the right-multiplication by a constant group element, i.e.~$g(x) \r g(x) g_0$ with $g_0 \in G$. However, this continuous symmetry of the low-energy supergravity theory is broken to its maximal discrete subgroup in the full string theory.

Particularising the discussion to our case of interest, we have $G = SO(P,Q)$, $H=SO(P) \times SO(Q)$. %and we parametrize the elements of $G$  by
%
%Following \cite{}, we will  treat all these cases in a unified manner. 
% It is very convenient to describe the scalars that parametrize the coset  $(S)O(P,Q)/O(P) \times O(Q)$ in a ``vielbein'' formalism, whereby one introduces 
% $(P+Q)^2$ fields $X_{i\L}, X_{I\L}$ that satisfy/
It is customary to parametrise the $SO(P,Q)$ group elements %\footnote{It is in fact fully possible to parametrize them by just $X^i_\L$, with $P^2$ constraints + gauge-invariances.} \emph{(is there some choice of rep here?)} 
 as $X_{i\L}, X_{\bar \imath \L}$, where the lower case latin indices transform under the local $SO(P)$ or $SO(Q)$, while the upper case greek ones do so under the global $SO(P,Q)$; these fields are sometimes referred to as `vielbeine' \cite{Cremmer:1979up}. The $SO(P,Q)$ condition $g^T \eta g=\eta$ reads

\be \label{cosetcon}
X_{i\L} X_{i\Sigma} - X_{\bar \imath   \L} X_{\bar \imath  \Sigma} = \eta_{\L\Sigma}
\ee
where summation over the repeated indices is understood, and  $\eta_{\L\Sigma}$ is a diagonal matrix whose first $P$ entries equal $1$, while the remaining $Q$ ones are $-1$. In addition to these  $(P+Q)(P+Q+1)/2$ constraints on the original $(P+Q)^2$ fields, one needs to fix the gauge symmetry\footnote{For example, one may, via an $SO(P) \times SO(Q)$ transformation, make the  $ij$ and $\bar \imath \bar \jmath$ blocks of $g$ symmetric, as follows from  the polar decomposition of a real matrix.}, leaving a total of $P Q$ physical scalars. 
%
% {\color{ForestGreen}(we have $P^2+Q^2+2PQ$ fields and above only , so we need to impose extra constraints in order to get down to the dimension $PQ$ of the coset, right? One possibility is to impose also symmetry in the blocks which brings other $\frac{P(P+1)}{2}+\frac{Q(Q+1)}{2}$ constraints, I think this is what Ferrara does.)}.  %The index $\L$ transforms under the ``global'' $SO(P,Q)$ rotations of the physical fields, whereas $r, I$ transform under ``local'' $SO(P) \times SO(Q)$ rotations, which are gauged.   The matrix $X_{a\L}$ is a vielbein in the sense that it converts the ``curved'' $\L$ indices, transforming under the global $SO(P+Q)$ symmetry of the theory, into ``local'' $SO(P) \times SO(Q)$ indices, for the ``symmetry'' that is gauged. \emph{Need better wording!/Do we still need this wording?}  
%
 Of course, $g^T$ is also a symmetric matrix, which leads to 
\be
\label{eq: coset conditions 2}
X_{i\L } X_j{}^\L = \d_{ij} \;, \;\;\;\;\; X_{i\L} X_{\bar \imath}{}^\L =0 \;, \;\;\;\;\; X_{\bar \imath \L} X_{\bar \jmath }{}^\L = - \d_{\bar \imath \bar \jmath} \,,  
\ee
% {\color{ForestGreen}(the first and the last one are extra $\frac{P(P+1)}{2}+\frac{Q(Q+1)}{2}$ constraints to reach the dimension of the coset. The middle one should follow from them and 2.4 contracting the indices, which also implies that the sign should be $-\delta_{IJ}$ in the last one to have $P+Q=\delta^{\Lambda}_{\Lambda}$, right?)}
 %
where the  index $\L$ is raised and lowered with $\eta$.

So far for the scalar manifold.  As stated above, the theory contains also $P+Q$ three-form tensor fields, with field strengths $F^\L= d A^\L$, which satisfy %. One may write down an $SO(P,Q)$ - invariant pseudo-action \emph{Is this correct? I can't match scalar kinetic term.} {\color{ForestGreen}(not working yet for me either)}
%
%\be
%S = \frac{1}{16\pi G_6} \int d^6x \sqrt{-g} \left( R + \frac{1}{8} tr (\p \N^{-1} \p\N) - \frac{1}{3} \N^{-1} F^2 \right)
%\ee
%and impose post-factum 
the modified (anti)-self-duality condition
\be
\label{eq: N-self duality condition}
\N_{\L\Sigma} \star F^\Sigma = \eta_{\L\Sigma} F^\Sigma 
\ee
Here we have introduced the matrix $\N = g^T g$, which can be written as 

\be \label{matrixNkin}
\N_{\L\Sigma} = X_{i\L} X_{i\Sigma} + X_{\bar \imath \L} X_{\bar \imath \Sigma}
\ee
and can be thought of as a gauge kinetic term in a pseudo-action  used to derive a set of equations of motion that are consistent with the self-duality condition above for the three-forms, see e.g. \cite{skendy}  for more details.  Note that \eqref{eq: N-self duality condition} implies that the electric and magnetic charges of black strings computed using these field strengths are equal up to a sign. The charges are defined as  %\emph{Check normalisation!}
\be
\label{eq: def charges}
q^\L = \frac{1}{2\pi^2 \a'} \int_{S^3} F^\L 
\ee
where $\a'$ is a dimensionful scale related to the six-dimensional Newton constant as $G_6 = \frac{\pi^2}{2} \a'^2$, and are integer-quantized. On the other hand, the combinations of field strengths that appear in the supersymmetry variations are 

\be
\label{eq: def usual SD and ASD 3forms}
F^{+i} = X^i{}_\L F^\L \;, \;\;\;\;\; F^{-\bar \imath} = X^{\bar \imath}{}_\L F^\L 
\ee 
which are  (anti)-self-dual  in the usual sense -
as follows from the properties of the matrix $\N$ -  but  satisfy modified Bianchi identities \cite{romans}. While we use the same letter, $F$, to denote  both sets of three-form  fields, it should be clear from the index structure which one we are referring to. 
 %\emph{Check! Ok.} 
One may also introduce the central charges
\be
\label{eq: def central charges Z}
Z^i = X^i{}_{\L} q^\L \;, \;\;\;\;\;\; Z^{\bar \imath}=X^{\bar \imath}{}_{\L} q^\L \, ,
\ee
which can be related to the $S^3$ integrals of the (anti)-self-dual fields (see e.g. \cite{finn} for a discussion of their meaning in the case of non-spherical solutions). %\emph{Only for sph. symm. sols, right?} {\color{ForestGreen}(in 5d in Larsen there is an non-spherical symmetric analogue around eq 89)} $Z_r$ appears in the susy algebra. 
%
%Finally, it is interesting to define the connection  $dV V^{-1}$, which is a one-form in field space. Its off-diagonal terms 
%%%
%\be
%P^{r I} = \frac{1}{\sqrt{2}} X^{r\L} d X^I{}_\L
%\ee
%give the scalar kinetic terms $(P^{rI}_\mu)^2$, whereas the antisymmetric diagonal blocks yield the $SO(P)$ and $SO(Q)$ connections to be used in building the spinor covariant derivative. \emph{Fix normalization and write it in action upstairs!}
The norms of these central charges are denoted
\be
|Z_+| \equiv (Z^i Z^i)^{\frac{1}{2}}\;, \;\;\;\;\;\;\; |Z_-| \equiv (Z^{\bar \imath} Z^{\bar \imath})^{\frac{1}{2}} 
\ee
 and they satisfy $|Z_+|^2 - |Z_-|^2 = q^\L q_\L$, where we used \eqref{eq: coset conditions 2}. The $SO(P)$ central charge $|Z_+|$  corresponds to the  eigenvalue of the central charge of the supersymmetry algebra.

When $P = Q$, the numbers of self-dual and anti-self dual tensor fields are  equal, hence it is  possible to write down a standard action for the theory. % \emph{Any attempts?}{\color{ForestGreen}(Romans discusses this in the second page of the paper, he points out to an old paper by Marcus,Schwarz that I couldn't find)}
Denoting the standard kinetic term  for the $P$ three-form fields in the lagrangian by $\mathcal{G}$ and the axionic couplings by $\mathcal{B}$,  the relation with  $\N_{\L\Sigma}$ is given by \cite{ferrara}
\be \label{definitionmatrixC}
C^T \N C =   \left(\begin{array}{cc} 1 & \mathcal{B} \\ 0 & 1 \end{array} \right) \left(\begin{array}{cc} \mathcal{G} & 0 \\ 0& \mathcal{G}^{-1} \end{array} \right)  \left(\begin{array}{cc} 1 & 0 \\ -\mathcal{B} & 1 \end{array} \right)  \;, \;\;\;\;\; C = \frac{1}{\sqrt{2}} \left(\begin{array}{cc} 1 & 1 \\ - 1 & 1 \end{array} \right) \, .
\ee
While will always have $P \neq Q$ at the level of the full supergravity theory, this relation will nevertheless be useful as we will consider a (purely bosonic) consistent truncation that contains equal numbers of self-dual and anti-self-dual three-form components. 
% The fields that transform nicely under duality rotations are $F^{\pm \L}$ and $\N_{\mp \L \Sigma} F^{\pm\Sigma}$. \emph{Check and understand Gaillard-Zumino!}

The main focus of this article is  the $(2,0)$ theory  obtained by compactifying type IIB supergravity on K3, which has $P=5$ and $Q=21$. There are thus $PQ = 105$ masless scalars  in this compactification. Of these, $80$ correspond to the moduli of the metric and  the $B$-field on K3, %\footnote{These comprise the overall volume of K3, its $57$ complex structure moduli, which transform under $O(3,19)$, and the $22$ moduli  $b^\g$ coming from the $B$-field on the  2-cycles of the internal space. \emph{Keep?}},
 which  parametrize an $SO(4,20)/SO(4) \times SO(20)$ subspace of the moduli space. The remaining 25 moduli correspond to the compactification  of the RR two-form field $C_2$ on the $22$ 2-cycles of the K3
 %  $c^\g$, coming from , 
  the axion $\chi_1$, the dilaton, and the RR four-form compactified on the entire (unit) internal manifold, here denoted as $\chi_2$. In a convenient basis, the $SO(5,21)$ vielbein %/coset representative takes the form/ 
can be (schematically) written in terms of these fields as %\com{Include $\Omega_4?$}
%
%Finally, one may wonder what is the parametrization of the $6d$ coset in terms of the 10d physical fields, e.g. for type IIB/K3. {\color{ForestGreen}The type IIB action is
%\begin{align}
%S&=\frac{1}{2\kappa^2_{10}}\int d^{10}x\sqrt{-\hat{g}}\bigg\{e^{-2\hat{\Phi}}\bigg(\hat{R}+4(\partial\hat{\Phi})^2-\frac{1}{12}\hat{H}^2_3\bigg)-\frac{1}{2}\hat{F}_1^2-\frac{1}{12}\hat{F}_3^2-\frac{1}{480}\hat{F}_5^2\bigg\}-\frac{1}{4\kappa^2_{10}}\int d^{10}x \hat{C}_4\wedge \hat{H}_3\wedge \hat{F}_3
%\end{align}
%where the field strengths for $p>1$ are defined as
%\begin{align}
%\hat{F}_{p+1}&=d\hat{C}_{p}-\hat{H}_3\wedge\hat{C}_{p-2}
%\end{align}
%and the hats stand for 10d. Upon reduction on K3, one obtains a pseudoaction of the form
%\begin{align}
%S&=\frac{1}{2\kappa_6^2}\int d^6x\sqrt{-g}\bigg\{ e^{-2\Phi}\bigg(R+4(\partial\Phi)^2+\frac{1}{8}tr(\partial M^{-1}\partial M)\bigg) + \frac{1}{2}\partial l^a M^{-1}_{ab}\partial l^b -\frac{1}{3}G^A_{MNP}\mathcal{M}_{AB}^{-1}G^{BMNP} \bigg\}
%\end{align}
%where $G^A$ are the 3-forms satisfying the usual Bianchi identities, $\mathcal{M}$ is the $SO(5,21)$ matrix of scalars, $M$ is the $SO(4,20)$ part of it and $l^a$ is the vector of RR scalars (see Skenderis). A particular parametrization of the vielbeine given in \cite{} 
\begin{align}
\label{eq: full X matrix}
g&=\Omega_4^T\begin{pmatrix}
e^{-\phi} & 0 & 0 & 0 & 0\\
-e^{\phi}(\chi_1 \chi_2-\frac{1}{2}c^\g c_\g)  & e^{\phi} & -e^{\phi}\tilde{\chi}_2 & -e^{\phi}\chi_1 & e^{\phi}\tilde{c}_{\delta}\\
e^{-\rho}\chi_1 & 0 & e^{-\rho} & 0 & 0 \\
e^{\rho}\chi_2 & 0 & \frac{1}{2}e^{\rho}b^2 & e^{\rho} & e^{\rho}b_\delta \\
\tilde{V}_{\delta\gamma}c^{\gamma} & 0 &\tilde{V}_{\delta\gamma}b^{\gamma} & 0 & \tilde{V}_{\gamma\delta}
 \end{pmatrix}\Omega_4 \, , 
%\quad 
% \Omega_4 = \frac{1}{\sqrt{2}}
% \begin{pmatrix}
% 1 & 0  & 0  & 1 \\
% -1 & 0  & 0  & 1 \\
% 0 & 1  & 1  & 0 \\
% 0 & -1  & 1  & 0 
% \end{pmatrix}
\end{align}
where  $\phi$ is the six-dimensional dilaton, $e^{-2\rho}$ is the string frame volume\footnote{Note our conventions differ by a sign and factor of $2$ as compared to  those of \cite{skendy}.} of the K3,  $\tilde{V}_{\alpha\beta}$ is the $SO(3,19)$ vielbein on the unit K3,  $b^\g$, $c^\g$ are the components of the two-form $B$ and $C$-field on the internal manifold,   while the tildes mean that the fields are shifted in various ways, as explained in detail in %[and the various scalars come from the reduction of the ten-dimensional fields as explained in]
\cite{skendy}.  The matrix $\Omega_4$ %is also given in appendix B.2 of Skenderis  and 
is the  $SO(5,21)$ analogue of  the matrix $C$ in \eqref{definitionmatrixC}, which takes into account the fact that there is no standard action for the type IIB self-dual five-form. %Note that the matrix $g$ is only defined up to an $SO(5) \times SO(21)$ rotation. 
 The above parametrization reveals the $ SO(4,20)/SO(4) \times SO(20) \subset SO(5,21)/SO(5)\times SO(21)$  structure of the moduli space. 

In the main part of this article, we will mostly work with an $SO(2,2)$ consistent truncation \cite{Duff:1998cr} of the type IIB action on K3 that only keeps the upper left $4 \times 4$ block of the above matrix. This truncation is discussed in detail in appendix \ref{sec: Appendix B - consistent truncation}. Nonetheless, \eqref{eq: full X matrix} makes it clear how  the $SO(4,20,\mathbb{Z})$ action of T-duality  will generate other solutions.

\subsection{Warm-up: half-BPS black string solutions in $(1,0)$ supergravity}
\label{section22n1}
Let us now review the construction of the half-BPS, spherically symmetric black string solutions in $(1,0)$ six-dimensional  supergravity coupled to $n$ tensor multiplets. While these solutions are extremely well-known  (see  \cite{HetLam:2018yba}  for much more general BPS  solutions of this theory) and they are simply related to attractor solutions of five-dimensional $\mathcal{N} =2$ supergravity coupled to vector multiplets \cite{finn}, we find this exercise  useful as a warm-up for the explicit construction of the attractor solutions in $(2,0)$ supergravity,  addressed in the next subsection.  

In this theory, 
%In  $(1,0)$ $6d$ supergravity, the gravity multiplet consists of the graviton, two left gravitini and one self-dual three-form field, while the tensor multiplets consist of one anti-self-dual tensor, 2 right ``tensorini'' and one scalar. 
the two chiral gravitini and fermions  in the tensor multiplets are related by a symplectic Majorana condition. However, since the R-symmetry $USp(2)$ action is trivial, one may equally well replace them by a single,  unconstrained left gravitino and an unconstrained right tensorino (with $8$ real components each) as in \cite{romans} and omit writing  the symplectic indices on the fermionic felds. 
The supersymmetry variations of the fermions then read
\be
\delta\psi_{\mu}=\nabla_{\mu}\varepsilon+\frac{1}{4}F^+_{\mu a b}\gamma^{ab}\varepsilon    \;,\;\;\;\;\;\;\;
\delta\chi^{\bar \imath}=\frac{1}{\sqrt{2}}\gamma^{\mu}P_{\mu}^{\;\;\bar \imath}\varepsilon - \frac{1}{12}\gamma^{abc}F_{abc}^{-\;\bar \imath}\;\varepsilon \label{kseqn1} \, ,
\ee
where $\varepsilon$ is a left-handed Weyl spinor (with $8$ real, or $4$ complex components), $\mu$ and $a$ are six-dimensional spacetime and, respectively,  tangent space indices %namely $\g_7 \varepsilon = - \varepsilon$ 
and, because $P=1$ in this theory, the covariant derivative contains only the usual spacetime spin connection. In fact, since the $i$ index only takes one value, we will omit writing it altogether, thus writing $P_\mu^{i \bar \imath} \r P_\mu^{\bar \imath}$.
 % $P_{\mu}^{iI} \r P_\mu^I$ for this case is defined in \eqref{}. \com{necessary?}
%The chirality condition reduces the number of complex nonzero components of $\varepsilon$ to 4.
Our conventions for the $\gamma$ matrices are spelled out in appendix \ref{sec: AppA - 6d susy solutions}. 
%{\color{blue}Here, because we only have 1 SD 2-form, the covariant derivative $D_{\mu}\varepsilon$ contains only the spacetime spin connection (the $Q$ part is 0, in Romans notations). \emph{Explain previously!} In our conventions, the chirality condition on the susy parameter is $\g_7 \varepsilon = - \varepsilon$.  }
%{\color{ForestGreen}(maybe say that since $P=1$, we will not write this SD index anymore)}

We will be interested in black string solutions that preserve an $SO(4) \times SO(1,1)$ symmetry. %\emph{Do we need to assume $2d$ Lorentz?}
 Supersymmetry then implies we can make the following  Ansatz for the metric  %\emph{Understand how much needs to be assumed!}
\begin{align}
\label{eq: metric Ansatz}
ds^2&=e^{2U(r)}(-dt^2+d\s^2)+e^{-2U(r)}(dr^2+r^2 d\Omega_3^{2}) \, .
\end{align}
The three-form field strengths must take the form  %\emph{Careful normalisations $\om_{S^3}$!}
\be
\label{eq: (1,0) 3-forms generic}
F^\L = q^\L \a' \om_{S^3} +  \a' \eta^{\L \Sigma} \N_{\Sigma P}\, q^P \star \om_{S^3} \, ,
\ee
where the first term follows from the definition of the charge, and the second from  the modified (anti)-self duality conditions \eqref{eq: N-self duality condition}. The parameter $\a'$, introduced above  by dimensional analysis, has dimensions of $(length)^2$ and is related to the six-dimensional Newton's constant via  $G_6 = \pi^2 \a'^2/2 $.  The %\old{exactly  (anti-)self-dual three-form  field strengths} 
three-forms \eqref{eq: def usual SD and ASD 3forms} that enter the supersymmetry variations then take the form   
%. Clearly, integrating on $S^3$, we should obtain the central charges, which fixes the sphere part of all the field strengths. Clearly, the rest should be its Hodge dual/minus Hodge dual
%
\be \label{formsusyvar}
F^+=Z \a' (\omega_{S^3}+\star\, \omega_{S^3})\;, \;\;\;\;\;\;
F^{-\,\bar \imath }=Z^{\bar \imath} \a' (\omega_{S^3}-\star \, \omega_{S^3}) \, ,
\ee
where we have used  the proprerties of the matrix $\N$ given in \eqref{eq: N-self duality condition}. 
%\old{that the integral of $F^\L$  over a sphere of radius $r$ is $q_\L$ and that the scalar fields entering the definition \eqref{} are constant over such a sphere} \new Eqs.~\eqref{eq: def charges}-\eqref{eq: def central charges Z}. 
 Plugging these expressions into the spinor variations and using the chirality of $\varepsilon$,  the  tensorini equations reduce to %\emph{Check factors of $U$!!}
%
%, we note that $P_\mu \neq 0$ only if $\mu = r$. Using chirality, the equation can be written as 
%
\begin{align}\label{tens1}
\bigg(\frac{1}{\sqrt{2}} P^{\bar \imath}_r  -\frac{\a' Z^{\bar \imath} e^{2U}}{r^3}\gamma^0\gamma^1\bigg)\varepsilon=0 \, .
\end{align}
%\old{using the fact that only the $r$ component of $P_\mu$ is non-vanishing.} 
Since this is an algebraic equation for $\varepsilon$, %and the gamma matrices square to one],]
 it can only have non-trivial solutions if %\emph{Check factors U!!}
\be
\label{eq: (1,0) projection on vep}
\frac{1}{\sqrt{2}}P^{\bar \imath}_r = \pm \a'\frac{Z^{\bar \imath} e^{2U}}{r^3} \, ,\hspace{1cm} \gamma^0\gamma^1\varepsilon= \pm \varepsilon \, ,
\ee
provided $Z^{\bar \imath}\neq 0$. %, otherwise the equation is trivially satisfied for $P_\mu=0$). 
As it turns out, it is the upper solution that is physical, in the sense of leading to positive central charge. %\emph{Why is $Z>0$ desirable? How about the horizon area?}  %{\color{ForestGreen}(in the notes is the other way round, $Z>0$ for the plus sign)}  {\color{ForestGreen} The extra projection reduces further the number of nonzero components of $\varepsilon$ to 2 complex, thus 4 real. In the particular case $Z^I=0$, which is what happens in the near horizon (we will see this later), the supersymmetry is enhanced to 8, as the projection is no longer necessary.}
Using the explicit expression \eqref{Pcosetrep} for  $P_\mu^{\bar \imath}$ in terms of the coset representatives% defined in the previous section, namely 
\begin{align}
P_{\mu}^{\bar \imath}&=\frac{1}{\sqrt{2}}\eta^{\Lambda\Sigma}X_{\Lambda}dX^{\bar \imath}_{\Sigma}=-\frac{1}{\sqrt{2}}\eta^{\Lambda\Sigma}dX_{\Lambda} X^{\bar \imath}_{\Sigma} \, ,
\end{align}
where $\L$ runs over $n+1$ values and ${\bar \imath}$ over $n$, 
this immediately translates into a flow equation for the scalar fields in the tensor multiplets.

The gravitino equations fix the coordinate dependence of $\varepsilon$, which turns out to be $t$- and $\s$-independent in this coordinate system, provided the additional projection \eqref{eq: (1,0) projection on vep} is obeyed; see appendix \ref{sec: AppA - 6d susy solutions} for the  explicit solution. In addition, they lead to an equation for the metric as a function of the scalar fields
\be \label{functionmetric}
\p_r U = \a' \frac{Z e^{2U}}{r^3}
\ee
The equations \eqref{eq: (1,0) projection on vep} and \eqref{functionmetric} can be easily manipulated to yield an explicit solution for the radial dependence of the scalar fields $X^\L, U$ ($X^{{\bar \imath} \L}$ is then entirely determined by the coset conditions). It is convenient to introduce a new radial coordinate $\tau \equiv 1/r^2$, in terms of which %one has \com{Do we want these vertical spaces?}

\be
\p_\tau X^\L \, X_{{\bar \imath}\L} = \a' e^{2U} q^\L X_{{\bar \imath}\L} \;, \;\;\;\;\;\p_\tau\, e^{-2U} = \a' Z \, 
\ee
%
%
%\be
%{\color{blue}\frac{1}{2} X^\L \p_r X_{I\L} =} - \frac{1}{2} \p_r X^\L  X_{I\L} = \frac{e^{2 U}}{r^3} q^\L X_{I\L}
%\ee
Multiplying both sides of the first equation by $X_{{\bar \imath}\Sigma}$, summing over ${\bar \imath}$ and using \eqref{cosetcon}, we find

\be
\p_\tau X_\L =  \a' e^{2 U} (q_\L - X_\L \, q_\Sigma X^\Sigma) \, 
\ee
%where we have introduced the more convenient variable $\tau = 1/r^2$.  In terms of it, \eqref{functionmetric} becomes $\p_\tau e^{-2U} = Z$, and 
This leads to 
\be
\p_\tau (e^{-2 U} X_\L) = q_\L \a' \, 
\ee
which immediately implies the solution is given in terms of harmonic functions

\be
e^{-2 U} X_\L = H_\L = \frac{q_\L \a'}{r^2} + c_\L \, 
\ee
Remembering the $X^\L$ obey the constraint \eqref{eq: coset conditions 2}, namely $X^\L X_\L =1$, one finally obtains

\be
e^{-2 U} = \sqrt{ H_\L H^\L} \;, \;\;\;\;\;\;\;\; X^\L = \frac{H^\L}{\sqrt{ H_\L H^\L} } \, 
\ee
In the near-horizon limit, i.e.~the small $r$ region, the geometry becomes AdS$_3 \times S^3$, of squared radius $ \alpha' q_\L q^\L$,  while the scalar fields $X^\L$ approach constant values. The latter are entirely determined by the charges of the black string, irrespective of their values at infinity. This is the well-known  attractor mechanism \cite{Ferrara:1995ih}. Note that the asymptotic value of the $n$ moduli, which correspond to the $n$ massless scalars in the tensor multiples, only  
depends on ratios of the $n+1$ constants $c_\L$, leading to the correct parameter/moduli count. The overall scale of the coefficients $c_\L$ is usually fixed by requiring that the asymptotic metric takes the standard flat form, which normalises $c_\L c^\L =1$; we will however not make this assumption in this work.  

 %model contains whose asymptotic value is determined by the $n+1$ constants  $c_\L$, only up to an overall rescaling. \new{More precisely, } \emph{Write better!} 

The supersymmetry central charge is given by 
\be
Z = q_\L X^\L = \frac{q_\L H^\L}{\sqrt{H_\L H^\L}} %\, .\old{\;, \;\;\;\; |Z_-| = \ldots}
\ee
The constraint $Z^2 - Z^{\bar \imath} Z_{{\bar \imath}} = q^\L q_\L$, obtained by contracting \eqref{cosetcon} with $q^\L q^\Sigma$, then implies that  $Z_{\bar \imath} \r 0$ in the near-horizon limit. Consequently, each term in \eqref{tens1} vanishes separately and the second projection in \eqref{eq: (1,0) projection on vep} is no longer needed, leading to the well-known supersymmetry enhancement in the near horizon region (see also \cite{Raeymaekers:2006np} for a clear exposition in the D1-D5 case). Thus, while at generic $r$ one has 
$4$ real supercharges, obtained by imposing the projection \eqref{eq: (1,0) projection on vep} on the initial six-dimensional Weyl spinor (with $8$ real components), this projection drops
%
%starting with a $6d$ spinor (with $8$ complex components), which are cut in half due to the Weyl condition, and further reduced by a factor of $2$ due to the projection \eqref{eq: (1,0) projection on vep}. 
in the near-horizon limit, and the supersymmetry is enhanced to $8$ real supercharges. The explicit solutions for the additional supersymmetry generators are provided in appendix \ref{sec: AppA - 6d susy solutions}.

Finally, the solution for the three-form fields is obtained from \eqref{eq: (1,0) 3-forms generic} and can  be more explicitly written as 
\be
F^\L =  \a' q^\L  (\om_{S^3} + \star \om_{S^3}) + 2  \a' X_{{\bar \imath}}^\L Z^{\bar \imath} \star \om_{S^3} \,,
\ee
which shows that all the fields $F^\L$ become self-dual  in the near-horizon limit.  

\subsection{Half-BPS black strings in $(2,0)$ supergravity}
\label{section23n2}

We would now like to follow the same steps to find the explicit solution for the $(2,0)$ case. This solution has been previously constructed in \cite{mikhailov}; here we include a parametrisation and  more details that are useful for our purposes.

%derivation has previously been sketched in \cite{ferrara}, but with fewer details given.  

%In $(2,0)$ supergravity, the gravity multiplet contains the graviton, four gravitini and five self-dual three-form fields, whereas the tensor multiplet contains one anti-sef-dual tensor, 4 fermions and five scalars 

The supersymmetry variations of the fermions are 
\begin{align}
\delta\psi_{\mu A}&=\nabla_{\mu}\varepsilon_A+\frac{1}{4}H_{\mu\nu\rho}^{+i}\gamma^{\nu}\gamma^{\rho}(\Gamma^i)_{A}^{\;B}\varepsilon_B
\,, \\
\delta\chi^{\bar \imath}_A&=\frac{1}{\sqrt{2}}\gamma^{\mu}P_{\mu}^{i \, \bar \imath}(\Gamma^i)_A^{\;B}\varepsilon_B-\frac{1}{12}\gamma^{\mu}\gamma^{\nu}\gamma^{\rho}H^{-\bar \imath}_{\mu\nu\rho}\varepsilon_A \,
\end{align}
where $A$ is a spinor index for the R-symmetry group, while the  covariant derivative now also contains the $SO(5)$ connection and is explicitly given in \eqref{generalcovariantder}.   The spinors satisfy a Weyl and a symplectic Majorana condition. We refer the reader to  appendix \ref{sec: AppA - 6d susy solutions} for all relevant notations and conventions.
%{\color{blue}\emph{Move to first section or appendix. Is this so that $(D_\mu g)_{ij} =0$?}
%\begin{align}
%D_{\mu}\varepsilon_A&=\bigg(\partial_{\mu}+\frac{1}{4}\omega_{\mu}^{\;ab}\gamma_a\gamma_b \bigg)\varepsilon_A-\frac{1}{4}Q_{\mu ij}(\Gamma^i\Gamma^j)_A^{\;B}\varepsilon_B
%\end{align}
%Clearly, only $Q_{r}$ is nonzero.

The Ansatz for the metric and three-form field strengths are still given by \eqref{eq: metric Ansatz}, upon  adding an  $SO(5)$ index $i$ to $F^+$ and $Z$.
The supersymmetry equations $\delta\psi_{\mu A}=0,\delta\chi^{\bar \imath}_A=0$ are somewhat harder to solve due to the more complicated R-symmetry structure. Guided by the $(1,0)$ results, 
we will start by looking for time-independent solutions for the Killing spinors, which we expect to corresponded to the Poincar\'e supersymmetries preserved throughout the bulk. Using this assumption, the time component of the gravitino  equation reduces to

\be
\left( \p_r U -  \a' \frac{Z^i \Gamma^i e^{2U}}{r^3} \g^0\g^1 \right) \varepsilon =0 \, \label{tcompgraveq} 
\ee
As before, this implies a projection
\be
\label{eq: (2,0) projection on vep}
\hat Z^i \Gamma^i   \g^0\g^1  \varepsilon = \pm  \varepsilon  \;, \;\;\;\;\;\;\; \hat Z^i \equiv \frac{Z^i}{|Z_+|} \, 
\ee
As in the previous section, we focus on the case with a plus sign.  In addition,  \eqref{tcompgraveq} yields an equation for $U(r)$, which reads 
\be
\label{eq: (2,0) der Exp2U}
\p_\tau \, e^{-2U} =  \a' |Z_+| %\;, \;\;\;\;\; |Z_+| \equiv \sqrt{Z^i Z_i}
\ee
where we have again introduced $\tau=1/r^2$. 
Using the above projection, the remaining gravitino equations simply determine the spacetime dependence of the Killing spinor, see appendix \ref{sec: AppA - 6d susy solutions} for more details. The only additional consistency requirement is that
%Demanding that the projection above be consistent with the $r$ component of the Killing spinor equation requires that 
%
\be
D_r \hat Z^i =0
\ee
where the covariant derivative $D_\mu$ has been defined in \eqref{eq: def covariant derivative}. 
%\textcolor{blue}{a condition that the scalar fields will need to satisfy}. \emph{This statement is wrong, correct?} %Note that the $6d$ Majorana-Weyl condition on the $6d$ spinors yields 16 real spinor components, {\color{ForestGreen}which get reduced to 8 
%upon imposing the projection, in agreement with the fact that the string is half-BPS.} %\emph{If $\hat Z^i \Gamma^i$ is diagonalized first, then it looks like $1/2$, which are too many susies. } {\color{ForestGreen}(don't we have 8 susy in the $N=(2,0)$ case, enhanced to 16 in the near horizon?)} \emph{Yes, I guess the string is half-BPS.}

Next, we look at  the tensorini equations. Plugging in the Ansatz for $F^{-{\bar \imath}}$ and using the chirality of the spinor and the projection \eqref{eq: (2,0) projection on vep}, one arrives at the following equation
\begin{align}\label{tensori2}
%\left( \frac{1}{\sqrt{2}}P_{r}^{i{\bar \imath}}\Gamma^i- \a' \frac{Z^{\bar \imath} e^{2U}}{r^3}\gamma^{0}\gamma^{1 \right)\varepsilon=
 \left(\frac{1}{\sqrt{2}}P_{r}^{i{\bar \imath}}- \a' \frac{Z^{\bar \imath} e^{2U}}{r^3}\hat Z^i \right) \Gamma^i\varepsilon=0
\end{align}
%where in the second line we have used the projector. 
Since the $\Gamma^i$ are linearly independent, the scalars are constrained to satisfy
% the term multiplying it should be set to zero. {\color{ForestGreen}This yields the following equation for the scalars 
%
\be
\frac{1}{\sqrt{2}}P_{r}^{i {\bar \imath}} = - \frac{1}{2} X^{{\bar \imath}\L} \p_r X^i_\L = \a' \frac{e^{2U} \hat Z^i}{r^3} q_\L X^{{\bar \imath}\L} \, 
\ee
Multiplying again by $X_{{\bar \imath}\Sigma}$, summing over ${\bar \imath}$ and using the coset conditions we find

%, and then choosing the connection $Q^{ij}_\tau = X^{j\L} \p_\tau X^i_\L$ \emph{Is this a choice/ an obligation by the e.o.m/correct?} we find

\be
\label{eq: (2,0) cov der of X}
D_\tau X^i_\L = \a' e^{2U} \hat Z^i (q_\L - Z_j X^j_\L) \, 
\ee
Note that this immediately implies the covariant constancy of $\hat Z^i$, upon contraction with $q^\L$ and using \eqref{eq: def central charges Z}. Putting together Eqs.~\eqref{eq: (2,0) der Exp2U} and \eqref{eq: (2,0) cov der of X}, we find 

\be
D_\tau (e^{-2U} X^i_\L ) = \a' q_\L \hat Z^i + \a' |Z_+| (X^i_\L - \hat Z^i X^{||}_\L) \;, \;\;\;\;\;\;\; X^{||}_\L \equiv \hat Z^i X_{i\L} \, 
\ee
By contracting the condition on the left with $\hat Z^i$, we find that, similarly to the (1,0) case, 
\begin{equation}
\p_\tau (e^{-2 U}X^{||}_\L) = \a' q_\L \label{solXparall}
\end{equation}
hence the corresponding solutions are written in terms of $26$ harmonic functions $H_\L$. Introducing the perpendicular components 
\be
X^{\perp i}_\L \equiv  X^i_\L - \hat Z^i X^{||}_\L \, 
\ee
one can immediately show that they are covariantly constant, $D_\tau X^{i\perp}_\L =0$. We conclude that the general solution for the scalars takes the form 

\be \label{generalsolutionscacoset}
X^i_\L = \hat Z^i H_\L e^{2U} + \bar g^i_\L \;, \;\;\;\;\;\;\; H_\L = \frac{\a' q_\L}{r^2} + c_\L
\ee
where
\be
D_r \bar g^{i\L} = D_r \hat Z^i =0  \;, \;\;\;\;\;\hat Z_i \bar g^{i\L} =0 \, \label{gperpzh}
\ee
%where the $\bar g^i_\L$ are covariantly constant. 
and the expression for $U$ can be obtained from the constraint $X^{i\L} X_{j\L} =\d^i_j$, yielding as before

\be
e^{-2 U} = \sqrt{H_\L H^\L}  \,
\ee
The solution for the central charges is 
\be \label{centralchargesmod}
|Z_+| = \frac{q_\L H^\L}{\sqrt{H_\L H^\L}} \;, \;\;\;\;\;\; |Z_-| = \sqrt{\frac{(q\cdot c)^2 -q^2 c^2}{H_\L H^\L}} \, 
\ee
where $q^2 \equiv q_\L q^\L$, $c^2 \equiv  c_\L c^\L$ and $q\cdot c\equiv q_\L c^\L$. This again shows that $|Z_-|$, and thus the individual $Z_{\bar \imath}$, vanishes in the near-horizon region. 

The equations \eqref{gperpzh} can be easily solved in a gauge in which the $SO(5)$  connection $Q^{ij} =0$, case in which $\bar g_{i\L}$ and $\hat Z^i$ are simply constants. This gauge condition  still allows for constant $SO(5)$ gauge rotations%\footnote{The gauge connection introduced around  \eqref{eq: def covariant derivative} can be set to zero via a gauge transformation with parameter $h$ satisfying $\p_\mu h h^{-1} = Q_\mu$. Note, however, that this leaves residual gauge transformations with a constant parameter, which can be used to rotate $\hat Z^i$ to a particular direction. \emph{Keep?}}
, which can be used to align $\hat Z^i$ to a particular direction, say $1$. The orthogonality conditions in \eqref{gperpzh} then simply set $g^{1\L} =0$. 

Let us now  compute the number of independent parameters of the solution. The constraint $X^{i\L} X_{j\L} = \d^i_j$ reads
\vskip-1mm
\be
\hat Z^i \hat Z^j + \bar g^{i\L} \bar g^j_\L + \frac{H_\L}{\sqrt{H\cdot H}} (\hat Z^i \bar g^{j\L} + \hat Z^j \bar g^{i\L}) =\d^{ij}
\ee
Working in the  gauge  $Q^{ij}=0$, the $\bar g^i_\L$  and  $\hat Z^i $ are constant. 
%which can further be chosen to point in a particular direction , say $Z^i=\delta^{i1} Z^1$. %\emph{Correct?}{\color{ForestGreen}(I think yes, we solved it in the TsT case for arbitrary angle and got the same counting)}
 Requiring that the $r$-dependent and $r$-independent terms vanish separately leads, under the  assumption that  the $c_\L$ are generic, to  
 %
%  and that the $q_\L $ and $c_\L$ are generic, we have that 
 % $H_\L \bar g^{i\L} =0$, or
%
\be
\bar g^{i\L} \bar g^j_\L = \d^{ij} -\hat Z^i \hat Z^j \;, \;\;\;\;\;\; q_\L \bar g^{i\L} =0 \;, \;\;\;\;\; c_\L  \bar g^{i\L} =0 \,  \label{constrcg}
\ee
The first two conditions, together with the orthogonality  relation from \eqref{gperpzh}, simply reduce $\bar g_{i\L}$ from an a priori $SO(5,21)$  matrix to an $SO(4,21)$ one. Note that the way the $SO(4,21)$ is embedded into the larger coset depends on the charges of the configuration \cite{Dijkgraaf:1996hk}. The last condition should be read as four constraints on the $c^\L$, reducing the number of independent such constants to $22$. These   constants parametrize the remaining $21$ asymptotic moduli in a `projective'  fashion, in the sense that the physical scalar fields do not depend on the overall scale of these coefficients, at least as long as $c_\L c^\L \neq 0$; the overall scale itself appears in the asymptotic metric, which we will not be setting to one.
The final solution thus has $84+22 = 106$ parameters. % Of these, 105 appear in the asymptotic values of the scalar fields, , and one in t 
% \new{Additionally, we have $\hat{Z}_{i} \hat{Z}^{i} = {\bar g}_{i\L} {\bar g}^{i\L} = 1$.} 
%Starting with the original $26 \times 5 = 130$ scalars \com{values of $g^{i\L}$?}, orthogonality to $\hat Z^i$ subtracts $26$, while that to the charge - another $4$. Finally, we subtract $16$ other components, representing the $SO(P)$ constraints and gauge freedom, to be left with the expected number of $84$. % {\color{ForestGreen}(it's not very clear. I would say the charge condition is $-5$ and the $\hat Z^i \hat Z^j + \bar g^{i\L} \bar g^j_\L=\delta^{ij}$ is $-15$, in total $5\times 26 - 26 - 5 -15=84$)}. 
%We also have $4$ constraints on $c_\L$, yielding only $22$ independent \old{constraints} \new{constants} that we will denote $c_+ \sim q_\L c^\L/|q|$ and $c_{\bar I}$. Note these constraints depend on the ``hypermultiplet'' moduli. Given that the physical scalar fields at infinity are ratios of these constants, we obtain the correct number of $21$ attracted moduli. \emph{What is the physical significance of these conditions, in terms of \underline{physical} restrictions on the fields?}
%
%
%\bi
%\item \textcolor{red}{Are all the $q_\L$ allowed by supersymmetry?}
%\item check whether the leading departure of the $F^\L$ from  the near-horizon  only has $F^{+\L} \propto q^\L$ 
%\ei
For the solution written directly in terms of the physical fields, see \cite{mikhailov}.

\subsection{Mapping to irrelevant deformations}
\label{section24}
Let us now review the interpretation of this solution in the D1-D5 CFT. To read off the dual interpretation of the various parameters, we
 will  mostly concentrate on the near-horizon attractor region, where we can use the standard AdS$_3$/CFT$_2$ correspondence. 

The   moduli space of the near-horizon AdS$_3 \times S^3$ region of the black string, parametrized by the  $\bar g_{i\L}$, is locally an $SO(4,21)/SO(4) \times SO(21)$ coset and is  identified with the moduli space of the D1-D5 CFT, i.e. its  set of exactly marginal deformations. % of this CFT. %[, which . ] 
It is important to note that, even though the $SO(4) \times SO(21)$ subgroup of $SO(4,21)$ is gauged from the perspective of the low-energy supergravity coset action, % \emph{Does this fact change in gauged sugra?} 
the CFT operators do carry $SO(4) \times SO(21)$ labels. These are not symmetry indices, but rather indicate which fields can mix with each other under flow along the moduli space, as dictated by the reduced  holonomy of the connection on such a homogenous space   %symmetric manifold 
\cite{deBoer:2008ss}. 

%As discussed at length in \cite{}, the $\N = (4,4)$ CFT operators will in general mix under flow on the moduli  space. While the general holonomy of a general $4n$ - dimensional conformal manifold is $SO(4n)$, the special symmetric coset structure of the manifold we have makes the holonomy reduce to $SO(4) \times SO(21)$. Consequently, CFT operators carry a label corresponding to representations of these groups, even though they do not correspond to a symmetry of the theory. From the supergravity perspective, this label matches that of the $SO(4)$ subgroup of R-symmetry and the $SO(21)$ rotations of the tensor multiplets, even though these do not correspond to symmetries of the theory (which is, as discussed, $SO(5,21)$).     \emph{Move up?}

%The result of the above analysis  is that the most general half-BPS black string solution in type IIB supergravity compactified on K3 has $84+22 = 106 $ independent parameters that, in our notation and in an appropriate gauge, are given by the  constants $\bar g^{i}_\L$ and $c_\L$,  subject to the restrictions \eqref{constrcg}, \eqref{gperpzh} and $\bar g^{i}_\L \bar g^{j\L} = \d^{ij} - \hat Z^i \hat Z^j$.  

%The other $22$ independent constants   parametrize the remaining $21$ moduli in a `projective'  fashion, in the sense that the physical scalar fields do not depend on the overall scale of these coefficients, at least as long as $c_\L c^\L \neq 0$. 

%As extensively discussed in \cite{deBoer:2008ss}, 

The $22$ independent constants in the harmonic functions, taken infinitesimal, precisely correspond to the set of scalar single-trace maximal supersymmetry - preserving irrelevant  deformations of the D1-D5 CFT, which have dimension $(2,2)$ in the supergravity limit. In the linearised analysis of \cite{Deger:1998nm}, these operators correspond to perturbations of the three-form gauge fields\footnote{The reason  we dropped the `$i$' index on $F^{+i}$ is that the self-dual field strengths are by definition proportional to $Z^i$, so there is no such  perturbation along the orthogonal directions in $SO(5)$. } $F^{+}$ and $F^{-\bar \imath}$ that carry no spin on $S^3$; these perturbations are linked, via the linearised equations of motion, to those of the trace of the AdS$_3$ (and $S^3$)  metric perturbation and, respectively, those of the matter central charge\footnote{This agrees with the linearised fields \cite{Deger:1998nm} identify with these deformations, for their specific choice of conventions (aligning the charge with a particular direction and identifying   the $i, \bar \imath$ and $\L$ indices).% are identified, but this looks like a reasonable proposal that has the correct labels and matches their proposal for their spacific choices.
 } $Z_{\bar \imath}$, thus making the scalar nature of the corresponding operators manifest. The first operator, denoted $\O^+$, is an $SO(21)$ singlet, while the remaining ones, $\O^{\bar \imath}$, transform as a   vector under $SO(21)$.  %[This can be easily checked by matching the leading perturbation around AdS$_3$   of our solutions to the tables of \cite{Deger:1998nm} ; $X^1_I$, all of which correspond to fields of $m^2=8$, or $\Delta=4$; All of these fields belong to the $\ell=0$ sector and they correspond to the scalars (whose derivative is $X_\mu^I$) dual to the $5$ and $r$ components of the $B$- field turned on along the AdS$_3$ directions; Looking at the tables in \cite{sezgin}, we find one dimension $(2,2)$ operator that is an $SO(4) \times SO(21)$ singlet, as well an $SO(21)$ multiplet, for a total of 22.]
The change in the D1-D5 CFT action that corresponds to turning on these deformations % $c_\L$
 infinitesimally takes the form %\emph{Notation!}

\be \label{deformedaction}
S=S_{CFT} +  \l_{+} \O^{+}_{(2,2)} + \l_{\bar \imath} \, \O^{\bar \imath}_{(2,2)}
\ee
The coefficient $\l_+$ is given by the leading departure of the black string metric from the near-horizon AdS$_3 \times S^3$ and can be read off from its explicit form 

\be
ds^2 = \frac{1}{\sqrt{H_\L H^\L}} (-dt^2+ d\s^2) + \sqrt{ H_\L H^\L} (dr^2 + r^2 d \Omega_3^2) \label{6dmetsol}
\ee
while $\l_{\bar \imath}$ is extracted from the leading non-zero term\footnote{For this, it is useful to note that $ e^{-2U} X^{||}_\L X^\L_{\bar \imath}= \frac{\a'}{r^2} Z_{\bar \imath} + c_\L X^\L_{\bar \imath} =0$, where we used \eqref{eq: coset conditions 2} and \eqref{solXparall}.} in $(-) Z_{\bar \imath}$.    
%where the coefficients of the operators are related to the parameters of the solutions via %\emph{Can we do better than $\propto$?}
%{\color{ForestGreen}(this fixes the relative normalization, but we can still have an overall normalization)}
We obtain  %\emph{\textcolor{red}{Normalisation??}}{\color{ForestGreen}(sign?)}
\be
\l_{+} \sim \a' \frac{ c_\L q^\L}{\sqrt{q_\L q^\L}}  \;, \;\;\;\;\; \l_{\bar \imath} \sim \a'  X_{\bar \imath \L} c^\L
\label{irrcouplgen}
\ee
where the precise numerical coefficient depends on the normalisation of the operators, which we have not worked out in detail. %{\color{ForestGreen}(since all our solutions go into $AdS_3\times S^3$ in the near horizon, can't we fix the overall normalization from here?)}
% \;, \;\;\;\;\; H_\L = \frac{q_\L \a'}{r^2} + c_\L
%where the first can be read off from the metric, while
% The second set corresponds, as explained, to the leading contribution to $ Z_{\bar \imath}$ away from the horizon. %, which directly affects the three-form fields $F^{-\bar \imath}$.  \emph{How does this fit with the fact that the $SO(21)$ is gauged in the coset?} 
% [We note that the condition $c_\L \bar g^\L{}_{i_\perp} =0$ automatically sets the coefficients of the self-dual deformations that are not parallel to the background to zero.\emph{???}]
%
Note that, while there is no obvious restriction on the deformation parameters $\l_+, \l_{\bar \imath}$ from the point of view  of linearised conformal perturbation theory, well-behavedness of the metric in the supergravity solution \eqref{6dmetsol} imposes the non-linear constraint%\footnote{We will not be considering the possibility of ``negative strings'' \cite{}, even though they may behave better than naively expected.} %\emph{Are $q^2 >0$ and $q \cdot c >0$ already required by susy?}
\be
c_\L c^\L \geq 0
\ee
With the normalisation above, and using \eqref{centralchargesmod}, this immediately translates into $\l_+^2 - \l_{\bar \imath} \l_{\bar \imath} \geq 0$. 
It would be very interesting, of course, if this constraint could be seen to emerge from second order conformal perturbation theory, but it is not clear this should be the case.

Let us now review a few more field-theoretical details about these operators. 
The $(2,2)$ operators above  are supersymmetric descendants of the $22$ $(1,1)$ chiral primaries in the D1-D5 CFT, which we  denote as $\mathcal{X}_{(1,1)}^{+}$ and $\mathcal{X}_{(1,1)}^{ \bar \imath}$ %\emph{Notation!} 

\be \label{22operator}
\O_{(2,2)}^{+/\bar \imath} = G^{-+}_{-\frac{1}{2}} G^{--}_{-\frac{1}{2}} \tilde G^{-+}_{-\frac{1}{2}} \tilde G^{--}_{-\frac{1}{2}} \mathcal{X}_{(1,1)}^{+/\bar \imath}
\ee
where the first label on the supersymmetry generators is the R-symmetry index, while the second  corresponds to the $SO(4)_{outer} \sim SU(2) \times SU(2)$ remnant of the six-dimensional R-symmetry in the near-horizon region that we have been discussing.  The chiral primaries are annihilated by $G_{-1/2}^{+\pm}$ and $\tilde G_{-1/2}^{+\pm}$, in addition to all the positive modes of the supercurrents. Given the above structure, the resulting  operator has left/right conformal dimensions $(2,2)$, zero spin on the $S^3$ and is an $SO(4)_{outer}$ singlet.  
%From the point of view of the $SO(21)$ outer automorphism of the sugra solutions, the perturbation of the $S^3$ size is a singlet, while the rest form a vector under $SO(21)$. 
 Interestingly, exactly the same chiral primary operators control the $(1,2)$ deformations of the D1-D5 CFT of dipole and Kerr/CFT type \cite{El-Showk:2011euy}.

As already noted, $\O_{(2,2)}^+$ transforms as a singlet under the  $SO(21)$ holonomy, while the rest transform as a vector. As explained in \cite{deBoer:2008ss}, only the singlet operator can be unambiguously identified across the moduli space of the D1-D5 CFT, while the remaining operators will generically mix among themselves along flow in the moduli space, in a way that in principle depends  on the path taken. The explicit map between the $SO(21)$ singlet $(1,1)$ chiral primary operator and the corresponding operator in the D1-D5 orbifold CFT can be found in \cite{El-Showk:2011euy} and involves both untwisted and twisted sector contributions. 
%[The chiral primaries can be identified, using supersymmetry, to those in the D1-D5 CFT, \emph{Copy from paper with Sheer!} where the first  is the singlet under $SO(21)$, and the remaining ones the vectors. Note the latter can mix under flow in the moduli space, and there is no unambigous identification for them. ]  

The action \eqref{deformedaction}  corresponds to the most general  maximally supersymmetric  single-trace irrelevant deformation of the D1-D5 CFT that is visible within supergravity.  It is interesting to ask whether there could be other operators consistent with the symmetries, at least at this level.  It is easy to show that in the supergravity limit, there is no other higher-dimensional scalar operator of R-symmetry spin zero, as descendants of chiral primaries (be them multitrace)  must have $|h-j| \leq 2$.  There are, however, double-trace operators of the same dimension, $(2,2)$,  that can and do mix with them \cite{Taylor:2007hs}, which are supersymmetric descendants of the double-trace $(1,1)$ chiral primaries that appear in the $SO(21)$ singlet and vector  representations one obtains from the tensor product of two $SO(21)$ $(1/2,1/2)$ chiral primaries. %In fact, one should expect mixing with such operators, as far as the identification of the CFT operators is concerned (Marika). 
%
%In the supergravity limit, these deformations must start as $(2,2)$. However, as is standard for such deformations \cite{}, in the supergravity limit the deforming operator is exact,  
Other than these,  only operators in long representations of the supersymmetry algebra may in principle appear, but those correspond to string modes. Thus, at the supergravity level, the deformation \eqref{22operator} may turn out to be (almost) exact, similarly to the  irrelevant deformation of $\N=4$ SYM studied in \cite{Intriligator:1999ai,Caetano:2020ofu} and the $(1,2)$ deformations of \cite{El-Showk:2011euy,Kraus:2011pf}. We will therefore be labeling it by the corresponding coefficient in \eqref{deformedaction}. Even if there are additional corrections to the operator, it appears safe to assume they will all depend on the same coefficient $\l_+$ or $\l_{\bar \imath}$ (i.e., the irrelevant coefficients are highly fine-tuned) and thus we can still use the corresponding label.

% This is very similar to other such cases.   It is however not clear how useful the sugra exactness of the operator  is in computations (see \cite{} for previous attempts in $d=4$). 
 
%[The only other maximally supersymmetric single-trace operators have dimension $(1,1)$, transform in the $(4,n)$ of $SO(4)\times SO(n)$, and correspond to the moduli of the theory. \emph{Holonomy discussion?}] 

It is an interesting exercise to  isolate each deformation in part and understand whether it may lead to a tractable theory. For this, it is useful to frame the discussion inside string theory. Assuming that the background AdS$_3$ is supported by purely NS-NS (self-dual) flux, the operators \eqref{22operator} may be labelled by the type of three-form field strength that they turn on at infinitesimal level around the near-horizon AdS$_3 \times S^3$

\be
\O^+_{(2,2)}\;\; \leftrightarrow \;\; \O_{H_3^+} \;, \;\;\;\;\;\;\;\; \O^{\bar \imath}_{(2,2)}\;\; \leftrightarrow \;\; \O_{H_3^-}\;,  \O_{F_3^-}^S  \label{sugraidop}
\ee  
where the index $S \in \{1 , \ldots 20\}$ runs over  the anti-self-dual piece of the RR three-form flux, $F_3^-$,  and over the $19$ anti-self-dual three-form fields, $F_3^{-\hat S}$ obtained by compactifying the self-dual type IIB five-form on the $19$ anti-self-dual cycles of the K3. The $F_3^{-S}$ transform under each other under the $SO(4,20,\mathbb{Z})$ T-duality group.
The $H_3$ and $F_3$ labels on the above operators are interchanged, as expected, if we perturb instead around the pure RR background. 
 
Let us now discuss the supergravity description of the various deformations. The self-dual deformation will correspond to the non-linear solution  with purely self-dual $H_3$-flux  throughout. Since, by the self-dual choice, $Z_{\bar \imath} =0$, one immediately notes from \eqref{centralchargesmod} that this solution is necessarily asymptotically flat, as $c^2 = (q\cdot c)/q^2 >0$. In string theory language,  it is trivial to see that this noting but the standard NS5-F1 solution  \eqref{nsextrmet} with the dilaton at infinity already set to its attractor value, so it does not flow.% [ It would be very interesting if the $SO(21)$ singlet condition it satisfies could be enough to follow it in conformal perturbation theory on the field-theory side (note the deformation involves a twist-three operator). \emph{Move up?}] We further study the properties of this solution in section \ref{}. 

%\emph{Explain better: }

According to \eqref{sugraidop}, the remaining $21$ deformations will correspond to turning on anti-self-dual $H_3^-$ or anti-self-dual RR 3-form flux $F_3^{-S}$. The  background associated to the $H_3^-$ deformation simply corresponds to the general NS5-F1 solution. The  $F_3^-$ deformation can be constructed by applying an STsTS  solution generating transformation to the pure NS background, where the TsT acts along the black string directions. It is easy to see that in the S-dual RR frame, the TsT  generates a $B$ - field  along the common D1-D5 directions, which  then turns into an RR two-form potential under the final S-duality.  The remaining backgrounds with a particular $F_3^{-\hat S}$ turned on can be simply obtained from the $F_3^{-}$ one by acting with two T-dualities along the compact manifold. The properties of the resulting theories are thus closely related
%\footnote{\textcolor{blue}{For the `isolated' deformations only.} } 
to those of the $F_3^-$ - deformed one, which justifies concentrating most of our attention on the latter.
%
%All the solutions can be modelled by just turning on $H_3$ and/or $F_3$ on top of the AdS solution supported by pure NS flux, and then acting with T-dualities on specific cycles of K3 to turn the $F_3$ into $F_5$ - the anti-self-duality will be guaranteed by the fact that solution generating techniques such as TsT automatrically preserve the equations of motion. The general solution has both self-dual and anti-self-dual $H_3$. The solution with anti-self-dual $F_3$ can be obtained by   performing TsT on D1-D5, and then S-dualising. 
This procedure for generating the solutions  is  directly analogous   to that used in \cite{Bena:2012wc} to generate all the backgrounds corresponding to the $(1,2)$ irrelevant deformations of the D1-D5 system analysed in \cite{El-Showk:2011euy}. 

Note that, while the $\l_+$ deformation can be turned on independently of the others, the same is not true of the $\l_{\bar \imath}$ ones, as 
%Note, however,  that all these solutions with $\l_{\bar \imath}\neq 0$ must also have some self-dual $H_3$ flux turned on (in addition to the background), as
 reality of the metric \eqref{6dmetsol} imposes $c_\L c^\L = \l_+^2-\l_{\bar \imath} \l_{\bar \imath}\geq 0$, which corresponds to a lower bound on $\l_+$ when the $\l_{\bar \imath}\neq 0$. %\textcolor{blue}{The existence of a non-zero  lower bound on $\l_+$ can be easily extracted from the constraints \eqref{constrcg} that the $c_\L$ obey and the fact that $\l_+ \sim c_\L q^\L$.} 
 %
%[ Let us use the $SO(5,21)$ symmetry to rotate the charge into the vector $q_\L = q \d^1_\L$, which agrees with the positive norm. We can also use the independent $SO(5)$ symmetry to rotate the vector $\hat Z^i = \d^i_1$. Then, the conditions \eqref{} on $\bar g_{i\L}$ read: $\bar g_{1\L} = \bar g_{i 1} =0 $. This basically deletes the first row and column from $\bar g_{i\L}$, leaving an $SO(4,21)/SO(4) \times SO(21)$ matrix $\bar g_{i',\L'}$, where $i', \L'$ run over $4$ and respectively $25$ values. 
 %
%This can be easily seen if the index $i_\perp$ runs over a single value, $c_\perp = c_1$. Then  \emph{Notation!!}
%
%\be
%c_\L c^\L = c_+^2 + c_1^2 - \sum_I c_I^2 \;, \;\;\;\; c_1 = \frac{g_{I} c_I}{g_{11}} \;, \;\;\;\; g_{11} = \sqrt{1+ g_I^2} \;, \;\;\; g_I = g_{1I}
%\ee
%where we have already used the $q_\L g^{i \L} =0$ constraint to set to zero one value of $\L$. Therefore
%
%\be
%c_1^2 - c_I^2 = \frac{(g_I c_I)^2}{1+g_I^2} - c_I^2 \leq \frac{g_I^2 c_I^2}{1+g_I^2} - c_I^2 = - \frac{c_I^2}{1+g_I^2} < 0
%\ee
%\emph{We basically need a matrix generalisation of this argument}. We conclude that $c_\L c^\L >0$, as required by the supergravity solution, necessitates $c_+ >0$, and bounded below by the $c_I$. \emph{In which way?} [Write down infinitesimal def action with normalised coefficients. Note that $c_\L c^\L$ is probably a sugra artifact that we cannot currently see in field theory.
%
%Unlike in the $(1,2)$ case, \emph{Check!} it is not possible to turn on the ``pure'' anti-self-dual (or $SO(21)$ vector) deformations, as some $c_+$ must always be turned on. ]
%
It is interesting to ask what happens when we turn on the ``minimum'' allowed amount of $\l_+$, which happens for $c_\L c^\L=0$. We immediately note from \eqref{6dmetsol} that for $c_\L c^\L =0$ with $c_\L q^\L$ fixed, the asympotics of the solution degenerate. In the pure NS5-F1 system, which is our guiding example,  this degeneration (which corresponds to `dropping the one in the NS5 harmonic function' - a procedure that we will explain more carefully in the next section)  yields an asymptotically linear dilaton (ALD) spacetime, which is well-known to correspond to the NS5 decoupling limit. The dual theory in this case is little string theory \cite{Seiberg:1997zk}, a strongly-coupled six-dimensional string theory that is decoupled from gravity.

 The main goal of the next section is to show that  the more general backgrounds \eqref{6dmetsol} with $c_\L c^\L =0$, in which the metric  degenerates to what we will be calling an ``ALD-like'' space-time,  also correspond to known decoupling limits of string theory. Thus, our claim is 

%\bigskip
%
%\parbox{0.89\linewidth}{\emph{All backgrounds with $c_\L c^\L =0$ \;\;\;\;  $\leftrightarrow$ \;\;\;\; Known decoupling limits of string theory}}
%
%\bigskip

\vskip4mm
\begin{center}
\emph{All backgrounds with $c_\L c^\L =0$ \;\;\;\;  $\leftrightarrow$ \;\;\;\; Known decoupling limits of string theory}
\end{center}
%
%\vskip1mm
%
 %As we will try to argue in the next section, the answer is yes. 
%
\noindent Before ending this section, we would like to translate the $c_\L c^\L =0$ degeneration condition on the asymptotic metric into a condition on the asymptotic moduli of the theory, which \emph{a priori} depend on ratios of the coefficients $c_\L$. 
%
%let us note that, while the condition $c_\L c^\L =0$ looks like  a codimension one subspace of the moduli space at infinity, this statement is not very precise because the moduli only depend on ratios of the $c_\L$.
For this, we note that the $SO(5)$ - invariant combination $|Z_+| $ given in \eqref{centralchargesmod}, which can be expressed unambiguously in terms of the values of the moduli, also degenerates in this limit - more precisely, it becomes infinite. The matter central charge $|Z_-|^2=|Z_+|^2-q^2$ also becomes infinite, as does the hyperbolic angle defined in \cite{mikhailov}. %and this  (using $2 |Z|^2 = q^\L \N_{\L\Sigma} q^\Sigma + q^2$), 
This degeneration corresponds on a single restriction on the asymptotic moduli,   and thus the theories dual to the set of all ALD-like backgrounds live on a codimension one  subspace of the moduli space at infinity. %, which does appear to enjoy $SO(5,21)$ ``symmetry''.  
%
%.  Specifically in type IIB, it is xxx. \emph{Compute this!} This seems like a necessary, but not sufficient, condition on having a decoupling limit.
%
 %[It would be very interesting if the holonomy became $SO(5) \times SO(21) $ upon including the irrelevant deformations, even in the decoupling limits, which correspond to an $SO(5,21)$ invariant condition. Note then that we would mix max-susy $\O_{(2,2)}$ and $\a'^{-1} \O_{(1,1)}$ operators. \emph{What could this possibly mean?}]

\section{A family of decoupled backgrounds}
\label{section3}

Having worked out the general supergravity solutions, we found that the asymptotics change drastically when $c_\L c^\L =0$. In this section,  we would like to show that %in the specific example of type IIB/K3, 
this condition precisely corresponds to various decoupling limits of five-brane theories, implying that %. More precisely,  we propose that in the limit $c_\L c^\L \to 0$ 
the supergravity solutions obtained above are holographically dual to a family of six-dimensional, non-gravitational string theories compactified on K3. 

We start by reviewing the known decoupling limits associated with NS5- and/or D$p$-branes, possibly in the presence of various background fields. 
We then move to the gravitational description of these decoupled theories at strong coupling and show that, when written in the six-dimensional supergravity language   employed above, the $c_\L c^\L \to 0$ condition lands on the relevant limit for all cases under consideration. As anticipated above, it will be both useful and sufficient to work in the $SO(2,2)$ consistent truncation of \cite{Duff:1998cr}, for which several details are provided in appendix \ref{sec: Appendix B - consistent truncation}.

%how to go from the standard Lagrangian to the $F^\L$ description  

\subsection{Degeneration and decoupled six-dimensional string theories}
\label{section31rev}
One usually thinks of superstrings as describing ten-dimensional theories of extended objects, which include gravitational interactions. However, there exist a number of non-gravitational string theories in lower dimensions. They are obtained in certain parameter regimes, which we collectively refer to as \textit{decoupling limits}, and live on  different types of $p+1$-dimensional branes, with $p \leq 5$. These are string theories in the sense that they include stringy towers of states and stringy dynamics, as well as other non-local properties such as T-duality. Moreover, their entropy is of the Hagedorn type \cite{Atick:1988si} at high energies. We now briefly review how these models arise from the brane perspective. 

\bigskip

\noindent $\diamond$ \emph{Little string theory}

\vskip2mm

\noindent The first example, known as Little String Theory \cite{Seiberg:1997zk} (see also {\cite{Losev:1997hx,Berkooz:1997cq,Dijkgraaf:1996hk,Dijkgraaf:1996cv}}), corresponds to the effective description of the dynamics on a stack of $k$ NS5-branes in the limit of small string coupling $\tilde g_s$ and fixed  string tension, set by\footnote{Whenever a  confusion is possible, we will be denoting the NS frame parameters with a tilde.} $\tilde \alpha'$. While the  bulk modes decouple as we take $\tilde g_s\to 0$, the theory on the five-branes remains interacting. 

In the type IIB picture this is most easily argued by going to the S-dual frame. The low-energy fluctuations of the resulting D5-branes are characterized by a six-dimensional gauge theory with coupling  $g_{\rm D5}^2= g_s \ap$ (we will be ignoring numerical factors). Under  S-duality we have $g_s \to \tilde g_s= g_s^{-1}$ and $\ap \to  \tilde \a' =g_s \alpha'$, implying that $g_{\rm NS5}^2 = \tilde \alpha'$, hence the NS5 coupling remains finite in the above decoupling limit. For large $k$, the perturbative gauge theory description is valid up to energies of the scale set by the inverse `t Hooft coupling, $E \sim (k \tilde\alpha')^{-1/2}$, where additional UV degrees of freedom come into play. The energy estimate suggests that these are  non-perturbative string-type excitations corresponding to instantons in the six-dimensional  gauge theory, which are known as (closed) \textit{little strings}. The IIB NS5 worldvolume theory is then a non-critical   six-dimensional string theory with (1,1) supersymmetry \cite{Seiberg:1997zk}. Its BPS spectrum does not, however, contain a spin-2 massless state. 

The low-energy analysis is slightly more involved in the type IIA case. Here 
the supersymmetry is  (2,0) and there is a non-trivial IR fixed point, namely  the well-known (2,0) SCFT. The stringy degrees of freedom then correspond to (the endpoints of) D2-branes stretched between the five-branes. Given that it commutes with the $g_s \to 0$ limit,  
LST inherits the T-duality of the underlying critical strings. It follows that type IIA and type IIB LST are related upon compactification. For more details, see \cite{Kutasov:2001uf,Aharony:1999ks}.

\bigskip

\noindent $\diamond$ \emph{Non-commutative open strings}

\vskip2mm

\noindent Another type of decoupled string theories are non-commutative open string theories (NCOS), first studied in \cite{Seiberg:2000ms,Gopakumar:2000na}. They are  obtained by considering a stack of  D$p$-branes in a background configuration with $g_{\mu \nu} = g \eta_{\mu\nu}$ and $g_{i j}=g_\perp \delta_{ij}$ where $\mu,\nu=0,1$ and $i,j=2,\dots,9$, and with a constant electric $B$-field, $B_{\mu\nu} = E\, \vep_{\mu \nu}$. The constant $B$ - field can be traded for an electric field along the D$p$-brane worldvolume, and its effect is to lower the effective tension of a string oriented along the $E$-field to %\emph{Check! Factors!}
\be
\frac{1}{4 \pi \a'_{eff}} = \frac{g}{2\pi \a'} - E \label{alpeff}
\ee
Unlike for the magnetic case, there is a maximum  value of the electric field, $E_{\rm crit} = g /(2\pi \alpha')$, beyond which the system becomes unstable. The relation between open string quantities and closed string ones parallels that in the magnetic case \cite{Seiberg:1999vs}. In the electric case one obtains %gets \emph{Check power on $G_s$}
\begin{equation}
\label{openstring quantities Seiberg}
    G = g (1-\tilde{E}^2) \, , \quad \;\;
    G_\perp = g_\perp \, , \quad  \;\;
    \theta = \frac{\tilde{E}}{E_{\rm cr}(1-\tilde{E}^2)} \, , \quad \;\;
   G_s =  g_s \sqrt{1-\tilde{E}^2} \, 
\end{equation}
where  $\tilde{E} \equiv E/E_{\rm crit} \leq 1$. 
%\textcolor{blue}{ and, if $N$ D$p$-branes are present, then the effective open string coupling is $N G_s$}.
 These quantities %in \eqref{openstring quantities Seiberg}
 enter the open string propagator along the boundary of the disc (with parameter $\tau$), namely 
\begin{equation}
    \langle x^{M}(\tau)x^{N}(\tau)\rangle = - \ap G^{MN} \log \tau^2 + \frac{i}{2} \theta^{MN} \mbox{sgn} (\tau) \, 
\end{equation}
with $M=0,\dots,9$, and where $G_{\mu \nu} = G \eta_{\mu \nu}$, $G_{ij} = G_\perp \delta_{ij}$, while for the space-time non-commutative parameter  we have $\theta^{\mu \nu} = \theta \, \vep^{\mu \nu}$, $\theta^{ij} = 0$. 
% This can be estimated by computing the effective tension of an open string stretched in the preferred direction, which is given by
% \be
% \frac{g}{2\pi \a'} - E = \frac{1}{4\pi \a'_{eff}}
% \ee
Rewriting the third relation in  \eqref{openstring quantities Seiberg}  as 
\be
\a' G^{-1} = \frac{\th}{2\pi \tilde E} \geq \frac{\theta}{2\pi} \label{eqapGth}
\ee
shows that, unlike in the magnetic case, it is not possible to take the limit $\a' \r 0$ with $G$ fixed if one would like the non-commutativity parameter $\theta$ to remain finite. There still exists a scaling limit where the bulk string modes decouple, although - importantly - the brane dynamics remain both stringy and space-time non-commutative, describing the so-called non-commutative open string (NCOS) theory. Unlike in field theory, the presence of spacetime non-commutativity does not lead to causality violations \cite{Seiberg:2000gc}.

 This decoupling limit corresponds to tuning the electric field to its critical value,  $\tilde E \to  1 $. Note,  from \eqref{eqapGth},  that the non-commutativity and string scales are equal. The open strings remain dynamical and interacting as long as we scale the closed string quantities appropriately at the same time. 
 More precisely, if we define $\vep = 1- \tilde{E}^2$ and take 

\be
g_s \sim \frac{1}{\sqrt{\vep}} \;, \;\;\;\; g \sim \frac{G}{\vep} \;, \;\;\;\; g_\perp\sim \vep^a \, g\;,\;\;\;\mbox{with} \;\; \a' G^{-1} \;\;\mbox{fixed}
\ee 
with $a>0$ (usually, $a=1$), then the open strings decouple from closed ones. Indeed,  the dispersion relation for a putative closed string excitation that could be emitted/absorbed by the branes into/from the bulk reads   %\emph{\textbf{Check! } $g_\perp^{-1}$ diverges. }
\begin{equation}
     g^{-1} p_0^2  = g^{-1} p_1^2 + g_\perp^{-1}p_i^2 + m^2
\end{equation}
In the above  limit, this cannot be satisfied for states with finite energy $p_0$, showing that bulk strings effectively decouple. %\new{A more precise computation  \cite{Gopakumar:2000na} shows  this conclusion holds for $p\leq 5$. }

 The relative scaling of $G$ with $\vep$ is not fixed  - indeed, various scalings %\cite{Gopakumar:2000na, Gopakumar:2000ep, Harmark:2000wv, Harmark:2000ff,Berman:2000jw, Larsson:2001ug}
  exist in the literature. %\footnote{They correspond to $\a' \sim \sqrt{\vep}$ or \emph{What??}.}),
 This in turn implies that the scaling of $\a'$ with $\vep$ can differ. A choice that is standard in the supergravity literature \cite{Maldacena:1999mh,Gopakumar:2000na,Harmark:2000wv} is 
 
\begin{equation}
\label{eq: NCOS limit 1}
    \ap \sim G \sim G_\perp \sim \sqrt{\vep} \,  \quad  \Rightarrow \;\;\;
      g \sim 1/\sqrt{\vep} \;, \;\;\; g_\perp \sim \sqrt{\vep} 
\end{equation}
with all other quantities fixed (including the coordinates $x^M$). %[This ensures that all components in $\ap G^{MN}$ and $\theta^{MN}$ remain fixed, and so does $G_s$. ]
The theory still contains fundamental open strings, which remain light due to the presence of the near-critical background electric field. Their effective tension is given by $T_{\rm NCOS \, F1} \sim (\a'_{eff})^{-1}$, defined in \eqref{alpeff}.  If we fix $G_{MN} = \sqrt{\vep} \, \eta_{MN}$, then $\a'_{eff}$ is related to the original $\alpha'$ by  $\ap = \a'_{eff} \sqrt{\vep}$. %Note this  coincides with the effective non-commutativity scale, since \eqref{openstring quantities Seiberg} implies  \textbf{\emph{Check!}}
%\begin{equation}
%    \theta =  
%    \frac{2\pi \ap \tilde{E}}{g(1-\tilde{E}^2)} \to 2\pi \apt \, .
%\end{equation}
%
Similarly, one can identify $\sqrt{\vep}\, g_s$ with the the (squared) open coupling, $G_s$. 

For the purpose of comparing with the  supergravity analysis, to be performed below, we briefly discuss the scaling of the transverse metric $g_\perp$. In the  AdS/CFT limit it is useful to think of a coordinate $r$ parametrizing the separation of the branes. As we bring the branes together, we usually keep the mass $m = r/\ap$ of the fundamental strings stretched between them fixed at low energies \cite{Maldacena:1997re}, leading to the rescaling $r = \ap \tilde{r}$. In the NCOS case, one keeps the open strings dynamical by fixing their mass fixed  \textit{in string units}, hence the radial scaling should be $r = \sqrt{\ap} \tilde{r}$, which indeed results in $g_\perp \sim \ap$ as we take $\ap \to 0$ \cite{Maldacena:1999mh,Berman:2000jw,Harmark:2000wv}. %[Moreover, while in the bulk we can gauge away the constant $B$-field, in the presence of the D-branes the worldvolume field which is actually gauge invariant is $B + 2\pi \ap {\cal{F}}$, hence this induces an constant ${\cal{F}}_{01}$. The dual solutions can then be thought of as F1-D$p$ bound states.]

%%[Let us stress that the NCOS limit is not uniquely defined, in the sense that one can reach the same theory by using alternative scalings to the one presented in Eq.~\eqref{eq: NCOS limit 1}. More explicitly, one can scale $\ap$ and the closed string metric elements $g$ and $g_\perp$ in different ways in terms of our scaling parameter $\vep$ as long as $\ap G^{MN}$ and $\theta^{MN}$ remain fixed, and so does the relative scaling of $g/g_\perp$. As a result, various equivalent definitions for the NCOS theories can be found in the literature \cite{Gopakumar:2000na, Gopakumar:2000ep, Harmark:2000wv, Harmark:2000ff,Berman:2000jw, Larsson:2001ug}. For instance, instead of \eqref{eq: NCOS limit 1} one could use 
%\begin{equation}
%\label{eq: NCOS limit 2}
%    \ap \sim g_\perp \sim \vep \, , \quad 
%    g_s \sim 1/\sqrt{\vep}  \, , \quad g \sim 1\,.
%\end{equation}
%This particular scaling will prove useful later on. ]

\bigskip

\noindent $\diamond$ \emph{$6d$ NCOS and the OD1 theory}

\vskip2mm

\noindent As it turns out, six-dimensional NCOS (corresponding to D5-branes in a critical electric field) contains, in addition to the light  open strings, closed D1-brane excitations, which represent the string-type instantons of the underlying low-energy Yang-Mills description. Their tension is given by $T_{\rm NCOS}^{ D1} = (g_s \a')^{-1}$ (up to an  factor of $k$) where $g_s \a' = G_s \a'_{eff}$.  Thus, in the standard weakly-coupled NCOS description, where $G_s$ is small, these branes are very heavy. 

In order to interpret this, we go back to the bulk point of view. The bulk ten-dimensional dilaton diverges in the  NCOS limit \eqref{eq: NCOS limit 1}, implying  that a reliable description can only be obtained through S-duality. In the dual NS5 frame the string length is fixed, while the dilaton goes to zero. This is reminiscent of the LST limit considered above, hence we interpret the NCOS closed (D1) string modes as the  excitations S-dual to the LST little strings. However,  we now have a critical RR 2-form potential in the background. Taking into account the scaling of the metric elements, this indicates that the resulting (type IIB) NS5 theory also contains \textit{open} D1-branes (with tension $\a'_{eff}$), which remain light. Of course, these are nothing but the NS5 version of the open strings of the S-dual NCOS. This theory is known as the OD1 theory \cite{Gopakumar:2000ep,Harmark:2000wv,Harmark:2000ff}, an abbreviation for open D1-brane. 

In this duality frame, we can write the relevant parameters in terms of the $6d$ NCOS ones introduced above as the coupling $\tilde G_s= G_s^{-1}$ and the effective string scale $\apt= g_s \a' = G_s \a'_{eff}$, which is the LST scale.  $\a'_{eff} = \tilde G_s \tilde \a'$ is the `spacetime noncommutativity' scale. If we  take $\a'_{eff} \to 0$ with $\apt $ fixed in addition to the OD1 limit, %[defined by S-duality from the $6d$ NCOS limit in Eq.~\eqref{eq: NCOS limit 1},] 
the open D1-branes become very heavy and decouple, while the `non-commutativity' becomes negligible, so that we are left simply with  type IIB LST. 
%It was further conjectured in \cite{Gopakumar:2000ep} that the $6d$ NCOS/OD1 theory  inherits the self-duality property of critical type IIB string theory. %\emph{Also Shatashvili?}

\bigskip

\noindent $\diamond$ \emph{OM theory, other NCOS and OD$p$ theories}

\vskip2mm

\noindent The strong coupling limits of the remaining $p+1$-dimensional NCOS theories with $p<5$ can then be studied by going to the relevant bulk duality frame, which depends on the dimensionality (even or odd) of the D$p$-brane under study. For instance, for $p=3$ we are still in type IIB;  the authors of \cite{Gopakumar:2000na} used the self-duality of D3-branes under type IIB S-duality, which acts as  electric-magnetic duality on the branes,  to argue that the strong coupling limit of  $4d$ NCOS is  four-dimensional  non-commutative $\N=4$ SYM. This relation will be particularly useful when discussing the decoupling limit of the D1-D5 NCOS.

On the other hand, for type IIA cases such as $p=4$, one ends up in M-theory. The strongly coupled  description then corresponds to the so-called OM theory \cite{Gopakumar:2000ep,Bergshoeff:2000ai}, describing the  decoupled M5-brane dynamics in the presence of a critical 3-form potential. 
This M2-M5 bound state reduces to the F1-D4 one associated to $5d$ NCOS upon compactification along an electric circle. Compactifying on a magnetic circle instead leads to a D2-NS5 configuration. This theory, which contains closed little strings as well as light open D2-branes, is known as OD2, and is related to OD1 simply by T-duality. At small effective coupling it reduces to the type IIA LST discussed above.   

One can define the rest of the OD$p$ theories by means of T-duality. Note that these are always six-dimensional. The corresponding decoupling limits   can be obtained by starting with the  $6d$ NCOS limit%in %\eqref{eq: NCOS limit 2}
, performing an S-duality (together with a rescaling of the coordinates to absorb the resulting $g_s$ factor in the metric), followed by  $p-1$ T-dualities along directions parallel to the NS5-branes, but orthogonal to the critical background potential. This leads, in particular, to a scaling of $g_s \sim \vep^{\frac{3-p}{4}}$ \cite{Gopakumar:2000ep}.
%
%\begin{equation}
%\label{eq: NCOS limit 2}
%    \ap \sim \sqrt{g_\perp} \sim \sqrt{\vep}  \, , \quad  
%    g_s \sim \vep^{\frac{3-p}{4}}
%    \,, \quad g \sim 1\,.
%\end{equation}
Of particular interest is the OD3 theory, where $g_s$ is kept finite. It has been proposed \cite{Gopakumar:2000ep} that this is related to $6d$ spatially non-commutative SYM by S-duality, another relation that will be very useful in our supergravity decoupling analysis in section \ref{od3section}. %\new{Is this proposal correct?} 
Since the latter is non-renormalizable, the OD3 theory may be interpreted as providing a suitable  UV completion. 

This concludes our briew review of non-critical string theories in $d\leq 6$ dimensions. For further details and a more complete account, see \cite{Gopakumar:2000ep,Harmark:2000av,Larsson:2001ug} and references therein. We now move to the supergravity description of these decoupled theories.

\subsection{Warm-up: pure NS \emph{or} RR backgrounds}

The simplest case to consider is that of a  pure NSNS %\com{I changed this from NS-NS to be consistent with RR}
 background, for which the decoupling limit of interest yields the gravitational dual of LST. While this setup has been extremely well-studied, in particular from the recent perspective of the single-trace $T\bar T$ deformations, 
we will still treat it in detail, mainly to fix the notation. We then discuss the  decoupling limit in the corresponding S-dual solution with purely RR three-form flux. 

In ten-dimensional string frame, the half-BPS black string solution supported by pure NSNS flux is given by
\be
ds^2 = \frac{1}{f_1} (-dt^2 + d\s^2) + f_5 (dr^2 + r^2 d\Omega_3^2) + ds^2_{K3} \;, \;\;\;\;\; e^{2\Phi} = g_s^2  \frac{f_5}{f_1} \,
\label{nsextrmet}
\ee
%and \com{Check H in AdS}

\be \label{H3purensns}
H_3 = 2 k \a' \omega_{S^3} -\frac{2 p g_s^2 \alpha'}{v r^3 f_1^2}dr\wedge d\sigma \wedge dt \, 
\ee
where $\Phi$ is the ten-dimensional  dilaton and  the harmonic functions are, in the standard convention in the literature %\emph{Do we already want arbitrary constants? Notation!}
\be
f_1 = \frac{\a' g_s^2 p}{r^2 v}+1 \;, \;\;\;\;\;\; f_5 = \frac{\a' k}{r^2} + 1  \, ,
\ee
Here,  $k$ is the number of NS5 branes, $p$ - the number of F1 strings, and $v$ is the string-frame volume of $K3$ in units of $(2\pi)^4 \a'^2$. %\footnote{In the notations from Skenderis, we have $\rho=-\log v$ and $\Phi_{6d}=\Phi_{10d}+\frac{\rho}{2}$. \emph{Remove.}}.
 Note the equations of motion are still solved if we put arbitrary constants in the harmonic functions, but here we stick to the standard conventions. 
 %\com{In this sense, we can also consider $c_5\neq 1$, still with $c_1=1$. For this we need to rescale $r \to \sqrt{c_5} r$ and we can take $g_s^2 = c_5$. The limit with $c_5 \to 0$ and $r,t,\sigma$ fixed is then more similar to an NCOS/OD1 Seiberg-type limit where the coordinates remain fixed but the metric elements (at infinity) do not. Useful? }

In this frame, the six-dimensional dilaton, denoted as $\phi$,  is an attracted scalar, which approaches a ratio of integers at the horizon
 
\be
e^{2\phi} = g_6^2 \frac{f_5}{f_1}   \;,\;\;\; \;\;\;\;\;\;\;\;\; \lim_{r\r 0} e^{2\phi} = \frac{k}{p} \, , 
%= \frac{\frac{\a' k}{r^2} + 1}{\frac{\a' p}{r^2} + \frac{v}{g_s^2}}{\color{ForestGreen}
%\xrightarrow[\text{}]{r\r 0}\frac{k}{p}
\ee
independently of its asymptotic value $g_6^2 \equiv g_s^2/v$ at large distances.  
As reviewed, the NS5 decoupling limit is $g_s \r 0$ with $\a'$ and $r/g_s$ fixed (we have temporarily dropped the tildes from the NS-frame parameters). The constant term in the $f_5$ harmonic function drops out in this limit, although - importantly - this does not happen for $f_1$.   
The asymptotic value of the  dilaton tends to zero, but its value at the horizon remains fixed. %We also need to . \com{This can be argued from S-duality of the NCOS type limit. Comment later?} 

We would now like to discuss this solution and its five-brane decoupling limit from the six-dimensional perspective of section \ref{sugrareview}. This simple exercise is performed in detail in appendix %\footnote{\old{The pure NS solution can be  fit inside the more general truncation studied therein   by additionally setting $F_3$ and the axions $\chi_{1,2}$ to zero.} \com{Not necessary?}}
 \ref{appendixB2}. %Clearly there is a single three-form field, which contains both a self-dual component and an anti-self-dual one.  
 The only non-zero fields $F^\L$   are  %. We simply start with the $6d$ pure NS action, and split the $H_3$ into two components that satisfy the modified (anti)-self-duality conditions \eqref{}. These are
%\emph{ Can you fix the $1/4$ from the action?} {\color{ForestGreen}(we can fix the closed forms as those that give $k,p$ and then take the "SD/ASD" combinations such that their sum/difference gives the closed forms) }

\be
F^1 = \frac{1}{4} (H_3+ e^{-2\Phi} \star H_3) \;, \;\;\;\;\; F^{\bar 1} =  \frac{1}{4} (-H_3+ e^{-2\Phi} \star H_3) \, ,
\ee
where we have denoted the $\L$-indices associated to the negative eigenvalues of $\eta_{\L\Sigma}$  by a bar and the factor of $1/4$ is due to the normalisation of $H_3$ in \eqref{H3purensns}. It is easy to check these satisfy the  modified (anti)-self-duality conditions \eqref{eq: N-self duality condition}; as a result,   the `electric' charges associated to these combinations of field strengths equal plus or minus  the `magnetic' ones. More precisely, if the standard magnetic and electric charges with respect to $H$  are  $q_{\rm NS5} = k$ and   $q_{\rm F1} = p$,  then the `magnetic' charges $q^\Lambda$ associated  to the above three-forms are 

\be
q^1 = \frac{p+k}{2} \;, \;\;\;\;\;\;\; q^{\bar 1} = \frac{p-k}{2} \, ,
\ee
while the `electric' ones are $q_\Lambda = \eta_{\L\Sigma} q^\Sigma$.  The harmonic functions that parametrise this solution thus take the form 
\be
H^1 = \frac{\a'(p+k)}{2\tilde r^2} + c^1 % = f_1 + g_6^2 f_5
  \;, \;\;\;\; H^{\bar 1} = \frac{\a'(p-k)}{2\tilde r^2} + c^{\bar 1} % =  f_1 - g_6^2 f_5
  \label{harmns5f1}
\ee
where we have in principle allowed for a rescaled radial coordinate, $\tilde r$ and for arbitrary constants in the harmonic functions.  The relation between these harmonic functions and $f_{1,5}$ is obtained by bringing  the metric  \eqref{nsextrmet} to six-dimensional Einstein frame, which yields

%Let us now check this solution matches indeed the general one, \eqref{}. The metric computed from \eqref{} is 
%
\be
ds_6^2 = \frac{1}{g_6 \sqrt{f_1 f_5}} (-dt^2 + d\s^2) + \frac{\sqrt{f_1 f_5}}{g_6} (dr^2 + r^2 d\Omega_3^2) \, , \label{6defmet}
\ee
%with $g_6 = g_s/\sqrt{v}$.
This matches the general solution  \eqref{6dmetsol},  provided that 
\be
H^\L H_\L =  g_6^2 f_1 f_5 \;, \;\;\;\;\;\;\; r = g_6 \, \tilde r \label{Hsqf1f5}
\ee
Matching also the scalar coset, the relation between the harmonic functions is found to be %{\color{ForestGreen}%(we defined them with indices down in the solution, we also need a better notation)
 %\emph{Is notation ok now?} (yes, I think it's ok to keep H for harmonic functions and F for forms)}
\be
H^1 = \frac{f_1 + g_6^2 f_5}{2} %=\frac{\a'(p+k)}{2\tilde r^2} + \frac{1+g_6^2}{2}:=\frac{q_1}{\tilde{r}^2}+c_1 
 \;, \;\;\;\; H^{\bar 1} =  \frac{f_1 - g_6^2 f_5}{2}%=\frac{\a'(p-k)}{2\tilde r^2} + \frac{1-g_6^2}{2}:=\frac{q_2}{\tilde{r}^2}+c_2 
\ee
%\com{Comment on alternative parametrization? discuss}
from which one immediately reads off  the constants that appear in \eqref{harmns5f1}.  The combination 

%\be
%c^1 = \frac{1}{2} (1+g_6^2) \;, \;\;\;\;\;\;\; c^{\bar 1} = \frac{1}{2} (1-g_6^2) 
%\ee
\be
\left. c_\L c^\L\; \right|_{\rm NS5-F1} = g_6^2
\ee
becomes zero, indeed, in the NS5 decoupling limit. Thus, our decoupling criterion  is met.

We may also check that the central charge diverges in this limit.  The  non-zero components of the supersymmetry and matter  central charges \eqref{eq: def central charges Z} % \com{according to Ferrara for (1,0) and (2,0) the ASD one is not a central charge} are % \emph{Check!} {\color{ForestGreen}(I get overall $\frac{1}{2}$ factors, which agree with the fact that $(Z^1)^2-(Z^{\bar{1}})^2=kp$)}
\be
Z^1 = \frac{1}{2}( k e^{-\phi} + p e^{\phi} ) \;, \;\;\;\;\; Z^{\bar 1}= \frac{1}{2}( -k e^{-\phi}+  p e^{\phi}  )
\ee
%\be
%Z^{i,I} = X^{i,I}{}_\L q^\L = \left(\begin{array}{cc} \cosh \Phi_6 & \sinh \Phi_6 \\ \sinh \Phi_6 & \cosh \Phi_6 \end{array}\right)\left(\begin{array}{c} p+k \\ p-k \end{array}\right) = \left(\begin{array}{c} k e^{-\Phi_6} + p e^{\Phi_6} \\  - k e^{-\Phi_6} + p e^{\Phi_6}   \end{array}\right)
%\ee
We immediately note that $Z^{\bar \imath}$ vanishes in the near-horizon limit, as expected. At infinity we have $|Z_+| \to |Z_+^\infty|=\frac{1}{2}| k/g_6 + p g_6|$, which indeed diverges if $g_6 \r 0$ (including when $p=0$), as expected from our general arguments. %Note also $Z_-$ vanishes at the horizon, due to the attractor mechanism.

The leading irrelevant deformations away from AdS$_3$ are given by the linear perturbations in $Z^{\bar 1}$ and the metric%\footnote{More precisely, $\l_+$ is read off from the metric perturbation, upon rescaling $r^2$ by a factor of $|q| = \sqrt{q_\L q^\L}=\sqrt{kp}$, while $\l_-$ is read  from $Z^{\bar 1}$, with the same rescaling. We ignore  the relative normalisation of $\l_\pm$ by factors of $2$, etc. %\emph{\new{Ok?}}
% {\color{ForestGreen} (for $Z^{\bar{1}}$ I obtained $\frac{\alpha'p}{\sqrt{p k}}(g_6^2-\frac{k}{p})$ and for the metric  $\frac{r^2}{\sqrt{k p}}-\frac{\alpha'p}{2(kp)^{3/2}}(g_6^2+\frac{k}{p})r^4$)}
%}%\com{We could say $Z^1$, it's the same right? NOO!}
, which lead to %\emph{\textcolor{red}{Careful normalisation! Same as before?}} %\new{There seems to be an upper bound on $g_6$?? Understand!}
\be \label{irrdefpurensbackgr}
\l_\pm  = \a' k \left(1 \pm \frac{g_6^2}{(g_6^*)^2} \right) \;, \;\;\;\;\;\; g_6^* = \sqrt{\frac{k}{p}}
\ee
where we have normalised the overall coefficient\footnote{ More precisely, we have rescaled the irrelevant couplings by a factor of $2\sqrt{kp}$ with respect to \eqref{irrcouplgen}. } so that it equals the little string tension when $g_6 =0$. %{\color{red}\emph{Factor k correct? Relative normalisation?}} \com{We should remove the $k$}
 As expected, the irrelevant operator that turns on the anti-self-dual perturbation is only present if the value of $g_6$ at infinity differs from its attractor value, $g_6^*$. %\com{Say that on the other hand one cannot turn on only the ASD deformation by itself? We should also comment on absorbing the $g_s$ in the metric, because we do it here but not in the TsT solutions.}

A very similar analysis applies to the S-dual background, which is supported by a purely RR three-form field. The string-frame metric now reads %\emph{Factor $g_s$ in F?}{\color{ForestGreen}(I think that there is no factor of $g_s$ in $F$, I checked eom. We can  put the $g_s$ that comes in front of the metric from S-duality and $g_s$ in $F$, but we don't want any overall $g_s$ in the metric we don't have $g_s$  in $F$ like in the stringy exclusion paper)} \com{I agree. Also, we should write $F$ as the $H$ in the NSNS solution, not with the dilaton, agree?} \new{Ok, so do we drop the $g_s$?}
\be \label{pured1d5metric}
ds^2  =\frac{1}{\sqrt{f_1f_5}}\left( - dt^2 +d\s^2 \right)+\sqrt{f_1 f_5}\bigg(dr^2+r^2 d\Omega^2_3\bigg)  +\sqrt{\frac{f_1}{f_5}}\, ds^2_{K3} \, 
\ee
and 
\be \label{pured1d5}
F_3=2k \alpha'\omega_{S^3}+\frac{2 p\alpha'}{v} e^{-2\Phi}\hspace{0.01cm}\star \omega_{S^3} \;, \;\;\;\;\; e^{2\Phi } = g_s^2 \frac{f_1}{f_5} \,
\ee
where now $g_s, \a'$ are the corresponding S-dual parameters\footnote{Under S-duality, $g_s \r \tilde g_s = \frac{1}{g_s}$ and $\a' \r \tilde \a' = \a' g_s$, implying that $v \r \tilde v = \frac{v}{g_s^2} = \frac{1}{g_6^2}$, and vice-versa. },  in terms of which the 
 harmonic functions $f_{1,5}$ read
\be \label{functionsd1d5}
f_1 = \frac{g_s \a' p}{v r^2} + 1% = \frac{p}{g_s \a' \tilde r^2} +1 
\;, \;\;\;\;\; f_5 = \frac{g_s \a' k}{r^2} + 1 \,  %= \frac{v k}{g_s \a' \tilde r^2} + 1
\ee
From the six-dimensional perspective, it is now the string-frame volume of the K3 in string units, denoted $e^{-2\rho}$,  that is an attracted scalar
\be
 e^{-2\rho} = v \frac{f_1}{f_5} \;, \;\;\;\;\;\;\; \;\;\lim_{r \r 0} e^{-2\rho} = \frac{p}{k}
\ee
while the six-dimensional dilaton is now an arbitrary constant, $ e^{2\phi} = g_s^2/v = g_6^2$ throughout. This is of course consistent with S-duality.  The six-dimensional Einstein frame metric is again given by \eqref{6defmet}, but with $g_6 \r g_6/g_s = 1/\sqrt{v}$, where the additional factor of $g_s$ comes from the fact that under S-duality, the metric \eqref{nsextrmet} turns into \eqref{pured1d5metric} multiplied by an overall factor of $g_s$. Consequently, the line element  reproduces the general form \eqref{eq: metric Ansatz}, as it should,  with the identification
\be
H_\L H^\L = \frac{f_1 f_5}{v} \;, \;\;\;\;\;\;\; r = \frac{\tilde r}{\sqrt{v}} \label{idmetr}
\ee
% and so \eqref{Hsqf1f5} still holds \emph{True?} {\color{ForestGreen}(yes)}, as well as the fact that we need to rescale $r$ by a factor of $g_6$.
in full agreement with the transformation of $g_6$ into $v$ under S-duality. % since the latter is an $SO(5,21)$ transformation, hence it must leave $H_\L H^\L$ invariant. % 

Matching the rest of the fields, the  relation we find between the harmonic functions  is

 \be
 H^2 = \frac{1}{2} \left(f_1 + \frac{f_5}{v}\right) \;, \;\;\;\;\;\; H^{\bar 2}  = \frac{1}{2} \left(f_1 - \frac{f_5}{v}\right) \,
 \ee
The associated integer-quantised charges are $(p\pm k)/2$. %\old{The reason that their coefficient is $g_s \a'$ can be traced back to the fact that, due to the additional factor of $g_s$ obtained in S-duality, the $10d$ Newton's constant ($\propto \a'^4$) is rescaled by a factor of $g_s^4$.} \emph{Correct? Check!!} {\color{ForestGreen}(we have $G_N=g_s^2\alpha'^4$ which is invariant under S-duality. The coordinates are all rescaled with $\sqrt{g_s}$ to eat the factor in front of the metric)} 

The coefficients of the harmonic functions can be readily read off, %( $c_{2,\bar 2} = (1\pm v)/2v$) 
and satisfy 

  \be
 \left. c_\L c^\L \; \right|_{\text{D1-D5}} = \frac{1}{v} \, .
 \ee
This shows that the decoupling limit of the D5 branes,  which is  $\a' \r 0$,  $g_s\to \infty$ with $\a'g_s $ and the volume $v \a'^2$ of the K3 fixed, does indeed correspond to the  $c_\L c^\L \r 0$  limit, since $v \to \infty$. Note that, under $S$-duality, $\a' g_s$ becomes the little string tension, so this is exactly the same limit as the previous one. Since the dilaton diverges in this limit, it is more natural to study it in the S-dual NS5 frame, as is usually done.

 %
% We will also rescale it by a factor of $\a'$, in order to make contact with the AdS$_3$ decoupling limit. 
%Going to $6d$ notation the (linear combinations of) constants in the harmonic functions are $g_s \a' c_1$ and $g_s \a'/v c_5 = g_s \a'^3/V_{K3}$, where $c_{1,5} \sim \O(1)$. In the D5 decoupling limit $\a' \r 0$ with $g_s \a'$ and $V_{K3}$ fixed (as we would like to ``scale'' the K3 coordinates in the same way as $t,\s$ - i.e., not at all), the first constant survives, whereas the second goes to zero \cite{}. 
We have thus explicitly shown that  $c_\L c^\L = 0$  corresponds to a decoupling limit also in the D5 frame, in full agreement with our expectations. We may check the decoupling criterion also at the level of the central charge, whose only non-zero components read

% The central charges are \emph{Check!} {\color{ForestGreen}(checked) (should we say in these simple cases that $|Z_+|=Z^2 etc?$)}

\be \label{d1d5centralcharges}
Z^2 = \frac{1}{2}  (k e^{-\rho} + p\, e^\rho) \;, \;\;\;\;\;\; Z^{\bar 2} = \frac{1}{2} (-k e^{-\rho} + p\, e^\rho) %\pm? (k e^{-\rho} - p\, e^\rho)
\, 
\ee
We  note that $|Z_+^\infty| = \frac{1}{2} |k \sqrt{v} + p\,/\sqrt{v}|$ , which diverges as expected in the D1-D5 decoupling limit, $v \r \infty$. 

 The coefficients of the irrelevant deformations  can be computed, as before, from the metric and the matter central charge %{\color{ForestGreen}(normalization here also in case we fix 2.51)}%{\color{red}\emph{Normalisations?}} %{\color{ForestGreen}(I'll look for references, I don't know how to fix these normalisations, but I checked thar I obtained the result below if we fix the normalisation for the NS as above)}
 \be \label{d1d5irrcof}
\l_{\pm} =  g_s \a' k  \left(1  \pm \frac{v_*}{v}\right) \;, \;\;\;\;\;\;\;\; v_* = \frac{p}{k}
 \ee
where $v_*$ is the attractor value for the K3 volume, and we have again rescaled \eqref{irrcouplgen} by a factor of $2\sqrt{kp}$.  This entirely parallels the NS analysis, and matches \eqref{irrdefpurensbackgr} under the S-duality exchange $g_6 \leftrightarrow 1/\sqrt{v}$. Note that in the NS frame, $\l_\pm$ were the irrelevant deformations associated to the $\O_{H_3^\pm}$ operators, whereas in the RR frame, they correspond instead to $\O_{F_3^\pm}$.

\subsection{Backgrounds with both NS \emph{and} RR three-form fields}\label{sectiontstextremal}

Let us now study a slightly more complicated case, where we have both NS-NS and RR three-form fields. Starting from the pure RR D1-D5 solution \eqref{pured1d5metric} - \eqref{pured1d5}, we would like to turn on an irrelevant  deformation that starts as the $(2,2)$ deformation corresponding to $\O_{H_3^-}$ ($\O_{F_3^-}$ in the S-dual frame, to which the discussion around equation \eqref{sugraidop} refers). 

The simplest way to generate the solution\footnote{{As we will see, this  way to generate the solution does not provide a useful parametrisation at finite temperature.}} is via a TsT transformation with parameter $\g$ along %\footnote{More precisely, we perform a T-duality on $\s$, a shift $t \r t + \g \a' \tilde \s$, followed by a T-duality back on $\tilde \s$. \new{\emph{Check $\a'$!}} 
%}
 $t,\s$. The string-frame metric, B-field and dilaton are given by 
 %\emph{Factors $\a'$! Notation!}  {\color{ForestGreen}(I think it's ok, $R\rightarrow \frac{\alpha'}{R}$ at T-duality, if we replace $\gamma\rightarrow \gamma\alpha'$ then the metric is what we wrote, but the B-field gets a $\frac{1}{\alpha'}$)}{\color{red}(edit: I looked in becker becker schwarz 6.97-6.98 and I don't think there should be any $\alpha'$ in Buscher rules; also, if the metric (and dilaton) is the same and the B-field is changed by an multiplicative factor, the dilaton eom is not satisfied anymore)}
 \cite{Berman:2000jw}

\be
ds^2=\frac{\sqrt{f_1 f_5}(-dt^2+d\sigma^2)}{f_1f_5-\gamma^2}+\sqrt{f_1f_5}(dr^2+r^2d\Omega_3^2)+\frac{\sqrt{f_1}}{\sqrt{f_5}} ds^2_{K3} \label{d1d5tstmet}
\ee

\be \label{10ddilaton}
\a' B=\frac{\gamma}{f_1f_5-\gamma^2}dt\wedge d\sigma\;, \;\;\;\;\;
e^{2\Phi}=\frac{ g_s^2 f_1^2}{f_1f_5-\gamma^2}
\ee
and RR fields  
\be \label{axionsd1d5}
C_0 = \frac{\g}{g_s f_1} \;, \;\;\;\;% C_2 = \ldots \; \;\;\; 
C_4 = - \frac{v \g}{ g_s f_5} \omega_{K3} + \ldots
\ee
%{\color{red}(effect on the RR fields of the $\alpha'$ in the TsT??)} 
where the $\ldots$ stand for the self-dual completion of $C_4$, which is given explicitly, along with the remaining RR field $C_2$, in appendix \ref{appendixB2}. %{\color{ForestGreen}(mention that we can also add $s_0,s_4$, which will be important for the S-dual?)}
 The parameter $ g_s$ is the same one that appeared in the undeformed D1-D5 solution and, in particular, in the definition of $f_{1,5}$ in \eqref{functionsd1d5}. While it no longer corresponds to the value of the dilaton at infinity, which is now $g_s^\infty = g_s/\sqrt{1-\g^2}$, it still turns out to be a useful parameter. %It does correspond to the open string coupling. 

Bringing the metric to  six-dimensional Einstein frame, one finds 
\be
ds^2_{6}=\frac{g_s}{g_6} \left[ \frac{-dt^2+d\sigma^2}{\sqrt{f_1f_5-\gamma^2}}+\sqrt{f_1f_5-\gamma^2}(dr^2+r^2d\Omega_3^2) \right]
\ee
in agreement with  the general form \eqref{eq: metric Ansatz} of the supersymmetric solutions, with 

\be
H_\L H^\L = \frac{f_1 f_5 - \g^2}{v} \;, \;\;\;\;\; r = \frac{\tilde r}{\sqrt{v}}
\ee
We consequently read off 
\be
\left. c^\L c_\L \right|_{TsT \; D1-D5}  = \frac{1-\g^2}{v} 
\ee
The six-dimensional solution thus degenerates if $ \g \r 1$, even for $v$ finite,  as does the ten-dimensional one. In terms of ten-dimensional string frame fields, we can easily identify the $\g \r 1$ limit, in which %\emph{Factors?} 
\be
%G_{\mu\nu}=\eta_{\mu\nu},\;\;\;  
g_{\mu\nu}  =\frac{\eta_{\mu\nu}}{1-\g^2}\;, \;\;\;\;\;g_{ij} = \d_{ij} \;, \;\;\;\;\;\;g_s^\infty = \frac{ g_s}{\sqrt{1-\g^2}}  \;, \;\;\;\;\;\; \a' B_\infty = \frac{\g}{1-\g^2} =  g
\ee
with  the NCOS decoupling limit \eqref{openstring quantities Seiberg} for $\vep=1-\g^2$ and  $G_{MN}=\eta_{MN}$, again showing  that $c_\L c^\L \r 0$ corresponds to a known decoupling limit of string theory. 

The  decoupling limit at the level of the supergravity solution proposed in \cite{Maldacena:1999mh,Gopakumar:2000na,Harmark:2000wv}  places some of the singular scaling inside the coordinates, rather than  the metric components\footnote{ It effectively makes $G_{MN} \propto \sqrt{\vep}\,  \eta_{MN}$ for fixed coordinates but, as discussed, this still leads to the same NCOS theory. Note our scaling of $\s,t$ differs from that of \cite{Harmark:2000wv} by a factor of $\sqrt{1-\g^2}$, as our parametrisation is  different.} %.  It reads%\footnote{ %\emph{Copy footnote from section $4$!}
%} % \cite{}
%
%\com{This is a bit confusing as written because with this rescaling the $B$-field doesn't diverge at infinity. That's of course fine because we have decoupled the asymptotic region in this way, but it should be clear that it's equivalent to "take $\ap$ and all coordinates constant while scaling $\gamma \to 1$" and to "set $\gamma=1$ while scaling $\ap \to 0$ with the indicated rescalings" since the full solution gets simply rescaled by a factor of $\ap$, which drops out. See 0103188, footnote 3 of page 5.}
%
\be
\g \r 1 \;, \;\;\;\; \a' = \sqrt{1-\g^2}\,  b \; ,\;\;\;\;r = \sqrt{\a'} \tilde r \;, \;\;\;\; \s,t = \sqrt{\frac{(1-\g^2) b}{\a'}} \hat \s, \hat t \;, \;\;\;\;\; x^i \r \sqrt{\frac{\a'}{{b}}} \, \hat x^i \label{ncosharm}
\ee
with $b, g_s$ %{\color{ForestGreen}($\bar{g}_s$  because we want $\alpha' g_s$ to be fixed for the S-dual frame, right?)}
 and all the hatted quantities fixed. {Interpreting the rescaling factor of the coordinates (squared) as a non-trivial open string metric, we have $\a' G^{-1} = b $, which should therefore be identified with the spacetime non-commutativity parameter, $\a'_{eff}$.} Note also that $g_s^\infty \a' = g_s b$ is fixed, and will become the little string length in the S-dual frame; $g_s$ itself  is identified with the open string coupling $G_s$ of section \ref{section31rev}, which is held fixed. %, while $b$ is identified with $\a'_{eff}$.
 
Note  the  limit \eqref{ncosharm} does not affect the functions $f_{1,5}$; %, it just multiplies the metric \eqref{d1d5tstmet} by an overall factor of $\a'$. 
  %{\color{blue} The $-dt^2+d\s^2$ factor in the metric becomes (up to the overall $\a'$ factor) %{\color{ForestGreen}(also the factors of $b$ in the parallel part?)}. 
 %
%\be
%-dt^2 + d\s^2 = \frac{(1-\g^2)b}{\a'^2} (-d\tilde t^2+d\tilde \s^2) = \frac{1}{b} (-d\tilde t^2+d\tilde \s^2)
%\ee 
%}
 % \emph{Check also the B-field, etc!} {\color{ForestGreen}(the rescaling of coordinates cancels the $\alpha'$ factor and the result is finite except at infinity) }
thus, {even though  {\cite{Harmark:2000wv}} proposed this limit for extremal D5-branes alone,} the presence of the  F1 strings does not affect it, {as also remarked in \cite{Berman:2000jw}}. %\textcolor{red}{\emph{Check! Is this what we get from the $r_0=0$ limit of the NCSYM way of doing things?}} {\color{ForestGreen}(the D1s do not affect the decoupling limit in the extremal case)} %This solution  has been previously written down in {\color{red}\cite{}REF}. 
 {As we will see, this statement will no longer be true at finite temperature.} 
  %\emph{\textcolor{red}{Agree?}}
  
 { The best way to understand where the above scalings originate from is to T-dualize the D1-D5 system to one of intersecting D3-branes, and use the fact that the D3 NCOS is S-dual to the \emph{spatially} non-commutative $\N=4$ SYM \cite{Gopakumar:2000na}, for which the decoupling limit is very well understood \cite{Maldacena:1999mh,Seiberg:1999vs}. Moreover, the NCSYM perspective appears to be essential for understanding the decoupling limit at finite temperature;  we thus analyse it in full detail in appendix \ref{appendixdecoupling}. }
%
 %[While oftentimes one also takes the limit $\a' \r 0$ on the gravity side
%\footnote{Such a solution is written in \cite{maldarusso}. By taking $\theta=0$ in their eqn (2.5) and adapts it to D5, then $f_M \r f_5$ and $H_M = (f_5-\g^2)/(1-\g^2)$. Consequently, the $0,1$ part of the metric is multiplied by an extra $1-\g^2$ factor, which is there identified with $\a'^2$, in agreement with the proposed scaling of $g_s$. \emph{Check!} The scaling of the various spacetime coordinates is then adjusted such that the metric scales with an overall $\a'$. \emph{Keep?}}

Since the solution \eqref{d1d5tstmet} - \eqref{axionsd1d5} falls within the consistent truncation Ansatz of \cite{Duff:1998cr}, we can readily write it in six-dimensional language.  The six-dimensional dilaton, string frame volume and axions are 

\be \label{fieldstrtst}
e^{2\phi} = g_6^2 \frac{f_1 f_5}{f_1f_5-\g^2} \;, \;\;\;\;\; e^{-2\rho} = v \frac{f_1}{f_5} \;, \;\;\;\;\; \chi_1 = \frac{\g}{ g_6 \sqrt{v} f_1}% +s_0
 \;, \;\;\;\; \chi_2 = - \frac{\g \sqrt{v}}{g_6 f_5} %+ s_4
\ee
%where, for generality, we have also included two constant shifts of the axions, which clearly still solve the equations of motion {\color{ForestGreen}(again maybe mention quantities are in the NS frame?)}. 
The parameter $g_6=g_s/\sqrt{v}$ is the one before the deformation. While it is no longer the asymptotic value of the six-dimensional dilaton, it is a useful parameter for our solution,  representing the value of the $6d$ dilaton at the horizon, which needs to stay  finite (as it is part of the near-horizon D1-D5 moduli).% In fact, $g_6$ is the open string coupling in the theory, which is to be held finite.]

 The  charges of the black string, which are given by integrals of $H_3, F_3$ and their duals (given in \eqref{generalFerraraformsNS} - \eqref{generalFerraraformsR}) over the $S^3$ are still 

\be \label{chargess0s4}
q_{D5} = k \;, \;\;\;\; q_{NS5} = 0 \;, \;\;\;\;\; q_{D1} =   p \;, \;\;\;\;\; q_{F1} = 0 % k s_4-p s_0
\ee
%\textcolor{red}{Is the F1 charge compatible with susy?} 
%\emph{It would be nice to come up with an argument why the TsT solution is susic.} 
%{\color{ForestGreen}Just like in the previous section, we will set $q_{F1}=0$.}
Linear combinations of these  charges correspond to sources for the harmonic functions characterising this solution, which are fully worked out in appendix 
\ref{appendixB2}.  A difference with the previous cases is that now the $SO(2) \subset SO(5)$ connection $Q_{ij} \neq 0$, so we need to perform an $SO(2)$ rotation, whose general expression is given in {\eqref{anglegeneralrot}} %appendix \ref{sec: Appendix B - consistent truncation}
, to make it vanish and to align the central charge with a fixed direction. The rotation parameter, given in  {\eqref{rotangletstd1d5}}, approches zero near the horizon, and $\pi/2$ near infinity. %\new{\emph{Check!}}.
 %After bringing the vielbein matrix to the ``standard form'' \eqref{standardformXmat}, 
We can then easily read off  the coefficients $c_\L$ in the harmonic functions and the moduli $\bar g_{i\L}$, {which are given in \eqref{harmonictstd1d5} and \eqref{modulitstd1d5}, respectively}. We find the latter are identical to those in the undeformed  D1-D5, as the TsT parameter $\g$ entirely drops out of the near-horizon geometry.

 The central charge \eqref{eq: def central charges Z} %is invariant under the shifts of the axions, so we can simply compute it for   $s_0=s_4=0$. It  
 reads 
 
 \be \label{centralchdiv}
 |Z_+| =  \sqrt{\frac{f_1 f_5}{f_1 f_5 - \g^2}} \,  |Z_+|_{\g=0} \;, \;\;\;\;\; |Z_+|_{\g=0} = \frac{1}{2} ( k e^{-\rho} + p e^\rho)
 \ee
Its value at infinity 
\be
|Z^+_\infty|= \frac{k \sqrt{v} + p/\sqrt{v}}{2\sqrt{1-\g^2}} \label{ccncos}
\ee
 diverges as $\g \r 1$; note that the previous $v \r \infty$ limit is no longer needed to achieve this.  However, this limit is still allowed, in the sense that
$v$ does not appear among the near-horizon moduli, which would need to stay finite. 

%  an analysis of the full solution in the $6d$ frame shows that only $g_6$ and a certain combination of the axions are moduli in the near-horizon D1-D5 CFT limit. This will correspond to the LST limit of the (S-dual to) NCOS. \emph{Agree?} To turn off the LST deformation, we simply set $v = v_\star$.
The coefficients of the irrelevant deformations  can again be inferred from the components of the metric and the matter central charge. If we normalise $\l_\pm$ (which, in this frame, are the couplings associated with $\O_{F_3^\pm}$) to be given by the LST coupling $ g_s \a'$ as $ v \r \infty$ (via the - by now, standard - rescaling by $2\sqrt{kp}$), then%\footnote{\textcolor{magenta}{Note these values are read off in the unrescaled coordinates $t,\s,r$. As discussed at length in appendix \ref{}, the natural coordinates in the decoupling limit differ from these by a rescaling factor. \emph{More?}}\textcolor{red}{Keep?}} %\emph{Re-re-check! Notation!} {\color{ForestGreen}(I checked that I obtain the same results with this  normalisation established)}
%\emph{\textcolor{red}{Do we get the same result if we use the NCSYM parametrization?}}{\color{ForestGreen}(no, the normalization is affected by the rescaling of the coordinates, I had the results in the Decoupling limit note; it was actually interesting that in the hatted parametrization the 6d metric does not depend on the TsT parameter(s))}
%
\be \label{irredeftstsol}
\l_\pm =  g_s \a' k \left(1 \pm \frac{v_*}{v}\right) \;, \;\;\;\;\;\; \l_-' = 2\g g_s \a' k \sqrt{\frac{v_*}{v}} 
\ee
%

%
%\begin{align}
%\lambda_{\pm}=\tilde{\alpha'} \frac{p \pm k v}{2 v \sqrt{p k}}=\frac{\tilde{\alpha'} k}{2 \sqrt{p k} }\bigg(1 \pm \frac{v_*}{v}\bigg)\hspace{1cm}\lambda'_-=-\tilde{\alpha'}\frac{\gamma}{\sqrt{v}}
%\end{align}
where, as before,  $v_* = p/k$ is the attractor value for the volume modulus and $\l_-'$ is the coupling associated with $\O_{H_3^-}$ which, according to our previous discussions, should be identified with the NCOS coupling $b \leftrightarrow \a'_{eff}$.  

As can  be seen from the central charge \eqref{ccncos}, % Note that, with the conveniently-chosen normalisations above, the coefficients of the three irrelevant deformations at play satisfy
%
%. Remembering that $\l_R \sim b \sim 2\g  k \tilde \a'/\tilde g$ \emph{Is $k$ scaling correct?} we identify the NCOS coupling 
%
%\be
%\tilde g = \sqrt{v/v_*}
%\ee
%We thus have three independent irrelevant parameters: $\tilde \a' k, \tilde \a k/v$ and $\g \tilde \a' k \sqrt{v_*/v}$, with
%
% $\l_+^2-\l_-^2 -(\l_-')^2 = 4 (1-\g^2) (\tilde \a' k)^2 v_*/v $, so
  the degeneration limit requires either $\g \r 1$ or $v \r \infty$. {Note that $\lambda_+^2-(\lambda_-^2+\lambda'^2_-)=({2}g_s \a' k)^2 (1-\gamma^2) v_\star/v$  vanishes in either limit.} The non-commutativity can be turned off by sending $\g \r 0$. Then, the only way to achieve decoupling is $v \r \infty$. If we insist on keeping $v$ fixed, then the only other way to achieve decoupling is $\g \r 1$. This is the NCOS limit. %\textcolor{blue}{The effective NCOS coupling constant is given in \eqref{effncosc} above. If we send it to infinity, we recover LST (by switching to the dual frame)}. 
  {Conversely, if we send $v \r \infty$, then the non-commutativity is automatically turned off, and we recover LST (by switching to the dual frame), which has $\l_-=\l_+$. }   But, we  need not do this, and we can in particular switch off the $\l_-$ coupling by tuning $v \r v_*$, which implies that the `non-commutativity' scale $\l_-'$ is of the same order of magnitude as the LST one. %\emph{Does this agree with dual picture?}
%A weakly coupled description doe
 Note that, since we are in the supergravity regime, the NCOS description is strongly coupled.

One may consider the same background in the ``NCSYM parametrisation'', given in  \eqref{newparnonext} and discussed at length in appendix \ref{appendixdecoupling}. The irrelevant couplings \eqref{irrcoefnewpar} one reads off from it are - naturally - the same as the ones above,  upon using the appropriate identifications of the parameters and the coordinates. A bonus feature of that parametrisation is that in the decoupling limit, the $SO(21)$ singlet coupling is identified with the LST one, $\l_+ =  \hat g_s \a' k$.  %, which is also the irrelevant coupling that enters the entropy formula. {\color{ForestGreen}(a bit too early to mention the entropy?)}}
% An advantage of that alternate parametrisation is that, in it, we simply find $\l_+ = \tilde \a' k$.  } \emph{\textcolor{red}{So, can we actually fix the normalisation?}}

Since the ten-dimensional asymptotic dilaton \eqref{10ddilaton} diverges  in the $\g \r 1$ limit, it is preferable to change to the S-dual frame. For this, we must first perform a constant shift by $s_0 = -\frac{\g}{g_s}$ in the axion \eqref{axionsd1d5} to make it vanish at infinity; otherwise the S-dual dilaton is still divergent. For reasons that will soon be clear, we also consider a constant shift in the axion associated with $C_4$ on K3;  all this simply introduces two constants, $s_0$ and $s_4$, in the last two fields in \eqref{fieldstrtst}. 
The shifts in the axions induce an F1 charge

\be
q_{F1} = k s_4-p s_0
\ee
Since we would like the charge to be quantized (and, in this particular case, exactly zero), the above shift in $s_0$
% Note this leads to a contribution to the charge proportional to $-\g g_s$. {\color{ForestGreen}This also 
 sets $s_4=-\frac{\g p}{k g_s}$. %[Given though that the dilaton  We find that the dilaton is well-behaved in this case only if $\chi_1^\infty =0$. ]
With these choices, the S-dual axion-dilaton is%\footnote{\textcolor{blue}{We denote the string coupling in this frame by $\tilde{g}_s$ in the rest of the section to avoid confusion.}} %\emph{Can we still add a constant term to $C_0$? Factors!!!}
\be \label{C0sugrans}
e^\Phi = \tilde{g}_s \frac{f_5 + \g^2 (f_1 -2)}{\sqrt{f_1 f_5 - \g^2}} \;, \;\;\;\;\; C_0 = \frac{\g (f_1-1)}{\tilde{g}_s[f_5 + \g^2 (f_1-2)]}%\propto \sqrt{1-\g^2}
\ee
where $\tilde g_s = 1/g_s$ is the new string coupling parameter.
The resulting string frame metric and RR fields read %\emph{Is this more like Om or like Harmark?}

\be
ds^2 = \sqrt{ \frac{f_5+\g^2 (f_1 -2)}{f_5}} \left( \frac{f_5(-dt^2+d\s^2)}{f_1 f_5-\g^2} + f_5 (dr^2 + r^2 d\Omega_3^2) + ds^2_{K3}\right) \label{sdtstmet}
\ee

\be
C_2 =  - \frac{\g}{f_1 f_5 - \g^2} dt \wedge d\s \;, \;\;\;\; C_4 = \bigg(- \frac{\g \tilde{v}}{ \tilde{g}_s f_5} -\frac{\g p}{\tilde{g}_s k}\bigg)\omega_{K3}  
\ee
%{\color{ForestGreen}(if the B-field from the other frame had an overall factor)}
%\textcolor{magenta}
{where $\omega_{K3}$ is the unit-normalised K3 volume form}. The formula for the B-field can be found in appendix \ref{appendixB2}.  {%\color{ForestGreen}(In principle, one could also add new constants in $C_0,C_4$, which will be constrained by charge quantisation; not interesting, we will not do this) }\emph{Is there also $C_4$?}

%{\color{red}(I arrived here with the checks)}

One immediately notices that in the $\g \r 1$ limit, the asymptotic values of the supergravity fields scale as expected in the OD1/OBLST  theory, namely  %\cite{} \emph{Mention differences with Om}

\be
g_{\mu\nu} \sim \frac{\eta_{\mu\nu}}{\sqrt{\varepsilon}} \;, \;\;\;\; g_{ij} \sim  \sqrt{\varepsilon}\, \d_{ij} \;, \;\;\;\; g_s^\infty \sim \sqrt{\varepsilon} \;, \;\;\;\; C_2 \sim \frac{1}{\varepsilon} \;, \;\;\;\; C_4 \sim \O(1)
\ee
%scales in the expected way for an $OD1$/$(1,1)$ OBLST  theory,
 where  $\varepsilon = 1-\g^2 $. The decoupling limit is word-for-word the S-dual of the one above.
%
% \emph{Work out explicitly!} \emph{Does whether $C_4$ vanishes or not matter?} The string frame volume and $6d$ dilaton computed from the above read
%
One may also easily work out the six-dimensional fields using the consistent truncation presented in the appendix, and check that the six-dimensional solutions take the expected form. The coefficients of the irrelevant couplings are given by the S-dual expression to \eqref{irredeftstsol}, namely

\be
\l_\pm =  \tilde \a' k \left(1 \pm \frac{\tilde{g}_6^2}{(\tilde{g}_6^*)^2}\right) \;, \;\;\;\;\;\; \l_-' = 2\g \tilde \a' k \frac{\tilde{g}_6}{\tilde{g}_6^*} 
\ee
where now $\l_\pm$ correspond to the $\O_{H_3^\pm}$ operator deformations, while $\l_-'$ is associated with $F_3^-$. The LST limit is $\tilde{g}_6 \r 0$ with $\tilde \a'$ fixed, which also shuts off the `non-commutative' deformation.

\subsection{Backgrounds with non-zero five-form flux}\label{od3section}

%
%\bi
%\item Understand  OD3! (T-duality on (12) or (12)-(34)?)
%\item is anti-self-duality automatic, or one needs to impose it for supersymmetry?
%\ei
%
%{\color{ForestGreen}
%(sketch, probably we don't need most of the formulae here)

In this section, we perform two additional T-dualities on the internal space, which should have the effect of mapping the OD1 decoupling limit to the OD3 one. To simplify our task, we temporarily replace the K3 by $T^4$, %\textcolor{magenta}
{with coordinates $x^2, \ldots, x^5$}. 

Starting from the NS-frame solution \eqref{sdtstmet}, we perform two  T-dualities along the $(23)$ directions of the torus. The B-field is not affected, whereas the string frame  metric and the dilaton change as follows: %\textcolor{red}{Recheck and factors $\tilde v$! Dilaton looks weird.}  %\emph{\textcolor{red}{Notation torus coord!}}
\be
ds^2=\sqrt{\frac{f_5+\gamma^2(f_1-2)}{f_5}}\bigg[\frac{f_5}{f_1 f_5-\gamma^2 }(d\sigma^2- dt^2)+f_5\big(dr^2+r^2 d\Omega^2_3\big)+ \frac{f_5 (dx^2_2+dx^2_3)}{f_5+\gamma^2(f_1-2)} +\, dx^2_4+dx^2_5\bigg] \nonumber
\ee

\be
e^{\Phi}=%\tilde{g}_s\frac{\sqrt{f_5}}{\sqrt{\tilde{v}}}\frac{\sqrt{f_5+\gamma^2(f_1-2)}}{\sqrt{f_1 f_5-\gamma^2}}=
{\frac{\tilde{g}_s \tilde \a'}{R_2 R_3}}\sqrt{\frac{f_5(f_5+\gamma^2(f_1-2))}{f_1 f_5-\gamma^2}} \label{od3soltdualities}
\ee
where $R_i$ denote the radii of the corresponding circles of the $T^4$.  For de OD3 decoupling limit, we are  particularly interested  in the behaviour of the four-form potential 
\be
C_4 = -\frac{\gamma}{f_1 f_5-\gamma^2} dt\wedge d\sigma \wedge dx_2\wedge dx_3 + {\frac{\tilde{v}\gamma(1-\gamma^2)}{\tilde{g}_s  R_4 R_5(f_1 f_5-\gamma^2)}dt\wedge d\sigma \wedge dx_4\wedge dx_5+...}
\ee
%{\color{ForestGreen}(we can absorb in $\tilde{g}_s$ the radii after T-dualities, case in which the factor of $\tilde{v}$ simplifies in the forms)}
where the dots are additional components on $S^3$ necessary in order to render the type IIB five-form self-dual.
%
%
%in the effect of T-dualities on the components of $C_4$ on $T^4$ (recheck (34)):
%\begin{align}
%C_4^{(12)}&=-\frac{\gamma}{f_1 f_5-\gamma^2} dt\wedge d\sigma \wedge dx_1\wedge dx_2\\
%C_4^{(12)}&=\frac{v\gamma(1-\gamma^2)}{g_s(f_1 f_5-\gamma^2)}dt\wedge d\sigma \wedge dx_3\wedge dx_4
%\end{align} 
The two-form RR field is given by

\be
C_2=C_0 dx_2\wedge dx_3 -{\frac{\tilde{v}\gamma}{\tilde{g}_s R_4 R_5 f_5}dx_4\wedge dx_5}
\ee
where $C_0$ is the value of the axion before the T-dualites, given in \eqref{C0sugrans}. 

In the decoupling limit $\gamma\rightarrow 1$, the component of the $C_4$ potential that lies along the  $t , \, \s,2,3$ directions diverges as $1/\vep$, while its  $t,\s,4,5$ component and the $C_2$ field stay finite. The asymptotic dilaton also stays finite, while 
%
%part on $(12)$ diverges like $\frac{1}{1-\gamma^2}:=\frac{1}{\epsilon^2}$, while the part on $(34)$ stays finite. The 6d dilaton equals the 10d one and stays finite in the decoupling limit, as it should for OD3. 
the components of the metric along the open D3 directions all scale as %$\frac{1}%%{\sqrt{1-\gamma^2}}=\
$\frac{1}{\sqrt{\varepsilon}}$, while the rest all scale as %$\sqrt{1-\gamma^2}=\epsilon$.
$\sqrt{\vep}$. This is precisely the OD3 limit described in \cite{Gopakumar:2000ep}, up to the fact that we have chosen a slightly different (overall, but not relative) scaling of the metric components. The $6d$ Einstein frame metric is simply 
\begin{align}
ds^2_{6}&=\frac{1}{\sqrt{\tilde{g}_6^2(f_1 f_5-\gamma^2 )}}\big(d\sigma^2- dt^2\big)+\sqrt{\tilde{g}_6^2(f_1 f_5-\gamma^2 )}\bigg(dr^2+r^2d\Omega_3^2\bigg)
\end{align}
which leads to the same degeneration condition as in the OD1 case; note, however, that now a different irrelevant deformation %\footnote{\textcolor{magenta}{To preserve supersymmetry, the cycles of the $T^4/K3$ upon which the two T-dualities are to be performed should be chosen appropriately. {\color{ForestGreen}(If the Killing spinors do not depend on the directions on which we do T-duality, than susy is preserved. In our case nothing depends on the $T^4$ coordinates, so probably any cycle is equally good)} \emph{Correct??}}}
 is turned on, namely one associated to one of  the $\O_{F_3^{-\hat S}}$. Were both the OD1 and OD3 deformations turned on, then we would naturally  expect the dimensionality of the  parameter space where the degeneration condition is obeyed to increase. 

{It is interesting to note that the same solution, \eqref{od3soltdualities}, can also - and more simply - be generated by using the proposed S-duality between the OD3 theory and spatially non-commutative $6d$ SYM \cite{Gopakumar:2000ep,Alishahiha:2000pu}. 
%\textcolor{red}{\emph{More cites?}}. {\color{ForestGreen}(cite \cite{Alishahiha:2000pu} also earlier?)} 
{ While their proposal was for NS5 branes alone, we find that the presence of the F1 strings does not affect the overall conclusion. }

To generate the background this way, we simply apply a TsT transformation, with parameter $\l$, along the torus directions (45) of the D1-D5 solution, and then S-dualize. We include hats on all parameters and coordinates of the D1-D5 seed, as the two final backgrounds are the same only  up to coordinate  rescalings and parameter redefinitions.  We obtain  %\emph{\textcolor{red}{More fields?}}
\be
ds^2 = \sqrt{\frac{\hat{f}_5+\lambda^2 \hat{f}_1}{\hat{f}_5}} \bigg[\frac{1}{\hat{f}_1}(-d\hat{t}^2+d\hat{\sigma}^2)+\hat{f}_5 \bigg(d\hat{r}^2+\hat{r}^2 d\Omega_3^2\bigg)+(d\hat{x}_2^2+d\hat{x}_2^3)+ \frac{\hat{f}_5}{\hat{f}_5+\lambda^2 \hat{f}_1}(d\hat{x}_4^2+d\hat{x}_5^2)\bigg] \nonumber \ee

\be
e^{2\Phi} = \hat{\tilde{g}}_s^2\frac{f_5+\lambda^2 f_1}{f_1}
\ee
%\textcolor{magenta}{where we have included, for generality, a blackening factor $\hat f = 1 - \hat r_0^2/\hat r^2$.}
As it turns out, this background is identical to  \eqref{od3soltdualities}, upon certain coordinate rescalings  and parameter redefinitions that we discuss at length in the appendix, including in the more non-trivial finite temperature case;   the relevant equations are \eqref{relationcoord} -  \eqref{alitoalh}. 

%. \emph{\textcolor{red}{Want to give concrete eqn numbers, or not?}}{\color{ForestGreen} (yes, maybe it's useful) (The coordinate rescalings are given in , the relation between TsT parameters in \ref{relationparam1}, between the rest of the parameters in \ref{relationparam2},\ref{alitoalh} ) }
% We chose to write the nonextremal version because the identification of the backgrounds is true also at finite temperature. 

Upon performing two T-dualities along {(23)} and one further S-duality, one finds precisely the {(pre-decoupling)}  extremal NCOS background  in the ``NCSYM'' parametrisation {briefly mentioned in section \ref{sectiontstextremal} above, and discussed at length in appendix \ref{appendixdecoupling}} (its non-extremal version can also be found in \eqref{newparnonext}). Interestingly, in both cases, the six-dimensional  metric only depends on  the  harmonic functions in the hatted parametrisation, 
$\hat f_{1,5}$  

\be
ds_6^2 = \frac{1}{\hat{\tilde{g}}_6\sqrt{\hat{f}_1\hat{f}_5}}(-d\hat{t}^2+d\hat{\sigma}^2) + \frac{\sqrt{\hat{f}_1\hat{f}_5}}{\hat{\tilde{g}}_6}\bigg({d\hat{r}^2}+\hat{r}^2d\Omega_3^2\bigg)
\ee
a feature that we discuss at length in appendix \ref{appendixdecoupling} for the NCOS/OD1 case. Note this metric is the same as the $6d$ metric \eqref{6defmet} obtained from the pure NS5-F1 solution. Since the decoupling limit in the hatted parametrisation is identical to the standard D5/NS5 decoupling limit %($\a' \r 0, \hat{g}_s \r \infty$ with $\hat{g}_s \a'$ fixed)
(plus a scaling limit on the TsT parameter $\l$, which only affects the other fields in the background), we immediately find that the self-dual irrelevant parameter will be $\l_+ = k \tilde \a' $ - the inverse little string tension -  in the  appropriate normalisation. 

%\textcolor{magenta}
{To summarize, in this section we have shown that for each representative choice \eqref{sugraidop}   of $(2,2)$ irrelevant operator  to turn on,   the  $c_\L c^\L =0$  asymptotic degeneracy condition corresponds to a known decoupling limit of string theory: LST, OD1 and, respectively, OD3. We presented two different ways to generate the OD$p$ backgrounds: either via TsT along the common space-time D1-D5 directions followed by S and T-dualities, or via a purely \emph{spacelike} TsT in a different duality frame.
%
%To sum up, we can obtain the OD1 background by S-duality from the (5+1) NCOS background, which can be obtained either by performing a TsT transformation on D1-D5 along the directions $t,\sigma$ or by the more complicated chain of dualities explained in appendix \ref{appendixdecoupling}. From  OD1, one can reach OD3 with two extra T-dualities. Alternatively, one can obtain OD3 by starting from D1-D5 and doing TsT along two spatial directions along D5 and then S-dualizing. 
It would be interesting to generalize this procedure for the case where more than two independent deformations are turned on simultaneously.
}

\section{Non-extremal solutions and their properties}
\label{section4}

We would now like to discuss the thermodynamic properties of the corresponding non-extremal (or nearly extremal) black string backgrounds. We will start with the NS5-F1 solution and its decoupling limit, and then move on to the more general decoupled backgrounds, which also include RR fields.

\subsection{Thermodynamics of the general non-extremal NS backgrounds}
\label{section41thermo}

Let us, for simplicity, start  by considering the non-extremal NS5 - F1 string frame solution of type IIB string theory, with $n$ units of momentum along the common $S^1$ \cite{Maldacena:1996ky,Chakraborty:2020swe}

\be
ds^2 = \frac{1}{f_1} \left[  d\s^2 - dt^2 + \frac{r_0^2}{r^2} (\cosh \a_n dt + \sinh \a_n d\s)^2 \right]  +f_5 \left( \frac{dr^2}{f} + r^2 d \Omega_3^2\right) + ds^2_{K3} %\nonumber
%\ee
%
%\be
%ds^2 = - \frac{f}{f_1 f_n} dt^2 + \frac{f_n}{f_1} \left(d\s +  \frac{r_0^2 \sinh 2 \a_n }{2 f_n r^2} dt\right)^2 +f_5 \left( \frac{dr^2}{f} + r^2 d \Omega_3^2\right) + \sum_{i=1}^4 dx_i^2
%\ee
%\be
%H = \frac{ g_s^2 \a' p}{v} \,  d\s\wedge dt \wedge d \left(\frac{1}{ f_1 r^2} \right)+ 2 \a' k \, \omega_{S^3}
\;, \;\;\;\;\;
 e^{2\Phi} = g_s^2 \,\frac{ f_5}{f_1 } \label{bckgndsol}
\ee
%{\color{ForestGreen}(also mention the 3-form field which is the same as in the extremal zero momentum case discussed previously)}
where, in the non-extremal case, %$d\Omega_3^2$ and $\om_{S^3}$ are the metric and, respectively, the volume form of the unit $S^3$, and  
the various harmonic functions are given by
\be \label{functionsnonextremal}
f= 1 - \frac{r_0^2}{r^2} \;, \;\;\;\;\; f_{i} = 1 + \frac{r_0^2 \sinh^2 \a_{i}}{r^2}
\ee
with
\be \label{parametersns5f1}
 \sinh 2 \a_1 = \frac{2 g_s^2 \a' p}{v r_0^2} \;, \;\;\;\;\; \sinh 2 \a_5 = \frac{2 \a' k}{r_0^2} \;, \;\;\;\;\; \sinh 2 \a_n = \frac{2 g_s^2 \a'^2 n}{R^2 v r_0^2}
\ee
%{\color{ForestGreen}(check that we have the definitions of $v,R$ in the previous sections)}
The ADM mass  of this black string  is

\be
M  = \frac{R v r_0^2}{2 \a'^2 g_s^2} (\cosh 2 \a_1 + \cosh 2 \a_5 + \cosh 2 \a_n) 
\ee
and its entropy 

\be \label{entropyexpression}
S = \frac{2\pi R v r_0^3}{g_s^2 \a'^2} \, \cosh \a_1 \cosh \a_5 \cosh \a_n
\ee
It is useful to split the mass into a ground state contribution - the energy at extremality $E_{extr}$, and a purely non-extremal piece, $E$ -  the  energy above extremality. They are  given by
\be
E_{extr} = \frac{R }{ \a' } \left( p + \frac{k v}{g_s^2} \right)\;, \;\;\;\;\;\; E = \frac{R v r_0^2}{2 \a'^2 g_s^2} (e^{-2\a_1} + e^{- 2 \a_5} + \cosh 2 \a_n) \label{energyexpr} 
\ee
In all the well-known decoupling limits (the standard AdS decoupling with $\a' \r 0$, $g_s$ fixed, or the LST limit with $g_s \r 0$, $\a'$ fixed) the extremal energy diverges, whereas the energy above extremality stays finite, provided $r_0$ is rescaled by an appropriate factor of $\a'$ or, respectively, $g_s$. In the first case, both $\a_{1,5} \r \infty$, and thus the entropy $S \propto r_0 \propto \sqrt{E}$ has  Cardy behaviour\footnote{The energy always scales as $E \sim  r_0^2$ in the various decoupling (or non-decoupling) limits, albeit with different numerical coefficients.}; in the second limit, only $\a_5 \r \infty$, and thus $S \propto r_0^2 \propto E$, leading to  Hagedorn behaviour instead. Taking  the zero-momentum limit for simplicity, the exact entropy-to-energy relation one finds is %\emph{Put it at zero momentum}
\begin{align}
S = 2 \pi \sqrt{ 2  p k E R + \a' k E^2 } \label{ldentropy1}
\end{align}
%
%\be
%S = 2 \pi \sqrt{  p k E_L R + \a' k E_L E_R }+ 2 \pi \sqrt{  p k E_R R  + \a' k E_L E_R } \label{ldentropy}
%\ee
which, as interestingly remarked in \cite{Giveon:2017nie}, is identical to the entropy of a symmetric product orbifold of $T\bar T$ - deformed CFTs.

For the general backgrounds with $\a_{1,5}$ finite, the entropy scales as $S \propto r_0^3 \propto E^{3/2}$ at large $r_0$, which leads to the negative specific heat characteristic of asymptotically flat black holes. Moreover,
no decoupling limit from the asymptotically flat geometry is known, though one option worth exploring may be the  the strict $g_s =0$, $N = \infty$ limit, with $g_s N$ fixed \cite{Gubser:1998iu,Danielsson:2000ze}.  However, it turns out that if one works in the canonical ensemble, the large black strings with this entropy-to-energy relation do not dominate the partition function, as stable small black holes of lower free energy coexist with them at the same temperature. 

%even in these cases, one may exhibit  an intermediate Hagedorn-like regime in the canonical ensemble, which lies near the boundary between the stable black holes localised in the AdS region and the thermodynamically unstable ones lying in the asymptotically flat one. The  two solutions coexist for temperatures smaller than a maximum one. 

%{\color{red}
%\begin{itemize}
%\item with the definition of $\xi$ below, I get $\frac{1}{\sqrt{\xi}}$ instead of $\frac{1}{\xi}$ in 4.11. This affects also 4.12 in the same way.
%\item the rescaling in order to have $r_0=\hat{r}_0'\sqrt{\alpha' k}$ is for me $\hat{r}'_0=\sqrt{\xi}\hat{r}_0$ and not $\hat{r}'_0=\xi\hat{r}_0$. Then the growth I obtain at small $\xi$ is $\xi^{1/3}$ (with the rescaling $\hat{r}'_0=\xi\hat{r}_0$ I get $\xi^{2/3}$, closer to what is written below). 
%\item I also get that the free energy is an increasing function of $r_0$
%\end{itemize}
%
%
%}
To show  this, we  closely follow the analysis 
of  \cite{kutd3} for non-extremal D3-branes.  We start  by computing the entropy (at zero momentum, for simplicity) as a function of the horizon size, $r_0$%\emph{Check! I'm not getting the factor of 2 downstairs}%{\color{ForestGreen}(checked, no factor of 2 downstairs)}
\be
S =  \frac{2 \pi R v r_0^3}{g_s^2 \a'^2} \cosh \a_1 \cosh \a_5  = \frac{\pi R  r_0}{g_6^2 \a'^2} \sqrt{\left(r_0^2 +  \sqrt{r_0^4+(2\a' k)^2}\right) \left(r_0^2 + \sqrt{r_0^4 + (2\a' p g_6^2)^2}\right)}
\ee
Letting $r_0 = g_6 \sqrt{\a'  p} \,\hat r_0$ %(see below without the $g_6$ factor)
, this can be written more compactly as %{\color{ForestGreen}(I would write it directly in terms of $\xi$ or at least $g_6^*$ since we define them here)}
%
%\be
%S = \frac{\pi R g_6 \a' \tilde r_0}{2} \sqrt{\left(\tilde  r_0^2 + 2  \sqrt{\tilde  r_0^4+\left(\frac{2 k}{\a' g_6^2}\right)^2}\right) \left(\tilde  r_0^2 + 2 \sqrt{\tilde  r_0^4 + (2 p/\a')^2}\right)}
%\ee
%
%
%
%
%{\color{ForestGreen}Letting $r_0 = \a' \hat r_0$, this can be written more compactly as
%
%\be
%S = \frac{\pi R \a' \tilde r_0}{2 g_6^2} \sqrt{\left(\tilde  r_0^2 +   \sqrt{\tilde  r_0^4+\left(\frac{2 k}{\a' }\right)^2}\right) \left(\tilde  r_0^2 + \sqrt{\tilde  r_0^4 + \bigg(\frac{2 p g_6^2}{\alpha'}\bigg)^2}\right)}
%\ee
%
%}
%
%{\color{blue}To get a dimensionless variable, let us further define $\tilde r_0 = \hat r_0 \sqrt{p/\a'}$. Then
%
%\be
%S = \frac{\pi R g_6 p^{3/2} \hat r_0}{2 \sqrt{\a'}} \sqrt{\left(\hat  r_0^2 + 2  \sqrt{\hat r_0^4+\left(\frac{2 k}{p g_6^2}\right)^2}\right) \left(\hat  r_0^2 + 2 \sqrt{\hat  r_0^4 + 4}\right)}
%\ee
%where we note that the attractor value of $g_6^2$ is $k/p$. }
%
%
%{\color{ForestGreen}To get a dimensionless variable, let us further define $\tilde r_0 = \hat r_0 \sqrt{p/\a'}$. Then

\be
S = \frac{\pi R g_6 p^{3/2} \hat r_0}{\sqrt{\a'}} \sqrt{\left(\hat  r_0^2 +   \sqrt{\hat r_0^4+\frac{4}{\varkappa^4}}\right) \left(\hat  r_0^2 +  \sqrt{\hat  r_0^4 + 4 }\right)} \;, \;\;\;\;\;\; \varkappa \equiv \frac{g_6}{g_6^*}
\ee
%
%\be
%S = \frac{\pi R g_6 p^{3/2} \hat r_0}{\sqrt{\a'}} \sqrt{\left(\hat  r_0^2 +   \sqrt{\hat r_0^4+\left(\frac{2 k}{p g_6^2}\right)^2}\right) \left(\hat  r_0^2 +  \sqrt{\hat  r_0^4 + 4 }\right)} \;, \;\;\;\;\;\; \varkappa \equiv \frac{g_6}{g_6^*}
%\ee
where $g_6^*$ is the attractor value of $g_6$, $\sqrt{k/p}$. Thus, unlike in the D3-brane case, now the entropy also depends on the dimensionless coupling $\varkappa$. 
%
%
%
%\emph{I seem to be missing a factor of $2$ in this older result. Please check!}
%{\color{ForestGreen}(is this for the decoupled ALD?)}
%\be
%S =  \frac{2 \pi R v r_0^3}{g_s^2 \a'^2} \cosh \a_1 \cosh \a_5  = \pi R \, g_6 \, \a' \tilde r_0 \left(\sqrt{\tilde r_0^4 + 4 p^2/\a'^2} + \tilde r_0^2 \right)
%\ee

The expression for the energy as a function of the same variables is
%
%
%Remembering that $\a_1=\a_5$ in our case, we have (remember $r_0^2 \sinh 2\a_5 = 2 \a' k$)
%
%\be
%E = \frac{R v}{g_s^2 \a'^2} (r_0^2 e^{-2\a_5} + L_u + L_v) = \frac{R v}{g_s^2 \a'^2}  \left( L_u + L_v - 2 \a' k + 2 \sqrt{\a'^2 k^2 + 4 L_u L_v}\right)
%\ee
%%It is also interesting to compute $S(E)$, here for $\a_n=0$ \emph{Generalise to $\a_n \neq 0$!} We have
%{\color{blue} (I wrote it below directly in terms of the rescaled one)
%\be
%E = \frac{R}{2} \left(\tilde r_0^2 +  \sqrt{\tilde  r_0^4+\left(\frac{2 k}{\a' g_6^2}\right)^2}  - \frac{2k}{\a' g_6^2} + \sqrt{\tilde  r_0^4 + \left(\frac{2 p}{\a'}\right)^2} - \frac{2p}{\a'}\right)
%\ee
%In terms of the rescaled variable $\hat r_0$
%
%\be
%E = \frac{R p}{2\a'} \left(\hat r_0^2 +  \sqrt{\hat  r_0^4+\left(\frac{2 k}{p g_6^2}\right)^2}  - \frac{2k}{p g_6^2} + \sqrt{\hat  r_0^4 + 4} - 2\right)
%\ee
%}
%
%{\color{ForestGreen}
%
%In terms of the rescaled variable $\hat r_0=\frac{r_0}{\sqrt{p \alpha'}}$
%{\color{ForestGreen}
\be
E = \frac{R p}{2\a' } \left(\hat r_0^2 +  \sqrt{\hat  r_0^4+\frac{4}{\varkappa^4}}  - \frac{2}{\varkappa^2} + \sqrt{\hat  r_0^4 + 4 } - 2\right)
\ee
%
%}
%\be
%E = \frac{R p}{2\a' } \left(\hat r_0^2 +  \sqrt{\hat  r_0^4+\left(\frac{2 k}{p g_6^2}\right)^2}  - \frac{2k}{p g_6^2} + \sqrt{\hat  r_0^4 + 4 } - 2\right)
%\ee
%Thus, in terms of the rescaled variable $\hat r_0$, the energy and entropy only depend on the ratio of $g_6$ to its attractor value, up to some overall factors.
 It is interesting to understand how the temperature $T = (\p S/\p E)^{-1}$, given below  %\emph{Check!Coefficient!} {\color{ForestGreen}(checked, but I get in front $\pi g_6 \sqrt{\alpha' p}$ which is $\pi\sqrt{\alpha' k} \sqrt{\xi}$ with out notation $\xi=\frac{g_6^2}{(g_6^*)^2}$)}

\be \label{tempfunctionr0}
T= \frac{1}{\pi \varkappa \sqrt{\a' k}}\frac{ \hat r_0}{\left[\left(\sqrt{\hat r_0^4+4}+\hat r_0^2\right) \left(\sqrt{\frac{4}{\varkappa^4}+\hat r_0^4}+\hat r_0^2\right)\right]^{1/2}}
\ee
behaves as a function of $\hat r_0$, for different values of $\varkappa =  g_6/g_6^*$.  
The plot of $T$ as a function of $\hat r_0$ is given in the left figure \ref{Tvsr0h}, for various values of the parameter $\varkappa$. We note there is a maximum value of the temperature, $T_{max}$, above which no black string solutions exist; this is reminiscent of Hagedorn behaviour. For temperatures below $T_{max}$, there are always two black string solutions that coexist: one small, which has positive specific heat, and one large, which is thermodynamically unstable. As in \cite{kutd3}, the heat capacity is discontinuous at the transition point between the two branches. The free energy  is an increasing function of $\hat r_0$, and thus the small black string always dominates the thermal ensemble.   

%\com{Can you put both plots on the same line?} {\color{ForestGreen}(done below)}

\begin{figure}[h] 
    \centering
     \captionsetup{width=.8\linewidth}
    \subfloat{{\includegraphics[width=5.7cm]{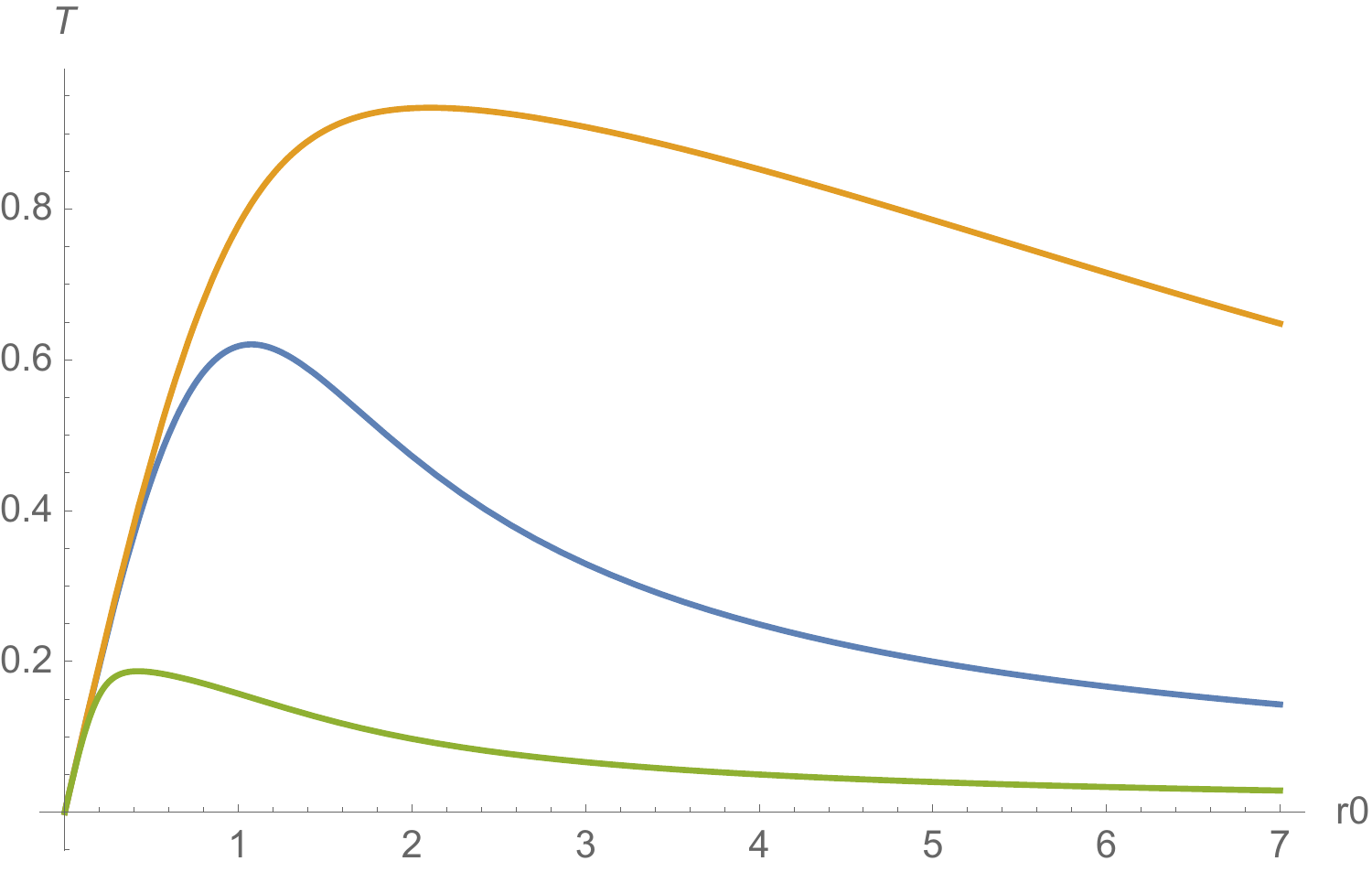} }}%
    \qquad\qquad\qquad
    \subfloat{{\includegraphics[width=5.7cm]{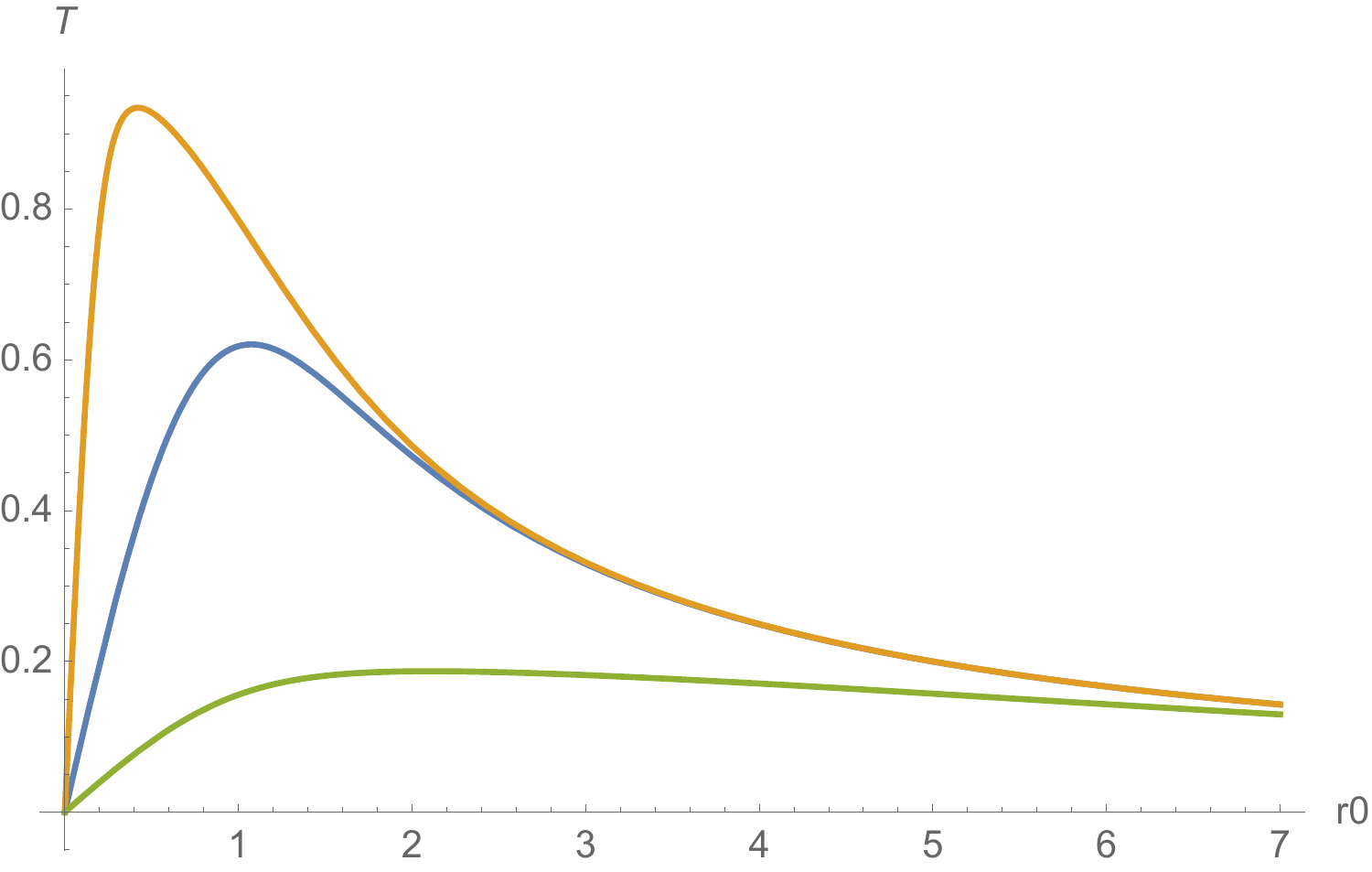} }}%
    \caption{
\footnotesize{  $T$ as a function of $\hat r_0$ (left) and  $\hat r_0' = \varkappa \,\hat r_0 $ (right) for $\varkappa=5$ (green), $\varkappa=1$ (blue) and $\varkappa = 0.2$ (orange). }}%
    \label{Tvsr0h}%
\end{figure} 

%
%
%\begin{figure}[!htb]
%   \begin{minipage}{0.45\textwidth}
%     \centering
%      \captionsetup{width=.7\linewidth}
%     \hbox{\hspace{1cm}\includegraphics[width=.9\linewidth]{playbhthermov2roh}}
%     \caption{\small $T$ as a function of $\hat r_0$ for $\varkappa=5$ (green), $\varkappa=1$ (blue) and $\varkappa = 0.2$ (orange). }\label{Tvsr0h}
%   \end{minipage}
%   \begin{minipage}{0.45\textwidth}
%     \centering
%      \captionsetup{width=.7\linewidth}
%     \hbox{\hspace{0.9cm}\includegraphics[width=.9\linewidth]{playbhthermov2rohp}}
%     \caption{\small $T$ as a function of $\hat r_0/\varkappa$ for $\varkappa=5$ ( green), $\varkappa=1$ (blue) and $\varkappa = 0.2$ (orange).}\label{Tvsr0hp}
%   \end{minipage}
%\end{figure}

%\begin{figure}[h]
%\centering
%\includegraphics[height=4.5cm]{playbhthermo}
%\caption{$T$ as a function of $\hat r_0$ for $\xi=10$ (upper, green), $\xi=1$ (blue) and $\xi = 0.1$ (orange). }
%\end{figure}

\noindent From the left plot, we can observe that the value $\hat r_0^*$ at which the temperature reaches its maximum value increases with decreasing coupling. In particular, $\hat r_0^*$
%
% [Plotting the max, the value of $\hat r_0 $ decreases very mildly with $\xi$ for large $\xi$. {\color{ForestGreen}(we can see this nicely in a 3d plot)}] Note $\hat r_0^*$ 
 diverges as $\varkappa \r 0$, indicating the Hagedorn regime is reached near infinity in this rescaled variable, as expected. 

It is also interesting to understand how the maximum small black string horizon size scales with $g_6$. Then, the plot of $T$ as a function of $\hat r_0 '= \varkappa \, \hat r_0$ (right), so that $r_0 = \hat r_0' \sqrt{\a' k}$, is more appropriate. We find that near $\varkappa \approx 0$, $(\hat r_0')_* \sim \varkappa^{0.66}$, whereas for very large $\varkappa$, the scaling is $(\hat r_0')_* \sim \varkappa^{0.34}$. 
{Consequently, the size of the small, stable black string can be made much larger than that of the AdS$_3$ throat, by tuning $\varkappa $ to be large.}  This can be easily achieved for  $p>>k$, even if $g_6$ small, as required by reliability of the solution. Note this separation of scales is made possible by the existence of an extra parameter, $\varkappa$, with respect to the D3-brane case (which is quite analogous to the self-dual solution with $\varkappa =1$). %\textcolor{red}{ We would thus expect a clean separation between the Hagedorn regime that is presumably associated with $T_{max}$ and the Cardy one when $\varkappa$ is large.  } \emph{I think it's this statement that's not true, as Hagedorn requires $\a' E/Rp >>1$, which translates into $\hat r_0'^2 >> \varkappa^2$. \emph{Check!}} {\color{ForestGreen}(agree)}

 The scaling of $T_{max}$ with $\varkappa$ is% \emph{Are we sure?} {\color{ForestGreen}(checked)}

\be
T_{max} \sim \left\{\begin{array}{ccc} \frac{1}{2\pi \varkappa \sqrt{\a' k}} & \;\;\;\mbox{for}& \varkappa >>1 \\[2mm] \frac{1}{2\pi \sqrt{\a' k}} & \;\;\;\mbox{for}& \varkappa <<1 \end{array}\right.
\ee
which interpolates between the LST Hagedorn temperature 
at small $g_6$ and  a new, coupling dependent scale that roughly corresponds to $\l_+ - \l_- $ in \eqref{irrdefpurensbackgr}, the coefficient of the operator that turns on the deformation to flat space. 
%the $OD1$/OBLST one \eqref{} at $g_6 >> g_6^*$.  }

The results of this section suggest that, since large black holes never dominate the thermal ensemble, only the entropy of small black holes can be meaningfully matched to a computation in the holographic dual. It is interesting to compare this to the situation in global AdS$_5$, where the large black holes are stable, while the small black holes are  unstable %\footnote{Here we mean thermodynamically unstable. Small black holes in AdS are also dynamically unstable \cite{Hubeny:2002xn}, but the alternate stable solution is still subdominant with respect to the large black holes. }
 and never dominate the canonical ensemble. %This comparison suggests a (vague) analogy between small black holes in AdS and our large black strings, with possibly similar prospects for understanding the microscopic explanation of their entropy \cite{}. 

\subsection{Thermodynamics of the non-extremal NCOS  $\&$ ODp backgrounds}
\label{section42ncoshag}
Let us now add RR flux to this background. In fact, following the steps of  section \ref{section3}, we will start from the S-dual D1-D5 non-extremal solution,  perform a TsT transformation and then, eventually, S-dualize again. For simplicity, we set the momentum to zero, though adding it in  is straightforward. The NS-NS sector fields read 
%\emph{Figure out the $\a'$ in B}

\be \label{nsnonextremalbackgr}
ds^2=\frac{\sqrt{f_1 f_5}}{f_1 f_5 - \gamma^2 f} \big(d\sigma^2 - f dt^2\big) +\sqrt{f_1 f_5}\big(\frac{dr^2}{f}+r^2d\Omega^2_3\big)+\sqrt{\frac{f_1}{f_5}}\, ds^2_{K3}
\ee
\be
B=\frac{\gamma f}{f_1 f_5-\gamma^2 f}dt\wedge d\sigma\;, \;\;\;\;\;
e^{\Phi}=g_s\frac{f_1}{\sqrt{f_1 f_5-\gamma^2 f}} \nonumber
\ee
while the RR ones are given in \eqref{rrtst02} - \eqref{rr4formtst} in the appendix. The functions $f_{1,5}$ and $f$ still take the form \eqref{functionsnonextremal}, except that now the relation between the angles and the quantized charges is given by  %\emph{Notation?}
\be \label{standardquantcond}
r_0^2 \sinh 2 \a_1 = 2 \frac{g_s \a'}{v} p \;, \;\;\;\;\;\; r_0^2 \sinh 2 \a_5 = 2 \a' g_s k
\ee
as compared to their expression \eqref{parametersns5f1} in the NS case. 
Note that the asymptotic value of the dilaton is $g_s/\sqrt{1-\g^2}$, but we use this parametrisation because $g_s$ enters the quantisation condition above. %\eqref{standardquantcond}.

{ The NCOS decoupling limit of the finite-temperature background \eqref{nsnonextremalbackgr} is quite subtle.} 
 %{\color{red}(cite both Harmark papers?)} 
For the case of  pure non-extremal D5-branes in a background critical $B$-field (obtained by setting $\a_1 =0$, or $f_1 =1$ in the solution above),  the limit  proposed in \cite{Gopakumar:2000na,Harmark:2000wv,Harmark:2000ff} scales, when translated to our parametrisation\footnote{Note that our coordinates $\s,t$ differ from the initial coordinates of \cite{Harmark:2000wv} by a rescaling  $\s, t = \sqrt{1-\g^2}\, \s_H, t_H$, which is responsible for the slightly different scaling of the coordinates above. Also note that our $g_s = g_s^H \sqrt{1-\g^2}$,  $\cosh^2\th_H = (1+ \g^2/\sinh^2 \a_5)/(1-\g^2)$,  and we omitted some factors of the fixed scale $b$ from the rescaled quantities. {In the notation of appendix \ref{appendixdecoupling}, $\cosh^2 \th_H = 1+\l^2$. }}
%{\color{ForestGreen}(the scaling in 30 from 0006023 of the parallel coordinates is with $b$, not $\sqrt{b}$, shouldn't we have the same?)We are using the 0007 paper. }
\be \label{coordscal}
\a' \r 0 \;, \;\;\;\;\;r = \sqrt{\a'} \, \tilde r \;, \;\;\;\;\; r_0 = \sqrt{\a'}  \, \tilde r_0  \;, \;\;\; \s, t = \sqrt{\frac{(1-\g^2) b}{\a'}} \, \hat \s, \hat t
\ee
and the K3 coordinates as $\sqrt{\a'}$, keeping fixed %{\color{ForestGreen}(is it obvious from the text that we set $\gamma=1$?)}

\be \label{keptctdec}
\tilde \ell_s^2 = \frac{g_s \a'}{\sqrt{1-\g^2}} \;, \;\;\;\;\; b = \frac{\a'}{\sqrt{1-\g^2}} \coth \a_5
\ee
{Note that the scaling of $r_0$ above implies that both $\a_1$ and $ \a_5$ are finite, which is rather non-standard for a decoupling limit.}  The ratio of the two fixed scales above defines an effective coupling
\be \label{effectivecouplingpured5}
\tilde g = \frac{\tilde \ell_s^2}{b} = g_s \tanh \a_5
\ee
which is to be held fixed in the decoupling limit. Note that in terms of it, the relation between  $\a_5$ and the quantized NS5 charge becomes %\textcolor{red}{\emph{Is this exact in this case?}}  {\color{ForestGreen}(yes, the way it is written is ok because it just uses 4.18)}

\be \label{quantcond2}
\tilde r_0^2 \sinh^2 \a_5 = \tilde g k
\ee
One may  rightfully worry that the explicit factor of $\coth \a_5$ above  makes the scaling of the coordinates energy-dependent. The solution is to work in terms of coordinates that differ by a factor of $\sqrt{1-\g^2}$ from the $\s ,t$ coordinates above, which is what \cite{Harmark:2000wv} does. In this case, $\g$ and $g_s$ are state-dependent, but $\a', \tilde g, b$ and the coordinates are not.
%
%{\color{blue} The $\s,t$  coordinates can nevertheless be made to scale by a \emph{fixed} factor of $\epsilon$ if

%\be \label{scalingxtcoord}
%\frac{(1-\g^2) b}{\a'} = \sqrt{1-\g^2} %\coth \a_5 \equiv \epsilon
%\ee
%so one may envisage a state-independent rescaling if one takes the TsT parameter to approach $1$ in an energy-dependent fashion, $\g \sim 1 - \frac{1}{2} \e^2 \tanh^2 \a_5$, and also $\a' =\e \, b \tanh^2 \a_5$. The fact that $g_s$ must depend on the energy as in \eqref{effectivecouplingpured5} then simply follows from the definition of the fixed scale $\ell_s$ in terms of the actual value of the dilaton at infinity. These  energy-dependent scalings of $1-\g$ and $\a'$ appear essential  for  obtaining the desired Hagedorn behaviour.% working out \emph{Check therwise it just doesn't work!!}.  We have no further  comment on this. 
%
At the level of the metric, one obtains 

\be \label{pured5nonextrmet}
ds^2 = \a' \left[ \frac{\sqrt{ f_5} \tilde r^2}{ \tilde r_0^2 \cosh^2 \a_5} \cdot \frac{(1-\g^2) b}{\a'^2} \big(d\hat \sigma^2 - f d\hat t^2\big) +\sqrt{ f_5}\big(\frac{d\tilde r^2}{f}+\tilde r^2d\Omega^2_3\big)+\sqrt{\frac{1}{f_5}}\, d\tilde s^2_{K3}\right]
\ee
with 
\be
f_5 = 1 + \frac{\tilde r_0^2 \sinh^2 \a_5}{\tilde r^2} = 1 + \frac{\tilde g k}{\tilde r^2} \;, \;\;\;\;\; f = 1 - \frac{\tilde r_0^2}{\tilde r^2}
\ee
Using  \eqref{keptctdec} and \eqref{quantcond2}, the prefactor of the first term becomes $\sqrt{f_5\, }\tilde r^2/(\tilde g k b)$, so the $\a_5$ dependence entirely drops out. {It was shown in \cite{Harmark:2000wv} that the entropy-to-energy relation in this background is Hagedorn.

The peculiar scalings of the coordinates and the parameters in the D5 NCOS decoupling limit above can be more easily understood by mapping the system, using T- and S-dualities, to a \emph{spatially} non-commutative D3-brane (related via S-duality to the D3 NCOS \cite{Gopakumar:2000na}), whose decoupling limit is quite clear \cite{Maldacena:1999mh} - in particular, the corresponding angle is sent to infinity. This mapping is reviewed in appendix \ref{appendixdecoupling}. Upon adding the D1-branes, the \emph{only} way to make sense of the decoupling limit at finite temperature is to go to the spatially non-commutative SYM frame.

 More concretely, one notes that there is another way to arrive at the background \eqref{nsnonextremalbackgr} starting from the undeformed D1-D5 system, which uses two T-dualities to map this initial system to one of intesecting D3-branes, then uses \emph{spatial} TsTs to generate   magnetic B-fields on them (see figures \ref{figgduality}, \ref{figdual2} in the appendix). By appropriately adjusting the ratio of the  parameters $\l, \mu$ of these TsT transformations one may, upon a further S-duality, arrive at a system of intersecting D3 branes in a purely \emph{electric} B-field. The corresponding D1-D5 system in an electric B-field is obtained via two further T-dualities.  More details about this chain of (pseudo)dualities can be found in appendix \ref{appendixdecoupling}.
 
The NS-sector fields one finds after this set of transformations are
\begin{align}
ds^2=\sqrt{(\hat{f}_5+\lambda^2 \hat{f}_1)(\hat{f}_1+\mu^2 \hat{f}_5)}\bigg(\frac{-\hat{f} d\hat{t}^2+d\hat{\sigma}^2}{\hat{f}_1 \hat{f}_5} +\frac{d\hat{r}^2}{\hat{f}}+\hat{r}^2 d\Omega_3^2\bigg)+\sqrt{\frac{\hat{f}_1+\mu^2 \hat{f}_5}{\hat{f}_5+\lambda^2 \hat{f}_1}}\, \sum_{i=2}^5 d\hat{x}_i^2 \nonumber
\end{align}

\be
B= \hat{\alpha'}\hat{g}_s \left[\frac{\cos\theta}{4} \left(\mu k+ \frac{\lambda p}{\hat{v}}\right)d\psi\wedge d\phi - \bigg(\frac{\mu p}{\hat{v}\hat{r}^2 \hat{f}_1}+\frac{\lambda k}{\hat{r}^2 \hat{f}_5}\bigg) d\hat{t}\wedge d\hat{\sigma} \right] \nonumber
\ee

\be
e^{2\Phi}=\hat{g}_s^2 \frac{(\hat{f}_1+\mu^2 \hat{f}_5)^2}{\hat{f}_1 \hat{f}_5} \label{newparnonext}
\ee
where the hatted functions and coordinates are those of the seed D1-D5 we have used, which differ from the parameters and coordinates of the seed D1-D5 solution used to generate \eqref{nsnonextremalbackgr} via space-time TsT. Nonetheless, the two backgrounds (including the RR fields) are identical if we identify the hatted and unhatted coordinates and parameters in a particular way, which we describe in full in appendix \ref{appendixdecoupling}. In particular, the relationship between the angles is
\be
\sinh \a_5 =\sqrt{\frac{\sinh^2 \hat \a_5 + \l^2  \sinh^2 \hat \a_1}{1+\l^2} } \;, \;\;\;\;\; \sinh \a_1 = \sqrt{\frac{\sinh^2 \hat \a_1 + \mu^2 \sinh^2 \hat \a_5}{1+ \mu^2} } \;, \;\;\;\;\; \mu = - \frac{\l p}{k \hat v}
\ee
In the decoupling limit, $\hat \a_{1,5}$ have the standard LST scaling $\hat \a_5 \r \infty$  and $\hat \a_1$ fixed. $\l$ is related to the NCSYM spatial non-commutativity parameter, and scales as $\l \sim 1/\a'$. Together with $\sinh \hat \a_5 \sim 1/\a'$ and $\hat v \sim 1/\a'^2$, we note that both $\a_{1,5}$ will be finite in the decoupling limit. However, their values are correlated in a subtle way, as there is only one independent parameter at fixed charges and moduli. Clearly, the hatted parametrisation  makes the decoupling limit look much more straightforward. 

The energy of the background \eqref{newparnonext} can be computed using e.g. the covariant phase space formalism. The answer is $\l$ - independent and thus coincides with the energy of the hatted undeformed D1-D5 background\footnote{One may also compute the energy in the unhatted coordinates \eqref{nsnonextremalbackgr}. While the metric contribution is the same upon using the map \eqref{relationparam1} of the parameters, the na\"{i}ve contribution from the B-field differs in an important way. This difference can be traced to the energy-dependent relation between the hatted and unhatted coordinates, and the correct result corresponds to the one where the former coordinates are held fixed. More discussion of this point can be found in the appendix.}
\be
\hat M = \frac{\hat{R}\hat{v}}{2{\alpha}'^2 \hat{g}_s^2}\hat{r}_0^2(\cosh 2\hat{\alpha}_1+\cosh 2\hat{\alpha}_5+1)
\ee
%{\color{ForestGreen}(maybe we can remove the hat on $\alpha'$)} 
The entropy in hatted variables is also $\l$-independent
\begin{align}
{S}&=\frac{2\pi \hat{R}\hat{v}}{{\alpha}'^2 \hat{g}_s^2}\hat{r}_0^3 \cosh\hat{\alpha}_1 \cosh\hat{\alpha}_5
\end{align}
 Since the energy above extremality is defined in exactly the same way as in the undeformed D1-D5 system, the entropy-to-energy relation in the decoupling limit is given by the same $T\bar T$ - like formula as in the decoupled D1-D5 background without a B-field. The
 $T\bar T$ coupling is identified with $\hat g_s \a'$,  where $\hat g_s$ is the value of the dilaton at infinity in the decoupled geometry,  which becomes the LST coupling $\tilde \a' = \a' \hat g_s$ upon one further S-duality   %\textcolor{red}{\emph{Correct? Write it down!}}
\begin{align}
{S}=2\pi\sqrt{2kp \hat{R} \hat{E} + k{\alpha}' \hat{g}_s \hat{E}^2}
\end{align}
It is interesting to note that the entropy still only depends on the LST parameter $\tilde \a' = \hat g_s \a'$, despite the fact that we now have two independent irrelevant parameters \eqref{irredeftstsol}   depending also on the internal space volume $v$, which is finite in the decoupling limit. In fact, in the decoupling limit we should be working in terms of the hatted coordinates, in terms of which the irrelevant couplings read\footnote{Since the couplings are irrelevant, they naturally change under a rescaling of the spacetime coordinates by an overall factor, which is how \eqref{irrcoefnewparlim} differs with respect to 
  \eqref{irredeftstsol}. Note they still satisfy  $\hat{\lambda}_+^2=\hat{\lambda}_-^2+\hat{\lambda}'^2_-$.
%
%, which is given by \eqref{coordscal},\eqref{keptctdec} in the unhatted parametrisation, and by \eqref{decouphat},\eqref{decoupsc2} in the hatted one. The two formulae differ due to the , 
%
}

\be
\hat \l_+ =  \tilde \a' k \;, \;\;\;\;\;\;\;\;\hat \l_- = % \tilde \a' k \, \frac{p-\mu^2 k \hat v}{p+ \mu^2 k \hat v} = 
\tilde \a' k \, \frac{1-v_*/v}{1+ v_*/ v} 
\;, \;\;\;\;\;\; \hat \l_-' = %- 2 \tilde \a' k  \l \frac{\sqrt{p/k\hat v} }{1+ \l^2 p/k\hat v} =
 - 2 \tilde \a' k \, \frac{\sqrt{v_*/v} }{1+ v_*/v}
\label{irrcoefnewparlim}
\ee
They were obtained by taking the decoupling limit ($\hat v \r \infty$) of the formulae  \eqref{irrcoefnewpar}  and then re-expressing finite quantities such as  $\hat v/\l^2 \sim p^2/(k^2 \hat v \mu^2)\sim v$  in terms of the  physical volume of the K3.
%
%the   relation \eqref{relationparam1} between the TsT parameters was also used.
%where we have used  {\color{ForestGreen}()} and the relation between the internal space volume, $v$, in the final frame, and that in the initial D1-D5 hatted seed, $v = \hat v/\l^2$ in the decoupling limit.
% It is not hard to check that in this limit,  $\hat{\lambda}_+^2=\hat{\lambda}_-^2+\hat{\lambda}'^2_-$. 
 It is  interesting to note that with this parametrisation (which also affects the energy), the irrelevant coupling that appears in the entropy formula is simply $\hat{\lambda}_+$.

%\textcolor{magenta}
{Thus, we have been able to show that, with an appropriate scaling, the entropy of black strings in the decoupled NCOS background takes  precisely the Cardy $\r$ Hagedorn form characteristic of single-trace $T\bar T$ - deformed CFTs. The irrelevant parameter that controls the Hagedorn behaviour is still the LST tension, and the entropy looks insensitive to the fact that non-commutativity has been turned on. In the hatted parametrisation we have been arguing for, the irrelevant parameter also appears to coincide with the value of the $SO(21)$ singlet irrelevant coupling $\hat{\lambda}_+$ one reads off from the supergravity solution. These conclusions trivially extend to the OD$1$ and OD$3$ backgrounds (for a single  $F_3^{\hat S}$ component turned on), which are simply related via S- and T-dualities to the background discussed above.    }

\subsection{Three-dimensional truncation of the self-dual black string}

Even though this article is mostly concerned with the decoupled asymptotic linear-dilaton-like black string backgrounds, it is interesting, for comparison, to also study the asymptotically flat six-dimensional backgrounds from the same perspective. We have already drawn a parallel between the entropies of black holes in these spacetimes in section \ref{section41thermo}. Now, we would like to compare the asymptotic symmetries of the ALD and the asymptotically flat black string spacetimes. 

The asymptotic symmetries of the (pure NS) linear dilaton background were analysed in detail in \cite{Georgescu:2022iyx}. Since  the dual LST theory  lives only along the black string directions ($\s,t$ and the internal manifold) it is justified in this case, in order to understand the ``spacetime'' symmetries of the dual field theory, to only concentrate on the $\s,t$ dependence of the asymptotic symmetry generators, ignoring their dependence on the $S^3$ factor, which is associated to  the R-symmetry directions.   To simplify the analysis, \cite{Georgescu:2022iyx} thus used a spherically-symmetric consistent truncation to three dimensions, containing only the $3d$ metric and the dilaton. Also, they made central use of the known black hole solutions in the ALD background, as the asymptotic symmetry generators were fixed by requiring that their symplectic product with the metric perturbations that simply changed the black hole parameters vanish. Interestingly, \cite{Georgescu:2022iyx} found that the asymptotic symmetry generators are precisely the field-dependent generalisation of two-dimensional conformal transformations that appears in single-trace $T\bar T$ - deformed CFTs, and the non-trivial symmetry algebra matched exactly to the order it was computed. 

We would now like to study the asymptotic symmetries of the asymptotically flat black string backgrounds from the same viewpoint, which focuses on the dependence 
of the asymptotic symmetry generators on $\s,t$ and ignores to first approximation the $S^3$ dependence. 
%Note 
% namely the one in which the dual theory lives along the black string directions, $\s, t$. In particular, this suggests that the asymptotic celestial $S^3$  simply corresponds to the R-symmetry directions,  
%is \emph{not} part of the dual theory, 
%this is rather different from the celestial perspective \cite{} on  flat space holography. 
In this section, we prepare the ground for this computation, by writing down a simple, spherically symmetric consistent truncation of these backgrounds, and describing the black string solutions from this three-dimensional point of view.

%This viewpoint motivates concentrating, in an  asymptotic symmetry group analysis of flat space,  on those diffeomorphisms that depend on the black string directions $\s, t$. To simplify the problem, we ignore for now the part of the ASG that lies on the $S^3$, though its structure is certainly very interesting \cite{}. 

% We will in particular be interested in how the thermodynamics of the asymptotically flat black holes compares with that of the ALD ones \cite{}, as well as the asymptotic symmetries. 
 
% We remind the reader that in the ALD background, the asymptotic symmetry algebra turned out to be \emph{identical} to that of single-trace $T\bar T$.

We concentrate on the simplest asymptotically flat background, which has self-dual pure $H_3$ flux and is given in \eqref{H3purensns}.  The dilaton is a constant. Since, as explained, we are less interested in what happens on the $S^3$, we will concentrate on the following consistent truncation of the six-dimensional theory to three dimensions, which only keeps the spherically symmetric modes  %\emph{Notation!}

\be
ds_6^2  = e^{-6F} ds_3^2 + e^{2F} d\Omega_3^2 \;, \;\;\;\;\; H = 2k \a' (\omega_{S^3} +\star_6 \omega_{S^3} )  = \star H \;, \;\;\; e^{2\phi}= \frac{k}{p}=const
\ee
where $ds_6^2$ is the six-dimensional string frame metric, $d\Omega_3^2 , \omega_{S^3}$ are the metric %{\color{red}(the metric was denoted $d\Omega_3^2$ in the rest of the paper)}
 and, respectively, the volume form on the  unit three sphere (they are given explicitly in \eqref{volumeforms3},\eqref{metrics3}), and $F$ only depends on the three-dimensional coordinates. The fact that the six-dimensional dilaton is constant implies that 
 its value at infinity equals its attractor value, $g_6^2=k/p$.  The normalisation of the three-dimensional Einstein metric was chosen so it be in Einstein frame.  The six-dimensional equations  of motion are %\emph{Check!}
\be
R_{MN} + 2 \nabla_M \nabla_N \phi = \frac{1}{4} H_{MPQ} H_N{}^{PQ} \;, \;\;\;\; \Box \phi - 2 (\p\phi)^2= - \frac{1}{12} H^2 \;, \;\;\;\;\; d (e^{-2\phi} \star H) =0
\ee
The last two of them  are trivially satisfied by our Ansatz (as $H^2 = H \wedge \star H = H\wedge H =0$), while the spherical and non-compact $3d$ components of the Einstein equations lead to the following constraints on the scalar field $F$  and the  $3d$ Einstein frame  metric % {\color{red}\emph{Write in terms of $R_{\mu\nu}$}}

\be
\Box  F = 2 e^{-8 F} - \frac{a^2}{2} \, e^{-12 F} \;, \;\;\;\;\;\; G_{\mu\nu} = 12 \p_\mu F \p_\nu F - 6 g_{\mu\nu} (\p F)^2 + g_{\mu\nu} \left(3 e^{-8F} - \frac{a^2}{2} e^{-12F}\right)
\ee
with $a = 2 k \a'$,  which can be easily derived from the three-dimensional action 

\be
S = \int d^3 x \sqrt{g} \left( R - 12 (\p F)^2 + 6 e^{-8F} - a^2 e^{-12 F}\right) \label{3dact}
\ee
Let us now try to understand how the NS5-F1 solution \eqref{bckgndsol} fits into this. One immediately reads off 
\be
e^{2F} = f_5 r^2= r_5^2 +  r^2 \;, \;\;\;\;\;\;r_5^2 \equiv r_0^2 \sinh^2\a_5
\ee
Note that $r_5^2 (r_0^2 +r_5^2) = (\a' k)^2$. The scalar $F$ diverges logarithmically in this limit; note, however, that its potential is well-behaved. The $6d$ dilaton is 

\be
e^{2\phi} = g_6^2 \frac{f_5}{f_1} = \frac{k}{p} \;\;\;\; \Rightarrow \;\;\;\; f_1=f_5\;, \; g_6^2 = \frac{k}{p}
\ee
The three-dimensional Einstein frame metric reads %\emph{Already remove $f_1$? Check!}

\be
ds_3^2 =  f_5^2 r^6  \left[-dt^2 + d\s^2 + \frac{r_0^2}{r^2} \left(\cosh \a_n dt +\sinh \a_n d\s\right)^2 \right] + \frac{f_5^4 r^6 }{f} dr^2 
\ee
with $f, \a_n$ given in \eqref{functionsnonextremal}, \eqref{parametersns5f1}.
%\be
%f = 1 - \frac{r_0^2}{r^2} \;, \;\;\;\;\;\sinh 2\a_n = ...
%\ee
%Since the six-dimensional dilaton is constant, from \eqref{} we must have $r_1=r_5$, where $r_1^2 (r_0^2 +r_1^2) = (g_6^2 \a' p)^2 $. This implies that $g_6^2 = \frac{k}{p}$ throughout this solution, i.e. its attractor value at the horizon. 
The asymptotics of the Einstein frame are conformally flat. 
%
%\be
%ds^2 = r^6 (-dt^2 + d\s^2) + r^6 dr^2 \sim 
%\ee
%which differs from that of the ALD background by a factor of $r^2$. In a certain  sense, the holographic coordinate here is $r$, while there it was $\log r$. Otherwise, this is not dissimilar from the $3d$ Einstein frame metric in that paper.

Let us now study the black hole solutions in this background. Following \cite{Georgescu:2022iyx}, it is convenient to rescale the coordinates as 

\be
r= g_6 \a' \tilde r \;, \;\;\;\;\; r_0 = g_6 \a' \tilde r_0
\ee
and  introduce the parametrization %\emph{Note $r$ and $\rho$ are switched w.r.t previous paper}

\be
\tilde r = \rho + \frac{\tilde r_0^2}{4\rho} \;, \;\;\;\;\;\;\; L_{u,v} = \frac{\tilde r_0^2 e^{\pm 2\a_n}}{4} \;, \;\;\;\;\;\; U,V= \s \pm t
\ee
Then 
\be
e^{2 F} = r^2 + r_5^2 = (g_6 \a')^2 \left( \rho^2 + \frac{\tilde r_0^2}{2} \cosh 2 \a_5 + \frac{\tilde r_0^4}{16 \rho^2}\right) \equiv (g_6 \a')^2  e^{2 \tilde F}
\ee
where
\be
e^{2\tilde F} = 
 \rho^2 + \sqrt{4 L_u L_v + \frac{ k^2}{(g_6^2 \a')^2}} + \frac{L_u L_v}{\rho^2}  =  \rho^2 + \sqrt{4 L_u L_v + \frac{ p^2}{\a'^2}} + \frac{L_u L_v}{\rho^2} 
\ee
The metric can be written as
\be
ds^2 = (g_6\a')^6 e^{4 \tilde F} \left[  \rho^2 dU dV  + L_u dU^2 + L_v dV^2 + \frac{L_u L_v}{\rho^2} \, d U dV \right] + (g_6\a')^8 e^{8 \tilde F} \, \frac{d\rho^2}{\rho^2}
\ee
which is extremely similar to the ALD metric of \cite{Georgescu:2022iyx} in the same parametrization. 
The quantities $L_{u,v}$ are related to the energy above extremality \eqref{energyexpr}  and angular momenta of the black holes as % \cite{Georgescu:2022iyx} 
%\emph{Check and shorten!}
%
\be \label{energyafsol}
 E = %\frac{R v r_0^2}{2 g_s^2 \a'^2} (e^{-2 \a_1} + e^{-2 \a_5} + \cosh \a_n) = R (\tilde r_0^2 e^{-2\a_5} + L_u + L_v) = 
 R \left(L_u + L_v - \frac{2p}{\a'} + 2\sqrt{4 L_u L_v + \frac{p^2}{\a'^2} }\right)
 \ee
where we used the fact that $g_6^2 = k/p$, implying that $\a_1=\a_5$. 
This highly resembles the formula in \cite{%Chakraborty:2020swe,
Georgescu:2022iyx}, up to the factor of $2$, which reflects the different scaling of $\a_5$.  The momentum is simply

\be  \label{momentumafsol}
   P = \frac{n}{R} = \frac{R v r_0^2 \sinh 2\a_n}{2 g_s^2 \a'^2} = R (L_u-L_v)
\ee
We will now use this parametrisation of the black string solutions to work out the asymptotic symmetries of this spacetime, in order to facilitate the comparison with the corresponding formulae in the ALD case.  
%and then compute the conditions on the allowed diffeos from the symplectic form. \emph{Which gauge is best?}

\subsection{Asymptotic symmetries of the self-dual black holes}

We would now like to find the asymptotic symmetries of the self-dual black holes   within this truncation. Since we do not know, \emph{a priori}, which boundary conditions to impose, we will be following the same strategy that we used in \cite{Georgescu:2022iyx} to find the asymptotic symmetries of the ALD background, namely to require that the asymptotic symmetry generators have zero symplectic product with the metric and scalar field perturbations that simply change the mass and angular momentum parameters of the black holes -  namely, the parameters $L_{u,v}$ in the solutions above - as the latter should clearly be allowed perturbations of the spacetime. 
%Note that while this vanishing condition is necessary, but not sufficient, we would probably need a more complicated analysis to fully fix the 

To find the asymptotic symmetries, we first 
%
%We proceed the study of the asymptotic symmetries by looking for the AKV. Our strategy is the following, inspired by the ALD analysis in \cite{Georgescu:2022iyx}: we 
choose a gauge that fixes the radial dependence of the diffeomorphism. We will thus be working in radial gauge
%
% %and then we compute the symplectic form between a variation given by $\delta L_u,\delta L_v$ and a variation generated by the corresponding diffeomorphism. For a consistent phase space, we require this to vanish, which  provides constraints on our AKV. N
%
%We start with an arbitrary vector field:
%\begin{align}
%\xi&=F_{\rho}(\rho,U,V)\partial_{\rho}+F_U(\rho,U,V)\partial_U+F_V(\rho,U,V)\partial_V
%\end{align}
%which we choose to be  radial gauge
\begin{align}
(\mathcal{L}_{\xi}g)_{\rho\mu}&=0
\end{align}
The resulting gauge-fixed diffeomorphisms take the form 
%\new{I rescaled $\b$ by $\a'$, please check (checked)}
\begin{align}
\xi&=\frac{\rho^5}{(\rho^4+\rho^2\beta/\a'+L_u L_v)^2}F_{\rho}(U,V)\partial_{\rho}+\bigg(F_U(U,V)+\frac{(\alpha' g_6)^2(\rho^2\partial_v F_{\rho}-L_v\partial_u F_{\rho})}{ \rho^4-L_u L_v }\bigg)\partial_U+\nonumber\\
&\;\;\;\; +\, \bigg(F_V(U,V)+\frac{(\alpha' g_6)^2(\rho^2\partial_u F_{\rho}-L_u\partial_v F_{\rho})}{\rho^4-L_u L_v }\bigg)\partial_V
\label{xirad}
\end{align}
where $F_{U,V,\rho}$ are \emph{a priori} arbitrary functions of their arguments and we introduced the notation 
\be
\beta=\sqrt{p^2+4 \alpha'^2 L_u L_v}
\ee %(if we want to compare with ALD there is a difference of $\alpha'$ in the notation because in ALD we denoted $\beta=\sqrt{p^2+4 \alpha'^2 L_u L_v}$)
This is very similar to the corresponding radial diffeomorphism in the ALD spacetime, except for the fall-off of the radial component. 

%The radial component goes asymptotically as $\rho^{-3}$, compared to the ALD where we had $\rho^{-1}$. In the other components, the radial dependence is just like in ALD.

The symplectic form receives contributions from both the metric and the scalar

\be
\omega=\frac{1}{16\pi G_3}(\omega_{g}+\omega_{sc})
\ee
For an action of the form \eqref{3dact}, 
%\begin{align}
%S&=\frac{1}{16\pi G_3}\int d^3x \sqrt{-g}\bigg(R-\frac{1}{2}f_{ab}\partial_{\mu}\phi^a\partial^{\mu}\phi^b +V(\phi)\bigg)
%\end{align}
%for which 
the general expressions for these contributions, for two arbitrary variations $\delta_1,\delta_2$ of the fields are \cite{Compere:2014bia}
%\new{\cite{Compere:2014bia}: Re-check eq (checked)}
\be
\omega^{\mu}_g=\frac{1}{2}\bigg( (2\nabla_{\lambda}h_{1\rho}^{\mu}-\nabla^{\mu}h_{1\lambda\rho})h_2^{\lambda\rho} -\nabla_{\rho}h_1 h_2^{\rho\mu}+h_1(\nabla_{\rho}h_2^{\rho\mu}-\nabla^{\mu}h_2) - (1\leftrightarrow 2) \bigg)\ee
\be
\omega^{\mu}_{sc}=\frac{\kappa}{2}\bigg(-\delta_1(\nabla^{\mu}F)\delta_2 F-\frac{1}{2}h_1 \nabla^{\mu}F \delta_2 F - (1\leftrightarrow 2)\bigg)
\ee
where $\kappa/2$ is the coefficient of the scalar kinetic term in the action ($\kappa=24$ in our case) and we use the notation $h^i_{\mu\nu}=\delta_i g_{\mu\nu}$ and $h_i=g^{\mu\nu}\delta_i g_{\mu\nu}$. The falloff conditions on the symplectic form that ensure it is conserved and normalisable are, in our coordinates 
\begin{align}
\omega_{UV}=o(\rho^0)\hspace{1cm}\omega_{\rho U,\rho V}=o(\rho^{-1})
\end{align}
where $\omega_{\mu\nu}=\epsilon_{\mu\nu\alpha}\omega^{\alpha}$ and the notation $o(\rho^a)$ signifies that the falloff needs to be faster than $\rho^a$. %For us there is only one scalar, $F$, and $f_{FF}=24$. 
We evaluate the symplectic form on a perturbation that simply changes the black hole parameters
\be
\delta_1=\delta L_u\frac{\partial}{\p L_u}+\delta L_v \frac{\partial}{\p L_v}
\ee
 and one that corresponds to the action of a diffeomorphism of the form \eqref{xirad}, $\delta_2=\mathcal{L}_{\xi}$.

Requiring that the coefficients of $\delta L_u,\delta L_v$  vanish independently, the asymptotic falloff condition on  $\omega_{UV}$ translates into the following two constraints 
\begin{align}
\partial_U F_U+\partial_V F_V&=\frac{\beta}{2 \a' L_v}\partial_V F_U=\frac{\beta}{2 \a' L_u}\partial_U F_V
\end{align}
These equations, up to a factor of 2 difference, are the same as in the ALD background.  The solutions to the equations are given in terms of two arbitrary functions $f,\bar f$
\begin{align}\label{functionfdep}
F_U(U,V)=f(u)+\frac{\beta-\tilde{\beta}}{4 \a' L_u}\bar f(v)\hspace{1cm}F_V(U,V)=\bar f(v)+\frac{\beta-\tilde{\beta}}{4 \a' L_v}f(u)
\end{align}
of the following field-dependent coordinates:
\begin{align}
u=U+\frac{4 \a' L_v}{\beta+\tilde{\beta}}\, V\hspace{1cm}v=V+\frac{4 \a' L_u}{\beta+\tilde{\beta}}\, U
\label{fdepcoord}
\end{align}
where we have introduced the additional  notation
\begin{align}
\tilde{\beta}=%\sqrt{\beta^2-16 L_u L_v}=
\sqrt{p^2-12 \alpha'^2 L_u L_v}
\end{align}
The field-dependent coordinates \eqref{fdepcoord} reduce to $U$ and, respectively, $V$  in the $L_u,L_v\rightarrow 0$ or $\a' \r 0$ limit, while $F_{U,V}$ become purely holomorphic functions of these fixed coordinates. % $F_U(U,V)=f_u(U),F_V(U,V)=f_v(V)$. 
Another interesting fact to note is that the expression for $\tilde \b$ breaks down at an energy or, correspondingly, horizon radius that \emph{precisely} coincides  with the maximum value of the radius of the small black strings we discussed in section \ref{section41thermo} %for $\varkappa=1$

\be
(L_u L_v)_* = \frac{(\tilde r_0^*)^4}{16} = \frac{p^2 (\hat r_0^*)^4}{16 \a'^2} = \frac{p^2}{12 \a'^2}
\ee
%
%Using $r_0=g_6\sqrt{\alpha' p}\hat{r}_0=g_6\alpha' \tilde{r}_0$ and $L_u L_v=\frac{\tilde{r}_0^4}{16}$, one can check that $\tilde{\beta}$ vanishes for
 where $\hat{r}_0^*=\sqrt[4]{\frac{4}{3}}$ is the value of the rescaled radius that corresponds to $T_{max}$ for $\varkappa=1$. 
 
% , which is precisely the value for which the maximum temperature is obtained in \ref{tempfunctionr0}, for $\varkappa=1$ since we are at the attracted value.
%
The $\omega_{\rho U,\rho V}$ components of the symplectic form contain terms that behave have both as $\rho$ and $\frac{1}{\rho}$ at infinity. However, these terms  are total derivatives, in the sense that $\omega_{\rho\sigma}=\partial_{\sigma}(...)$, which vanishes upon integration on the $\sigma$ circle. The radial function $F_{\rho}$ is not fixed by these constraints but, just like in the ALD background, it is fixed by the symplectic form evaluated on variations generated by two different diffeomorphisms of the form above, which yield the constraint 
\begin{align}
L_u\partial^2_V F_{\rho} +L_v\partial^2_U F_{\rho} =\frac{\beta}{2\;\alpha'}\partial_U\partial_V F_{\rho}
\end{align}
The solution to this equation takes the general form
\begin{align}
F_{\rho}(U,V)&=f_{\rho}(u)+\bar{f}_{\rho}(v)
\end{align}
for two new functions of the field-dependent coordinates.

We would now like to compute the charges associated to these asymptotic diffeomorphisms and show they are non-vanishing. Since our background depends on constant parameters, the only nontrivial charges will correspond to the energy and momentum, already computed in \eqref{energyafsol}, \eqref{momentumafsol}. To obtain non-trivial charges associated with the general functions $f, \bar f$ in \eqref{functionfdep}, we first perturb 
the background using an allowed diffeomorphism, denoted $\eta$ and  parametrized by $h(u),\bar h(v), h_{\rho}(u),\bar{h}_{\rho}(v)$,  and then compute the charge difference associated to another asymptotic diffeomorphism,  $\xi$, parametrized by $f(u), \bar f(v), f_{\rho}(u),\bar{f}_{\rho}(v)$. The result takes the form
\begin{align} 
\delta_{\eta}Q_{\xi}&=\frac{g_6^2\alpha'^2}{4\pi G_3}\frac{\tilde{\beta}}{\beta+\tilde{\beta}}\oint d\sigma \bigg[r_u\bigg(2L_u f(u)h'(u)+g_6^2\alpha'^2(f(u)h''_{\rho}(u)-h(u)f''_{\rho}(u))\bigg)-\nonumber\\
&-r_v\bigg(2L_v \bar f(v) \bar h'(v)+g_6^2\alpha'^2(\bar f(v)\bar{h}''_{\rho}(v)-\bar h(v)\bar{f}''_{\rho}(v))\bigg)+\nonumber\\
&+\frac{1}{2}\bigg(r_u f'_{\rho}(u)\bar h'(v)+r_v f_{\rho}(u) \bar h''(v)-r_v \bar{f}'_{\rho}(v)h'(u)-r_u \bar{f}_{\rho}(v)h''(u)\bigg)\bigg]
\end{align}
where $r_{u,v}$ are the normalised radii of the field-dependent coordinates \eqref{fdepcoord} 
\begin{align}
r_u:=\partial_{\sigma}u=\frac{4\alpha'L_v+\beta+\tilde{\beta}}{\beta+\tilde{\beta}}\hspace{2cm}r_v:=\partial_{\sigma}v=\frac{4\alpha'L_u+\beta+\tilde{\beta}}{\beta+\tilde{\beta}}
\end{align}
It is clear that the last line vanishes after integration by parts. We are left with the following result %that is almost the same as in ALD:
\begin{align}\label{chargevar3d}
\delta_{\eta}Q_{\xi}&=\frac{g_6^2\alpha'^2}{4\pi G_3}\frac{\tilde{\beta}}{\beta+\tilde{\beta}}\oint d\sigma\; r_u\bigg(2L_u f(u)h'(u)+g_6^2\alpha'^2(f(u)h''_{\rho}(u)-h(u)f''_{\rho}(u))\bigg) + RM%\left\{\begin{array}{ccc}f,h &\r& \bar f, \bar h \\ u & \r &v\end{array}\right.
\end{align}
where `RM' stands for the expression obtained by replacing $f,h$ etc. by $\bar f, \bar h$ and $u \r v$. 
%right-movers (we call ``left" what depends on $u$ only and ``right" what depends on $v$ only). The result goes to 0 in the limit $L_u,L_v\rightarrow 0$. 
%
This result is extremely similar with the analogous expression in the ALD spacetime. Setting, for simplicity, the radial piece of the diffeomorphism to zero, the difference simply amounts to replacing $\tilde \b$ by $p$ (the same holds at the level of the field-dependent coordinates), which becomes negligible at low energies.%, the difference between the two is negligible. 
%
%Let us set the radial functions and the RM to 0 and compare with ALD for the same choice, ignoring overall factors:
%\begin{align}
%\delta_{\eta}Q^{AF}_{\xi}&=\frac{\tilde{\beta}}{\beta+\tilde{\beta}}\oint d\sigma\; r_u L_u f_u(u)h'_u(u)\hspace{0.6cm}\delta_{\eta}Q^{ALD}_{\xi}=\frac{p}{p+\beta}\oint d\sigma\; r_u L_u f_u(u)h'_u(u)
%\end{align}
%The periodicities that appear are also different, since the field-dependent coordinates are different. In AF $r_u=\frac{4L_v+\beta+\tilde{\beta}}{\beta+\tilde{\beta}}$, while in ALD we have $r_u=\frac{\beta+p+2L_v}{2p}$. The formulae above suggest that one needs to replace $\tilde{\beta}$ by $p$, but this statement is not precise. At very large $p$, the results should be similar and it would be interesting to explore the large $p$ limit of the ALD background.

To conclude, it is rather remarkable that i) the asymptotic symmetries that we have obtained in this simple manner are so similar to the symmetries of single-trace  $T\bar T$ (identified with the asymptotic symmetries of the ALD spacetime) and ii) they appear to know about the thermodynamic stability of the backgrounds that they depend on, in the sense that the symmetry generators become imaginary precisely when one reaches the large, unstable black string region. It would be very interesting to better understand this connection between asymptotically flat backgrounds and $T\bar T$.

\section{Discussion}
\label{discussion}
In this article, we have argued that turning on a particular $21$-dimensional subset of the $22$ maximally supersymmetric $(2,2)$ single-trace irrelevant  deformations of the D1-D5 CFT for a \emph{finite} amount leads to a UV-complete theory decoupled from gravity, which is in general  a deformation of little string theory, compactified on K3. Our argument was largely based on the supergravity description of these deformations, which are characterised by the degeneration of the flat asymptotia to linear dilaton-like ones in specific ways that we could identify with  known decoupling limits of string theory.  By studying the behaviour of the entropy in these theories - inferred from that of the holographically dual black holes -  we also argued that the %se  (\emph{a priori} intractable) 
resulting two-dimensional theories  should admit a $T\bar T$ - like effective description.  Finally, we put forth evidence that, for low enough energies,  asymptotically flat black strings also  admit a similar description, as suggested by the resemblance between the asymptotic symmetry generators we found for flat space and the  field-dependent coordinate transformations characteristic of $T\bar T$. 

If this global picture is correct, there are many interesting future directions to explore. 

One such direction is to gain a  better technical understanding of the general ALD-like spacetimes we discussed and their associated holographic dictionary. One may, for example, study the asymptotic symmetries of the general such spacetimes and check whether they, too, display $T\bar T$ - like features.  Another interesting question is whether the behaviour of the entropy and symmetries changes if  more than two $\l_{\bar \imath}$ deformations are turned on at the same time.
%, and whether there are differences between the different decoupled theories. 
One may also try to  understand the precise map between supergravity modes and their dual operators, by adapting the holographic renormalisation technique to this setup (see \cite{Marolf:2007ys} for some early work); the careful linearised analysis of the fields in  a consistent truncation of the standard ALD background, performed in \cite{Georgescu:2022iyx}, already showed several unusual features. %\textcolor{magenta}
{This holographic analysis should also shed light on how the holographic expectation values appear in the asymptotics of the metric and of the other fields, an issue that  our analysis of the NCOS backgrounds in appendix \ref{appendixdecoupling} indicated was rather subtle.} It is also interesting to understand how the matching of supersymmetry multiplets works in these cases; for example, whether fields in the same  multiplet acquire the same momentum-dependence in their radial fall-offs.

 All  these analyses would presumably provide important clues for what the properties of the dual theories should be and, in particular, which observables should be expected to be universal. In addition, this could prove to be a useful warm-up  exercise for flat space holography, with which the ALD-like backgrounds studied herein share many features \cite{Marolf:2006bk}, but without some of the puzzles related to decoupling and the black hole instability. 

Another possibility is to approach the same problem from the field-theory perspective, using conformal perturbation theory and/or supersymmetry. One may hope, for example, that for the case of the $SO(21)$ - singlet irrelevant deformation, its mixing with other operators  may be constrained enough to render the deformation solvable.  The conformal perturbation theory calculation may be performed at a generic point in moduli space, or at the orbifold point. In the latter case, it may be possible to understand the exact single-trace $T\bar T$ deformation in terms of the dual worldsheet theory and, to the extent this field progresses \cite{Gaberdiel:2023lco,Frolov:2023pjw}, study the other types of deformations by turning on RR fields in the bulk.

Still from the boundary perspective, it would be very interesting to further explore the properties of non-commutative open string theories, and  
%\textcolor{blue}{, in particular, understand the assumptions that underlie the particular energy dependence of the five-dimensional NCOS decoupling parameters, which is ultimately responsible for the observed Hagedorn behaviour in the supergravity dual.  It would of course be interesting if this could be} 
possibly  extend this to general OD$p$. 

%A better understanding of NCOS and OD$p$/OBLST theories would also be desirable, and in particular of the assumptions that underlie the finite-temperature decoupling limit reviewed in section \ref{section42ncoshag}.  

Finally, it would be interesting to explore whether the perspective of section \ref{section4} can shed light on  more realistic black holes, such as  Reissner-Nordstr\"{o}m or Kerr ones, %\textcolor{magenta}
{or more general ones in the string/M-theory context. For example, in  \cite{Manschot:2022lib} it has been observed that the spectrum of BPS black holes in Calabi-Yau compactifications of M-theory displays $T\bar T$ - like behaviour away from the strict decoupling limit; it would be interesting to understand what happens with generic states of this system. 
}

%
%\bi
%\item we argued there exist finite flows $\r$ UV-complete theories (LST-like)
%\item intractable, but effective $T\bar T$ behaviour (thermodyn)
%\ei

%This motivates a better technical understanding of the ALD-like theories, both in supergravity and on the dual side:  , ASG ODp, holographic renormalisation. Interesting to study the sugra description of the full moduli space of ALD-like backgrounds, which may have $SO(5) \times SO(21)$, which would be interesting to reproduce from dual perspective (maybe what one needs is to show that the mass of the $3d$ fields in the decoupling limit becomes zero asymptotically, and so they are equivalent to the moduli).  Loose ends: better understanding of the NCOS decoupling limit at finite temperature and see if correct IR behaviour can be reproduced, better understanding of holographic map between $c_\L$ and the coefficients of the irrelevant deformations (holonomy business). 
%
%
%In field theory, try understanding flow upwards a la $T\bar T$ e.g. for SO(21), holonomy, can one see quadratic constraint on $c_\L$ in CPT?
%
%How does this help? Maybe effectve $T\bar T$ theories have a universal symmetry definition, that would fix form of entropy and correlations. 
%
%\bi
%\item could check ASG of ODp, other universal observables?
%\item precision holo $\r$ correlation functions, understand stress tensor sector
%\item field theoretical analyses: what happens to chiral primaries?
%\item flow over moduli space?
%\item CPT, k=1?
%\ei

\subsection*{Acknowledgements}
We are grateful to Iosif Bena, Nikolay Bobev, Soumangsu Chakraborty, Ruben Minasian  and especially  Kyriakos Papadodimas for insightful conversations. The work of NK is supported by the ERC Consolidator Grant 772408-Stringlandscape. The work of SG is supported by the ERC Starting Grant 853507.  SG and NK  also thank CERN and EPFL, where part of this work was carried out, for their hospitality.

\appendix

\section{Details of the six-dimensional supersymmetric solutions}

In this appendix, we fill in many of the details of the supersymmetric solutions discussed in section \ref{sugrareview}, including conventions and the explicit form of the Killing spinors.

\subsection{Supersymmetry conventions}
\label{sec: AppA - 6d susy solutions}
%{\color{red}(Recheck consistency of notation with the whole paper, $\bar{\imath}$ instead of $I$ etc)}

In six dimensions, the R-symmetry group is 
%\cite{} {\color{ForestGreen}(see the table in van Proeyen's book pg 240, but he uses conventions in which $N=(1,0)$ corresponds to $N_L=2,N_R=0$; the $USp$ groups are only for even $N$s)} $\text{USp}(N_L)\times \text{USp}(N_R)$, \emph{where $N$ is what??!!} {\color{ForestGreen} (where $N=(N_L/2,N_R/2)$ counts the number of supersymmetries; I agree it's confusing, we can write directly the particular cases we are interested in)} so we have 
USp$(2)$ for $\N=(1,0)$ supergravity and USp$(4)$ for $\N=(2,0)$. These groups are isomorphic to spin groups \cite{romans}
\begin{align}
\text{USp}(2)\cong SU(2)\cong Spin(3)\nonumber\\
\text{USp}(4)\cong Spin(5)\label{Rsym1}
\end{align}
and we can construct their generators  in the standard way using gamma matrices in three and, respectively, five Euclidean dimensions. In $\N=(1,0)$ supergravity, the spinors are USp$(2)$ doublets of six-dimensional Weyl spinors satisfying the symplectic Majorana condition; in $\N=(2,0)$, they are  quadruplets of  six-dimensional Weyl spinors that transform under USp$(4)$ and  satisfy the corresponding symplectic Majorana condition.

In the following,  $\mu,\nu$ will denote curved spacetime indices, and $a,b$ 
will be six-dimensional tangent space indices. 
%\emph{Comment about how in $(1,0)$, this is trivial?} {\color{ForestGreen} (the spinors transform indeed under $SU(2)$ as doublets, but in eq 3.8 from Romans since the connection $Q=0$ just because it would be a 1d antisym matrix, the big Gamma matrices of $Spin(3)$ don't appear. Next, the index $i$ takes only one value, so $F^i$ is not a vector under $Spin(3)$ but a singlet, so in eq A.3, A.4 below we don't actually have the big Gamma matrix there. Hence, the components of the doublets are not mixed in the eqs and we can just solve for the components without writing indices. Is the explanation ok? I also had this question when I was solving these eqs)}
In $\N=(1,0)$ theories, the $USp(2)$ structure is trivial, as all bosonic quantities are singlets under it, and all fermionic ones are doublets; we will consequently drop all the associated indices. In $\N=(2,0)$ theories, given the correspondence between the fundamental and, respectively, the traceless antisymmetric representation of $USp(4)$ on the one hand and the spinor and, respectively, the vector representation of $Spin(5)$ on the other, we will be using $Spin(5)$ spinor indices  $A,B=\{ 1,\ldots, 4\}$ on the fermions, and vector indices 
$i,j=\{1,\ldots 5\}$ on the bosons. % transforming in the vector representation of $Spin(5)$, we will also need the indicestransform in the spinor representation of $Spin(5)$. % The flat $SO(P)\times SO(Q)$ metric is $(\underbrace{+,...,+}_{P},\underbrace{-,...,-}_{Q})$. 
%
%The indices $i,j$ will be taken to transform  in the vector representation of the R-symmetry group \eqref{Rsym1},
 Meanwhile, the indices  $\bar \imath,\bar \jmath=\{ 1, \ldots Q\}$ transform in the vector representation of $SO(Q)$.  We will be using the notation $\gamma^a$ for the six-dimensional Lorentzian gamma matrices, and $\Gamma^i$ for the five-dimensional Euclidean gamma matrices. 
The chirality matrix in six dimensions is defined as  
%\emph{Cite!} {\color{ForestGreen}(Polchinski vol 2 eq B.1.11 in $d=2k+2$ dimensions; however he works with mostly plus signature and I am not sure if the definition changes if we change signature, this would just switch left with right. Mikhailov worked in mostly minus signature and his spinors were left though. I checked that $(\gamma_7)^2=I$ as it should, the only question regards the minus in front.)}
\begin{align}
\gamma_7&=-\gamma^0...\gamma^5
\end{align}
and the chirality projectors are given by $P_{\pm}=\frac{I_7\pm\gamma_7}{2}$.
%{\color{ForestGreen} (I check that $P_+^2=P_+$ for this chirality matrix)}. 
The left and right spinors satisfy $P_{+}\varepsilon_+=\varepsilon_+$ and $P_{-}\varepsilon_-=\varepsilon_-$, respectively.

In $\mathcal{N}=(2,0)$ theories, the supersymmetry variations are given by 
%\emph{In both, or just $(2,0)$? Different product $\g$!} {\color{ForestGreen} (the big Gammas don't appear in the N=1 because $F^i$ are singlets; I am not sure how to write both cases in one eq; the small gammas are the same, wrt Romans I just traded the antisym versions by products because they are dotted into forms which impose the antisym)}
\begin{align}
\delta\psi_{\mu A}&=\nabla_{\mu}\varepsilon_A+\frac{1}{4}F_{\mu a b}^{+i}\gamma^{a}\gamma^{b}(\Gamma^i)_{A}^{\;B}\varepsilon_B\\
\delta\chi^{\bar{\imath}}_A&=\frac{1}{\sqrt{2}}\gamma^{\mu}P_{\mu}^{i\bar{\imath}}(\Gamma^i)_A^{\;B}\varepsilon_B-\frac{1}{12}\gamma^{\mu}\gamma^{\nu}\gamma^{\rho}F^{-\bar{\imath}}_{\mu\nu\rho}\varepsilon_A \label{tensorinoeq1}
\end{align}
where the covariant derivative contains both the spacetime spin connection, $\omega_\mu{}^{ab}$, and the $SO(P)$ 
%\emph{??} {\color{ForestGreen} (because $\varepsilon$ transforms as a spinor of $Spin(5)$ which we identify with $SO(5)$; $Spin(5)$ is the universal covering of $SO(5)$, but all the coset discussion is local, so we probably don't care about global properties (all these papers don't) )} 
one, $Q_\mu^{ij}$
\be \label{generalcovariantder}
\nabla_{\mu}\varepsilon_A=\bigg(\partial_{\mu}+\frac{1}{4}\omega_{\mu}^{\;ab}\gamma_a\gamma_b \bigg)\varepsilon_A-\frac{1}{4}Q_{\mu ij}(\Gamma^i\Gamma^j)_A^{\;B}\varepsilon_B
\ee
whose construction in terms of the decomposition of the Lie algebra element $dg\, g^{-1}$ was explained in section \ref{sugrareview} and is given by 
% {\color{ForestGreen} (it agrees with 3.6 in Romans and it's also that for this sign we get the covariant derivative appearing in our eqs)}
\be
Q^{ij}= (dg g^{-1})^{ij}=-X^{i\Lambda}dX^j_{\Lambda}\;, \;\;\;\;\;\;
Q^{\bar{\imath}\bar{\jmath}}= (dg g^{-1})^{\bar{\imath}\bar{\jmath}}=X^{\bar{\imath}\Lambda}dX^{\bar{\jmath}}_{\Lambda}
\ee
We have also included the $SO(Q)$ connection, for completeness. 
In $\mathcal{N}=(1,0)$ theories,  there is only one self-dual form, and so the index $i$ takes a single value. Consequently,  $Q=0$, case in which the equations simplify to \eqref{kseqn1}.

%In the scalar coset, by taking $V$ to be the coset representative, we look at $dV V^{-1}$ and expand it in terms of the generators (send to \cite{ferraraUduality} for details). In the matrix realization we get the expressions in Romans (more details?). We identify the vielbein:
%\begin{align}
%P^{iI}&=\frac{1}{\sqrt{2}}X^{i\Lambda}dX^I_{\Lambda}=-\frac{1}{\sqrt{2}}dX^{i\Lambda} X^I_{\Lambda}
%\end{align} 

\subsection{Details of the computation of the $(1,0)$ solutions}

In the $\N=(1,0)$ case, 
we can replace the doublet of spinors satisfying the Majorana condition by a single unconstrained spinor. 
Let us now explicitly solve the Killing spinor equations \eqref{kseqn1} in the $(1,0)$ theory.  Our derivation closely follows that of \cite{mikhailov,Raeymaekers:2006np} for $\mathcal{N}=(2,0)$. We choose (Hopf) coordinates on the sphere, denoted  $\{\psi,\theta,\varphi\}$
\begin{align} \label{metrics3}
d\Omega^2_3&=\frac{1}{4}(d\theta^2+d\psi^2+d\varphi^2 + 2\cos\theta d\varphi d\psi)
\end{align}
The orientation on $S^3$ is
\begin{align} \label{volumeforms3}
\omega_{S^3}=\frac{\sin\theta}{8}d\psi\wedge d\theta \wedge d\varphi%=\frac{e^{3U}}{r^3}e^3\wedge e^4\wedge e^5
\end{align}
and use the convention  $\epsilon_{r\s t\psi\theta\varphi}=+ \sqrt{g}$ when computing the Hodge dual. %{\color{red}(Should we also move this to the conventions subsection?)} 
The vielbein associated to the metric \eqref{eq: metric Ansatz} is given by 
\be \label{vielbein6dm}
e^{0}=e^U dt\;,  \hspace{0.3cm}e^{1}=e^U d\s \;, \hspace{0.3cm}e^{2}=e^{-U} dr\;, \hspace{0.3cm}e^{3}=\frac{r \, e^{-U}}{2} (d\psi+\cos\theta d\varphi)
\ee
\be e^{4}=\frac{r \, e^{-U}}{2}d\theta\;,  \hspace{0.3cm}e^{5}=\frac{r \, e^{-U}}{2}\sin\theta d\varphi
\ee
and the nonzero components of the spin connection $\omega_{\mu}^{\;ab}=e^{\;a}_{\nu}\nabla_{\mu}e^{\;\nu b}$ are
\begin{align} \label{spinconn}
\omega_{02}&=e^{U}U'e^0\;, \hspace{0.5cm}
\omega_{12}=-e^{U}U'e^1 \;, \hspace{0.5cm}\omega_{23}=\frac{e^{U}(1-rU')}{r}e^3 \;, \hspace{0.5cm}\omega_{24}=\frac{e^{U}(1-rU')}{r}e^4\\
\omega_{25}&=\frac{e^{U}(1-rU')}{r}e^5 \;, 
\hspace{0.5cm}\omega_{34}=\frac{e^{U}}{r}e^5\;, \hspace{0.5cm}\omega_{35}=-\frac{e^{U}}{r}e^4\;, \hspace{0.5cm}\omega_{45}=\frac{e^U}{r}(-e^3+2\cot\theta e^5)
\end{align}
where $U'=\p_r U$. In terms of the vielbeine, our Ansatz \eqref{formsusyvar} for the 3-forms is given by 
%{\color{ForestGreen}(maybe we can stress here that these are really SD/ASD, different from the forms $F$ in the main text which are closed and that this happens because the forms that appear in the susy variations are the SD/ASD ones)}
\begin{align} \label{formsansatz}
H^+&=\frac{Z e^{3U}}{r^3}\big(e^3\wedge e^4\wedge e^5-e^0\wedge e^1\wedge e^2\big)\\
H^{-\bar{\imath}}&=\frac{Z^{\bar{\imath}} e^{3U}}{r^3}\big(e^3\wedge e^4\wedge e^5+e^0\wedge e^1\wedge e^2\big)
\end{align}
Plugging this into the tensorino equations \eqref{tensorinoeq1} and using the chirality of the $6d$ spinor they can be reduced to 
\begin{align}
%\frac{1}{\sqrt{2}}\gamma^a P^I_a\varepsilon -\frac{1}{2}\frac{Z^{\bar{\imath}} e^{3U}}{r^3}(\gamma^0\gamma^1\gamma^2  +\gamma^3\gamma^4\gamma^5)\varepsilon=
\frac{1}{\sqrt{2}}\gamma^2  e_2^r P^{\bar{\imath}}_r\varepsilon -\frac{Z^{\bar{\imath}} e^{3U}}{r^3}\gamma^0\gamma^1\gamma^2\varepsilon=0
\end{align}
As explained in the main text, for $Z_{\bar{\imath}} \neq 0$ this leads to the projection 
\begin{align} \label{projs1}
\g^0\g^1 \varepsilon = \varepsilon
\end{align}
as well as the equation \eqref{tensori2} for $P_r^{\bar \imath}$. 

Let us now turn to the gravitino equations. Writing them explicitly in components, they read
\begin{align}
&e^{-U}\partial_t\varepsilon+\frac{1}{2}e^U U' \gamma^0\gamma^2\varepsilon-\frac{1}{2}\frac{Z e^{3U}}{r^3}\gamma^1\gamma^2\varepsilon=0\\
&e^{-U}\partial_\s\varepsilon+\frac{1}{2}e^U U' \gamma^2\gamma^1\varepsilon-\frac{1}{2}\frac{Z e^{3U}}{r^3}\gamma^2\gamma^0\varepsilon=0\\
&e^{U}\partial_r\varepsilon-\frac{1}{2}\frac{Z e^{3U}}{r^3}\gamma^0\gamma^1\varepsilon=0\\
&\frac{2 e^U}{r}\partial_{\psi}\varepsilon+\frac{1}{2}\bigg( \frac{e^U(1-rU')}{r}\gamma^2\gamma^3 -\frac{e^U}{r}\gamma^4\gamma^5 \bigg)\varepsilon+\frac{1}{2}\frac{Z e^{3U}}{r^3}\gamma^4\gamma^5\varepsilon=0\\
&\frac{2 e^U}{r}\partial_{\theta}\varepsilon+\frac{1}{2}\bigg( \frac{e^U(1-rU')}{r}\gamma^2\gamma^4 -\frac{e^U}{r}\gamma^3\gamma^5 \bigg)\varepsilon+\frac{1}{2}\frac{Z e^{3U}}{r^3}\gamma^5\gamma^3\varepsilon=0\\
&\frac{2 e^U}{r\sin\theta}(\partial_{\varphi}-\cos\theta\partial_{\psi})\varepsilon+\frac{1}{2}\bigg( \frac{e^U(1-rU')}{r}\gamma^2\gamma^5 +\frac{e^U}{r}\gamma^3\gamma^4 +\frac{e^U 2\cot\theta}{r}\gamma^4\gamma^5 \bigg)\varepsilon+\frac{1}{2}\frac{Z e^{3U}}{r^3}\gamma^3\gamma^4\varepsilon=0
\end{align}
Using the projection \eqref{projs1} inside these equations, one can easily factorize the $r$ dependence. Requiring that the $t$ (or $\s$) equation does not impose any additional projection on $\varepsilon$ sets \footnote{More generally, one easily sees $e^{2U} (U' - Z e^{2U}/r^3)$ is $r$ independent, as should be.} 

\be
U' = \frac{Z e^{2U}}{r^3}
\ee
which in particular implies that, in this coordinate system, $\p_t \varepsilon = \p_\s \varepsilon =0$. The radial and $S^3$ equations can then be trivially solved to give 
\begin{align}\label{solution1}
\varepsilon&=e^{\frac{U(r)}{2}}e^{-\frac{\theta}{2}\gamma^2\gamma^4}e^{-\frac{\varphi}{2}\gamma^2\gamma^3}\varepsilon_0
\end{align}
with $\varepsilon_0$ coordinate-independent and satisfying 
\begin{align}
\gamma^0...\gamma^5\varepsilon_0=-\varepsilon_0\hspace{1cm}\gamma^0\gamma^1\varepsilon_0=\varepsilon_0
\end{align}
which follow from the corresponding projections on $\varepsilon$.
The metric and scalar fields $X^\L$ obey \eqref{tens1}, \eqref{functionmetric}; how to solve these equations is explained in the main text. 

\subsection{Details on the computation of the $(2,0)$ solutions}

In the $\mathcal{N}=(2,0)$ case, there are four Killing spinors, $\varepsilon_A$, related via a symplectic Majorana condition. %\emph{(Why) can we ignore this thoughout?} {\color{ForestGreen}(I need to think about it. I was asking myself the same thing when I was solving the eqs.)} 
We will package them into a four-index object, on which the $\Gamma^i$ matrices act. %\emph{More details..} {\color{ForestGreen}(the spinors transform under $USp(4)$ can be read as "the fundamental representation of $USp(4)$ is the same as the spinor representation of $Spin(5)$, so the objects transforming under it are basically spinors". The spinors have 4 components, matching the number of components of Gamma matrices in 5d)}
The formulae for the vielbeine, spin connection and field strengths are the same as before \eqref{vielbein6dm}, \eqref{spinconn} and \eqref{formsansatz}. 
%
%
%and its inverse is given by $e^{\mu}_i=e^j_{\nu}g^{\mu\nu}\eta_{ij}$, so
%\begin{align}
%\hspace{-0.6cm}e_{0}=e^{-U} \partial_t \hspace{0.3cm}e_{1}=e^{-U} \partial_x\hspace{0.3cm}e_{2}=e^{U} \partial_r\hspace{0.3cm}e_{3}=\frac{2e^U}{r}\partial_{\psi}\hspace{0.3cm}e_{4}=\frac{2e^U}{r}\partial_{\theta} \hspace{0.3cm}e_{5}=\frac{2e^U}{r\sin\theta}(\partial_{\varphi}-\cos\theta\partial_{\psi})
%\end{align}
%
The gravitino equations, written explicitly in components are 
%\emph{Is it $D_r$ or $\nabla_r$?} {\color{ForestGreen}(here the covariant derivative contains spacetime spin connection and $SO(5)$ connection)}
\begin{align}
&2e^{-2U}\partial_t\varepsilon+\gamma^0\gamma^2\left(U' - \frac{Z^i e^{2U}}{r^3} \Gamma^i\gamma^0\gamma^1\right)\varepsilon=0\\
&\partial_\s\varepsilon-\gamma^0\gamma^1\partial_t\varepsilon=0\\
&2D_r\varepsilon-\frac{Z^i e^{2U}}{r^3} \Gamma^i\gamma^0\gamma^1\varepsilon=0\label{eqr}\\
&4\partial_{\psi}\varepsilon+(1-rU')\gamma^2\gamma^3\varepsilon -\left(1-r \frac{Z^i e^{2U}}{r^3}\Gamma^i\right)\gamma^4\gamma^5 \varepsilon=0\\
&4\partial_{\theta}\varepsilon+(1-rU')\gamma^2\gamma^4\varepsilon -(1+r \frac{Z^i e^{2U}}{r^3}\Gamma^i)\gamma^3\gamma^5\varepsilon=0\\
&4\partial_{\varphi}\varepsilon+\big(\cos\theta +\sin\theta\gamma^3\gamma^5 \big)\bigg((1-rU')\gamma^2\gamma^3+(1+r \frac{Z^i e^{2U}}{r^3}\Gamma^i)\gamma^4\gamma^5\bigg)\varepsilon=0
\end{align}
 Looking for time-independent solutions, we see from the first two equations that they also need to be $\s$-independent and satisfy
\begin{align}
U'=\pm \frac{|Z^+|e^{2U}}{r^3}\hspace{2cm}\hat{Z}^i\Gamma^i\gamma^0\gamma^1\varepsilon =\pm\varepsilon
\end{align}
We focus on plus sign. %\emph{Briefly explain why}. 
Plugging the projection into the above equations and using six-dimensional chirality, one recovers  exactly the same sphere equations as in the $(1,0)$ case, obviously with the same solution. After using the projection,  the $r$ equation reads
\begin{align}
2D_r \varepsilon =2\bigg(\partial_r\varepsilon-\frac{1}{4}Q_{rij}\Gamma^{ij}\varepsilon\bigg)=\frac{|Z^+| \hat{Z}^i e^{2U}}{r^3}\Gamma^i\gamma^0\gamma^1\varepsilon=U'\varepsilon
\end{align}
which fixes the $r$ dependence and yields the same solution as before, if we set $Q^{ij} =0$.

% In particular, we see that $k^i=U' \hat{Z}^i$.
  The solution for the Killing spinors is of the form
\begin{align}
\varepsilon(r,t,\s,S^3)&=e^{-\frac{\theta}{2}\gamma^2\gamma^4}e^{-\frac{\varphi}{2}\gamma^2\gamma^3}\varepsilon_0(r)
\end{align}
where $\varepsilon_0(r)$ is a $6d$ Majorana-Weyl spinor that depends on $r$ only. It needs to satisfy:
\begin{align}
D_r\varepsilon_0=\frac{U'}{2}\varepsilon_0
\end{align}
and the projection since the $\gamma^0\gamma^1$ commute with the exponentials from $S^3$:
\begin{align}
\hat{Z}^i\Gamma^i\varepsilon_0=\gamma^0\gamma^1\varepsilon_0
\end{align}
We can see that the covariant derivative on the projection gives:
\begin{align}
D_r(\hat{Z}^i\Gamma^i)\varepsilon_0+\hat{Z}^i\Gamma^i\frac{U'}{2}\varepsilon_0=\frac{U'}{2}\gamma^0\gamma^1\varepsilon_0=\frac{U'}{2}\hat{Z}^i\Gamma^i\varepsilon_0
\end{align}
which requires for consistency
\begin{align}
D_r \hat{Z}^i=0
\end{align}
%{\color{red}\emph{What happened to the symplectic projection? Does it commute with this?}}
%
The tensorini equations, after using the projection  $\hat{Z}^i\Gamma^i\varepsilon=\gamma^0\gamma^1\varepsilon$
give
\begin{align}
\bigg(\frac{1}{\sqrt{2}}P_{r}^{i\bar{\imath}}-\frac{Z^{\bar{\imath}} e^{2U}}{r^3}\hat{Z}^i\bigg)\Gamma^i\varepsilon_0=0
\end{align}
which implies that we have the following equations for the function $U(r)$
\begin{align}
U'=\frac{|Z^+|e^{2U}}{r^3} \;, \hspace{1cm}\frac{1}{\sqrt{2}}P_{r}^{i{\bar{\imath}}}=\frac{Z^{\bar{\imath}} e^{2U}}{r^3}\hat{Z}^i%=\frac{q^{\Sigma} e^{2U}}{r^3}\hat{Z}^i X^{\bar{\imath}}_{\Sigma}
\end{align}
In the second equation, after plugging in the definition of $P_r$  and $Z_{\bar \imath}$,  we obtain
\begin{align}
-\frac{1}{2}X^{\bar{\imath}}_{\Lambda}\partial_r X^{i\Lambda}=\frac{q^{\Lambda} e^{2U}}{r^3}\hat{Z}^i X^{\bar{\imath}}_{\Lambda}
\end{align}
We multiply by $X^{\bar{\imath}}_{\Sigma}$, sum over $\bar{\imath}$ and use $X^{\bar{\imath}}_{\Lambda}X^{\bar{\imath}}_{\Sigma}=X^j_{\Lambda}X^j_{\Sigma}-\eta_{\Lambda\Sigma}$:
\begin{align} \label{eqcovde5}
\partial_r X^{i}_{\Sigma}-X^j_{\Lambda}X^j_{\Sigma}\partial_r X^{i\Lambda}=2\frac{e^{2U}}{r^3}\hat{Z}^i (X^j_{\Sigma}Z^j-q_{\Sigma})
\end{align}
Then, using the $SO(5)$ connection $Q_{\mu}^{ij}=-X^{i\Sigma}\partial_{\mu}X^j_{\Sigma}$ and $X^i_{\Lambda}X^{j\Lambda}=\delta^{ij}$, we recognize on the left-hand side the covariant derivative
\begin{align}
D_{r}X^{i}_{\Lambda}&=\partial_r X^{i}_{\Lambda}-Q^{ij}_r X^{j}_{\Lambda}=\partial_r X^{i}_{\Lambda}+X^{i\Sigma}X^j_{\Lambda}\partial_r X^{j}_{\Sigma}
\end{align}
Hence, we can write
\begin{align}
D_r X^i_{\Lambda}=\frac{2e^{2U}}{r^3}\hat{Z}^i (X^j_{\Lambda}Z^j-q_{\Lambda})
\end{align}
Introducing $\tau=\frac{1}{r^2}$, we obtain the equations
\begin{align}
\partial_{\tau}(e^{-2U})=|Z^+|\hspace{1cm}D_{\tau}X^{i}_{\Lambda}=e^{2U}\hat{Z}^i(q_{\Lambda}-Z^j X^j_{\Lambda})
\end{align}
which imply that
\begin{align}
D_{\tau}(e^{-2U}X^i_{\Lambda})&=\hat{Z}^i(q_{\Lambda}-Z^j X^j_{\Lambda})+|Z^+|X^i_{\Lambda}=\hat{Z}^i q_{\Lambda}+|Z^+|(X^i_{\Lambda}-\hat{Z}^i \hat{Z}^j X^j_{\Lambda})
\end{align}
Multiplying by $\hat{Z}^i$ and summing over $i$ kills the second term on the left-hand side and, since $D_{r}\hat{Z}^i=0$, we can write
\begin{align}
D_{\tau}(e^{-2U}\hat{Z}^i X^i_{\Lambda})=q_{\Lambda}
\end{align}
where we used the normalization $\hat{Z}^i \hat{Z}^i=1$. Since the quantity above is a $SO(5)$ scalar, we can turn the covariant derivative into a partial derivative and since $q_{\Lambda}$ is constant, we can integrate:
\begin{align}
e^{-2U}\hat{Z}^iX^i_{\Lambda}=\tau q_{\Lambda}+c_{\Lambda}
\end{align}
We can now solve for the parallel component $X^i_{\Lambda}\propto \hat{Z}^i$, and obtain
\begin{align}
X^i_{\Lambda}&=\hat{Z}^i (\tau q_{\Lambda}+c_{\Lambda}) e^{2U}+\bar{g}^i_{\Lambda}
\end{align}
where we denoted by $\bar{g}^i_{\Lambda}$ the perpendicular part. From here we can extract the function entering the metric to be given by
\begin{align}
e^{-4U}=\big(q_{\Lambda}\tau+c_{\Lambda}\big)\big(q^{\Lambda}\tau+c^{\Lambda}\big)
\end{align}
The remaining details of the solution are given in the main text.

\section{Attractor solutions within a consistent truncation \label{attrsolsB}}

In this appendix, we explicitly work out the map between solutions to ten-dimensional supergravity whose bosonic part fits within a consistent truncation to six dimensions and the six-dimensional language of section \ref{sugrareview}. 

\subsection{A consistent truncation with $SO(2,2)$ symmetry}
\label{sec: Appendix B - consistent truncation}

We will be using a bosonic consistent truncation of ten-dimensional supergravity  that preserves an  $SO(2,2)$ subgroup of the original global $SO(5,21)$ symmetry group of this low-energy theory, which is given in\footnote{To match the standard conventions, we have flipped the sign of $\chi_1$ with respect to this reference. }  \cite{Duff:1998cr}
\begin{align} \label{lagtrunc}
 \mathcal{L} &= \frac{1}{8 \pi^3 \a'^2} \bigg(  R - \frac{1}{2} (\p\phi_1)^2 -  \frac{1}{2} (\p\phi_2)^2 - \frac{1}{2} e^{2\phi_1} (\p\chi_1)^2 - \frac{1}{2} e^{2\phi_2} (\p\chi_2)^2 - \frac{e^{-\phi_1-\phi_2}}{12}  H_3^2 -\nonumber\\
 &- \frac{e^{\phi_1-\phi_2}}{12}  (F_3 - \chi_1 H_3)^2 + \chi_2 H_3 \wedge F_3  \bigg)
\end{align}
where $\phi_1 +\phi_2 \equiv 2 \phi$, $\phi_1 - \phi_2 \equiv - 2\rho$ is the logarithm of the string frame volume of the K3, $\chi_1$ is the $10d$ axion and $\chi_2$, the part of $C^{(4)}$ that is proportional to the volume of K3. %{\color{red}\emph{Also comment on $G_6$ and how we do S-duality.}} 
 As discussed at length in \cite{Duff:1998cr} the global $SO(2,2) \sim SL(2,\mathbb{R}) \times SL(2,\mathbb{R})$ %(a subgroup of $SO(5,21)$ in the full type IIB) 
 acts on the scalar combinations
\begin{align}
\tau_1=\chi_1+i e^{-\phi_1} \hspace{1cm} \tau_2=\chi_2+i e^{-\phi_2}
\end{align}
via fractional linear transformations
%on which the two (independent) transformations act as:
\begin{align}
\tau \rightarrow \frac{a  \tau +b }{c \tau +d}\;, \;\;\; ad-bc = 1\hspace{1cm} a,b,c,d \in \mathbb{Z}
%\tau_2\rightarrow \frac{a_2 \tau_2+b_2}{c_2\tau_2 +d_2}
\end{align}
where $\tau \in \{\tau_1, \tau_2\}$. Note that the transformation of $\tau_1$ is nothing but type IIB S-duality.

%where $a_{1,2} d_{1,2}-c_{1,2} b_{1,2}=1$, meaning that the matrices below are in $SL(2,\mathbb{R})$:
%\begin{align}
%\Lambda_{1,2}:=\begin{pmatrix}
%a_{1,2} & b_{1,2}\\
%c_{1,2} & d_{1,2}
%\end{pmatrix}\in SL(2,\mathbb{R})
%\end{align}

We would now like to put this action in the general form described in section \ref{sugrareview}.

\subsubsection*{The $SO(2,2)$ vielbein}

As explained in the main text, the vielbein can be found from the matrix $\N = g^T g$, where $\N$ itself is constructed from the gauge field kinetic ($\mathcal{G}$) and axionic ($\mathcal{B}$) terms.  The gauge kinetic matrix is 
\be
\mathcal{G} = \left(\begin{array}{cc} e^{-\phi_1-\phi_2} + \chi_1^2 e^{\phi_1-\phi_2} & - \chi_1   e^{\phi_1-\phi_2}  \\ - \chi_1 e^{\phi_1-\phi_2} &  e^{\phi_1-\phi_2}   \end{array}\right)\;, \;\;\;\;\;\; \mathcal{B} =  \left(\begin{array}{cc} 0 & \chi_2 \\ -\chi_2 & 0 \end{array}\right)
\ee
From this, one can easily build the matrix $\N_{\L\Sigma}$  using its definition \eqref{matrixNkin}. Writing $\N = g^T g$, it  is easy to see that, up to an $SO(2)$ rotation, the 'vielbein' satisfies %{\color{ForestGreen}(maybe type, I get $\mathcal{E}^T\mathcal{E}=\mathcal{G}$ by plugging in the $\mathcal{E}$ below)}

\be
C^T g\, C =   \left(\begin{array}{cc}\mathcal{E} & 0  \\ 0 & (\mathcal{E}^T)^{-1} \end{array}\right)  \left(\begin{array}{cc}1 & 0 \\ -\mathcal{B} & 1 \end{array}\right) \;, \;\;\;\;\; \mathcal{E}^T  \mathcal{E} =\mathcal{G}
\ee 
for the matrix $C$ defined in \eqref{definitionmatrixC}. A convenient choice for the matrix $\E$ is, letting $\phi_1+\phi_2 = 2 \phi$ and $\phi_1-\phi_2 = -2\rho$ 
\be
\mathcal{E} = \left(\begin{array}{cc}e^{-\phi} & 0 \\ - \chi_1 e^{-\rho}  & e^{-\rho} \end{array}\right) 
\ee
which leads to the following choice of parametrisation of the vielbein in terms of the physical fields
\be \label{matrixofsc6dtrunc}
C^T g \, C =  \left(\begin{array}{cccc} e^{-\phi} &0 &0 &0 \\ -\chi_1 e^{-\rho}& e^{-\rho} & 0 & 0 \\ \chi_1 \chi_2 e^\phi & - \chi_2 e^\phi & e^\phi &  \chi_1 e^\phi\\ \chi_2 e^{\rho} & 0 & 0 & e^{\rho}\end{array}\right)
\ee  
Note this precisely coincides with the upper $4 \times 4$ block of \eqref{eq: full X matrix}, up to a transposition of the second and third row,  %the fact that  $\rho = \rho/2$,
 as well as a global $O(2,2)$ sign flip.  On this restricted subspace, $\Omega_4$ is exactly the same as $C$. Performing the multiplications by the $C$'s, the expression for $g$ reads %{\color{red}(need to recheck that I replaced everywhere $v$ with $\rho$)}

\be \label{cosetrep6dtruncation}
g = \frac{1}{2}\left(\begin{array}{cccc}  e^\phi+e^{-\phi} + \chi_1 \chi_2 e^\phi & e^\phi (\chi_1-\chi_2) & e^\phi-e^{-\phi} - \chi_1 \chi_2 e^\phi & e^\phi(\chi_1+\chi_2) \\
\chi_2 e^{\rho} -\chi_1 e^{-\rho} & e^{\rho}+e^{-\rho} & \chi_1 e^{-\rho} - \chi_2 e^{\rho} & e^{\rho}-e^{-\rho}\\
e^\phi-e^{-\phi}+ \chi_1 \chi_2 e^\phi & e^\phi (\chi_1-\chi_2) & e^\phi+e^{-\phi} - \chi_1 \chi_2 e^\phi & e^\phi (\chi_1+\chi_2) \\
\chi_1 e^{-\rho} + \chi_2 e^{\rho} & e^{\rho}-e^{-\rho} & -\chi_1 e^{-\rho} -\chi_2 e^{\rho} & e^{\rho}+ e^{-\rho} \end{array}\right)
\ee
Let us check that the scalar kinetic terms in \eqref{lagtrunc} are reproduced using the general construction reviewed in section \ref{sugrareview}. For this, we first compute 
\bea
\p_\mu g g^{-1}& = & C  \left(\begin{array}{cc} \p_\mu \mathcal{E} & 0  \\ - \p_\mu [(\mathcal{E}^{-1})^T \mathcal{B}] & \p_\mu (\mathcal{E}^T)^{-1} \end{array}\right)  \left(\begin{array}{cc}\mathcal{E}^{-1} & 0  \\ \B \E^{-1} & \mathcal{E}^T \end{array}\right) C^T = \\[3mm]
&  & \hspace{-2cm}=\; \frac{1}{2} \left(\begin{array}{cc} \p_\mu \mathcal{E} \E^{-1} +  \p_\mu (\E^{-1})^T \E^T - (\E^{-1})^T \p_\mu \B \E^{-1} &  (\E^{-1})^T \p_\mu \B \E^{-1} -\p_\mu \E \E^{-1} + \p_\mu (\E^{-1})^T \E^T  \\[2mm] - \p_\mu \mathcal{E} \E^{-1}  + \p_\mu (\E^{-1})^T \E^T - (\E^{-1})^T \p_\mu \B \E^{-1} & \p_\mu \mathcal{E} \E^{-1} +  \p_\mu (\E^{-1})^T \E^T + (\E^{-1})^T \p_\mu \B \E^{-1} 
\end{array}\right)  \nonumber
\eea
The elements in the upper and, respectively, lower block represent  $Q_\mu^{ij}$ and $Q^{\bar{\imath}\bar{\jmath}}_\mu$ in the gauge \eqref{matrixofsc6dtrunc}. To find the scalar kinetic term, we simply remove these blocks, and compute the trace of the remaining matrix squared. Using

%\be
%L_{scal} = \frac{1}{4} Tr[(\E^{-1})^T \p_\mu \B \E^{-1} (\E^{-1})^T \p_\mu \B \E^{-1}] - \frac{1}{4} Tr[(\p_\mu \mathcal{E} \E^{-1}  - \p_\mu (\E^{-1})^T \E^T )^2] 
%\ee
% Intermediate steps
%%
\be
\p_\mu \E \E^{-1} = \left(\begin{array}{cc} - \p_\mu \phi & 0 \\ - \p_\mu \chi_1 \, e^{\phi-\rho} & - \p_\mu \rho \end{array}\right) = - (\p_\mu (\E^{-1})^T \E^T)^T
\ee

\be
 (\E^{-1})^T  \p_\mu \B \E^{-1} = \left(\begin{array}{cc} 0 & e^{\phi+\rho} \p_\mu \chi_2 \\ - e^{\phi+\rho} \p_\mu \chi_2 & 0 \end{array} \right)
\ee
one may easily check the Lagrangian agrees with the scalar part of \eqref{lagtrunc}.

\subsubsection*{The basis of three-form fields}

Let us now write down explicitly the the three-form field strengths $F^\L$ that satisfy the modified self-duality condition \eqref{eq: N-self duality condition}. The general procedure for constructing them is outlined in \cite{Gaillard:1981rj}. Stating with a lagrangian $\mathcal{L}(F)$ for the three-form fields that takes the schematic form %\emph{Careful matrix multiplication!}
\be
\mathcal{L} = \mathcal{G}\,  F \wedge \star F + \B \, F \wedge F
\ee
one forms the linear combinations 
%{\color{red}\emph{Normalisation?}}
%
\begin{align}
F^\L \equiv  \frac{1}{\sqrt{2}}\; C \left(\begin{array}{c} F \\ \p_F \mathcal{L} \end{array}\right)= \frac{1}{2}\begin{pmatrix}
F+\p_F \mathcal{L}\\
-F+\p_F \mathcal{L}
\end{pmatrix}
\end{align}
where, for a Lagrangian of the general form we assume, 
\begin{align}
\p_F \mathcal{L} &=\mathcal{G} \star F+\mathcal{B}F
\end{align}
Using the relation  $C^T \eta\,  C = \s_1 \otimes I_2$, it is easy to check that the forms $F^\L$ so defined satisfy \eqref{eq: N-self duality condition}, with $\N$ given in \eqref{matrixNkin}. The charges $q^\L$  associated to the sphere integrals of $F^\L$ read, in terms of the usual electric and magnetic charges %{\color{red}%\emph{Normalisation?}(such that 2 times the charges are quantized? this is what we have)}

\be
q^\L = \eta^{\L\Sigma} q_\Sigma =\frac{1}{2} \left(\begin{array}{c} q_{el} + q_{mag} \\ q_{el} - q_{mag}\end{array} \right)  =\frac{1}{\sqrt{2}}\; C \left(\begin{array}{c} q_{mag} \\ q_{el} \end{array}\right)
\ee
To sum up, the 3-forms that satisfy \eqref{eq: N-self duality condition}, are the combinations $(\pm F +\p_F \mathcal{L})/2$. For the particular case of the $SO(2,2)$ truncation, these 3-forms are:
\begin{align} 
\frac{1}{2}\bigg[\pm H_3 + (e^{-\phi_1-\phi_2}\hspace{0.001cm}^*H_3-\chi_1 e^{\phi_1-\phi_2}\hspace{0.001cm}^*(F_3-\chi_1 H_3)-\chi_2 F_3)\bigg]\label{generalFerraraformsNS}\\
\frac{1}{2}\bigg[\pm F_3+(e^{\phi_1-\phi_2}\hspace{0.001cm}^*(F_3-\chi_1 H_3)+\chi_2 H_3)\bigg]\label{generalFerraraformsR}
\end{align}
where the combination $F_3-\chi_1 H_3$ is the modified field strength. We denote the charges associated to these forms by $(q_{F1} \pm q_{NS5})/2 $ and, respectively, $(q_{D1} \pm q_{D5})/2$.

\subsubsection*{A useful gauge rotation}

While the gauge \eqref{matrixofsc6dtrunc} allows for a simple relation between the vielbein and the physical fields, the gauge   $Q^{ij}_\mu =0$ is much more convenient for the study of the black string attractor solutions. To reach it, we note the gauge connection reads

\be
Q^{ij}_\mu = %\left(\begin{array}{cc} 0 &  \p_\mu \chi_1 \, e^{\phi-\rho} - \p_\mu \chi_2 \, e^{\phi+\rho} \\ - \p_\mu \chi_1 \, e^{\phi-\rho} + \p_\mu \chi_2 \, e^{\phi+\rho} & 0 \end{array} \right) = 
 i \s_2 (\p_\mu \chi_1 e^{\phi_1} - \p_\mu \chi_2 {e^\phi_2}) \equiv -i\s_2 \p_\mu \a
 \label{exprQ}
\ee
This can be set to zero via a gauge transformation that satisfies $h^{-1} \p_\mu h = - Q_\mu$, with solution $h = e^{i \a \s_2}$. One can also  set $Q_\mu^{\bar \imath \bar \jmath} =  i \s_2 (\p_\mu \chi_1 e^{\phi_1} + \p_\mu \chi_2 e^{\phi_2})$ to zero via a similar gauge transformation, $e^{i\sigma_2 \beta}$. Note that the constant piece of $\a ,\b$ is not fixed by the vanishing of $Q^{ij}$, and can be used to further rotate $\hat Z^i$ by a constant angle. 

 These gauge transformations act on the group element as

\be \label{anglegeneralrot}
g \r \left(\begin{array}{cc} R_\a & 0  \\  0& R_\b  \end{array}\right) g\;, \;\;\;\;\; R_\a = \left( \begin{array}{cc}  \cos \a & \sin \a \\ -\sin \a & \cos \a \end{array} \right) %\;, \;\;\;\;\;\; \p_\mu \a,\b =  \pm \p_\mu \chi_1 e^{\phi_1} \pm \p_\mu \chi_2 e^{\phi_2}
\ee
Choosing $\a$ to be given by \eqref{exprQ}, one can rotate the corresponding 
%
%What this does is to rotate the upper blocks of the matrix of scalars as 
%
%\be
%C^T g C \r  \frac{1}{2}\left(\begin{array}{cc} R_\a + R_\b & R_\a -R_\b \\ R_\a - R_\b & R_\a + R_\b    \end{array}\right) C^T g C
%\ee
%One can rotate by $\a\b$ tuned to these special values, so as to set both
 gauge connection to zero. %Or else, given that for our attractor solutions we only need $Q^{ij} =0$, one may rotate by $\a_{sp}$ and then choose $\b =\a_{s}$ {\color{red}(notation?)} so as to simplify the end form of $g$, which is what we are doing in \ref{appendixB2}. %\emph{Say we do this explicitly later.} 

\subsubsection*{The central charge}

It is not hard to see that 
\be
C^T Z = C^T g \, C \left(\begin{array}{c} q_{mag} \\ q_{el} \end{array} \right) = \left(\begin{array}{c} q_{NS5}\,  e^{-\phi} \\  e^{-\rho} (q_{D5} - \chi_1 q_{NS5}) \\ e^\phi (q_{F1} +\chi_1 \, q_{D1} -\chi_2 \, q_{D5} + \chi_1 \chi_2\,  q_{NS5}) \\ e^\rho (q_{D1} + \chi_2 \, q_{NS5}) \end{array} \right) 
\ee
We immediately find 
\begin{align}
|Z_+|^2 &= \big[ q_{NS5}\, e^{-\phi} +e^\phi (q_{F1} +\chi_1 q_{D1} -\chi_2 q_{D5} + \chi_1 \chi_2 \, q_{NS5})\big]^2 +\nonumber\\
&+ \big[ e^{-\rho} (q_{D5} - \chi_1 q_{NS5}) +  e^\rho (q_{D1} + \chi_2 q_{NS5})  \big]^2
\end{align}
in perfect agreement with the expression of {\cite{Larsen:1999uk}}, particularised to our case. %{\color{ForestGreen}(checked)}

The matter central charge $|Z_-|^2$ is given by the same expression with relative minus signs, and one can check that $|Z_+|^2 - |Z_-|^2$ is moduli-independent, as expected.

%
%
%
%\subsection{The pure NS solution}
%
%The type IIB NS-NS action, upon reduction to six dimensions,  reads
%%
%\bea
%S_{NS} &= & \frac{1}{{\color{ForestGreen}8\pi^3} \a'^2 } \int d^6 x \sqrt{g_S}\, e^{-2 \Phi_6} \bigg(R + 4 (\p\Phi_6)^2 - \frac{1}{2} |H_3^2|\bigg) \nonumber \\
%&=&  \frac{1}{2\kappa_6^2} \int d^6 x \sqrt{g_E}\, \left(R + \# (\p\Phi)^2 - \frac{e^{-2\Phi}}{2} |H_3^2|\right) 
%\eea
%, and in the second line we have passed from $6d$ string to $6d$ Einstein frame.  

\subsection{Putting various  string solutions in six-dimensional form}
\label{appendixB2}

In this appendix, we exemplify how we put the string solutions of section \ref{section3} in the form above. 

\subsubsection*{The pure NS5-F1 solution}

For this solution, written in the main text, the axions are zero, and so is $F_3$. The vielbein \eqref{cosetrep6dtruncation} splits as (up to a transposition of the second and the third row)

\be \label{purens5matrixsc}
g = \left( \begin{array}{cc} \cosh \phi & \sinh \phi \\ \sinh \phi & \cosh \phi \end{array} \right) \otimes  \left( \begin{array}{cc} 1 & 0 \\ 0 & 0 \end{array} \right) +  \left( \begin{array}{cc} \cosh \rho & \sinh \rho \\ \sinh \rho & \cosh \rho \end{array} \right) \otimes  \left( \begin{array}{cc} 0 & 0 \\ 0 & 1 \end{array} \right)
\ee
with the only non-trivial dynamics affecting  the six-dimensional dilaton

\be \label{solutionpurens5f1}
e^{2\phi} = g_6^2 \frac{f_5}{f_1} \;, \;\;\;\;\;\;\; e^{-2\rho} = v = const.
\ee 
Note that only $H_3$ contributes to the forms $F^\L$. Numbering the $\L$ index as $\L = \{1,2,\bar 1, \bar 2 \}$, where $1,\bar 1$ corespond to the combinations in \eqref{generalFerraraformsNS}, and $2,\bar 2$ to those in \eqref{generalFerraraformsR}, the non-zero field strengths that satisfy the modified (anti)-self duality conditions \eqref{eq: N-self duality condition} are 
%\emph{Factors!}
%
\be
F^1 = \frac{1}{4} (H_3+ e^{-2\phi} \star H_3) \;, \;\;\;\;\; F^{\bar 1} =  \frac{1}{4} (-H_3+ e^{-2\phi} \star H_3)
\ee
We immediately note that the charges of these combinations are integer quantized, being given by $(p \pm k)/2$. %\emph{Where do the dimensions go? Should we rather say integer $ \times \a'$, if we integrate oover the unit sphere?} 
The magnetic charge of $F^{1,\bar 1}$ equals plus or minus its electric charge, as expected. 
%{\color{ForestGreen}(we actually have on the sphere $H+e^{-2\Phi}*H=2\alpha'(k+p)\omega_{S^3}$, so I would say we need to divide by $\alpha'$ and also other numerical factors to get the quantized charges after integration on $S^3$, see for ex eq A.8 from 1911.12359)}

The combinations that appear in the supersymmetry variations are 
\be
F^+ = X^{1}_{\L} F^\L = \frac{e^{-\phi}}{4} (H_3+\star H_3) \;, \;\;\;\;\; F^-= X^{\bar 1}_{\L} F^\L = \frac{e^{-\phi}}{4} (-H_3 + \star H_3)
\ee
which are, indeed, properly (anti)-self-dual. The explicit expresion for the quantities in parantheses is 
\be
H_3 \pm \star H_3 = 2\a' \bigg(k \pm \frac{g_s^2 p f_5}{v f_1}\bigg) (\omega_{S^3} \pm \star \omega_{S^3}) 
\ee
Note that the anti-self-dual part of $H_3$ vanishes in the near-horizon limit.
%{\color{ForestGreen}(with the correct numerical factors in front, these $H^{\pm}$ are the 6d 3-forms from Skenderis and from the sphere part of the SD we can readoff $Z^+$)}.

Let us now check this solution matches indeed the general one, \eqref{eq: metric Ansatz}. The Einstein frame metric is 
\be
ds^2 = \frac{1}{g_6 \sqrt{f_1 f_5}} (-dt^2 + d\s^2) + \frac{\sqrt{f_1 f_5}}{g_6} (dr^2 + r^2 d\Omega_3^2) 
\ee
with $g_6 = g_s/\sqrt{v}$. From this, we read off  that 
\be
H^\L H_\L =  g_6^2 f_1 f_5 
\ee
and that we need to rescale $r$ by a factor of $g_6$: $r = g_6 \tilde r$. Note from this that $c_\L c^\L = g_6^2$,  which indeed degenerates in the decoupling limit.

The harmonic functions are given by identifying the various rows of \eqref{purens5matrixsc} with our general solution \eqref{generalsolutionscacoset}, leading to

\be
H^1 = \cosh \phi \, \sqrt{H_\L H^\L} =   \frac{1}{2} (f_1 + g_6^2 f_5) = \frac{\a' (p+k)}{2 \tilde r^2} + \frac{1}{2} (1+g_6^2)
\ee
\be
H^{\bar 1} = - \sinh \phi \, \sqrt{H_\L H^\L} =   \frac{1}{2} (f_1 - g_6^2 f_5) = \frac{\a' (p-k)}{2 \tilde r^2} + \frac{1}{2} (1-g_6^2)
\ee
The modulus $v$ simply enters the D1-D5 CFT moduli, via

\be
\bar g_{2\L} = \left( 0, \cosh v,0 \, \sinh v\right)
\ee
which is obviously orthogonal to the charge and the $c_\L$.  %Note that the rescaling of $r$ that has  been forced on us results in   $c^\L c_\L =g_s^2/v$, which indeed degenerates in the decoupling limit. %It would be interesting to understand whether this argument could be made more general {\color{ForestGreen}(in general for the NS frame? I think it should work for the TsT also)}. Note the constants can also be made arbitrary.

\subsubsection*{The pure D1-D5 solution}

This entirely parallels the previous discussion, we just concentrate on $\rho$ instead of $\phi$ and switch $H_3 \leftrightarrow F_3$. The solution for the scalars is now 

\be
e^{-2\rho} =  v \frac{f_1}{f_5} \;, \;\;\;\;\; e^{2\phi} =  g_6^2 = const.
\ee
where $ v, g_6$ are now the corresponding string frame K3 volume and six-dimensional string coupling in this frame. Note that  S-duality exchanges $\phi$ with $\rho$ and $g_6^2$ with $v^{-1}$, so the solution above is the same as \eqref{solutionpurens5f1}, with these identifications.

To find the map to our $H_\L$ we put the metric \eqref{pured1d5metric} in $6d$ Einstein frame. One should be careful that S-duality yields an overall factor of $ g_s$ (which we ommitted in \eqref{pured1d5metric}), and  thus the overall factor we obtain is $ g_s/ g_6 = \sqrt{v}$. We thus find the identification

\be \label{hLhLd1d5frame}
H_\L H^\L = \frac{f_1 f_5}{ v} \;, \;\;\;\; \tilde r = r \sqrt{ v}
\ee
The non-zero harmonic functions then are %\emph{Explain factor?}
\be
H^{2,\bar 2} = \frac{1}{2} (e^{-\rho} \pm e^\rho)\sqrt{H_\L H^\L} = \frac{1}{2} (f_1 \pm \frac{f_5}{ v}) = \frac{ g_s \a' (p\pm k)}{2 \tilde r^2} + \frac{1}{2} (1 \pm 1/ v)
\ee
Now, the horizon moduli are given by $ g_6$, which maps as expected to $v$ in the NS case. 

\subsubsection*{The D1-D5 solution with axionic moduli}
It is also interesting to understand how to turn on more general moduli in this solution. We will be working within the $SO(2,2)$ truncation described above, and choose the D1-D5 frame for concreteness. As discussed above, in this frame the volume of the K3 is an attracted scalar, while the $6d$ dilaton is unfixed  and constant throughout. We can trivially turn on two constant modes for $\chi_{1,2}$ via large gauge transformations; they are denoted as $s_0, s_4$. These shifts in the axions induce an F1 charge, as the charges of the solution are
\be
q_{NS5} = 0 \;, \;\;\;\; q_{D5} = k \;, \;\;\;\;\; q_{F1} = k s_4 -p s_0\;, \;\;\;\;\;q_{D1} = p 
\ee
 Note, however, that in the full string theory this charge is quantized. 
 
% At the level of the $10d$ solution this change is trivial. 
 At the level of the $6d$ fields, the scalars shift predictably%, while the $F^\L$ become {\color{red}(recheck notation for the charges)}

\be
e^\phi = g_6 \;, \;\;\;\;\; e^{\rho}  = \sqrt{\frac{f_5}{v f_1}}\;, \;\;\;\;\; \chi_1 = s_0\;, \;\;\;\;\; \chi_2 = s_4
\ee
and the scalar matrix is given by \eqref{cosetrep6dtruncation} with these replacements. 
%
%
%
%from which one can read off an electrical charge induced by this transformation.  with 
%
%Plugging in, we note it already takes the standard form
\begin{align}\label{matrixscaxionsd1d5}
g=\begin{pmatrix}
\frac{g_6^2(s_0 s_4+1)+1}{2 g_6} & \frac{1}{2}g_6(s_0-s_4) & \frac{g_6^2(1-s_0 s_4)-1}{2 g_6} & \frac{1}{2}g_6(s_0+s_4)\\
\frac{f_5 s_4-f_1 s_0 v}{2\sqrt{v f_1 f_5 }} & \frac{f_5+v f_1}{2\sqrt{v f_1 f_5 }} & \frac{f_1 s_0 v-f_5 s_4}{2\sqrt{v f_1 f_5 }}  & \frac{f_5-v f_1}{2\sqrt{v f_1 f_5 }} \\
\frac{g_6^2(s_0 s_4+1)-1}{2 g_6} & \frac{1}{2}g_6(s_0-s_4) & \frac{g_6^2(1-s_0 s_4)+1}{2 g_6} & \frac{1}{2}g_6(s_0+s_4)\\
\frac{f_5 s_4+f_1 s_0 v}{2\sqrt{v f_1 f_5 }} & \frac{f_5-v f_1}{2\sqrt{v f_1 f_5 }} & -\frac{f_1 s_0 v+f_5 s_4}{2\sqrt{v f_1 f_5 }}  & \frac{f_5+v f_1}{2\sqrt{v f_1 f_5 }}
\end{pmatrix}
\end{align}
{Comparing with \eqref{cosetrep6dtruncation} we note that the attracted scalars are $e^\rho$ and $\chi_1 e^{-\rho} - \chi_2 e^\rho$, in agreement with the conclusions of \cite{Larsen:1999uk} for the case $q_{F1} =0$. }

\subsubsection*{The D1-D5 frame TsT solution\footnote{{Since there will be several types of TsT transformations discussed in the following appendix,  what we mean by `TsT' here are transformations that act \emph{within} the six-dimensional consistent truncation, being performed along the common D1-D5 direction and time. }}}

We generate the (non)-extremal such solution by performing a TsT along the common  space and time direction of the general/extremal D1-D5 one.  In the following, we will be giving the expressions for the non-extremal solutions, generated from the seed %{\color{ForestGreen}(why did we write this with momentum?)}
%It is equally easy to treat the directly non-extremal case, where the seed is given by 
\begin{align}
ds^2 &= \frac{1}{\sqrt{f_1 f_5}} \left(  d\s^2 - f dt^2 \right)  +\sqrt{f_1 f_5} \left( \frac{dr^2}{f} + r^2 d \Omega_3^2\right) + \sqrt{\frac{f_1}{f_5}}ds^2_{K3}
\end{align}
where the function $f$ is the blackening factor, $f = 1-r_0^2/r^2$ and $f_{1,5}= 1+r_{1,5}^2/r^2$, with $r_1,r_5$ defined in \eqref{standardquantcond}. The dilaton and the 3-form field have the same expressions as in the extremal case, in terms of the non-extremal functions $f_1,f_5$. After  the TsT (followed by two trivial axionic shifts with parameters $s_0, s_4$, which will be useful later), we obtain:
\begin{align}\label{nonextremald1d5sol}
ds^2&=\frac{\sqrt{f_1 f_5}}{f_1 f_5 - \gamma^2 f} \big(d\sigma^2 - f dt^2\big) +\sqrt{f_1 f_5}\big(\frac{dr^2}{f}+r^2d\Omega^2_3\big)+\sqrt{\frac{f_1}{f_5}}\, ds^2_{K3}
\end{align}
\be
B=\frac{\gamma f}{f_1 f_5-\gamma^2 f}dt\wedge d\sigma\;, \;\;\;\;\;
e^{\Phi}=g_s\frac{f_1}{\sqrt{f_1 f_5-\gamma^2 f}}
\ee
and
\be \label{rrtst02}
%\;,\;\;\;\;\;
%F_1&=\gamma \frac{2 Q_1}{r^3f_1^2}dr\\
%F_3&=2Q_5\omega_{S^3}+\frac{2Q_1}{r^3f_1^2}\bigg(1+\frac{f\gamma^2}{f_1 f_5-\gamma^2 f}\bigg)dr\wedge dt\wedge d\sigma\\
%F_5&=-\gamma\frac{2Q_5}{r^3f_5^2}dr\wedge\omega_{K3}+2Q_5\frac{f\gamma}{f_1f_5-\gamma^2 f}dt\wedge d\sigma\wedge\omega_{S^3}\\
C_0=\frac{\alpha' p}{v r_1^2}\frac{\gamma}{f_1}+s_0\;, \;\;\;\;\;
C_2=-2\alpha' k \frac{1}{8}\cos\theta d\psi\wedge d\phi +\frac{\frac{\alpha' p}{v} f_5-s_0 r_1^2 \gamma f}{ r_1^2(f_1f_5-\gamma^2 f)}dt\wedge d\sigma\ee

\be \label{rr4formtst}
C_4=\bigg(-v\frac{\alpha'k}{r_5^2}\frac{\gamma}{f_5}+s_4\bigg)\omega_{K3}-\frac{\gamma \alpha'k f \cos\theta}{4(f_1 f_5-f\gamma^2)}dt\wedge d\sigma\wedge d\psi\wedge d\phi
\ee
%We also introduced the notations $Q_1=\frac{\alpha' g_s p}{v}, Q_5=\alpha' g_s k$. \emph{\textcolor{red}{Do we actually need it?}} 
To obtain the extremal solution, we simply send $r_0\r 0$, keeping $r_1^2=\frac{\alpha' g_s p}{v}, r_5^2=\alpha' g_s k$ fixed. Note this sets $f=1$.

The six-dimensional fields can be readily extracted from the above. We particularize to the extremal case, which is the only one we need to match. The scalars are 
\be
e^{2\phi} = g_6^2 \frac{f_1 f_5}{f_1 f_5-\g^2} \;, \;\;\;\;\; e^{2\rho} = \frac{f_5}{v f_1}\;, \;\;\;\; \chi_1 = \frac{\g}{g_s f_1} + s_0 \;, \;\;\;\; \chi_2 = - \frac{v \g}{g_s f_5} +s_4
\ee
and the Einstein frame metric  %{\color{red}(again notation metrics, is it consistent throughout the paper?)}
\be
ds^2_{6}=\frac{1}{\sqrt{\frac{1}{v}(f_1f_5-\gamma^2) }}\big(d\sigma^2- dt^2\big)+\sqrt{\frac{1}{v}(f_1f_5-\gamma^2)}\bigg(dr^2+r^2d\Omega_3^2\bigg)
\ee
where we have rescaled $r$ as in \eqref{hLhLd1d5frame}, and then dropped the tilde. The six-dimensional charges are obtained by integrating the field strengths \eqref{generalFerraraformsNS}, \eqref{generalFerraraformsR}, evaluated on this solution, %lead to the following six-dimensional charges 
%{\color{red}(normalisations? consistent with the rest of the draft if we divide by 2)}
on $S^3$
\be
q^\L = {\color{ForestGreen}}\frac{1}{2}\{k s_4 - p s_0, k+p, k s_4 - p s_0, p-k\}
\ee
Using the $6d$ solution and the above charges, we can compute the central charge vector. Its modulus is 
\be
|Z_+| = \frac{f_5 p + f_1 k v}{2 \sqrt{v} \sqrt{f_1 f_5 - \g^2}} \label{ztst}
\ee
and  the following rotation aligns it with the previous direction

\be \label{rotangletstd1d5}
\left( \begin{array}{cc} \cos \a & - \sin \a \\ \sin \a & \cos \a\end{array} \right) \left(\begin{array}{c} Z_1 \\ Z_2 \end{array} \right) = \left(\begin{array}{c}0 \\ |Z_+| \end{array} \right) \;, \;\;\;\;\; \sin \a = \frac{\g}{\sqrt{f_1 f_5}}
\ee
%aligns the central charge with the previous direction. Note we do not perform a rotation on $Z_{\bar \imath}$. 
Note the rotation angle vanishes at the horizon, while approaching $\pi/2$ at infinity for the decoupled NCOS solutions with $\g =1$.  

Upon this rotation, the  coset solution takes the standard form \eqref{generalsolutionscacoset}, and we read off 
\begin{align} \label{modulitstd1d5}
\bar{g}_{i\L} =\bigg\{ \frac{g_6^2(s_0 s_4+1)+1}{2 g_6},\; \frac{1}{2}g_6(s_0-s_4),\; \frac{g_6^2(1-s_0 s_4)-1}{2 g_6}, \; \frac{1}{2}g_6(s_0+s_4)  \bigg\} 
\end{align}
exactly as in the undeformed D1-D5 solution {\eqref{matrixscaxionsd1d5}}, and 
% \left(\begin{array}{cccc} \frac{\text{g6}^2 \text{s0} \text{s4}+\text{g6}^2+1}{2 \text{g6}}& \frac{1}{2} \text{g6} (\text{s0}-\text{s4})& \frac{\text{g6}^2 (1-\text{s0} \text{s4})-1}{2 \text{g6}} & \frac{1}{2} \text{g6} (\text{s0}+\text{s4}) 
%\end{array}\right)
\begin{align} \label{harmonictstd1d5}
H_{\Lambda}=\bigg\{&\frac{-f_1 g_6 v s_0+ g_6 f_5 s_4+\gamma \sqrt{v} (g_6^2(s_0 s_4+1)-1)}{2 g_6 v},\;\frac{v f_1+f_5+\gamma \sqrt{v} g_6(s_0-s_4)}{2 v} ,\;\nonumber\\
& \frac{f_1 g_6 v s_0-f_5 g_6 s_4+\gamma \sqrt{v}(g_6^2(1-s_0 s_4)+1)}{2 g_6 v}, \;  \frac{-v f_1 +f_5+\gamma \sqrt{v} g_s(s_0+s_4)}{2 v} \bigg\}
\end{align}
\subsubsection*{The D1-D5 frame TsT solution in the ``NCSYM" parametrization}

As explained at length in appendix \ref{appendixdecoupling}, the background \eqref{nonextremald1d5sol} - \eqref{rr4formtst}  can also be obtained via a chain of dualities that relates it to a  system of intersecting D3-branes with  \emph{spatial} B-fields turned on, labeled by two related (spatial TsT) parameters, $\l$ and $\mu$.  This alternative procedure yields the solution in a different parametrization, which we will call the ``NCSYM" one, as in the decoupling limit, the theory on one stack of D3 branes is simply spatially non-commutative SYM.  The full non-extremal solution is written in \eqref{chaindual} - \eqref{RRhatted}, and in appendix \ref{appendixdecoupling} we show that it is in fact identical to \eqref{nonextremald1d5sol} - \eqref{rr4formtst} upon a set of coordinate rescalings and parameter redefinitions given in \eqref{relationcoord} - \eqref{alitoalh}. }

 It is useful to write the solution  in this alternative parametrization in $6d$ language, since the $6d$ Einstein frame extremal metric is extremely simple
\begin{align}
ds^2_{6}&=\sqrt{\frac{\hat{v}}{\hat{g}_s^2}} \left[\frac{1}{\sqrt{\hat{f}_1 \hat{f}_5}}(-d\hat{t}^2+d\hat{\sigma}^2)+\sqrt{\hat{f}_1 \hat{f}_5}\bigg(d\hat{r}^2+\hat{r}^2 d\Omega_3^2\bigg)\right]
\end{align}
%\textcolor{magenta}{ 
 and entirely independent of the TsT parameters. The $6d$  scalar fields that enter the truncated action \eqref{lagtrunc} are given by

\begin{align}
e^{\phi_1}&=\hat{g}_s \frac{\hat{f}_1+\mu^2 \hat{f}_5}{\sqrt{\hat{f}_1 \hat{f}_5}}\;, \hspace{0.5cm}e^{\phi_2}=\frac{\hat{g}_s}{\hat{v}}\frac{\hat{f}_5+\lambda^2 \hat{f}_1}{\sqrt{\hat{f}_1 \hat{f}_5}}\hspace{0.5cm}\chi_1=-\frac{\mu \hat{f}_5}{\hat{g}_s(\hat{f}_1+\mu^2 \hat{f}_5)}\hspace{0.5cm}\chi_2=\frac{\hat{v}\lambda \hat{f}_1}{\hat{g}_s(\hat{f}_5+\lambda^2 \hat{f}_1)}
\end{align}
from which one may extract the $6d$ dilaton and volume of the internal manifold
\begin{align}
e^{2\phi}&=\hat{g}_6^2\frac{(\hat{f}_1+\mu^2 \hat{f}_5)(\hat{f}_5+\lambda^2 \hat{f}_1)}{\hat{f}_1\hat{f}_5}\hspace{1cm}e^{2\rho}=\frac{\hat{f}_5+\lambda^2 \hat{f}_1}{\hat{v}(\hat{f}_1+\mu^2 \hat{f}_5)}
\end{align}
The six-dimensional conserved charges are given by %\emph{\textcolor{red}{Signs?}}{\color{ForestGreen} (the signs agree with the definitions B.16-B.17 of the forms)}
\begin{align}
q^{\Lambda}=\frac{1}{2}\{0,k+p,0,p-k\}
\end{align}
The magnitude of the central charge vector and its angle with respect to the `$2$' direction are given by
\be
|Z_+|=\frac{\hat{f}_5 p + \hat{f}_1 k\hat{v}}{2\sqrt{\hat{f}_1 \hat{f}_5 \hat{v}}}\;, \;\;\;\;\;
\sin \, \alpha=\frac{\mu \hat{f}_5+\lambda \hat{f}_1}{\sqrt{(\hat{f}_1+\mu^2 \hat{f}_5)(\hat{f}_5+\lambda^2 \hat{f}_1)}} \label{Zhatted}
\ee
Note that, in this parametrisation, $|Z_+|$  is the same as in the pure D1-D5 case. %\textcolor{magenta}
{It can be explicitly checked, using the relations \eqref{relharm} and \eqref{relationparam1}  between the hatted and unhatted quantities, that $|Z_+|$ above precisely agrees with \eqref{ztst}. After the above rotation, one may read off the}
%
% {\color{ForestGreen}(extra check: if we plug in the relations between hatted and unhatted quantities in the central charge B.46 we obtain exactly B.54. We can choose the axionic moduli in pure D1D5 to get the hatted parametrization and since this is in $SO(2,2)$ it's normal that the central charge stays the same)} indicating the decoupling limit will be the same. % The rotation angle is given by:
 near-horizon moduli 
\begin{align}
\bar{g}_{i\Lambda}&=\bigg\{\frac{-\hat{v}+\hat{g}_s^2(-1+\lambda\mu)}{2\hat{g}_s\sqrt{\hat{v}}},\frac{\mu(p-k)\sqrt{\hat{v}}}{2p},\frac{-\hat{v}+\hat{g}_s^2(1-\lambda\mu)}{2\hat{g}_s\sqrt{\hat{v}}},\frac{\mu(p+k)\sqrt{\hat{v}}}{2p}\bigg\}
\end{align}
and  harmonic functions\footnote{
If we want to compare them with the ones obtained via TsT along $t,\sigma$, we need to rescale by a factor of $\hat{g}_s$, which was absorbed in the coordinates when performing the S-duality.} %that we obtain are:
\begin{align}
H_{\Lambda}&=\bigg\{\hat{g}^2_6\frac{\mu \hat{f}_5 +\lambda \hat{f}_1}{2},\hat{g}_s\frac{\hat{f}_5+\hat{v} \hat{f}_1}{2\hat{v}},\hat{g}_6^2\frac{\mu \hat{f}_5 +\lambda \hat{f}_1}{2},\hat{g}_s\frac{-\hat{v} \hat{f}_1+\hat{f}_5}{2\hat{v}}\bigg\}
\end{align}
%\textcolor{magenta}
{These quantities precisely agree with those of the pure D1-D5 system with axionic moduli \eqref{matrixscaxionsd1d5}, with the choice $s_0=\frac{\lambda \hat{g}_s}{\hat{v}} $ and $s_4=\frac{\lambda \hat{g}_s p}{k\hat{v}}$, so  that $k s_4 = p s_0$ and no F1 charge is induced. This  explains why the modulus of the central charge \eqref{Zhatted} is identical to that in pure D1-D5, as axionic shifts correspond to $SO(2,2)$ transformations that leave invariant the central charge.    }

\subsubsection*{The S-dual (NS5-F1 frame) TsT solution}
The non-extremal solution obtained by S-dualising \eqref{nonextremald1d5sol} is given by
\be
ds^2=\sqrt{\frac{f_1 f_5-\gamma^2 f +\left(\frac{\alpha' p\gamma}{v r_1^2}+f_1 g_s s_0\right)^2}{f_1 f_5}}\bigg[\frac{f_5}{f_1 f_5-\gamma^2 f}(d\sigma^2-f dt^2)+f_5\bigg(\frac{dr^2}{f}+r^2 d\Omega^2_3\bigg)+ds^2_{K3}\bigg]
\nonumber\ee

\be
e^{\Phi}= g_s\frac{f_1 f_5-\gamma^2 f+(f_1 g_s s_0 +\frac{\alpha' p\gamma}{v r_1^2})^2}{ f_1\sqrt{f_1 f_5-\gamma^2 f}} \;, \;\;\;\;\;
C_0= \frac{-f_1(\frac{\alpha' p\gamma}{v r_1^2}+f_1  s_0)}{f_1 f_5-\gamma^2 f+(f_1 g_s s_0 +\frac{\alpha' g_s p\gamma}{v r_1^2})^2}\ee

\be
B=
- \frac{\alpha' k g_s}{4}\cos\theta d\psi\wedge d\phi +\frac{\frac{\alpha' g_s p}{v} f_5-s_0 g_s r_1^2 \gamma f}{r_1^2(f_1f_5-\gamma^2 f)}dt\wedge d\sigma\;, \;\;\;\;\;
C_2=-\frac{\gamma f}{g_s(f_1 f_5-\gamma^2 f)}dt\wedge d\sigma \nonumber
\ee
where $s_0$ is the previously introduced  arbitrary shift of the S-dual axion.  The type IIB four-form potential can be obtained from the one in the D1-D5 frame by $\tilde{C}_4=C_4-B\wedge C_2$ and its part on $K3$ is not changed. 

In order for the  dilaton  to not  diverge at infinity, we  need to fix $s_0 =- \g \frac{\alpha' p}{g_s v r_1^2}$. Charge quantization then fixes $s_4$ in terms of $s_0$, as explained in section \ref{section3}.

The six-dimensional fields read (particularising again to the extremal solution)

\be
e^{-2\rho} = \frac{v}{g_s^2} \frac{f_5 + \g^2 (f_1-2)}{f_5} \;, \;\;\;\;\; e^{2\phi} = \frac{1}{v} \frac{f_5 (f_5 + \g^2 (f_1-2))}{f_1 f_5 - \g^2}
\ee
and axions
\be
\chi_1 = \frac{1}{g_s} \frac{\g (f_1-1)}{f_5+\g^2 (f_1-2)} \;, \;\;\;\;\;\; \chi_2  = - v\frac{\alpha' k}{r_5^2}\frac{\gamma}{f_5}+s_4
\ee
There is again an $SO(2)$ rotation that brings the solution to the standard form.

\section{Details of the NCOS decoupling limit} \label{appendixdecoupling}
\noindent In this appendix, we discuss the NCOS decoupling limit of the finite temperature D1-D5 system, singling out certain subtleties that are not fully emphasized in the older literature. We start by reviewing the decoupling limits for four-dimensional non-commutative $\N=4$ SYM and its S-dual D3 NCOS, which are related via T-dualities to the system we are interested in. 

\subsection{Review of the NCSYM and D3 NCOS decoupling limits \label{D3ncos}}

As discussed in detail in \cite{Seiberg:1999vs}, in the presence of a $B$-field along a brane's worldvolume, open and closed strings experience a different metric and coupling constants, related as 
\be
g+2\pi \a' B = \left(G+\frac{\theta}{2\pi \a'}\right)^{-1} \;, \;\;\;\;\; G_s = g_s \sqrt{\frac{\det G}{\det (g+2\pi  \a' B)}} \label{openclrel}
\ee
where lower case letters denote closed string quantities, and capitals denote open string ones. 

\subsubsection*{Non-commutative SYM decoupling}

The non-commutative $\N=4$ SYM limit is obtained by taking  $\a' \r 0$ keeping the open string quantities: $G, G_s \, \a'^{\frac{p-3}{2}}$ and $\theta$ fixed. Due to the various (anti)symmetry properties, this implies $B \sim 1/\theta$ is fixed in this limit, while the closed string metric $g \sim \a'^2/\theta^2$ along the non-commutative directions.  For D3-branes and a B-field of rank two,  we also  have $g_s \sim \a'$. %\emph{Position indices?} 

To implement this limit in holography, one first turns on a $B$ - field along the D3 brane directions using a TsT (T-duality, shift, T-duality) transformation.  More concretely, one starts from the non-extremal D3-brane solution 

\be
ds^2 = \frac{1}{\sqrt{f_3}} (-f dt^2 + dx_1^2 + dx_2^2 + dx_3^2) + \sqrt{f_3} \left(\frac{dr^2}{f}+r^2 d\Omega_5^2\right) \nonumber
\ee

\be
e^\Phi = g_s\;, \;\;\;\;\; F_5 = 16\pi \a'^2 N (\omega_{S^5} + \star \omega_{S^5}) \label{undefd3}
\ee
with

\be
f_3 = 1 + \frac{r_0^4 \sinh^2 \a_3}{r^4} \;, \;\;\;\;\;\; f=1-\frac{r_0^4}{r^4} \;, \;\;\;\;\;\; r_0^4 \sinh 2 \a_3 = 8\pi \a'^2 g_s N \label{chquantd3}
\ee
and performs a T-duality on $x_2$, followed by a shift $x_3 \r x_3 + \l\, \tilde x_2$ - where $\tilde x_2$ is the T-dual coordinate to $x_2$ -  and a T-duality back. The resulting string frame background is\footnote{By comparison, the background of \cite{Maldacena:1999mh} is generated via a $TrT$ transformation, where $r$ is a rotation  with angle  $ \theta = \arctan \l$. As a consequence, the noncommutative coordinates $x_{2,3}$ of \cite{Maldacena:1999mh} differ by a factor of $\cos \theta$ from ours. This factor is such that our coordinates need not be rescaled when taking the decoupling limit. }  

\be
ds^2 = \frac{1}{\sqrt{f_3}} \left( - f dt^2 + dx_1^2 + \frac{f_3}{f_3 +\l^2} (dx_2^2+dx_3^2) \right)  + \sqrt{f_3} \left(\frac{dr^2}{f}+r^2 d\Omega_5^2\right) \;, \;\;\;\;\;\;e^{2\Phi} = \frac{ g_s^2 f_3}{f_3+\l^2} \nonumber
\ee

\be
\a' B_{23}= \frac{\l}{f_3 + \l^2} \;, \;\;\;\;\;\; C_{01} = \lambda \frac{4\pi\alpha'^2 N}{r^4 f_3}\;, \;\;\;\;\;\;\;  F_5 = 16\pi \a'^2 N  (\omega_{S^5} + \star \omega_{S^5})   \label{ncd3} 
\ee
The $4d$ NCSYM decoupling limit is simply
\be
\a' \r 0 \;\;\;\;\mbox{with} \;\;\;\; \frac{r}{\a'}\;, \;\frac{r_0}{\a'}\;, \; g_s\;\; \mbox{and} \;\; \theta \equiv \l \, \a' \;\;\; \mbox{fixed} \label{ncdeclim}
\ee 
 We immediately note that the scaling of the asymptotic  closed string quantities (as $r \r \infty$) is exactly as expected from field theory, the TsT effectively implementing the transformation \eqref{openclrel}. In particular, $g_s$ - the value of the dilaton before the TsT - corresponds to the fixed open string coupling in the deformed theory.  In the decoupling limit, the constant term in $f_3$ can be dropped  and $\a_3 \r \infty$ as $\sinh 2 \a_3 \propto \a'^{-2}$. 

It is not hard to see that the $\l$ dependence entirely drops out from the Bekenstein-Hawking entropy, as well as the Hawking temperature; the energy follows from the first law \cite{Maldacena:1999mh}. This result can be reproduced by directly computing the energy of the background \eqref{ncd3}. For this, we use the covariant phase space formalism; the relevant formulae are detailed in appendix \ref{appendixcovphasesp}. % {\color{ForestGreen}(I found a nice paper with the expressions for type IIB surface charges, I'd like to cite it  1912.01030. I checked that their terms are the same as my terms up to numerical factors that I didn't check; their definition of $C_4$ differs from mine because they work with $F_5=dC_4- \# (H\wedge C_2 + F\wedge B) $ and I have $F_5=dC_4-H\wedge C_2$, but after taking this into account everything seems to agree with their eq 26.)}
We find the following expression for the energy difference between two backgrounds in our phase space %\,\textbf{Dimensions, overall factors, notation!! } 
%\emph{Notation various energies!}
%
\begin{align}
\delta M&=\frac{V_3 vol_{S^5}}{2^5\pi^7\alpha'^4 g_s^2}\bigg[ \delta\left(\frac{r_0^4}{4}(3+2\cosh 2\alpha_3)\right)+(\lambda r_0^4\sinh^2\alpha_3 + 4\pi \alpha'^2 N C^{(2)}_{\infty} )\delta B_{\infty} +12\pi\alpha'^2 N \delta C^{(4)}_{\infty}\bigg] \label{engnc}
\end{align}
%{\color{ForestGreen}(I am not sure about the factors of $g_s$. The computation should match to NCOS when we include also the factors of $\lambda$, this is one way to fix these factors maybe)}
where $V_3$ is the spatial volume of the D3-branes, $vol_{S^5}$ is the volume of the unit five-sphere, and we have allowed for arbitrary constant values of the RR potentials and $B$-field at infinity. %{\color{ForestGreen}(Remark: the last two terms, provided that the values at infinity don't depend on $r_0$ would drop anyway when we take the energy above extremality)}
 If these values are non-zero, but fixed (i.e., energy-independent) for given $\l$, the variations vanish, and we find the same result for the energy as in the $\l=0$ case. %\footnote{Up to an arbitrary constant shift that is allowed by the formalism, which will not enter the energy variation inside the first law, nor the energy above extremality. \textbf{\emph{Keep?}}} . 
   We take note that the value of the $B$ - field at infinity does not enter the calculation in any important way. It is in fact customary to 
%
%
%for the middle term I am not sure how to integrate it in general. In any case, if $B_{\infty}$ is energy-independent, we see that the term drops, the same for the constant in $C_4$ (probably we don't need to include it). We are then left with the first terms which come from the metric and give
%\be
%E = \frac{V_3 vol_{S^5}}{(2\pi)^7}r_0^4(5+4\sinh^2\alpha_3)
%\ee
%which is the pure D3 result.
%
%where $B_\infty$ is the value of the $B$ - field at infinity, while $C_\infty$ is the value of $C^{(2)}$ (\emph{We can also give the result for arbitrary $C^{(n)}$ if you have it}). The various contributions come from xxx, respectively. Note the energy is indeed the same as in the undeformed system \emph{provided} the $B$ - field at infinity is set to zero. 
%
%{\color{blue}One can do
 perform a gauge transformation to turn the constant B-field at infinity into a magnetic  worldvolume field strength, upon which the action of S-duality is simpler to understand. 
 
 % If we want to obtain an electric B-field by doing S-duality on the brane it is necessary to do this. When we shift the B-field by a constant we generate D1 charge (can we see this in our case?). According to Maldacena Russo, the constant flux induced for the gauge field on the worldvolume will change the solution (D1 dissolved into D3) (I still don't understand this part, but the fact that we need to have 0 B-field at infinity when we do S-duality it's ok)\emph{Transcribe relevant comments from the literature!} Once this is the case, \emph{Correct?} the result is independent of the value of the $C^{(2)}$ - field at infinity.
 
  This computation holds in the asymptotically flat background, before decoupling.   Since the decoupling limit is identical, for the quantities that appear, 
   to the standard D3 decoupling limit, it is obvious the $4d$ NCSYM thermodynamics is identical to the $\mathcal{N} =4$ SYM one. Concretely,
  after setting to zero the asymptotic values of the form fields, we obtain:
\begin{align}
M&=\frac{V_3 vol_{S^5}}{(2\pi)^7\alpha'^4 g_s^2}r_0^4(2\cosh 2\alpha_3 +3)=\frac{V_3 vol_{S^5}}{(2\pi)^7\alpha'^4 g_s^2}r_0^4\, [\underbrace{2\sinh 2\alpha_3}_{extremal} +\underbrace{3+2 e^{-2\a_3}}_{above\; extremality}]
\end{align}  
where in the second rewriting we have separated the answer into the extremal contribution, $E_{extr}$, which is proportional to the quantized charge using \eqref{chquantd3}, and the energy above extremality, $E$. 
%
%The extremal energy obtained for $r_0\rightarrow 0$ is given by:
%\begin{align}
%E_{ext}&=\frac{V_3 vol_{S^5}}{(2\pi)^7\alpha'^4g_s^2}4\times 4\pi g_s N\alpha'^2=\frac{V_3 vol_{S^5}}{(2\pi)^7\alpha'^4g_s^2}2r_0^4\sinh2\alpha_3
%\end{align}
%which indeed just turns the cosh into an exp:
%\begin{align}
%E-E_{ext}&=\frac{V_3 vol_{S^5}}{(2\pi)^7\alpha'^4g_s^2}r_0^4(3+2 e^{-2\alpha_3})
%\end{align}
%and the expression should be S-dual invariant, so we can put a hat everywhere. In the decoupling limit, the $\frac{r_0^4}{\alpha'^4}$ factor in front is fixed. 
 % {\color{ForestGreen} 
%   Inside the bracket, the second term drops, so we are left with:
%\begin{align}
%\mathcal{E}:=(E-E_{ext})_{dec}&=\frac{V_3 vol_{S^5}}{(2\pi)^7 g_s^2}3\bar{r}_0^4
%\end{align}
%where the $\bar{r}_0$ is fixed in the decoupling limit. 
The entropy  is given by
\be
S = \frac{r_0^5 V_3 vol_{S^5}\cosh \a_3}{2^5\pi^6 g_s^2 \a'^4}%=\frac{r_0^3 V_3 vol_{S^5}}{2^5\pi^6 g_s^2 \a'^4}\sqrt{\frac{r_0^4+\sqrt{r_0^8+(8\pi N g_s\alpha'^2)^2}}{2}}
\ee
In the decoupling limit \eqref{ncd3}, $\a_3 \r \infty$, and we are left with the standard thermodynamic relation in $\N=4$ SYM
%
%, the first term in the square root drops and we are left with:
%\begin{align}
%S&=\frac{\bar{r}_0^3 V_3 vol_{S^5}}{2^5\pi^6 g_s^2}\sqrt{4 \pi g_s N}
%\end{align}
%Extracting $\bar{r}_0$ from the energy expression, we get 
%(I hope I put the factors correctly)\emph{Can you check?}{\color{ForestGreen}(checked)}
\begin{align}
S&=(E V_3^{\frac{1}{3}})^{3/4}\sqrt[4]{\frac{2^5 \pi^2 N^2}{3^3}}
\end{align}
This expression is of course  S-duality invariant. %, so we expect to find the same from the NCOS perspective. 

% (which does nothing), this gives our entropy-energy relation in hatted variables and we should be able to reproduce the same result from Harmark.

  %}

\subsubsection*{The D3 NCOS}

So far for NCSYM. The D3 NCOS is obtained by turning on an \emph{electrical} $B$ - field instead. We may  similarly generate it in supergravity via a space-time TsT transformation, consisting of a T-duality along $x_1$, a shift $t \r t+ \g\, \tilde x_1$, and a T-duality back. We obtain
\be
ds^2 = \frac{\sqrt{f_3}}{f_3-\g^2 f} ( - f dt^2 + dx_1^2) + \frac{1}{\sqrt{f_3}} (dx_2^2+dx_3^2)   + \sqrt{f_3} \left(\frac{dr^2}{f}+r^2 d\Omega_5^2\right)  \;, \;\;\;\;\;\;e^{2\Phi} = \frac{g^2_s  f_3}{f_3-\g^2 f}  \nonumber 
\ee

\be
\a' B_{01}= \frac{\g f}{f_3 - \g^2 f} \;, \;\;\;\; C_{23} = -\gamma\frac{4\pi\alpha'^2 N}{r^4 f_3} \;, \;\;\; F_5 = 16\pi \a'^2 N  (\omega_{S^5} + \star \omega_{S^5}) \label{spttstd3}
\ee
The NCOS limit described in the main text can be easily noticed to correspond to the $\g \r 1$ limit. However, the scaling of the various parameters, especially at finite temperature,  is not entirely clear. To help out, it is useful  to notice that the background \eqref{spttstd3} is simply S-dual to \eqref{ncd3}. Thus, in this particular case, one may \emph{define} the (finite-temperature) D3 NCOS decoupling limit as being the S-dual to the decoupled NCSYM theory  and read off the appropriate scalings by rewriting \eqref{spttstd3} in terms of the S-dual variables \cite{Gopakumar:2000na}. 

To extract the D3 NCOS limit via this route, we should find the mapping of parameters between the background \eqref{spttstd3} and the S-dual to \eqref{ncd3}. More precisely, the background we compare \eqref{spttstd3} to is obtained by applying to \eqref{undefd3}, with   all the parameters and the coordinates hatted,  a (trivial) S-duality\footnote{So that  the string couplings we compare are proportional, rather than inversely proportional, to each other. }, then a TsT along $\hat x_{2,3}$ and an S-duality back.  We do indeed find that the two backgrounds are identical, provided we identify

\be
r=\hat r \, \sqrt[4]{1+\l^2} \;, \;\;\;\;\; x_{2,3} = \frac{\hat x_{2,3}}{\sqrt[4]{1+\l^2}}\;, \;\;\;\;\; x_{0,1} = \hat x_{0,1} \, \sqrt[4]{1+\l^2} \sqrt{1-\g^2} \label{coordrescd3}
\ee
with
\be
\g^2 = \frac{\l^2}{1+ \l^2 + (\sinh \a_3)^{-2}}  \;, \;\;\;\;\; (1+\l^2) \sinh^2 \a_3 = \sinh^2 \hat \a_3 \label{relparam}
\ee
and we remove the constant piece from  $B_\infty$. %\emph{Is this necessary, in the end?} {\color{ForestGreen}(It's not clear, maybe we need to turn it all into electric field in order to take the limit from Seiberg)}
 We are using $t$ and $x^0$ interchangeably.

The most important take-home message is that the relationship between $\l$ and $\g$ is energy-dependent, as 
 the parameter $\a_3$ is a direct measure of the energy above extremality %{\color{ForestGreen}(factors of $g_s$?)} 
 %\emph{$V_3$ or $\hat V_3$?}{\color{ForestGreen} ($\hat V_3$ and maybe we can also put a hat for the energy because it's computed wrt the hatted time)}

\be
\hat E=  \frac{\hat{V}_3\, vol_{S^5}}{(2 \pi)^7{\hat{g}_s^2}}\frac{\hat r_0^4}{\hat \a'^4} (3+2 e^{-2 \hat \a_3}) \xrightarrow[\hat  \alpha' \rightarrow 0]{\;}  \frac{\hat{V}_3 \pi^3}{(2 \pi)^7{\hat{g}_s^2}}\frac{3 \hat{r}_0^4}{\hat{\alpha}'^4}= \frac{3 \hat{V}_3 }{2^7 \pi^4 {\hat{g}_s^2}} \frac{8\pi \hat g_s N}{\hat \a'^2 \sinh 2 \hat \a_3} =  \frac{3 \hat{V}_3  N}{ 32 \pi^3 {\hat{g}_s}(\l \hat \a')^2 \sinh^2 \a_3}
\ee
which stays finite even after taking the decoupling limit \eqref{ncdeclim}, in which $\l \a'$ is held fixed. %(Here $V_3$ denotes the spatial volume of the D3 branes, while $vol_{S^5}$ is the volume of the unit five-sphere.) 
This implies a state-dependent relation between the time coordinates we chose. Thus, care must be exercised when taking the NCOS decoupling limit, as working in terms of coordinates that are rescaled from the fixed ones by energy-dependent factors can non-trivially affect the thermodynamics. The choice advocated in the literature is to hold the parameters and coordinates in the \emph{spatially} non-commutative SYM frame fixed, which implies, via \eqref{coordrescd3}, that $t,x_1$ in the NCOS frame are state-dependent. 
 To have the state-dependence appear only in the metric components, we can already pull out  a factor of $\sqrt{1-\g^2}$ from $t,x_1$ in \eqref{spttstd3}. This is precisely the choice of \cite{Maldacena:1999mh}, but now it becomes clear it is a compulsory one.  

The relation between the (finite) constants $g_s, \hat g_s$ we chose to parametrise the asymptotic string coupling with is
\be
 g_s = \hat g_s \sqrt{1-\g^2} \sqrt{1+\l^2}  = \hat g_s \sqrt{\frac{1+\l^2}{1+\l^2 \tanh^2 \a_3}}
\ee
which is also energy-dependent. This must be carefully taken into account when taking the decoupling limit. Note, however, that the asymptotic value of the dilaton in the NCOS frame,  $g_s/\sqrt{1-\g^2}$, is not energy-dependent, though it is divergent in the decoupling limit. 

The reason that we consider $g_s$, $\hat g_s$ instead of the respective asymptotic values of the dilaton is that these coupling parameters naturally enter the relation between the angles $\a_3, \hat \a_3$ and the quantized charge, $N$. Concretely, we have

\be
\hat r_0^4 \sinh 2 \hat \a_3 = 8 \pi \hat g_s N \hat \a'^2 \;, \;\;\;\;\; r_0^4 \sinh 2 \a_3 = 8 \pi g_s N \a'^2 \label{qcondd3}
\ee
Note we have allowed  $\a'$ and $\hat \a'$ to be different in principle.  Plugging in the expressions for $r_0, g_s, \a_3$ in terms of $\hat r_0,, \hat g_s, \hat \a_3$ in the second  relation one finds precisely the first one, provided that $ \a'= \hat \a'$. We have therefore succeeded in matching \emph{all} the parameters of the solution \eqref{spttstd3}  in terms of those of \eqref{ncd3} with hats added.

Note  so far our  match  is between  two asymptotically flat solutions, written in different parametrisations. To study the decoupling limit, we simply need to  
translate the NCSYM decoupling limit \eqref{ncdeclim} to unhatted variables. Note that   $\l = \theta/(\hat g_s\hat \a')$, where $\theta$ is the NCSYM non-commutativity parameter and the factor of $\hat g_s$ is due to the fact that the string length in the S-dual frame is $\hat g_s \hat \a'$; we will be writing $\l = b/\a'$, where $b\equiv  \th/\hat g_s$ is assumed to be fixed. Using this and the NCSYM scaling  $\hat r \sim \hat \a'$  in \eqref{coordrescd3}, we immediately obtain the NCOS decoupling limit %\textbf{Check!} {\color{ForestGreen}(checked)}
\be
\a' \r 0 \; \;\;\;\; \mbox{with} \;\;\;\; \frac{r}{\sqrt{\a'}} \;, \; \frac{r_0}{\sqrt{\a'}}\; ,  \;\; g_s \tanh \a_3 \equiv \hat g_s \;,\; \; \frac{(1-\g^2) \tanh^2 \a_3}{\a'^2}\equiv \frac{1}{b^2}  \;\;\;\mbox{fixed}
\ee
%where $b$ is defined in terms of the spatially noncommutative parametrisation as $b \equiv \l \a'$. 
It is not immediately obvious, from the point of view of the D3 NCOS, why $\g$ needs to approach $1$ in the above peculiar, energy-dependent way. This scaling is nontheless clear from  the spatially non-commutative point of view.

One oftentimes finds the decoupling limit written in terms of  mixed variables: $r_0, \a_3, \hat g_s, b$. In terms of these, the second  quantization condition in  \eqref{qcondd3} is modified to\footnote{The exact relation between $\a_3$ and $\bar r_0$ is  $
\sinh^2\alpha_3=\frac{1}{2(1+\lambda^2)}\bigg(\sqrt{1+\bigg(\frac{8\pi\hat{g}_s N(1+\lambda^2)}{\bar{r}_0^4}\bigg)^2}-1\bigg)$.} 

% In the above decoupling limit, letting $r_0 = \sqrt{ \a'} \bar r_0$ and $g_s = \hat g_s \coth \a_3$, we find that 

\be
\bar r_0^4  \sinh^2 \a_3 = 4 \pi \hat g_s N \left(1 - \frac{\a'^2}{{2}b^2 \sinh^2 \a_3} + \O(\a'^4) \right)
\ee
to leading order in the decoupling limit, where $\bar r_0 = r_0/\sqrt{\a'}$. This (including the first subleading contribution) affects the computation of the energy above extremality in these variables.
 %\textbf{Check first order is enough!}
%{\color{ForestGreen}(It works for small angles. I don't know how to solve the eq for arbitrary angle, but the corrections always come with extra powers of $\alpha'$, which don't matter in the decoupling limit)}%, to which we now turn. %. More about this in the D1-D5 case. \textbf{Please answer:}
%{\color{ForestGreen}
%The exact equation for the sinh
%\begin{align}
%\sinh^4\alpha_3+\frac{\sinh^2\alpha_3}{1+\lambda^2}-\bigg(\frac{4\pi \hat{g}_s N}{\bar{r}_0^4}\bigg)^2=0
%\end{align}
%with solution
%\begin{align}
%\sinh^2\alpha_3&=\frac{1}{2(1+\lambda^2)}\bigg(\sqrt{1+\bigg(\frac{8\pi\hat{g}_s N(1+\lambda^2)}{\bar{r}_0^4}\bigg)^2}-1\bigg)
%\end{align}
%}
%\bi
%\item who studied the D3 NCOS at finite temperature? Just Harmark?
%\item anything part. wrong with the naive termodynamic calculation in the unhatted variables?
%\ei

It is instructive to first perform the energy computation on the na\"{i}ve background \eqref{spttstd3}, i.e. assuming that the unhatted coordinates and $\g, g_s$ are fixed. 
%Let us now discuss the thermodynamics of this sytem. 
% The energy computation, performed using
  The covariant phase space formalism  on the na\"{i}ve background \eqref{spttstd3} yields the following %\textbf{Factors, notation!!}:
{
\begin{align}
\d M_{naive} &= \frac{V_3 \, vol_{S^5}}{2^5\pi^7 g_s^2\alpha'^4}\, \bigg[ \frac{r_0^3\delta r_0(5-\gamma^2)+4 r_3^3\delta r_3}{1-\gamma^2}-4B_{\infty}\gamma\big( r_0^3\delta r_0 + r_3^3\delta r_3 \big)+\nonumber\\
&\hspace{1.2cm} +4\pi N \alpha'^2 \left(\frac{\gamma}{1-\gamma^2}+\sqrt{1-\gamma^2}B_{\infty}\right)\delta C_{\infty}^{(2)}+12\pi N  \alpha'^2 \delta C_{\infty}^{(4)}  \bigg]
\end{align} 
where $r_3^4\equiv r_0^4\sinh^2\alpha_3$. Note that now, the value of the B-field at infinity does enter the energy computation in a non-trivial way: the energy contribution on the first line (the only one that survives if the values of the RR fields at infinity are energy-independent), reads

\be \label{naiveenerg}
M_{naive} = \frac{V_3 \, vol_{S^5} r_0^4}{(2\pi)^7 g_s^2\alpha'^4} \left[ 4 \cosh^2 \a_3 \left(\frac{1}{1-\g^2} - \g B_\infty \right) + 1 \right]
\ee
We note that for the value of $B_\infty$ obtained via TsT  in \eqref{spttstd3}, the $\g$ dependence in the total energy drops out.  Na\"{i}vely, this appears to fit in  nicely with the other thermodynamic quantities of this system - the Bekenstein-Hawking entropy and the temperature (viewed as the identification of the euclidean unhatted time  coordinate)  - which are  $\g$ - independent at fixed $g_s, V_3$, etc., due to  $f(r_0) =0$.
%
%
%, the $\g$ dependence drops out from the Bekenstein-Hawking entropy, and we obtain the same expression as in standard $\N=4$ SYM %\emph{Check!} {\color{ForestGreen}(checked)}
%%
%\be
%S = \frac{r_0^5 V_3 \cosh \a_3}{32\pi^3 g_s^2 \a'^4}
%\ee
%%Care must be taken, though, as $g_s$ is in fact energy-dependent, as is the radius of the $x^1$ circle.
%where $V_3$ is the volume in the unhatted coordinates. Also 
% the temperature (viewed as the identification of the euclidean unhatted time  coordinate) appears to be  $\g$ independent. %, but one needs to remember that the true time is rescaled with respect to $t$, as in \eqref{coordrescd3}.
Note, however, that  there is no parametric separation between the extremal contribution to the energy and the non-extremal one in the decoupling limit, as one would have expected.   

The computation above is incorrect because, as we have already argued, we are supposed to keep fixed the  hatted coordinates and vary $\g$ appropriately. In particular, the fact that 
% In this computation we didn't rescale the coordinates and didn't take into account that $\gamma$ also varies.
%t is interesting to note that, in this way of performing the calculation, 
the energy  depends non-trivially on the value of the B-field at infinity is strange, given that it did not \eqref{engnc} in  the S-dual background. To recover the correct result, it should be sufficient to rescale the coordinates in \eqref{spttstd3} by a factor of $\sqrt{1-\g^2}$, as \cite{Maldacena:1999mh}  does, and take into account the variation of $\g$. We find %\emph{Which $V_3$ are you using?}  {\color{ForestGreen} (I rescaled the coordinates with $1-\gamma^2$ factors, but not with $1+\lambda^2$. I also wrote explicitly the $1-\gamma^2$ dependence of $g_s$, so this $g_s$ is energy independent, although is not the hatted one.)}
\begin{align}
\delta \tilde M &=\frac{\tilde V_3 V_5}{2^5 \pi^7 \alpha'^4 \tilde g_s^2}\bigg(  \frac{r_0^3(5-6\gamma^2+\gamma^4)\delta r_0+2r_0^4 \gamma \delta \gamma +2r_3^3(2(1-\gamma^2)\delta r_3 + r_3\gamma\delta\gamma)}{(1-\gamma^2)^2} - \nonumber\\
&-B_{\infty} \frac{4r_0^3\gamma(1-\gamma^2)\delta r_0 + r_0^4(1+\gamma^2)\delta\gamma+r_3^3(4\gamma(1-\gamma^2)\delta r_3+r_3(1+\gamma^2)\delta\gamma)}{(1-\gamma^2)^2}+\nonumber\\
&\hspace{1.5 cm}+4\pi N\alpha'^2\big(\gamma+ B_{\infty}\big)\delta C^{(2)}_{\infty}+ 12\pi N \alpha'^2 \delta C^{(4)}_{\infty}\bigg)
\end{align}
where the tilde denotes quantities computed in the coordinates so rescaled and 
%
 %$r_3^4=r_0^4\sinh^2\alpha_3$ and
 $\tilde g_s = g_s/\sqrt{1-\g^2} $ is the asymptotic value of the dilaton at infinity. Upon using \eqref{relparam} and \eqref{qcondd3} to express the variation of $\g$ and $r_3$ in terms of that of $r_0$, we find that the coefficient of $B_{\infty}$ vanishes, as expected. 
%{\color{ForestGreen}To be more precise, we use that $\delta\bigg[\frac{r_3^4(r_3^4+r_0^4)}{1-\gamma^2}\bigg]=0$ by the quantization condition.} \emph{Keep?}
Under the reasonable assumption that the RR potentials at infinity are fixed, the total energy is given by integrating just the first term in field space, upon replacing $\d \g$ appropriately %\emph{Pseudo-hats? Does this agree with Harmark?}{\color{ForestGreen}(after taking into account the rescalings with $1+\lambda^2$ in $V_3,r_0,g_s$ and also the total $E$ vs $\hat{E}$, we get the same energy as in the other parametrization)}
\be
\tilde M = \frac{\tilde V_3 \, vol_{S^5}}{(2 \pi)^7 \alpha'^4 \tilde g_s^2} \left[5r_0^4+4 r_3^4(1+\lambda^2)\right] =   \frac{\tilde V_3 \,  vol_{S^5}}{(2 \pi)^7 \alpha'^4 \tilde g_s^2} \left[5r_0^4+4 r_0^4 \left(\frac{\cosh^2 \a_3}{1-\g^2}-1\right)\right]
\ee
Using \eqref{relparam}, we can trade the $\sinh \a_3$ inside $r_3$ above for a $\sinh\hat \a_3$, and the whole expression can be checked to agree with the NCSYM one, upon the appropriate rescalings by powers of $(1+\l^2)$ to reach the hatted varibles. Note, also, that our second rewriting in terms of $\g, \a_3$ agrees with the na\"{i}ve energy computed in \eqref{naiveenerg} (the rescalings of space, time and $g_s$ by $\sqrt{1-\g^2}$ cancel each other), \emph{provided} we set the $B$-field at infinity to zero. That the  metric contribution to the energy does not depend on the particular integration through field space is not surprising, given it can simply be extracted from the falloff of the metric components near asymptotically flat infinity. 

Obtaining the energy above extremality in the hatted NCSYM variables is trivial and explained in the previous subsection.  If, however, we insist  upon working  in terms of $r_0$ and $\a_3$, we need to use the appropriate quantization condition, including the subleading terms, in order to obtain the correct expression for the energy above extremality.

It is not hard to check that the entropy and temperature also come out correctly upon making the appropriate rescalings. The take-home message is that one needs to be very careful in which parameters one works in order to obtain the correct thermodynamic quantities and scalings in the decoupling limit. %; one may check whether requiring that the metric asymptote to a fixed Minkowski one is one consistency requirement. \textcolor{red}{ \emph{In which frame?}}

\subsection{The NCOS decoupling limit of the D1-D5 system}

We consider, for simplicity, the D1-D5 system on $T^4$. 
The simplest way to turn on a $B$-field along the common D1-D5 direction is, as in the main text, to perform a  spacetime TsT (T-duality on $\s$, shift $t\rightarrow t+\gamma \tilde{\s}$, T-duality on $\tilde{\s}$). The resulting background (also presented in appendix \ref{attrsolsB}) is given by
\begin{align} \label{spacetimeTsT}
ds^2&=\frac{\sqrt{f_1 f_5}}{f_1 f_5-\gamma^2 f}(-fdt^2+d\sigma^2)+\sqrt{f_1 f_5}\bigg(\frac{dr^2}{f}+r^2 d\Omega_3^2\bigg)+\sqrt{\frac{f_1}{f_5}}\sum_{i=2}^5 dx_i^2
\end{align}
\be
B=\frac{\gamma f}{f_1 f_5-\gamma^2 f}dt\wedge d\sigma\;, \;\;\;\;\; e^{\Phi}=g_s\frac{f_1}{\sqrt{f_1 f_5-\gamma^2 f}} \;, \;\;\;\;\nonumber
\ee
and RR field strengths
\be
 F_1=\gamma \frac{2 \alpha'p}{v r^3f_1^2}dr \;, \;\;\;\;\;
F_3=2\alpha' k\omega_{S^3}+\frac{2\alpha'p}{v r^3f_1}\frac{f_5}{f_1 f_5-\gamma^2 f}dr\wedge dt\wedge d\sigma \label{spttstRR}
\ee
\be
F_5=-\gamma\frac{2\alpha' k}{ r^3f_5^2}dr\wedge\omega_{T^4}+2\alpha'k\frac{f\gamma}{f_1f_5-\gamma^2 f}dt\wedge d\sigma\wedge\omega_{S^3} \nonumber
\ee
where the functions that appear are parametrised in the standard D1-D5 fashion
\begin{align}
f=1-\frac{r_0^2}{r^2}\hspace{1cm} &f_{1,5}=1+\frac{r^2_{1,5}}{r^2}\hspace{1cm}r_{1,5}^2=r_0^2\sinh \alpha_{1,5}^2\\
r_0^2\sinh 2\alpha_1&=\frac{2\alpha' g_s p}{v} \hspace{1cm}r_0^2\sinh 2\alpha_5=2\alpha' g_s k
\end{align}
However, as we just saw in the case of D3-branes, the natural parametrisation  obtained from performing spacetime TsT can be quite misleading - or, at least, unintuitive -  from the point of view of the thermodynamics and of which quantities should be fixed in the decoupling limit. It is therefore quite useful to try to map the decoupling limit of the D1-D5 NCOS at finite temperature to that of \emph{spatially} non-commutative D3 branes. To our knowledge, the finite-temperature limit has only been previously discussed from this point of view for pure D5 branes in \cite{Harmark:2000wv}; the NCOS limit of the D1-D5 system has only been written down for the extremal case in \cite{Berman:2000jw} but, as we will see, its NCOS limit at finite-temperature is significantly more subtle. %\textbf{Check and add refs!}

The goal of this appendix is to map the above D1-D5 system in a spatio-temporal $B$ - field to D3 branes in a purely spatial $B$ - field, and to use the non-commutative decoupling limit of the latter system to \emph{define} what we mean by NCOS  decoupling limit 
of  D1-D5. 
%In order to understand the scaling of different parameters in decoupling limit, it is again useful to note that there is another way to obtain this background, via D3-branes.
 The precise chain of dualities  we consider is $T^2 S(TsT)^2ST^2$, acting on a (hatted) D1-D5 solution. Concretely, we start with the D1-D5 system, with $k$ D5-branes lying along the $0,1,2,3,4,5$ directions, and $p$ D1-branes lying along  $0,1$. By 
   performing two T-dualities along the $4,5$ directions,
  we turn it into a stack of intersecting D3 branes lying   along $0,1,2,3$ and, respectively,  $0,1,4,5$, and smeared along the remaining two directions. As in the previous section, we further perform an S-duality for convenience, which does not affect the form of the solution. This preparatory phase of our system can be pictured as

\vskip4mm
  
  \tikzstyle{rect} = [draw, rectangle, minimum height = 4em, text width = 6.5em, text centered]
\tikzstyle{arrow} = [draw, -latex']

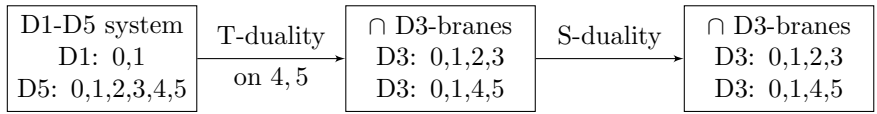
\begin{figure}[H]
    \begin{center}
        \begin{tikzpicture}[node distance = 4.5cm, auto]
        \small
        \node[rect](d1d5) {D1-D5 \text{system}\\ D1: 0,1\\
        D5: 0,1,2,3,4,5  };
        \node[rect, right of=d1d5](D3) {$\cap$ D3-\text{branes}\\
        D3: 0,1,2,3\\
        D3: 0,1,4,5};
        \node[rect, right of=D3](dd3) {$\cap$ D3-\text{branes}\\
        D3: 0,1,2,3\\
        D3: 0,1,4,5};

        \path[arrow] (d1d5) -- node [above]{T-duality} node[below]{on $4,5$}(D3) ;
        \path[arrow] (D3) -- node[above]{S-duality}(dd3);
        \end{tikzpicture}
        \caption{The chain of dualities taking the D1-D5 system to  intersecting D3-branes.}
        \label{figgduality}
    \end{center}
\end{figure}
\vskip-2mm
 \noindent Next, we make the theory on the D3-branes spatially non-commutative by turning on a B-field.  To generate the B-field, we perform two sets of TsT transformations along the spatial worldvolume directions of the D3s: a T-duality on $x_2$, followed by a shift $x_3\rightarrow x_3+\lambda \tilde{x}_2$ and a T-duality back on $\tilde{x}_2$, as well as a T-duality on $x_4$, a shift $x_5\rightarrow x_5+\mu \tilde{x}_4$, and a T-duality on $\tilde{x}_4$. The parameters of these two TsT transformations are related by the requirement that upon a further S-duality, one just obtains the system of intersecting D3 branes in a spacetime B-field with legs purely along $x^{0,1}$.    Next, we perform another S-duality and finally, two T-dualities on the directions $4,5$, in order to come back to a system of D5 and D1 branes.

\tikzstyle{rect} = [draw, rectangle, minimum height = 4em, text width = 6.5em, text centered]
\tikzstyle{arrow} = [draw, -latex']

\vskip4mm

\begin{figure}[H]
    \begin{center}
        \begin{tikzpicture}[node distance = 4.5cm, auto]
        \small
        \node[rect](dd3) {$\cap$ D3-\text{branes}\\
        D3: 0,1,2,3\\
        D3: 0,1,4,5};
        \node[rect, right of =dd3](tst) {NC \, D3 \\
        D3: 0,1,\textbf{2,3}\\
        D3: 0,1,\textbf{4,5}};
        \node[rect, right of =tst](sdu) {D3 NCOS\\
        D3: \textbf{0,1},2,3\\
        D3: \textbf{0,1},4,5};
        \node[rect, right of= sdu](final) {D1-D5 NCOS\\
        D1: \textbf{0,1}\\
        D5:\! \textbf{0,1},2,3,4,5};

        \path[arrow] (dd3) -- node[above]{\; TsT on 2,3 \;} node[below]{TsT on 4,5}(tst);
        \path[arrow] (tst) -- node[above]{S-duality} (sdu);
        \path[arrow] (sdu) -- node[above]{T-duality} node[below]{on 4,5} (final);
        
        \end{tikzpicture}
    \end{center}
    \caption{The chain of dualities relating spatially non-commutative D3 branes ($2^{nd}$ box) to the D1-D5 NCOS. The directions in bold represent the non-commutative directions.  }
       \label{figdual2}
\end{figure}
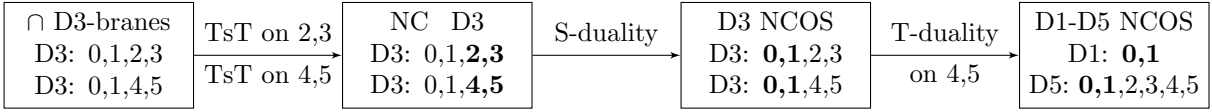

\noindent  Applying this chain of transformations to the undeformed D1-D5 system with hatted parameters, we obtain the following NS-NS fields
% \textbf{\emph{Replace $\hat Q_i$ by their expressions!}}
\begin{align}\label{chaindual}
ds^2&=\sqrt{(\hat{f}_5+\lambda^2 \hat{f}_1)(\hat{f}_1+\mu^2 \hat{f}_5)}\bigg(\frac{-\hat{f} d\hat{t}^2+d\hat{\sigma}^2}{\hat{f}_1 \hat{f}_5} +\frac{d\hat{r}^2}{\hat{f}}+\hat{r}^2 d\Omega_3^2\bigg)+\sqrt{\frac{\hat{f}_1+\mu^2 \hat{f}_5}{\hat{f}_5+\lambda^2 \hat{f}_1}}\, \sum_{i=2}^5 d\hat{x}_i^2 
\end{align}
\be
B= \hat{\alpha'}\hat{g}_s\frac{\cos\theta}{4}(\mu k+\frac{\lambda p}{\hat{v}} )d\psi\wedge d\phi -\hat{\alpha'}\hat{g}_s\bigg(\frac{\mu p}{\hat{v}\hat{r}^2 \hat{f}_1}+\frac{\lambda k}{\hat{r}^2 \hat{f}_5}\bigg) d\hat{t}\wedge d\hat{\sigma} \;, \;\;\;\;\;e^{2\Phi}=\hat{g}_s^2 \frac{(\hat{f}_1+\mu^2 \hat{f}_5)^2}{\hat{f}_1 \hat{f}_5} \nonumber
\ee
where, for illustation purposes, we have let the parameters $\l, \mu$ of the two TsT transformationsbe  arbitrary. Comparing with \eqref{spacetimeTsT}, it should be very clear that the two backgrounds can be the same only provided the first term in the B-field vanishes, which sets 
%\emph{Replace?}

\be
\mu k = - \frac{\l p}{\hat{v}} \label{relmul}
\ee
Thus, the two parameters are not independent of each other; however, we will oftentimes find it convenient to not plug in their relation above, so the formulae simplify and are more symmetric. The expressions for the RR fields read %\emph{Replace?}
\bea
F_3&=&\hat{\alpha'}\bigg[2k-\frac{2\mu \hat{f}_5}{\hat{f}_1+\mu^2 \hat{f}_5}\bigg(\frac{\lambda p}{\hat{v}}+\mu k\bigg)\bigg]\,\omega_{S^3}+\frac{2\hat{\alpha}'}{\hat{r}^3(\hat{f}_1+\mu^2 \hat{f}_5)}\bigg(\frac{p}{\hat{v}\hat{f}_1}-\frac{\mu\lambda k}{\hat{f}_5}\bigg)d\hat{r}\wedge d\hat{t}\wedge d\hat{\sigma} \nonumber \\
{F_1}&=&d\bigg(\frac{\mu \hat{f}_5}{\hat{g_s}(\hat{f}_1+\mu^2 \hat{f}_5)}\bigg)\;, \;\;\;\;\; {F}_5=-d\bigg(\frac{\lambda \hat{f}_1}{\hat{g_s}(\hat{f}_5+\lambda^2 \hat{f}_1)}\bigg)\wedge \omega_{T^4}+(...) \label{RRhatted}
\eea
where the dots represent the term that renders ${F}_5$ self-dual. Note the expression for $F_3$ simplifies upon using \eqref{relmul}. 
The functions labeling the solution are those of the hatted seed D1-D5 system (leftmost box in figure \ref{figgduality})
\begin{align}
\hat{f}=1-\frac{\hat{r}_0^2}{\hat{r}^2}\hspace{1cm} &\hat{f}_{1,5}=1+\frac{\hat{r}^2_{1,5}}{\hat{r}^2}\hspace{1cm}\hat{r}_{1,5}^2=\hat{r}_0^2\sinh \hat{\alpha}_{1,5}^2\\
\hat{r}_0^2\sinh 2\hat{\alpha}_1&=\frac{2\hat{\alpha}' \hat{g}_s p}{\hat{v}} \hspace{1cm}\hat{r}_0^2\sinh 2\hat{\alpha}_5=2\hat{\alpha}' \hat{g}_s k
\end{align}
Note that we have again allowed for the possibility that $\hat \a' \neq \a'$.

%{\color{ForestGreen}
%(details)
%After using the relation between $\lambda$ and $\mu$ we are left with
%\begin{align}
%\hat{F}_3&=2\alpha'k\omega_{S^3}+\frac{2\alpha' p}{ \hat{r}^3\hat{v}(\hat{f}_1+\mu^2 \hat{f}_5)}\frac{(\hat{f}_5+\lambda^2 \hat{f}_1)}{\hat{f}_1\hat{f}_5}d\hat{r}\wedge d\hat{t}\wedge d\hat{\sigma}
%\end{align}
%Now we just need to use the relations between hatted and unhatted variables and functions and we match the other $F_3$. For $F_1$ and $F_5$ the matching is more involved because we need to express $\gamma$ in terms of everything else, but nevertheless it works.
%
%}
Quite remarkably, we find that the backgrounds \eqref{chaindual} - \eqref{RRhatted} and \eqref{spacetimeTsT} -\eqref{spttstRR} are \emph{identical}, after imposing the relation \eqref{relmul} between the two TsT parameters $\lambda,\mu$, appropriately rescaling the coordinates and shifting the B-field  by a constant piece. Below, we first describe the map between the parameters and coordinates of the 
 two asymptotically flat backgrounds.  Then, we use this identification to derive the D1-D5 NCOS decoupling limit starting from the spacelike non-commutative D3 one.

\subsubsection*{Matching the parameters}

Upon closer inspection, the backgrounds  \eqref{chaindual} and \eqref{spacetimeTsT} are found to be exactly the same, including the RR fields,  upon
making the following identifications

\begin{align}\label{relationcoord}
r &= \sqrt[4]{(1+\l^2)(1+\mu^2)} \, \hat r\;, \;\;\;\;\; r_0 = \sqrt[4]{(1+\l^2)(1+\mu^2)} \, \hat r_0\; \nonumber\\[2pt]
x_{0,1} &= \sqrt[4]{(1+\l^2)(1+\mu^2)} \sqrt{1-\g^2}\, \hat x_{0,1}\;, \;\;\;\;\;\;
x_i = \sqrt[4]{\frac{1+\mu^2}{1+\l^2}} \ \hat x_i
\end{align}
where %{\color{blue}\emph{Notation! $r_0 \r 0$ limit?}}
\be \label{relationparam1}
1-\g^2 = \frac{(\hat r_5^2+\l^2 \hat r_1^2)(\hat r_1^2 + \mu^2 \hat r_5^2)}{\hat r_1^2 \hat r_5^2 (1+\l^2) (1+\mu^2)} \;, \;\;\;\;\;\; \mu k = - \frac{\l p}{ \hat v}% = - \l k \frac{\sinh 2 \hat \a_1}{\sinh 2 \hat \a_5} 
\ee
and $\hat r_i^2 = \hat r_0^2 \sinh^2 \hat \a_i$. 
Note, again, the energy-dependent rescaling of space and time. If one of $\hat r_{1,5}$ vanishes, then one may alternatively use the following  relation  between $\g$ and the angle parameters
\be \label{relnonextr0}
(1-\g^2) (\hat r_1^2+\hat r_5^2) - \g^2 \hat r_0^2 = \frac{\hat r_5^2 + \l^2 \hat r_1^2}{1+\l^2} + \frac{\hat r_1^2 + \mu^2 \hat r_5^2}{1+\mu^2}
\ee
%\textcolor{magenta}
{These relations are equally valid in the extremal limit, $\hat r_0 \r 0$; one simply needs to replace $\hat r_{1,5}$ by their expressions in terms of the quantised charges and the moduli, finding}
\begin{align}\label{extremalrelparam}
1-\gamma^2&=\frac{(1+\frac{\lambda^2 p}{k \hat{v}})(1+\frac{\mu^2 k \hat{v}}{p})}{(1+\lambda^2)(1+\mu^2)}% =\frac{(1+\frac{\lambda^2 p}{k \hat{v}})^2}{(1+\lambda^2)(1+\mu^2)}
\end{align}
%\textcolor{magenta}
{which may be rewritten in various ways using \eqref{relmul}.}
The relation between the  parameters labeling the string coupling and the internal space volumes is 
\be \label{relationparam2}
g_s = \hat g_s \sqrt{1-\g^2} (1+\mu^2) \;, \;\;\;\; v = \frac{1+\mu^2}{1+\l^2} \hat v
\ee
with the former relation energy-dependent. The relation between the various angles comes out to be 
\be 
\sinh \a_1 = \sqrt{\frac{\sinh^2 \hat \a_1 + \mu^2 \sinh^2 \hat \a_5}{1+\mu^2}} \;, \;\;\;\;\;\; \sinh \a_5 = \sqrt{\frac{\sinh^2 \hat \a_5 + \l^2 \sinh^2 \hat \a_1}{1+\l^2}} \label{alitoalh}
\ee
It is possible to  write the same relation in terms of the harmonic functions as
\begin{align}
f_1&=\frac{\hat{f}_1+\mu^2 \hat{f}_5}{1+\mu^2}\hspace{1cm}f_5=\frac{\hat{f}_5+\lambda^2 \hat{f}_1}{1+\lambda^2} \label{relharm}
\end{align}
Finally, let us look at the charge quantization conditions. Starting with

\be
r_0^2 \sinh 2 \a_1 = 2 \frac{\a' g_s}{v} p \;, \;\;\;\;\;\; r_0^2 \sinh 2 \a_5 = 2 \a' g_s k
\ee
and plugging in the expressions for $r_0, g_s,v$ in terms of their hatted counterparts, we find

\be \label{quantizeq}
\hat r_0^2 \sqrt{1+\mu^2} \sinh 2 \a_1 = \frac{2 \a' \hat g_s p}{\hat v} \, \frac{\sinh \a_1 \sinh \a_5 \sqrt{1+\l^2}}{\sinh \hat \a_1 \sinh \hat \a_5}
\ee
Using \eqref{alitoalh} and the relation  \eqref{relmul} between $\l$ and $\mu$, which implies that $\frac{\mu}{\lambda}=-\frac{p}{k\hat{v}}=-\frac{\sinh 2\hat{\alpha}_1}{\sinh 2\hat{\alpha}_5}$, we can write

\be
\cosh \a_1 = \sqrt{\frac{\cosh^2 \hat \a_1 + \mu^2 \cosh^2 \hat \a_5}{1+\mu^2}} = \frac{\cosh \hat \a_1 \sinh \a_5 \sqrt{1+\l^2}}{\sqrt{1+\mu^2} \sinh \hat \a_5} \label{relalalhat}
\ee
Plugging into \eqref{quantizeq}, we immediately find the other frame's quantization condition, thus impliying that $\hat \a' = \a'$. We have thus succeeded in fully matching the parameters of the two ways to write the solution. The only difference is a constant shift in the B-field (as visible at infinity) which can be absorbed into worldvolume flux via a gauge transformation.  
%
%
%{\color{ForestGreen}
%Finally, we note that, the B-field obtained by $T^2S(TsT)^2ST^2$ vanishes asymptotically, which means that it differs from the ones obtained by spacetime TsT by a constant piece $b_{01\infty}^2 = \gamma^2(1+\lambda^2)(1+\mu^2)$. 
%}

So far, we have described how to go from the hatted to the unhatted parameters. To go the other way, we can use
%
%Similarly, we may want to know how to go from the unhatted to the hatted variables in the general case: given $r_0, \a_i, g_s, \g$, how to construct $\l, \mu, \hat \a_i, \hat g_s$, etc.? We have
%
\be
\sinh \hat \a_1 \sinh \hat \a_5 = \frac{\sinh \a_1 \sinh \a_5}{\sqrt{1-\g^2}} \;, \;\;\;\;\;\;  \sinh^2 \hat \a_1 + \sinh^2 \hat \a_5  = \frac{\sinh^2 \a_1 +\sinh^2 \a_5 + \g^2}{1-\g^2} 
\ee
which yields $\hat \a_{1,5}$ in terms of $\a_{1,5}$ and $\g$.  In the  $\g \r 1$ limit that  will be of most interest to us, 
 the two solutions are, roughly
\be
\sinh^2 \hat \a_5 \approx \frac{ \sinh^2  \a_1 + \sinh^2  \a_5 +\g^2}{1-\g^2} \;, \;\;\;\; \sinh^2 \hat \a_1 \approx \frac{\sinh^2 \a_1 \sinh^2 \a_5}{\sinh^2 \a_1 + \sinh^2 \a_5 + \g^2} 
\ee
where we have simply called $\hat \a_5$ the divergent one.  The formulae for $\l,\mu$ can then be obtained from \eqref{alitoalh} 

\be
\l^2 = \frac{\sinh^2 \hat \a_5 - \sinh^2 \a_5}{\sinh^2 \a_5 - \sinh^2 \hat \a_1} \approx \frac{(\sinh^2 \a_1+\g^2 \cosh^2 \a_5)(\sinh^2 \a_1 + \sinh^2 \a_5 + \g^2)}{(1-\g^2) \sinh^2 \a_5 (\sinh^2 \a_5 + \g^2)} 
\ee
\be
 \mu^2= \frac{\sinh^2 \hat \a_1 - \sinh^2 \a_1}{\sinh^2 \a_1 - \sinh^2 \hat \a_5} \approx \frac{(1-\g^2) \sinh^2 \a_1 (\sinh^2 \a_1 + \g^2)}{(\sinh^2 \a_5+\g^2 \cosh^2 \a_1)(\sinh^2 \a_1 + \sinh^2 \a_5 + \g^2)} 
\ee
Again, from the perspective of the unhatted variables it is not at all obvious why  these particular combinations need to be fixed. In particular, the angles must be varied such that 

\be
\l^2 \mu^2 = \frac{(\sinh^2 \a_1 + \cosh^2 \a_5)\sinh^2 2\a_1}{(\sinh^2 \a_5 + \cosh^2 \a_1) \sinh^2 2 \a_5} \label{weirdratio}
\ee
is fixed, a condition that is entirely non-obvious. 

\subsubsection*{The decoupling limit}

As we saw above, the relation between the two parametrisations and coordinates is state-dependent. As discussed, the parameters that should be held fixed are the  NCSYM ones. We thus take 
\be
\a ' \r 0\;, \;\;\l = \frac{b}{\a'} \; \;\;\;\mbox{with} \;\;\;\; b = \frac{\theta}{\hat g_s^{(3)}} \,, \;\;  \hat g_s^{(3)} =  \frac{\hat g_s \a'}{\hat R_4 \hat R_5}\,, \; \; \frac{\hat r}{\a'}\, , \; \frac{\hat r_0}{\a'} \;\; \mbox{fixed}
\ee
%with $b, \hat \a_1, \hat g_s \a'$ fixed. Note $b = \theta/\hat g_s^{(3)}$, where $\hat g_s^{(3)} =  \hat g_s \a'/{\hat R_4 \hat R_5}$ 
where $\theta$ is the spatial non-commutativity parameter and $\hat g_s^{(3)} $
is the  initial SYM coupling constant  and $\hat{R}_{4,5}$ are the radii of the corresponding circles on $T^4$. In terms of the parameters of the original D1-D5 sytem, most scalings correspond to the D5 decoupling limit 
\be \label{decouphat}
\a' \r 0 \;\;\; \mbox{with} \;\;\;\; g_s \a' \,, \;\; \hat v \, \a'^2 \, , \;\;  \frac{\hat r}{\a'}\, , \; \frac{\hat r_0}{\a'} \;\; \mbox{fixed}
%\hat r_0 \propto \a' \;, \;\;\;\; \sinh \hat \a_5 \propto \frac{1}{\a'}, \;\;\;\; \hat \a_1 \;\;\emph{fixed} , \;\; \hat v \propto \frac{1}{\a'^2}   
\ee
that, upon an additional S-duality, lands precisely on the LST decoupling limit. This implies, as usual,  that 
\be \label{decoupsc2}
 \sinh \hat \a_5 \propto \frac{1}{\a'}, \;\;\;\; \hat \a_1 \;\;\emph{fixed}
\ee
in the decoupling limit. To this, we only need to add the scaling $\l = b/\a'$ with $b$ fixed.  

With this choice,  it is  interesting to note that the parameter $\mu$ of the other TsT scales as 

\be
\mu = - \frac{\l p}{k \hat v} \propto  \a' 
\ee
Thus, we are only taking the standard non-commutative decoupling limit for the stack of D3-branes that are related to the D5-branes via T-dualities, and not for those related to the D1-branes. We may consequently argue that the UV of the theory is dominated by the D5 NCOS, with its characteristic Hagedorn scaling \cite{Harmark:2000wv}, and not the D1 one, which has a rather different description \cite{Klebanov:2000pp}. This is just like in the NS5-F1 system, where  the UV is dominated by the NS5-branes. 

%This is how the symmetry between the two stacks is broken, and the UV is dominated by the D5 NCOS. \emph{Any way to make precise?}

Turning to the unhatted variables, the above scalings and \eqref{relalalhat} imply that both $\a_1$ and $\a_5$ are finite in the decoupled geometry, though they have relations among them. For example, when $\a_{1,5}$ vary, the ratio \eqref{weirdratio} must be held fixed if the theory is to be fixed. When both angles are non-zero, it is therefore best to work in terms of hatted variables to see the structure of the solution.

 Nonetheless, the $\hat \a_1 =0$ case (which implies  that $p=0$, so only D5-branes are present) is simple enough to treat in the unhatted variables. It is easy to check that $\mu =0$, which in turn sets $\a_1=0$, while 
 \be
 \sinh \a_5 = \frac{\sinh \hat \a_5}{\sqrt{1+\l^2}} \;, \;\;\;\;\; 1-\g^2 =  \frac{\cosh^2 \a_5}{\cosh^2 \hat \a_5}
 \ee
Plugging in this expression for $\g$ into the quantization condition yields

\be
r_0^2 \sinh 2 \a_5 = 2 \hat g_s \a' k \frac{\cosh \a_5}{\cosh \hat \a_5} \;\;\;\Rightarrow \;\;\; r_0^2 \sinh^2 \a_5 = \frac{\hat g_s \a' k}{\l}
\ee
in agreement with \cite{Harmark:2000wv}. We note again that such a simple condition does not occur when also D1 branes are present. If we wanted to go from unhatted to hatted variables, then we'd have

\be
\cosh \hat \a_5 = \frac{\cosh\a_5}{\sqrt{1-\g^2}} \;, \;\;\;\; \l = \frac{\g \coth \a_5}{1-\g^2}
\ee
with the latter assumed state-independent. 

Let us also comment on the decoupling limit in the extremal case. Then, \eqref{extremalrelparam} reduces to
\be
1-\gamma^2 \approx \frac{1}{\l^2} \left( 1+ \frac{\l^2 p}{k \hat v} \right)^2 \approx \frac{1}{\l^2} \left( 1+ \frac{v_*}{  v} \right)^2 
\ee
where we used that in the decoupling limit $\gamma\rightarrow 0$ and $\lambda\rightarrow\infty$, while $v\sim \hat v/\l^2$ is fixed. Since the second term is fixed, it is clear that we will have $\gamma\rightarrow 1$. %\textcolor{magenta}
{ The same will be true whether we have D1 branes or not, as setting their number to zero (i.e., setting $v_* =0$) only changes the value of the product $(1-\g^2) \l^2$, and not the fact that the first term must go to zero if the second goes to infinity.  All the angles involved are infinite in this limit.}
%On the other hand, we see that we can also set from the beginning $p=0,\gamma=0$ (pure D5) and the decoupling limit will not be affected.
%Thus, we conclude that in the extremal case the presence of D1s is not important for the decoupling limit.

%
%{\color{blue}
%
% Notice also that in the particular case $\a_1=\a_5$, the limit fixes $\coth 2 \a_5/\sqrt{1-\g^2}$, and not $\coth \a_5/\sqrt{1-\g^2} $, as in the $\a_1=0$ case.  \emph{Ok, but then why did the entropy look right? Cancelling mistakes?}
%
%}

\subsubsection*{Thermodynamics}

In the NCSYM parametrisation, the entropy is 
%{\color{ForestGreen}
\be
{S} = \frac{2\pi \hat v \hat R\hat r_0^3 }{\a'^2 \hat g_s^2} \sqrt{\hat{f}_1 \hat{f}_5} |_{\hat{r}=\hat{r}_0}= \frac{2\pi \hat v \hat R \hat r_0^3 \cosh \hat \a_1 \cosh \hat \a_5}{\hat g_s^2 \a'^2}
\ee
which is $\l$ - independent. One may check that the only contribution to the energy is from the metric, which is identical to the undeformed case

\be
\hat M = \frac{\hat R \hat v \hat r_0^2}{2\a'^2 g_s^2} (\cosh 2\hat \a_1+ \cosh 2\hat \a_5 +1) = \b_t \frac{R v r_0^2}{2\a'^2 g_s^2} \left(\frac{\cosh 2\a_1+ \cosh 2\a_5}{1-\g^2} +1 \right)
\ee
In the second equality, we have written the result of the same calculation of the metric contribution using the unhatted variables, where the factor $\beta_t$  accounts for the rescaling $\hat E = \beta_t E$ that follows from the rescaling \eqref{relationcoord} of the coordinates\footnote{%\textcolor{magenta}
{Nonetheless, as discussed in appendix \ref{D3ncos} for the D3 NCOS, if one na\"{i}vely performs the full energy calculation in unhatted variables, the B-field contribution will not be correct.}}. %\textcolor{magenta}
{Note that, in either parametrisation, the total energy in the decoupling limit is divergent, as it should. }

 %{\color{ForestGreen}(the only contribution is from the metric, no matter the value of the B-field at infinity)}Also, 
 
 Since the relative  part of the metric involving $\hat t, \hat r$ is unmodified, then 
the temperature is  the same as in the undeformed background. %\textcolor{magenta}
{Finally, given that the decoupling limit for the hatted parameters appearing above is identical to the D5 decoupling limit of the D1-D5 system, we trivially obtain the same LST thermodynamics as for $\l=0$. }

%
%{\color{blue}
%In the unhatted parametrisation, the entropy iis 
%
%\be
%S = \frac{r_0^3 v R \cosh \a_1 \cosh \a_5}{g_s^2 \a'^2}
%\ee
%The metric contribution to the energy is 
%
%
%using \eqref{}. . It is important, for the decoupling limit, that the relative size of the extremal and above extremality contributions is divergent (this wouldn't occur if we add the B-field contribution). With this, the energy above extremality also comes out correctly. 
%{\color{ForestGreen}(understand why do we need to set the B-field at infinity to 0)}

}

\subsubsection*{Irrelevant couplings}

As a final remark, let us also compute the coefficients  of the irrelevant deformations
present in the background \eqref{chaindual} - \eqref{RRhatted}  using the hatted parametrisation.  Upon the standard rescaling by $2\sqrt{kp}$ with respect to \eqref{irrcouplgen}, we find
% in this hatted parametrization. We obtain (put the hat also on $\alpha'$? we know it's the same):
\begin{align}\label{irrcoefnewpar}
\hat \lambda_+&=\hat{g}_s\alpha' \frac{p+k \hat{v}}{\hat{v}}\hspace{0.5cm}\hat \lambda_-=\frac{\alpha' \hat{g}_s (p-k\hat{v})(p-k\hat{v}\mu^2)}{\hat{v}(p+k\hat{v}\mu^2)}\hspace{0.5cm}\hat \lambda'_-=\frac{2\mu\alpha' \hat{g}_s \sqrt{pk}\, (p-k\hat{v})}{\sqrt{\hat{v}}(p+k\hat{v}\mu^2)}
\end{align}
where the hats indicate that they differ by a rescaling, associated to the rescaling \eqref{relationcoord} of the coordinates, from the irrelevant couplings \eqref{irredeftstsol} read off from the unhatted background. 

We may easily check  these couplings satisfy 
%\emph{\textcolor{red}{Hats!!}}
\begin{align}
\hat{\lambda}_+^2-\hat{\lambda}_-^2-\hat{\lambda}'^2_-&=\frac{4 \alpha'^2 \hat{g}_s^2 kp}{\hat{v}}
\end{align}
In the decoupling limit, since $\hat{g}_s\alpha'$ is fixed and $\hat{v}\rightarrow\infty$, we obtain that $\hat{\lambda}_+=\sqrt{\hat{\lambda}_-^2+\hat{\lambda}'^2_-}$. We also note that in the decoupling limit  %\emph{\textcolor{red}{Rescale and check signs!!}}
\begin{align}
\hat{\lambda}_+\rightarrow k \hat{g}_s \alpha'\hspace{0.5cm}\hat{\lambda}_-\rightarrow -\frac{\alpha' \hat{g}_s k(p-k\hat{v}\mu^2)}{(p+k\hat{v}\mu^2)}\hspace{0.5cm}\hat{\lambda}'_-\rightarrow -\frac{2\mu\alpha' \hat{g}_s k \sqrt{kp}\sqrt{\hat{v}}}{p+k\hat{v}\mu^2}
\end{align}
so the self-dual irrelevant coupling corresponds precisely to the LST one. The remaining couplings can be rewritten in terms of $v$ - see \eqref{irrcoefnewparlim} - which represents the physical size of the internal space  in string units. For this, one may use \eqref{relationparam2} and \eqref{relationparam1}, together with the decoupling limit.

\section{Covariant phase space expressions for the energy}
\label{appendixcovphasesp}
%{\color{red}(To check that all notations are consistent with the previous sections )}
In this appendix we list the formulae for computing the energy of the backgrounds discussed in section \ref{section3}, which fit into the six-dimensional  truncation  \eqref{lagtrunc}, as well as of the ten-dimensional backgrounds discussed in appendix \ref{appendixdecoupling}. We work in the covariant phase space formalism, following \cite{Compere:2018aar,Compere:2009dp}. In the ten-dimensional case, the charges were derived in \cite{Dias:2019wof}.

Let us consider an asymptotic Killing vector $\xi$ and a path in phase space between two solutions of the equations of motion associated to a $D$-dimensional action. The infinitesimal variation in phase space of the charge associated to $\xi$ is given by: 
\begin{align}
\delta\mathcal{Q}_{\xi}&=\frac{1}{2! (D-2)!}\int_{\Sigma_{D-2}}\epsilon_{\alpha_1\alpha_2...\alpha_{D-2}\mu\nu}\;k^{\mu\nu}_{\xi}dx^{\alpha_1}\wedge dx^{\alpha_2}\wedge...\wedge dx^{\alpha_{D-2}} \label{chvargen}
\end{align} 
where $\Sigma_{D-2}$ is a compact codimension $2$ spacelike surface and $\epsilon_{\alpha_1...\alpha_{D-2}\mu\nu}\, k^{\mu\nu}_{\xi}$ are the components of a $(D-2)$-form that receives contributions from the various fields of the theory.

For the particular case $\xi=\partial_t$ one obtains, after integration on the path in phase space, the energy difference between the two solutions, one of which is usually chosen to have zero energy.

\subsection*{Six-dimensional backgrounds}

We consider an action of the form: 
\begin{align}\label{laggeneral}
S_{6d}&=\frac{1}{16\pi G_6}\int d^6x \sqrt{-g} \bigg(R-\frac{1}{2}f_{ab}(\phi)\partial_{\mu}\phi^a\partial^{\mu}\phi^b-V(\phi)-\frac{1}{12}k_{ij}(\phi)F^{i}_{\mu\nu\rho}F^{j\mu\nu\rho}+\nonumber\\
&+\frac{1}{3!3!}h_{ij}(\phi)\epsilon^{\mu\nu\rho\sigma\alpha\beta}F^i_{\mu\nu\rho}F^j_{\sigma\alpha\beta}\bigg)
\end{align}
The $D-2$ - form $k_\xi$ entering the  charge variation \eqref{chvargen} receives 
contributions from the metric, the scalars and the three-form fields. The latter can be further split into a kinetic and a topological contribution. We thus write
\begin{align}
k^{\mu\nu}_{\xi}&=\frac{1}{8\pi G_6}\bigg(k^{\mu\nu}_{\xi,g}+k^{\mu\nu}_{\xi,sc}+k^{\mu\nu}_{\xi,kin}+k^{\mu\nu}_{\xi,top}\bigg)
\end{align}
The explicit expressions for each of these terms are written below, using the standard notation $h_{\mu\nu}=\delta g_{\mu\nu},h=g^{\mu\nu} \d g_{\mu\nu}$ and $(\delta f)^{\alpha_1...\alpha_{n}}=g^{\alpha_1\beta_1}...g^{\alpha_n\beta_n}\delta f_{\beta_1...\beta_n}$:
\bea
k^{\mu\nu}_{\xi,g}&=&\frac{1}{2}\bigg(\xi^{\mu}\nabla_{\alpha} h^{\alpha\nu}-\xi^{\mu}\nabla^{\nu}h +\xi_{\alpha}\nabla^{\nu}h^{\alpha \mu}+\frac{1}{2}h \nabla^{\nu}\xi^{\mu}-\frac{1}{2}h^{\nu\alpha}\nabla_{\alpha}\xi^{\mu}+\frac{1}{2}h^{\nu\alpha}\nabla^{\mu}\xi_{\alpha} - (\mu\leftrightarrow\nu)\bigg)\nonumber\\
k^{\mu\nu}_{\xi,sc}&=&\frac{1}{2}\bigg(\xi^{\nu}f_{ab}(\phi)\nabla^{\mu}\phi^a \delta\phi^b-(\mu\leftrightarrow\nu)\bigg)\nonumber\\
k^{\mu\nu}_{\xi,kin}&=&\frac{1}{4}\bigg[ -\partial_{\phi^a}k_{ij}(\phi)F^{j\mu\nu\lambda}\xi^{\alpha}C_{\alpha\lambda} \delta\phi^a+k_{ij}(\phi)\bigg(-(\delta F)^{j\mu\nu\lambda}+2h^{\mu\sigma}F^{j\;\;\nu\lambda}_{\;\sigma}+h^{\lambda\sigma}F^{j\mu\nu}_{\;\;\sigma}-\nonumber\\
&&-\frac{1}{2}h F^{j\mu\nu\lambda}\bigg)\xi^{\alpha}C^i_{\alpha\lambda}-k_{ij}(\phi)F^{j\mu\nu\lambda}\xi^{\alpha}\delta C^i_{\alpha\lambda}-k_{ij}(\phi)\xi^{\mu}F^{j\nu\alpha\lambda}\delta C^i_{\alpha\lambda}-(\mu\leftrightarrow\nu)\bigg]\nonumber\\
k^{\mu\nu}_{\xi,top}&=&\frac{1}{24}\bigg[\epsilon^{\mu\nu\alpha\beta\gamma\sigma}\bigg( (\partial_{\phi^a} h_{ij}(\phi))F^j_{\beta\gamma\sigma}\xi^{\rho}C^i_{\alpha\rho}\delta \phi^a+  h_{ij}(\phi)\delta F^j_{\beta\gamma\sigma}\xi^{\rho} C^i_{\alpha\rho}+h_{ij}(\phi) F^j_{\beta\gamma\sigma}\xi^{\rho}\delta C^i_{\alpha\rho}  \bigg)-\nonumber\\
&&-2h_{ij}(\phi)\xi^{\nu}\epsilon^{\mu\eta\alpha\beta\gamma\sigma}F^j_{\beta\gamma\sigma}\delta C^i_{\eta\alpha}-(\mu\leftrightarrow\nu)\bigg]
\eea
where $F^i=dC^i$. Note that the last two expressions %for \textcolor{red}{what?}{\color{ForestGreen}(the 3-forms, otherwise there are terms with the Lie derivative of the potentials $\mathcal{L}_{\xi}B$ etc)}
 are only valid when $\xi$ is an isometry; see  \cite{Compere:2009dp} for the most general expressions. 

The Lagrangian \ref{lagtrunc} is of the form of the Lagrangian of \ref{laggeneral}, so we can apply the formulae above, for the particular case $\xi=\partial_t$, in order to compute the energy of the backgrounds discussed in section \ref{section3}. The   integration  is over the compact surface  $\Sigma_4= S^1\times S^3$.

\subsection*{Ten-dimensional backgrounds}
In the case of the action of type IIB supergravity: 
\begin{align}
S_{IIB}^{Ein}&=\frac{1}{2\kappa_{10}^2}\int d^{10}x \sqrt{-g}\bigg(R-\frac{1}{2}(\partial\Phi)^2\bigg)-\frac{1}{4\kappa^2_{10}}\int d^{10}x\bigg( e^{-\Phi}H\wedge\ast H + e^{2\Phi}F_1\wedge \ast F_1 +\nonumber\\
&+ e^{\Phi}F_3\wedge \ast F_3+\frac{1}{2}F_5\wedge\ast F_5 +C_4\wedge H\wedge F_3\bigg)
\end{align}
the contributions to the infinitesimal charge variation are
\begin{align}
k^{\mu\nu}_{\xi}&=\frac{1}{8\pi G_{10}}\bigg(k^{\mu\nu}_{\xi,g}+k^{\mu\nu}_{\xi,\Phi}+k^{\mu\nu}_{\xi,B}+k^{\mu\nu}_{\xi,C_0}+k^{\mu\nu}_{\xi,C_2}+k^{\mu\nu}_{\xi,C_4}\bigg)
\end{align}
The explicit expressions for the terms coming from the metric, dilaton, 0-form and 4-form are
\begin{align}
k_{\xi,g}^{\mu\nu}&=\frac{1}{2}\bigg(\xi^{\nu}\nabla^{\mu}h-\xi^{\nu}\nabla_{\sigma}h^{\mu\sigma}+\xi_{\sigma}\nabla^{\nu}h^{\mu\sigma}+\frac{h}{2}\nabla^{\nu}\xi^{\mu}-h^{\rho\nu}\nabla_{\rho}\xi^{\mu}-(\mu\leftrightarrow\nu)\bigg)\\
k_{\xi,\Phi}^{\mu\nu}&=\frac{1}{2}\bigg(\xi^{\nu}\nabla^{\mu}\Phi\delta\Phi-(\mu\leftrightarrow\nu)\bigg)\\
k_{\xi, C_0}^{\mu\nu}&=\frac{1}{2}\bigg(-e^{2\Phi}\xi^{\mu}F_1^{\nu}\delta C_0-(\mu\leftrightarrow\nu)\bigg)\\
k_{\xi,C_4}^{\mu\nu}&=\frac{1}{16}\bigg[-(\delta F_5)^{\mu\nu\beta\rho\lambda}\xi^{\alpha}C_{4\alpha\beta\rho\lambda}-F_5^{\mu\nu\beta\rho\lambda}\xi^{\alpha}\delta C_{4\alpha\beta\rho\lambda}-\xi^{\mu}F_5^{\nu\alpha\beta\gamma\rho}\delta C_{4\alpha\beta\gamma\rho}+\bigg(-\frac{h}{2}F_5^{\mu\nu\beta\rho\lambda}+\nonumber\\
&+h^{\mu\sigma}F_{5\sigma}^{\;\;\;\nu\beta\rho\lambda}+ h^{\nu\sigma}F_{5\;\;\sigma}^{\;\mu\;\;\beta\rho\lambda}+3h^{\lambda\sigma}F_{5\;\;\;\;\;\;\;\sigma}^{\mu\nu\beta\rho}\bigg)\xi^{\alpha}C_{4\alpha\beta\rho\lambda}-(\mu\leftrightarrow\nu)\bigg]
\end{align}
It is easier to write the contributions associated with $C_2$ and $B$ using differential forms, which encode the charge variations of the eight-forms% are encoded in the following 8-forms:
\begin{align}
k_{\xi}^{(C_2)}&=\frac{1}{4}\bigg(-\delta\big(i_{\xi}C_2\wedge (2e^{\Phi}\ast F_3-C_4\wedge H)\big)+i_{\xi}\big(\delta C_2\wedge(2e^{\Phi}\ast F_3-C_4\wedge H)\big)\bigg)\label{formchC}
\end{align}
\begin{align}
k_{\xi}^{(B)}&=\frac{1}{4}\bigg(-\delta\big(i_{\xi}B\wedge (2e^{-\Phi}\ast H-2e^{\Phi}C_0\ast F_3 -C_2\wedge\ast F_5+C_4\wedge dC_2)\big)+\nonumber\\
&+i_{\xi}\big(\delta B\wedge(2e^{-\Phi}\ast H-2e^{\Phi}C_0\ast F_3 -C_2\wedge\ast F_5+C_4\wedge dC_2)\big)\bigg)\label{formchb}
\end{align}
Taking  the Hodge duals, we obtain
\begin{align}
k_{\xi,C_2}^{\mu\nu}&=\frac{1}{8}\bigg(t_1^{\mu\nu}+t^{\mu\nu}_2-(\mu\leftrightarrow\nu)\bigg)\hspace{0.8cm}k_{\xi,B}^{\mu\nu}=\frac{1}{8}\bigg(t^{\mu\nu}_3+t^{\mu\nu}_4+t^{\mu\nu}_5+t^{\mu\nu}_6-(\mu\leftrightarrow\nu)\bigg)
\end{align}
where each of the  $t^{\mu\nu}_i$ - which originate in the various eight-forms that appear in the expressions \eqref{formchC} - \eqref{formchb} above - %. All these terms can be written generically as
takes the generic form
\begin{align}
t^{\mu\nu}_i&=e^{\psi}\bigg[-F^{\mu\nu\lambda}\xi^{\alpha}C_{\alpha\lambda}\delta\psi-(\delta F)^{\mu\nu\lambda}\xi^{\alpha}C_{\alpha\lambda}-F^{\mu\nu\lambda}\xi^{\alpha}\delta C_{\alpha\lambda}-\xi^{\mu}F^{\nu\alpha\lambda}\delta C_{\alpha\lambda}+\bigg(-\frac{h}{2}F^{\mu\nu\lambda}+\nonumber\\
&+h^{\mu\sigma}F_{\sigma}^{\;\;\nu\lambda}+h^{\nu\sigma}F_{\;\;\;\sigma}^{\mu\;\;\lambda}+h^{\lambda\sigma}F_{\;\;\;\;\sigma}^{\mu\nu}\bigg)\xi^{\alpha}C_{\alpha\lambda}\bigg]
\end{align}
for different choices of the scalar $\psi$, three-form $F$ and two-form $C$.
The particular $\psi,F,C$ for which we obtain each contribution are listed below
\begin{align}
t^{\mu\nu}_1:&\;\psi=\Phi,\hspace{0.3cm}F=2F_3,\hspace{0.3cm}C=C_2\hspace{2cm}t^{\mu\nu}_2:\;\psi=0,\hspace{0.3cm}F=\ast(-C_4\wedge H),\hspace{0.3cm}C=C_2 \nonumber \\
t^{\mu\nu}_3:&\;\psi=-\Phi,\hspace{0.3cm}F=2H,\hspace{0.3cm}C=B\hspace{1.9cm}t^{\mu\nu}_4:\;\psi=\Phi+\log C_0,\hspace{0.5cm}F=-2F_3,\hspace{0.3cm}C=B\nonumber \\
t^{\mu\nu}_5:&\;\psi=0,\hspace{0.2cm}F=\ast(-C_2\wedge\ast F_5),\hspace{0.2cm}C=B\hspace{0.8cm}t^{\mu\nu}_6:\;\psi=0,\hspace{0.2cm}F=\ast(C_4\wedge dC_2),\hspace{0.2cm}C=B
\end{align}

\end{document}